\documentclass[printer]{aa} 
\usepackage{graphicx}
\usepackage{txfonts}
\usepackage{textcomp}
\newcommand{\degree}{\ensuremath{^\circ}}
\usepackage[english]{babel}
\begin{document}

\title{Distance to the northern high-latitude  HI shells
\thanks{Based on observations obtained using the Narval spectropolarimeter at the Observatoire du Pic du Midi (France), which is operated by the Institut National des Sciences de l'Univers (INSU)}}


\author{L. Puspitarini \inst{1}
	\and
	R.Lallement\inst{1}
	}

\institute{GEPI Observatoire de Paris, CNRS, Universit\'e Paris Diderot, 5 Place Jules Janssen, 92195, Meudon, France \\
	\email{lucky.puspitarini@obspm.fr}
	}
          
\date{Received;Revised}


  \abstract
   {A detailed three-dimensional (3D) distribution of interstellar matter in the solar neighborhood is increasingly necessary. It would allow a more realistic computation of photons and cosmic rays  propagation, which is of prime importance for microwave and gamma-ray background emission interpretation, as well as a better understanding of the chain of phenomena that gave rise to the main local structures.}
   {As part of a 3D mapping program, we aim at assigning a precise distance to the high-latitude HI gas, in particular the northern part (b$\geq$ 55$\degree$) of the shell associated with the conspicuous radio continuum Loop I. This shell is thought to be the expanding boundary of an interstellar bubble inflated and recently reheated by the strong stellar winds of the nearby Scorpius-Centaurus OB associations.}
  {We recorded high-resolution spectra of 30 A-type target stars located at various distances in the direction of the northern part of Loop I.  Interstellar neutral sodium and singly ionized calcium absorptions (NaI 5889-5895 and  CaII K-H 3934-3968 \AA) are modeled and compared with the HI emission spectra from the LAB Survey.}
  {About two-thirds of our stellar spectra possess narrow interstellar lines, while the remaining spectra show no absorption at all. Narrow lines are located at the velocity of the main, low-velocity Loop 1 HI shell ( [-6,+1] km.s$^{-1}$ in the LSR).  Using Hipparcos distances to the target stars, we show that the closest boundary of the b$\geq$+70$\degree$ part of this low-velocity Loop I arch  is located at of  98 $\pm$ 6 pc. The corresponding interval for the lower-latitude part (55$\leq$b$\leq$70$\degree$) is 95-157 pc. 
However, since the two structures are apparently connected, the lower limit is more likely.
At variance with this shell, the second HI structure, which is characterized by LSR Doppler velocities centered at -30 km.s$^{-1}$, is NOT detected in any of the optical spectra. It is located beyond 200 parsecs or totally depleted in NaI and CaII.}
{We discuss these results in the light of spherical expanding shells and show that they are difficult to reconcile with simple geometries and a nearby shell center close to the Plane. Instead, this high-latitude gas seems to extend the inclined local \textit{chimney}  wall  to high distances from the Plane.}


\maketitle

\section{Introduction}

   While emission surveys at various wavelengths are providing increasingly detailed maps of Galactic interstellar matter (ISM), there is no information on the distance to the emitting clouds.
Models of the 3D structure of ISM are generally based on the combination of  radio-emission spectral maps and distances assigned to the emitting gas clouds based on their velocities and a mean Galactic rotation curve. 
However, this leads to a poor description of the Sun's vicinity, while its emission dominates most of the sky away from the plane. 
Nowadays, there is an increasing need to improve the 3D distribution of interstellar (IS) gas and dust in our Galaxy, especially in the context of the Planck and Fermi missions, which intend to model and to subtract the Galactic dust and interstellar gamma rays that they map across the sky. 
In the first case, it is the computation of the interstellar radiation field that requires a realistic distribution of the ISM, and in the second case, the computation of the cosmic rays propagation and interaction with the IS gas. 
A precise 3D structure of the IS matter, and especially the dust, will also be mandatory for future analyses of the European Space Agency (ESA) cornerstone mission Gaia. 
Finally, a precise 3D distribution can be used in numerous and various ways, e.g., by providing context and foreground for specific objects, visibility, or detectability conditions.

    One way to obtain realistic 3D distributions of gas is to gather absorption data toward target stars located at known and widely-distributed distances and to invert the line-of-sight dasta. 
The first attempt to compute a 3D distribution of IS gas by inversion of line-of-sight data was made by Vergely et al, (2001), who applied a robust tomographic method based on the pioneering work of Tarantola and Valette (1982) to interstellar column densities from the literature. 
Lallement et al, (2003) then applied the same method to a larger neutral sodium data set gathered for that purpose. 
Such tomography provides the shape and size of the Local Bubble, the cavity that is supposedly filled with million K gas responsible for the diffuse soft X-ray background. 
More recently, Welsh et al, (2010) updated the maps and published a catalog of 2,000 NaI and CaII absorptions with new data from various observatories. 
In parallel, tools were developed by Vergely et al, (2010) and applied to extinction measurements.
About 6,500 stellar color excesses from Str\"{o}mgren photometry and Hipparcos distances were inverted to produce a 500 pc-wide-opacity cube, with the same inversion code being applied to the updated sodium dataset of 1,700 NaI columns.
The similarity between the locations of the major clouds deduced from the independent and very different sodium and extinction datasets demonstrates that the inversion method is efficient. However, the number of target stars is at present insufficient for obtaining a high spatial resolution, and allows mapping only the closest major features. 
Moreover, beyond 300 pc Hipparcos distance accuracy rapidly becomes poor, with errors of the order of 10\% at 300 pc.
More recently, Reis et al, (2011) also used a large set of Str\"{o}mgren individual photometric data to map the nearby gas opacity in the Galactic center direction with a different method.
For such methods using individual targets, great progress is expected on both target numbers and distances, thanks to Gaia.  
   On the other hand, 3D studies of dust based on stellar photometric surveys and photometric distances have been undertaken. 
Instead of individual measurements, they use statistical properties of a massive number of objects in various ways. 
Marshall et al, (2006) used the Two Micron All Sky Survey (2MASS) survey and the Besancon stellar population model (Robin et al, 2003) to select giant stars and produce 3D maps of the extinction for the entire Galactic disk.
This approach was successful in revealing the spiral arms. 
More recently, Gontcharov (2011) used 2MASS F dwarfs and the same method to map the more local dust, while Sale et al, (2009) undertook the construction of maps based on the Isaac Newton telescope Photometric H-Alpha Survey (IPHAS) database. 
Finally, Jones et al, (2011) used the Sloan Digital Sky Survey (SDSS) spectra of M stars to build local extinction maps. 
For all those studies, future cross-matching with Gaia will allow improved and refined mapping. 
However, these methods require a large number of stars to produce an average extinction as a function of distance, resulting in a limited resolution (of the order of 100 pc). As such, they are not appropriate for the local ISM as opposed to the inversion of individual data.
In addition, the approach based on individual gas absorption lines is the only one that provides kinematical information, thus making it possible to identify the clouds in the radio spectral cubes and benefit from their high precision. 
The price to be paid, however, is the requirement for huge sets of individual spectral observations. 
High-resolution spectroscopy is necessary and because it is time-consuming, only limited datasets have been built. 
The situation is currently changing, thanks to the new high-resolution, multi-object spectrographs (MOS). 
However, until wide-field MOS are built, no dedicated ISM surveys covering large fractions of the sky can be proposed because of the small field of view. 
This is why the new high-quality spectroscopic data with mono-object spectrographs are still, and will remain, very precious. 

   Our purpose here is to extend the spectroscopic database by adding high Galactic latitude targets that help constrain nearby, out-of-plane interstellar features. 
Such features are difficult to detect and position due to both the density fall-off with distance from the plane and the lack of suitable bright background stars, in particular early-type target stars required for absorption measurements because of their smooth continuum.
Nevertheless, as previously mentioned, these features are particularly important for current missions. 
Loop I is a Galactic giant radio continuum loop of 58$\degree$ $\pm$ 4$\degree$ radius, centered at \textit{l} = 329$\degree$ $\pm$ 1\degree.5 and \textit{b} = +17\degree.5 $\pm$ 3$\degree$ (\cite{Berkhuijsen}).
Several filamentary arches are visible in the radio maps.
A bright, conspicuous vertical structure at the bottom left of the main Loop I arch, centered at \textit{l} $\approx$ 30$\degree$ and extending between \textit{b} $\approx$ +5 and +40$\degree$ has been called the North Polar Spur (NPS).
HI shells have been observed at the periphery of Loop I (Colomb et al, 1980), and found to expand at low-velocity.
Estimations of this velocity are somewhat contradictory: Sofue et al, (1974) and Heiles (1984) found 19 and 25 km.s$^{-1}$ respectively, while Weaver (1979) gives a very low value of 2 km.s$^{-1}$ based on northern measurements. 
Ordered magnetic fields from radio and optical polarization data follow the continuum and HI shell orientation, however essentially for \textit{b} $\geq$ 30$\degree$, while a depolarization band characterizes the region between 0 and 30$\degree$ (see Wolleben, 2007). 
Finally, a region of enhanced X-ray emission corresponds to the interior of Loop I (Snowden et al, 1995). 
If X-rays and radio data are both related to the same exploding event, there is a controversy about its age: the relatively low expansion velocity of the neutral gas surrounding Loop I implies something of the order of a few million years, while X-Ray data suggest ten times less (Borken \& Iwan, 1977). 
In order to reconcile the various datasets, these authors and \cite{EggerAschenbach} proposed a multi-step scenario: Loop I is an old interstellar gas bubble initially blown up by the winds of early-type stars of the Sco-Cen association and recently reheated by a new series of star bursting. 
This scenario and the  resulting multi-phase structure, with Loop I and the Local Bubble being adjacent cavities, has been successfully modeled by \cite{deav12}. 
Wolleben (2007) also used the Northern Sky Polarization Survey and WMAP data to complete the picture. 
According to his scenario, two magnetic ionized gas shells generated by two distinct events are currently interacting, explaining the polarization pattern of the radio continuum. 

   The hot gas cavity associated with Loop I is thought to be centered around the Sco-Cen OB association, at about 140 pc, and thus the HI shell is generally believed to be closer, having already reached the Sun, according to Wolleben (2007). 
From observations of optical starlight polarization at low latitude, a distance of 100 $\pm$ 20 pc to the bottom of the NPS has been determined by Bingham (1967). 
Indeed, the geometrical association between the Loop I X-ray-emitting bubble, the nonthermal radio continuum, and the expanding HI shells, taken as a proof of their physical association, reinforces the assumption of proximity of the shells. 
It is, however, interesting to note that a very different estimate of the distance to the NPS was more recently proposed by Bland-Hawthorn and Cohen (2003), who argue that it could be part of the extraordinary bipolar structure detected in X-ray, microwave, and gamma ray full sky maps (Snowden et al, 1997; Finkbeiner, 2004; Sue et al, 2010) and found to be the manifestation of a powerful burst from the Galactic center. 

The high-latitude (\textit{b} $\geq$ 55$\degree$) HI shells have no known dust or absorbing gas counterpart that could confirm their distance.
The 3D maps of the nearby ISM should, in principle, reveal them, but due to the previously mentioned lack of high Galactic latitude data, this is not very clear yet.
The inverted 3D opacity distribution of Vergely et al, (2010) reveals a dust cloud within 5$\degree$ from the North Pole, which coincides in direction with the highest portion of the Loop I shell and the brightest area above 60$\degree$ latitude seen in Fig. 1, and is possibly associated to the HI emission feature. 
This cloud is visible, e.g., in Fig. 10 of Vergely et al, in the top part of the vertical half plane containing the Sun and sightlines with \textit{l} = -35$\degree$. 
Its closest part is at about 110-120 pc in this low-resolution map.
But this cloud has no detected counterpart in the current NaI maps, which may be due to the low number of high-latitude targets available and the absence of sightlines crossing the shells. 
On the other hand, the huge cavity supposedly bounded by Loop I has not been found in any inverted map yet, probably because it doesn't have enough spatial resolution and size. 

\begin{figure*}[ht!] 
\centering\includegraphics[width=\linewidth,height=70mm]{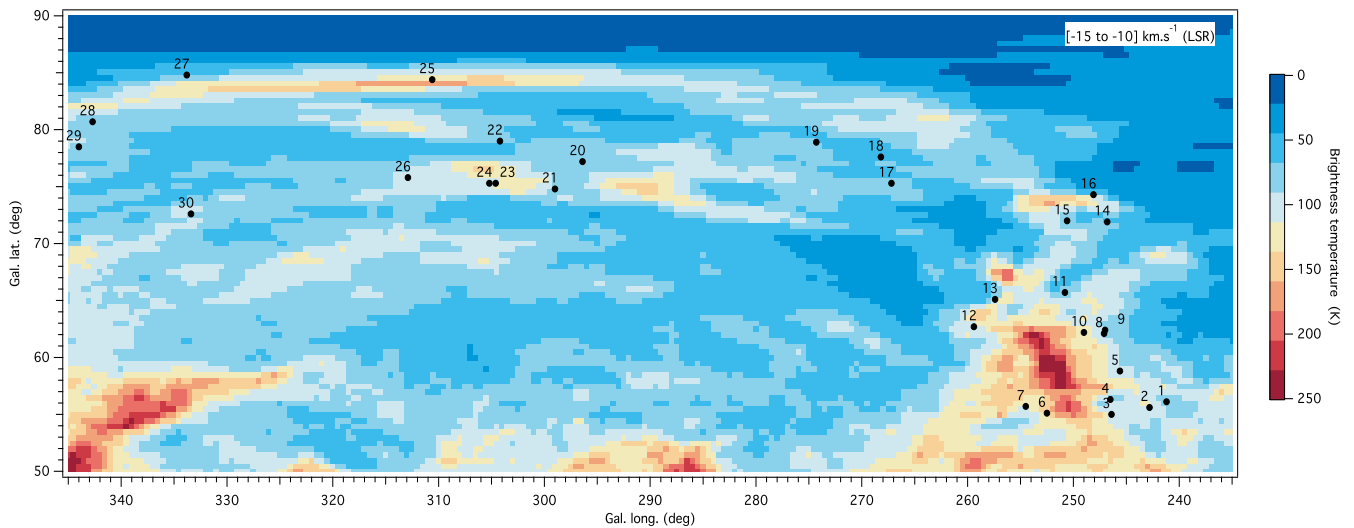} 
\caption{{Distribution} of the target stars observed, {superimposed on} HI emission map (-15\textless$v_{LSR}$\textless10 km/s). {The corresponding star numbers are given
 in Table \ref{table1}}.
} 
\label{HDSTAR_MAP} 
\end{figure*} 

In order to fill this gap and better represent the ISM distribution at high-latitude, we have recorded high-quality and high-resolution observations toward the Loop I arches, with the aim of locating the neutral sodium counterpart of the HI loop and determining whether or not this counterpart coincides with the high-latitude dust cloud mentioned above, which is revealed by the 3D extinction maps. 
Such observations also help complement the gas absorption database with the inclusion of targets located at  high northern latitude and at various distances. 
This should allow us later on to map the polar clouds that correspond to the HI shell. 
For those line-of-sight, we expect  low column densities, because the HI columns are only of the order of 10$^{20}$ cm$^{-2}$.
Using the quasi-quadratic relationship established by Welty and Hobbs (2001) between HI and NaI columns, this should correspond to log(NaI) =11 at maximum, i.e., a few mA absorption equivalent width.\\

In section 2, we describe the observations. In section 3, we describe the data analysis and derivation of the characteristics of the interstellar neutral sodium and ionized calcium lines. 
In section 4, we compare the results with the HI emission spectra and derive the constraints on the distance  to the NaI/ CaII  counterpart of the HI. We also discuss the results and consequences on the geometry.

\section{Observations}

  We selected a list of early-type, nearby target stars located in the direction of the Loop I northern shells and possessing a Hipparcos parallax distance. 
The target locations are shown in Fig. \ref{HDSTAR_MAP}, superimposed on the Leiden-Argentina-Bonn (LAB) HI map (Kalberla et al, 2005). 
The map here is restricted to local standard of rest (LSR) velocities between -15 and +10 km.s$^{-1}$ and clearly reveals the whole Loop I structure. 

The targets are of early A-type stars or later A stars that are fast rotators, allowing easier detection of the narrow interstellar absorption lines imprinted on the stellar lines. 
High signal-to-noise ratios ($\gtrsim$ 200) and high-resolution (R=75000, $\delta$v=4 km.s$^{-1}$) spectra of 30 targets were recorded with NARVAL, the spectro-polarimeter of the Bernard Lyot telescope (2m) at Pic du Midi Observatory. 
Here, we used NARVAL in its spectrometric mode.
Observations were distributed over 15 nights between December 2010 and February 2011 in order for each star to be observed under optimal conditions. 
The data are all of excellent quality and come from the standard reduction pipeline. 
The signal-to-noise ratio always exceeds 100. 
For stars which benefitted from several exposures, they were performed during the same night and their spectra were simply co-added. 
Exposure times range from 100s to two hours, according to brightness, spectral type, rotational velocity, and {potential constraints on the cloud distance (targets closer than 200 parsecs were given more weight)}.


\begin{table*}[!ht]

\caption{Target stars and NaI and CaII absorption measurements. {Equivalent widths ($EW$) are roughly proportional to the column densities: $NaI_{tot}=5.2 \times 10^{9} EW(NaI/D2)$) except for HD100974, $CaII_{tot}=1.150 \times 10^{10} EW(CaII/K) $.}
} 

\label{table1}

\begin{center}

\scriptsize

\begin{tabular}{ |l | c | c | c | c | c | c |c |c |c| c|}

\hline

No & HD No & Type & \textit{l} ($\degree$) & \textit{b} ($\degree$) & dist (pc) & NaI$_{tot}$ x$10^{10}$ & CaII$_{tot}$ x$10^{10}$ & NaI/CaII & Cloud's num. NaI ($v_{hel}$ in km/s) & idem for CaII \\ \hline

1 &  94194 & A2 & 241.2 & 56.1 & 192 $\pm$ 29 & 29 $\pm$ 2 & 12 :: & 2.4 :: & 1 (-6.13 $\pm$ 0.04) & 1 (-6.48 $\pm$ 0.48) \\ \hline

2 &  94266 & A2 & 242.8 & 55.6 & 481 $\pm$ 263 & 117 $\pm$ 7 & 67 $\pm$ 20 & 1.7 & 2 (-9.57 $\pm$ 0.02 \& {-1.81 $\pm$ 0.06}) &2 (-9.28 $\pm$ 0.26 \& -0.19 $\pm$ 0.21) \\ \hline

3 &  94766 & A2 & 246.4 & 55.0 & 143 $\pm$ 14 & 14 $\pm$ 2 & 10 $\pm$ 8 & 1.4 & 2 (-10.06 $\pm$ 0.09 \& {-0.48 $\pm$ 0.05}) & 2 (-7.83 $\pm$ 0.60 \& 0.31 $\pm$ 0.53) \\ \hline

4 &  95382 & A5III & 246.5 & 56.3 & 46 $\pm$ 1 & \textless 1.2 & \textless 6 & 0.2 :: & 0 & 0 \\ \hline

5 &  96398 & A2 & 245.6 & 58.8 & 826 $\pm$ 669 & 51 $\pm$ 7 & 12 :: & 4.3 :: & 1 (-3.51 $\pm$ 0.08) & 1 (-4.09 $\pm$ 0.86) \\ \hline

6 &  96399 & A2 & 252.5 & 55.1 & 251 $\pm$ 69 & 64 $\pm$ 4 & 61 $\pm$ 16 & 1.1 & 2 ({-5.83 $\pm$ 0.02} \& 2.49 $\pm$ 0.06) & 2 (-5.56 $\pm$ 0.29 \& 3.06 $\pm$ 0.23) \\ \hline

7 &  97230 & A3 & 254.5 & 55.7 & 204 $\pm$ 39 & 146 $\pm$ 3 & 78 $\pm$ 21 & 1.9 & 2 ({-1.84 $\pm$ 0.06} \& 11.5 $\pm$ 0.01) & (-2.30 $\pm$ 0.49\& 11.52 $\pm$ 0.20) \\ \hline

8 &  98377 & A2 & 247.1 & 62.1 & 239 $\pm$ 59 & 58 $\pm$ 1 & 43 $\pm$ 8 & 1.3 & 1 (-3.00 $\pm$ 0.01) & 1 (-3.61 $\pm$ 0.08) \\ \hline

9 &  98437 & A3 & 247.0 & 62.4 & 318 $\pm$ 148 & 36 $\pm$ 2 & 30 $\pm$ 9 & 1.2 & 1 (-3.31 $\pm$ 0.02) & 1 (-3.16 $\pm$ 0.18) \\ \hline

10 &  98747 & A2 & 249.0 & 62.2 & 73 $\pm$ 3 & \textless 1 & 0.2 :: & 5 :: & 0 & 0 \\ \hline

11 &  100518 & A2m & 250.8 & 65.7 & 103 $\pm$ 7 & \textless 10 & 1.2 :: & 8.3 :: & 0 & 0 \\ \hline

12 &  100974 & A2 & 259.4 & 62.7 & 170 $\pm$ 19 & 260:: (**)& 30 $\pm$ 14 & 8.7 & 1 (-5.04 $\pm$ 0.02) & 1 (-5.30 $\pm$ 0.20) \\ \hline

13 &  101470 & A2 & 257.4 & 65.1 & 102 $\pm$ 7 & \textless 1 & 6.2 :: & 0.2 :: & 0 & 0 \\ \hline

14 &  102660 & A3m & 246.8 & 71.9 & 63 $\pm$ 1 & 1 :: & \textless 12 & 0.1 :: & 0 & 0 \\ \hline

15 &  103152 & A2 & 250.6 & 72.0 & 93 $\pm$ 4 & \textless 17 & 2.3 :: & 7.3 :: & 0 & 0 \\ \hline

16 &  103877 & Am & 248.1 & 74.3 & 91 $\pm$ 4 & \textless 5 & 6.6 :: & 0.75 :: & 0 & 0 \\ \hline

17 &  106661 & A3V & 267.2 & 75.3 & 61 $\pm$ 1 & \textless 2 & \textless 3 & 0.7 :: & 0 & 1 \\ \hline

18 &  107612 & A2p & 268.2 & 77.6 & 103 $\pm$ 9 & 93 $\pm$ 8 & 9 :: & 10 :: & 2 (-8.33 $\pm$ 0.03 \& {-1.97 $\pm$ 0.04}) & 2 (-8.72 $\pm$ 0.61 \& -3.53 $\pm$ 1.63) \\ \hline

19 &  108714 & A0 & 274.3 & 78.9 & 184 $\pm$ 20 & 41 $\pm$ 3 & 15 :: & 2.7 :: & 2 (-7.92 $\pm$ 0.05 \& {-3.14 $\pm$ 0.06}) & 2 (-7.41 $\pm$ 1.35 \& -2.19 $\pm$ 1.74) \\ \hline

20 &  110932 & A0 & 296.4 & 77.2 & 230 $\pm$ 34 & 3 $\pm$ 2 & 2.8 :: & 1.1 :: & 1 (-8.18 $\pm$ 0.25) & 1 (-7.30 $\pm$ 0.50) \\ \hline

21 &  111164 & A3V & 299.0 & 74.8 & 83 $\pm$ 3 & \textless 2 & 0.5 :: & 4.0 :: & 0

& 0 \\ \hline

22 &  111893 & A7V & 304.2 & 79.0 & 111 $\pm$ 5 & \textless 2 :: & 6.1 :: & 0.3 :: & * & * \\ \hline

23 &  112002 & A2 & 304.6 & 75.3 & 177 $\pm$ 27 & 4 $\pm$ 1 & 1 :: & 4.0 :: & 1 (-6.81 $\pm$ 0.30) & 0 \\ \hline

24 &  112097 & A7III & 305.2 & 75.3 & 61 $\pm$ 3 & \textless 5 :: & 5.7:: & 0.9 :: & * & * \\ \hline

25 &  112197 & A2V & 310.6 & 84.4 & 361 $\pm$ 164 & 41 $\pm$ 3 & 12 :: & 3.4 :: & 1 (-6.14 $\pm$ 0.04) & 1 (-6.86 $\pm$ 0.50) \\ \hline

26 &  113124 & A1V & 312.9 & 75.8 & 306 $\pm$ 72 & 12 $\pm$ 1 & 22 $\pm$ 11 & 0.5 & 2 (-15.08 $\pm$ 0.52 \& {-7.36 $\pm$ 0.08 }) & 2 (-15.78 $\pm$ 0.26 \& -8.18 $\pm$ 0.38) \\ \hline

27 &  113365 & A0 & 333.8 & 84.8 & 172 $\pm$ 15 & 27 $\pm$ 2 & 8 $\pm$ 6 & 3.4 & 1 (-7.55 $\pm$ 0.03) & 1 (-8.18 $\pm$ 0.38) \\ \hline

28 &  115403 & A2 & 342.7 & 80.7 & 98 $\pm$ 6 & 6 $\pm$ 2 & 6 $\pm$ 5 & 1.0 & 1 (-9.83 $\pm$ 0.07) & 1 (-9.82 $\pm$ 0.47) \\ \hline

29 &  116379 & A2 & 344.0 & 78.5 & 127 $\pm$ 12 & 59 $\pm$ 4 & 16 :: & 3.6 :: & 1 (-9.88 $\pm$ 0.03) & 1 (-9.57 $\pm$ 0.36) \\ \hline

30 &  116960 & A0V & 333.4 & 72.6 & 266 $\pm$ 53 & 39 $\pm$ 1 & 32 $\pm$ 6 :: & 1.2 :: & 1 (-10.40 $\pm$ 0.02) & 2 (-25.71 $\pm$ 0.11 \& -9.88 $\pm$ 0.09) \\ \hline

\end{tabular}

\end{center}

\tablefoottext{*}{ambiguous: maybe stellar}

\tablefoottext{::}{uncertain}

\tablefoottext{**}{saturated: EW(NaI-D2)= 227 $\pm$ 7 m\AA\ }
\end{table*}


\section{Data Analysis}

We analyzed the interstellar NaI doublet absorption (5889-5895 \AA) in the 30 recorded spectra and derived Doppler velocities and column densities by means of classical profile-fitting methods.
Prior to the line-profile fitting, telluric water vapor absorption lines in the sodium doublet spectral region were removed on the basis of a synthetic telluric transmission spectrum, individually adjusted to each observation (see \cite{lall93}).
The spectra were then normalized, based on polynomial or Gaussian adjustments of the stellar continuum around the interstellar lines.  
After that, the absorptions were classically modeled  by products of Voigt functions representing the individual clouds, convolved by the instrumental response. 
{The cloud characteristics, i.e., radial velocities, column densities, and line broadening values (or b-values, which are combinations of thermal broadening and micro-turbulence) are free parameters and were derived from a simultaneous adjustment to the D2 and D1 transitions.}
The hyperfine structure has been taken into account, i.e., each D2 or D1 line is a product of six separate absorption lines (\cite{welty94}). 

We applied the same methods to the interstellar CaII absorption lines at  3933 and 3968 \AA, except for the telluric line correction and the hyperfine structure. 
Sodium and calcium were adjusted independently. 
As a matter of fact,  NaI and CaII absorption line strengths  depend on the physical conditions in the clouds, with NaI favored in dense-neutral gas (T \textless 1000 K) and CaII a tracer of both dense-neutral and warmer ionized gas (T  $\leq$ 10000 K). 
During the adjustment, we imposed an upper limit of 15,000 K to the effective temperature, which corresponds to a combination of a 10,000 K temperature and standard micro-turbulence. 
When absorptions are broader, we increase the number of clouds. 
It is well known that the bulk of the detected NaI and CaII may originate from different locations in the absorbing clouds.

\begin{figure*}[!h]
\begin{minipage}[t]{0.3\linewidth}
\centering
  	\includegraphics[width=1\linewidth]{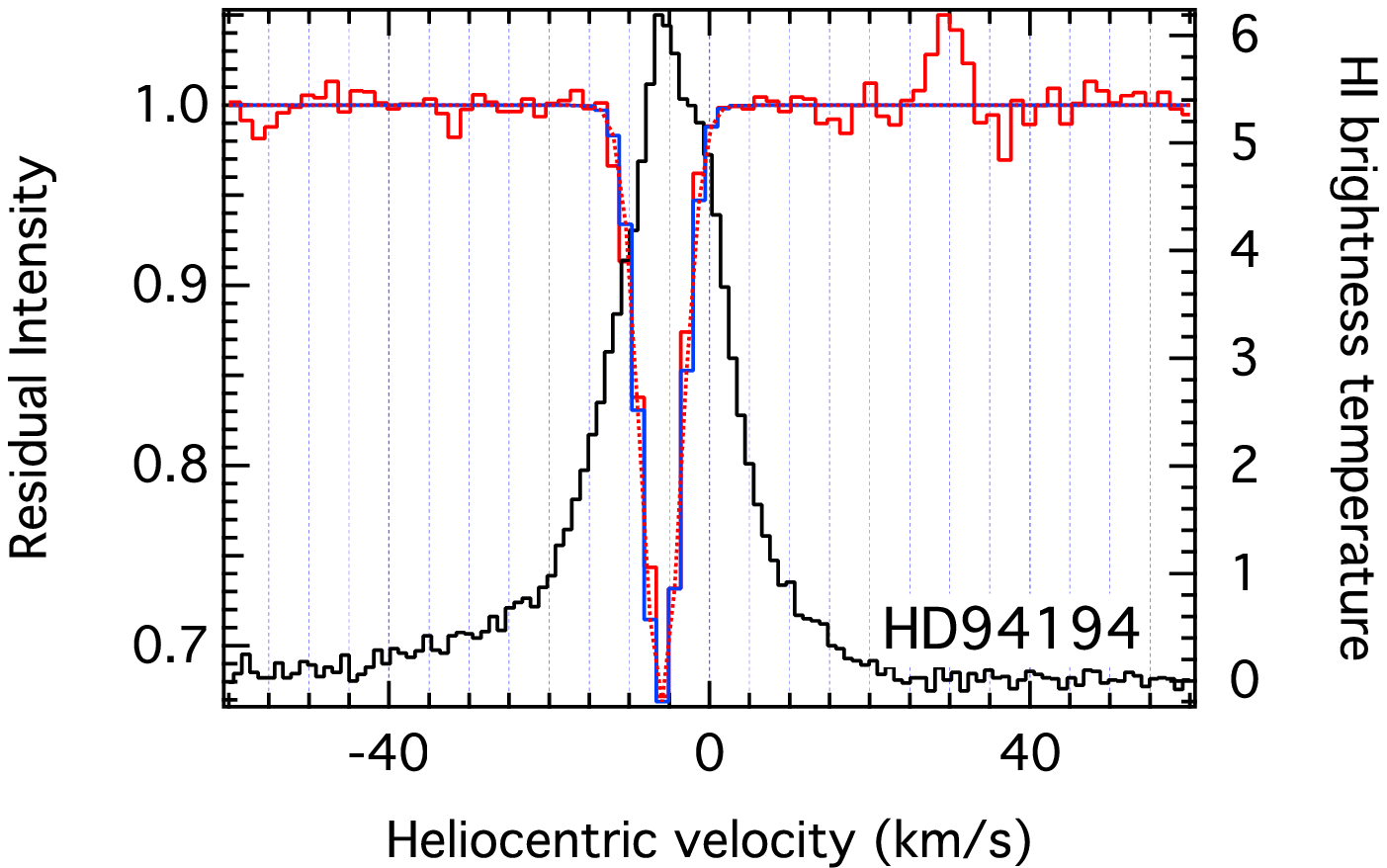}
  	\includegraphics[width=1\linewidth]{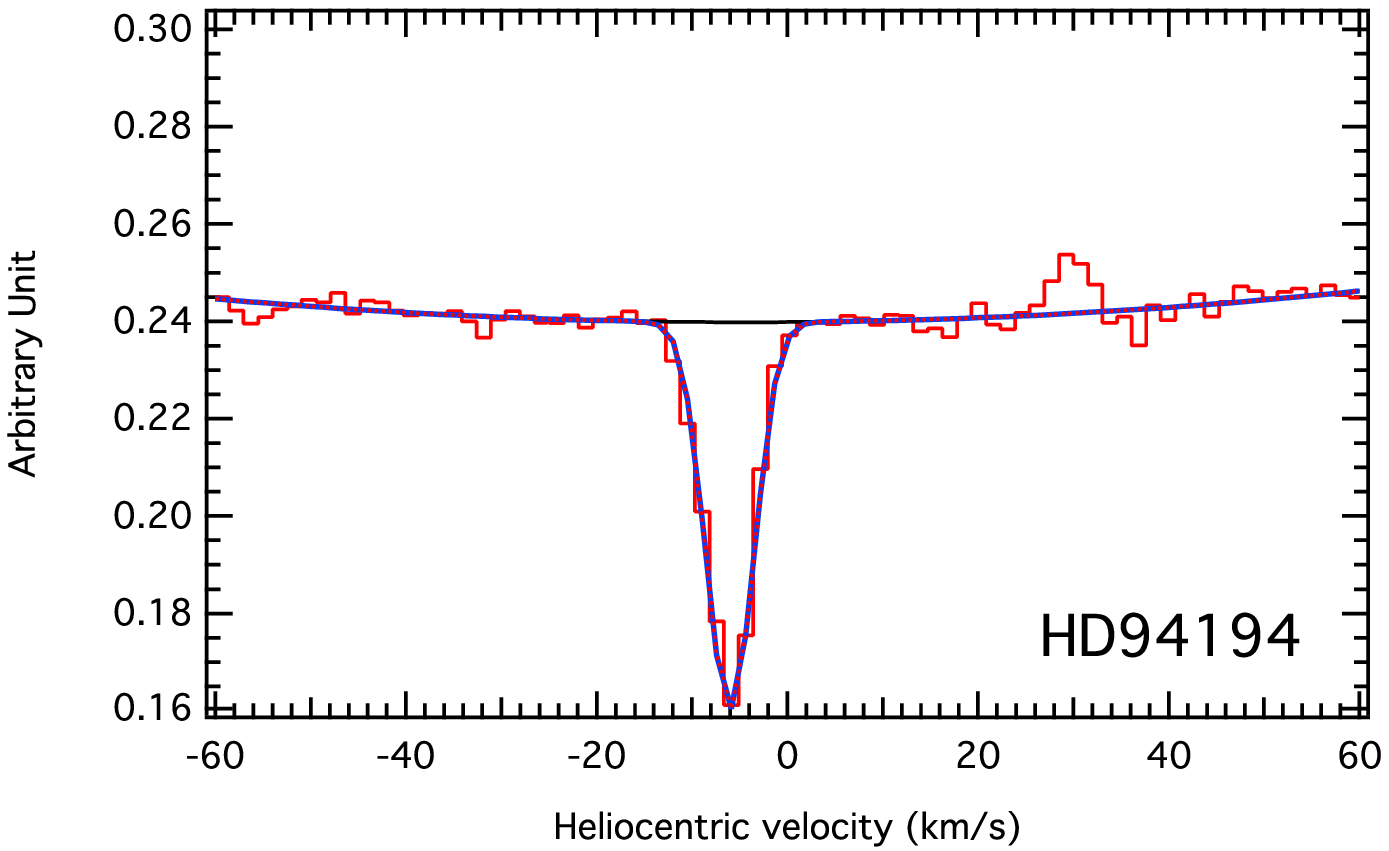}
  	\includegraphics[width=1\linewidth]{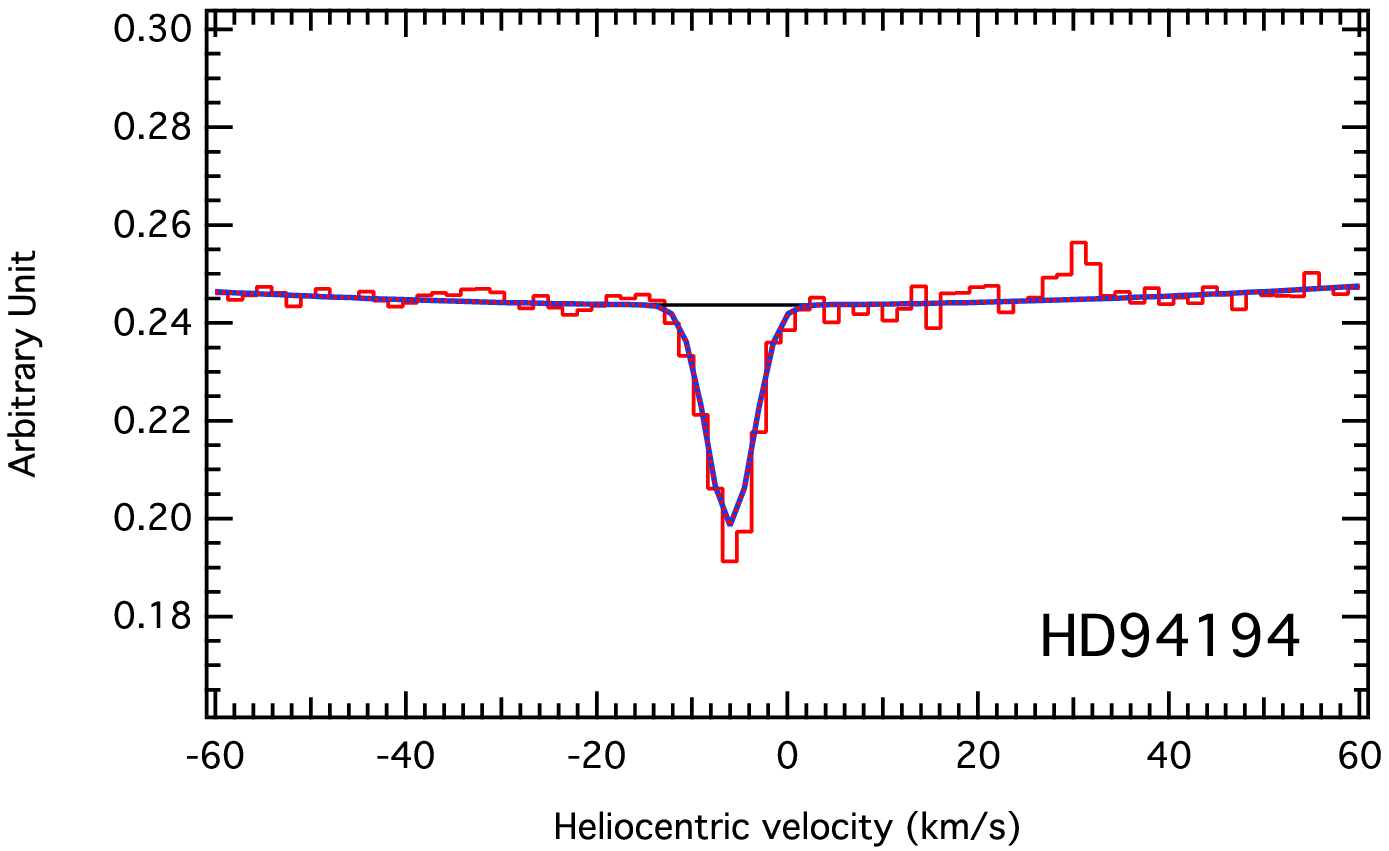}
\end{minipage}\hfill
\begin{minipage}[t]{0.3\linewidth}
\centering
  	\includegraphics[width=1\linewidth]{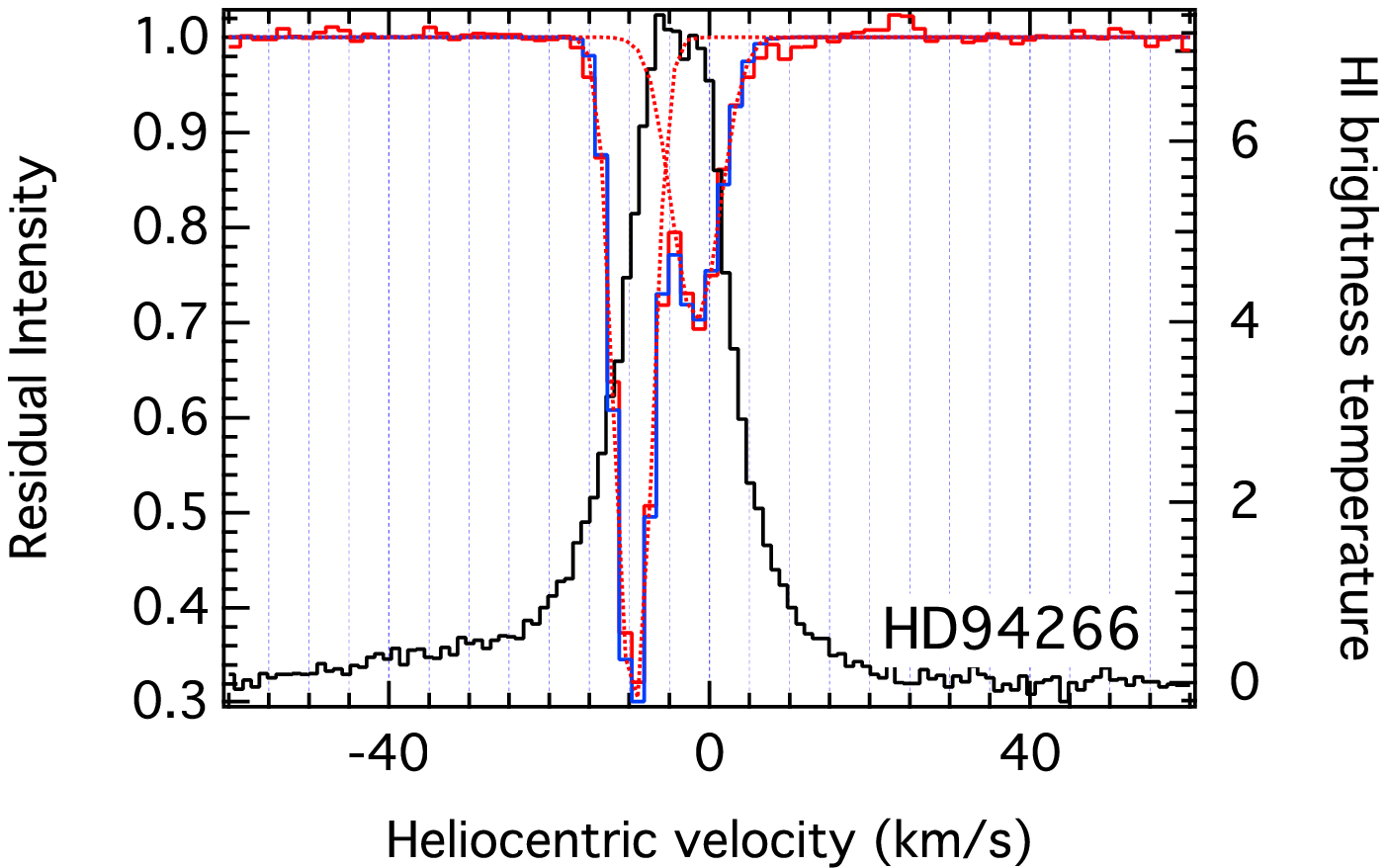}
  	\includegraphics[width=1\linewidth]{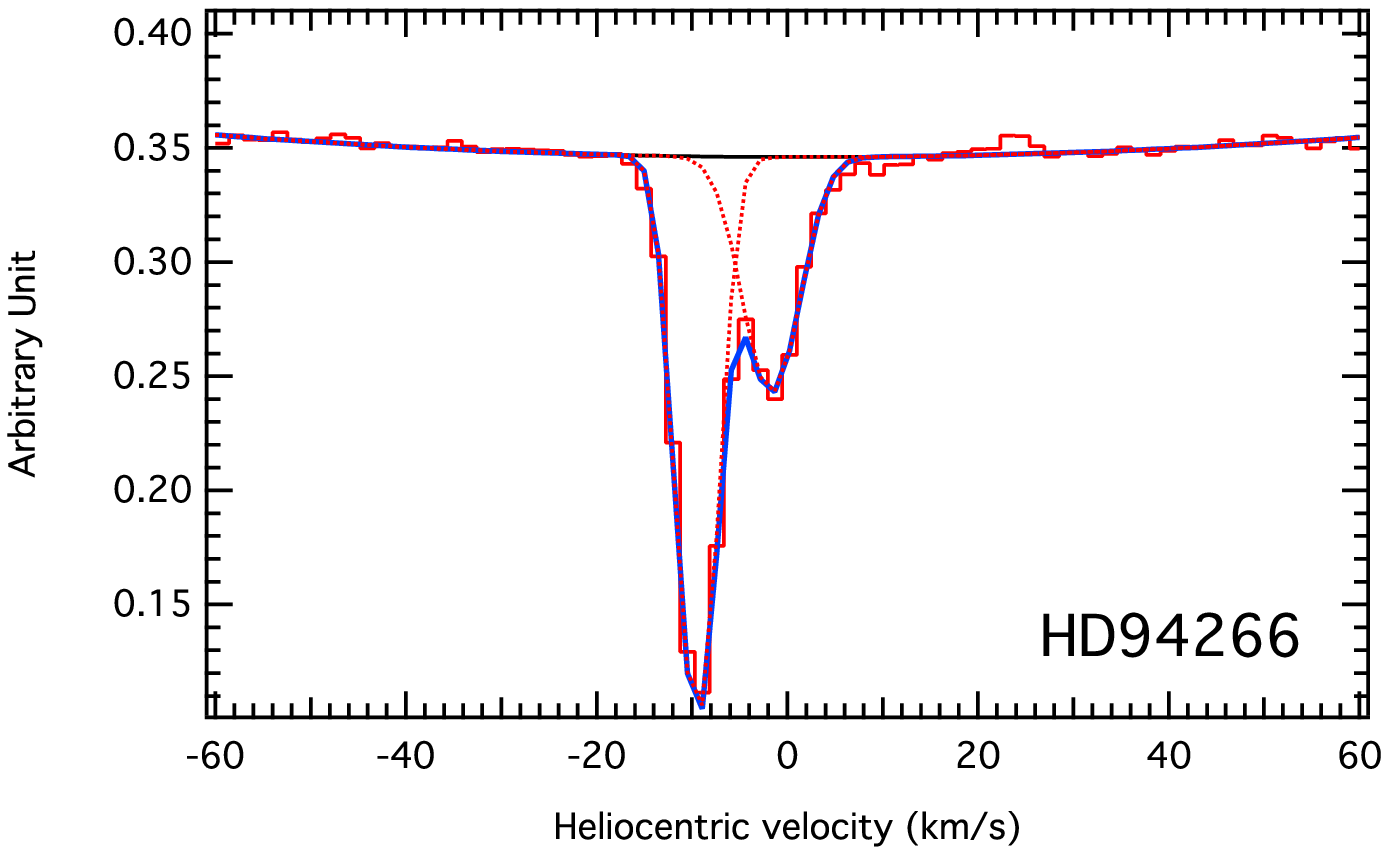}
  	\includegraphics[width=1\linewidth]{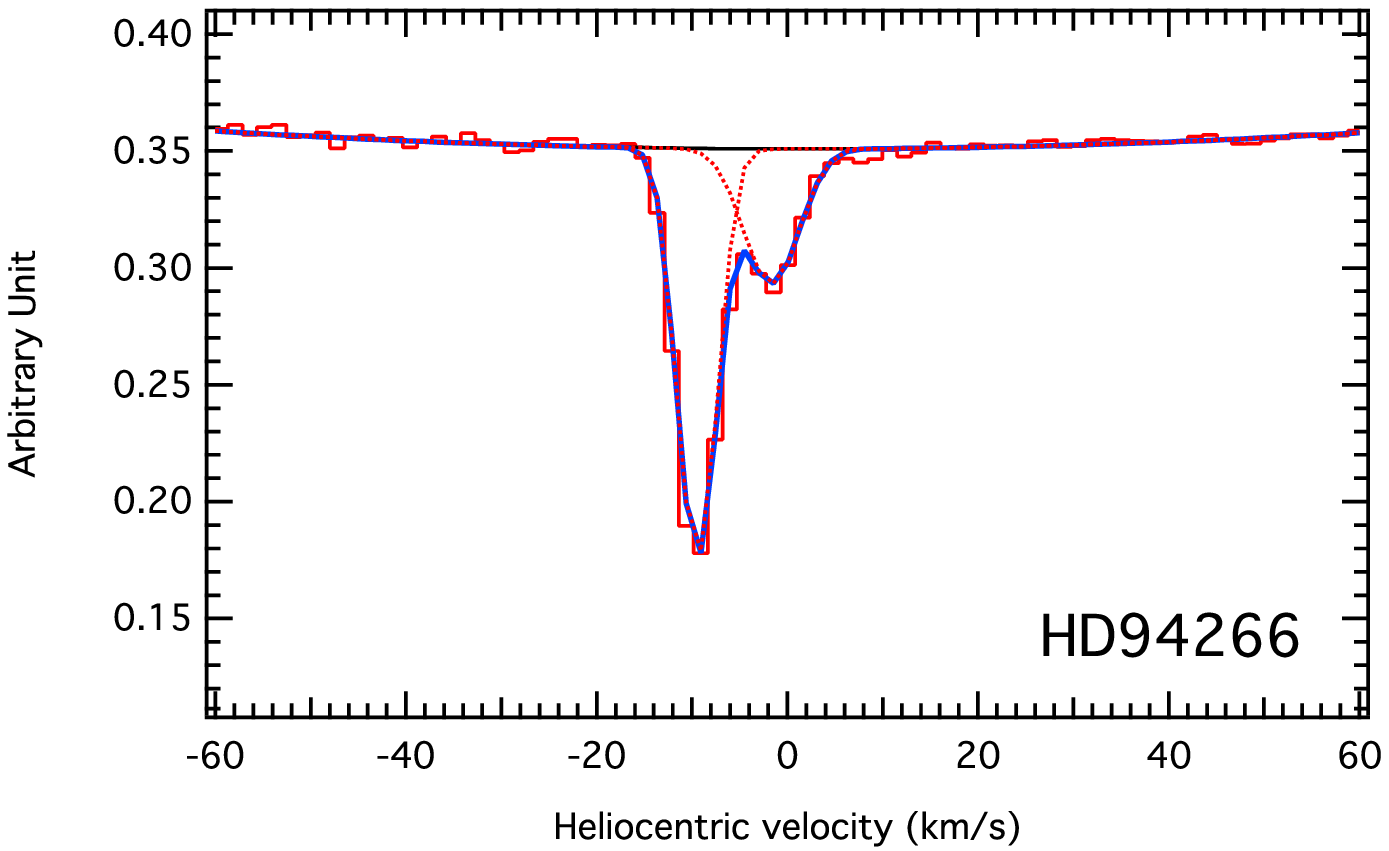}
\end{minipage}\hfill
\begin{minipage}[t]{0.3\linewidth}
\centering
  	\includegraphics[width=1\linewidth]{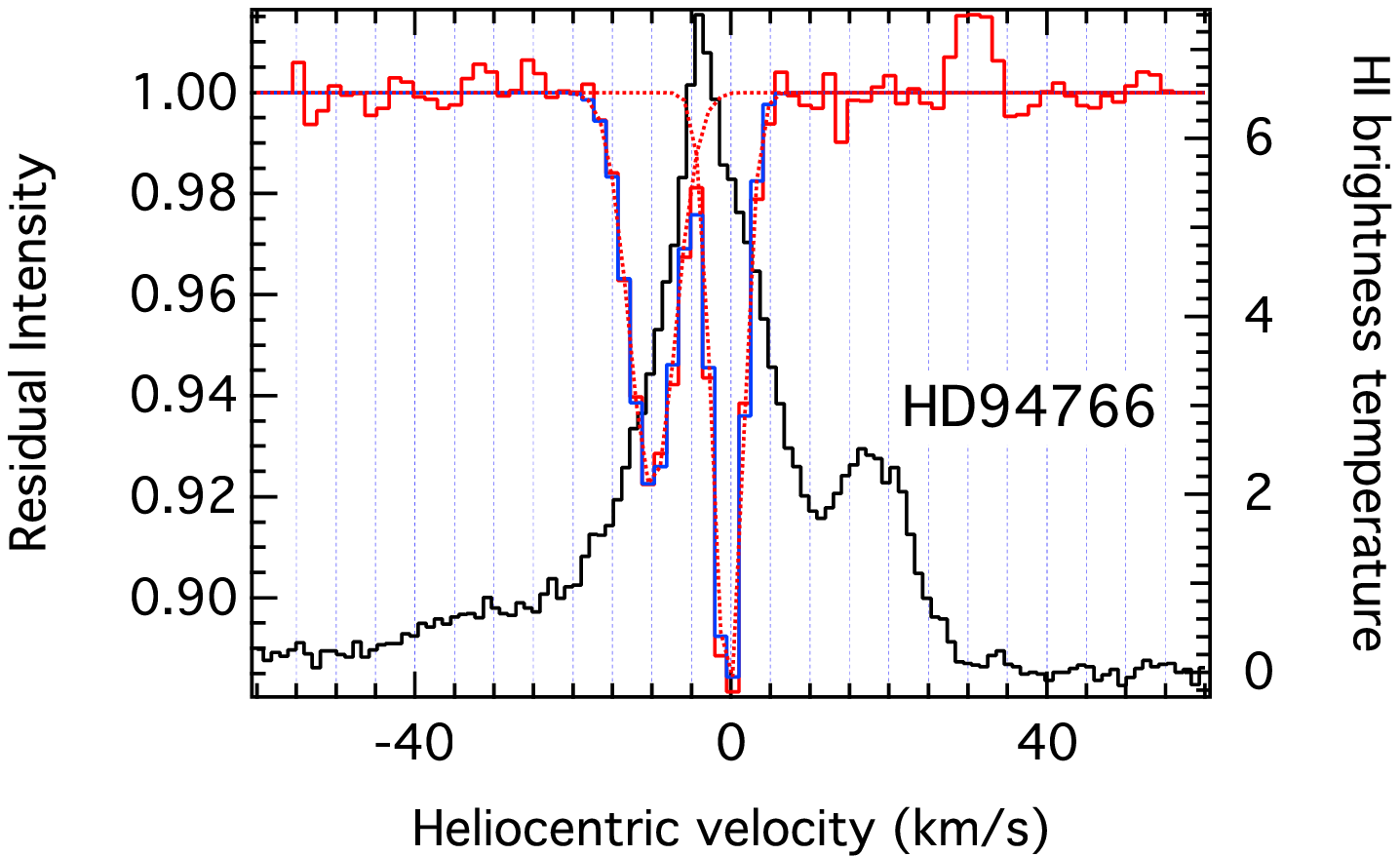}
  	\includegraphics[width=1\linewidth]{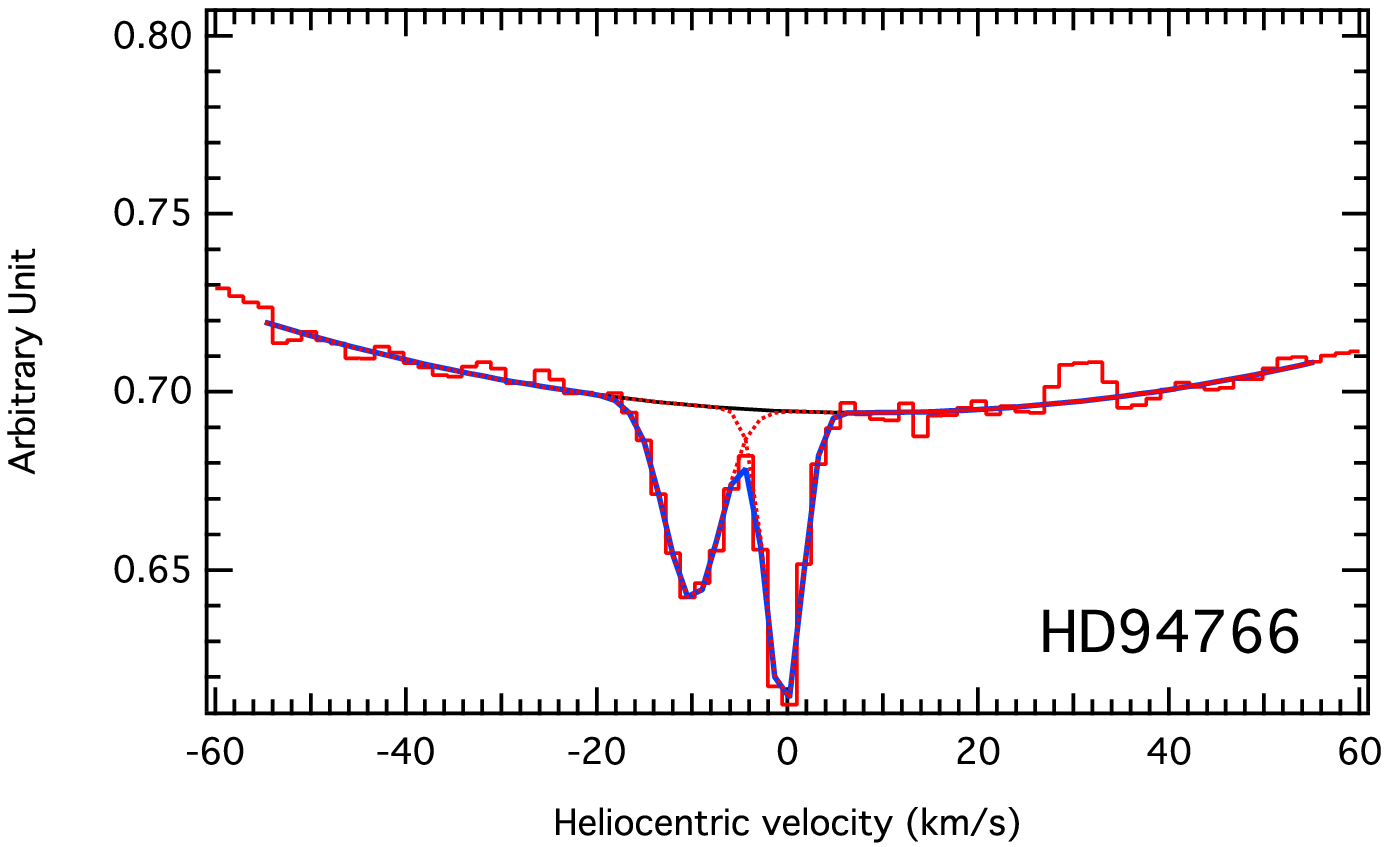}
  	\includegraphics[width=1\linewidth]{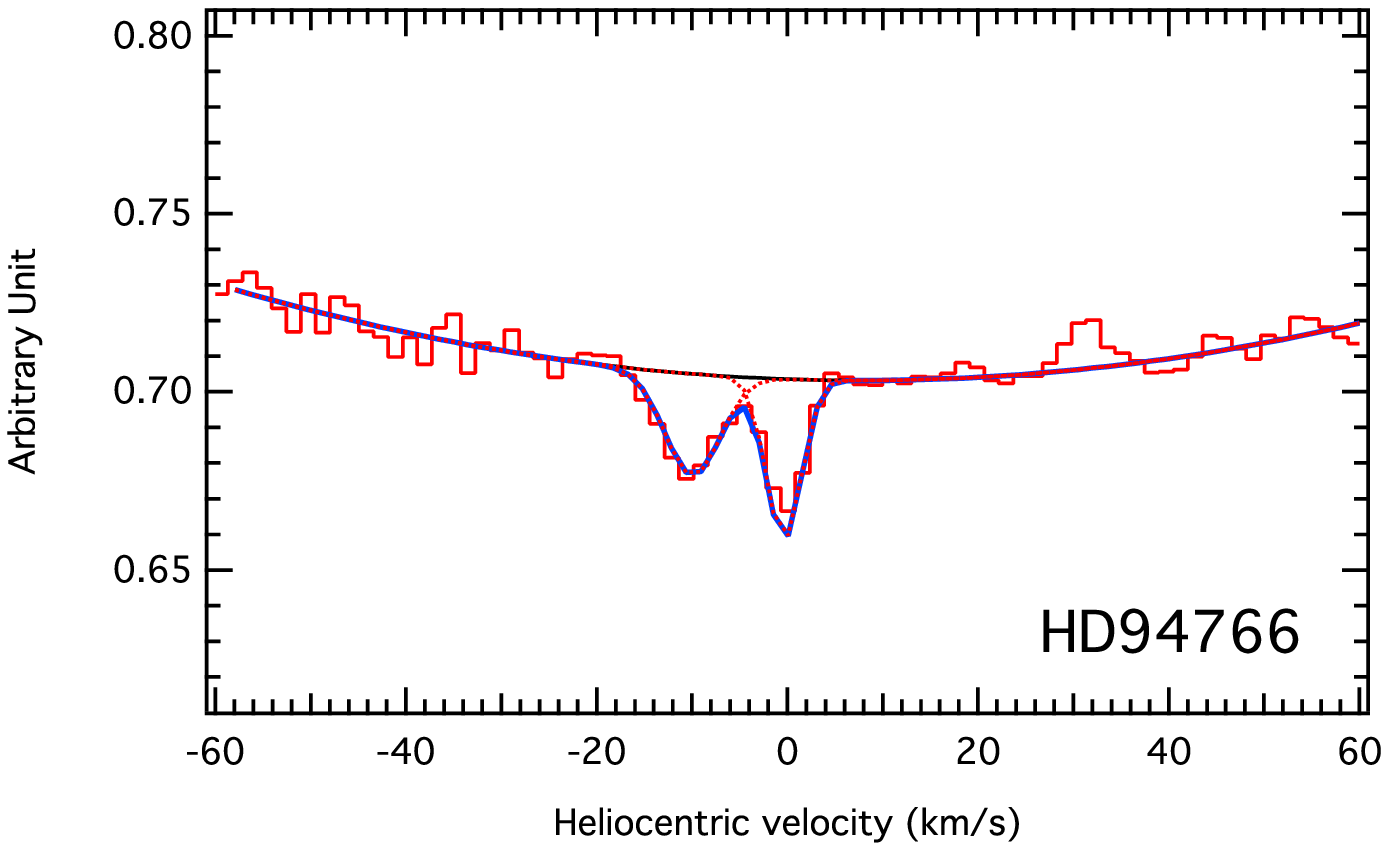}
\end{minipage}
\caption{Interstellar 5889  \AA\ NaI-D2 and 5895 \AA\  NaI-D1  absorption lines for the stars HD94194, HD94266, and HD94766. The first row shows the normalized interstellar NaI-D2 line-profile {(red histogram line}) superimposed on the HI emission spectrum ({black histogram line}). The second and hird rows show the unnormalized interstellar NaI ({D2 and D1, respectively}) line-profiles ({red histogram line}) superimposed on the fit model (blue line); and the {individual} model components (dotted red line). {For other target stars, line profiles and models can be found in the online appendix}.}
\label{HD94194} 
\end{figure*}

\begin{figure*}[!h]
\begin{minipage}[t]{0.3\linewidth}
\centering
  	\includegraphics[width=1\linewidth]{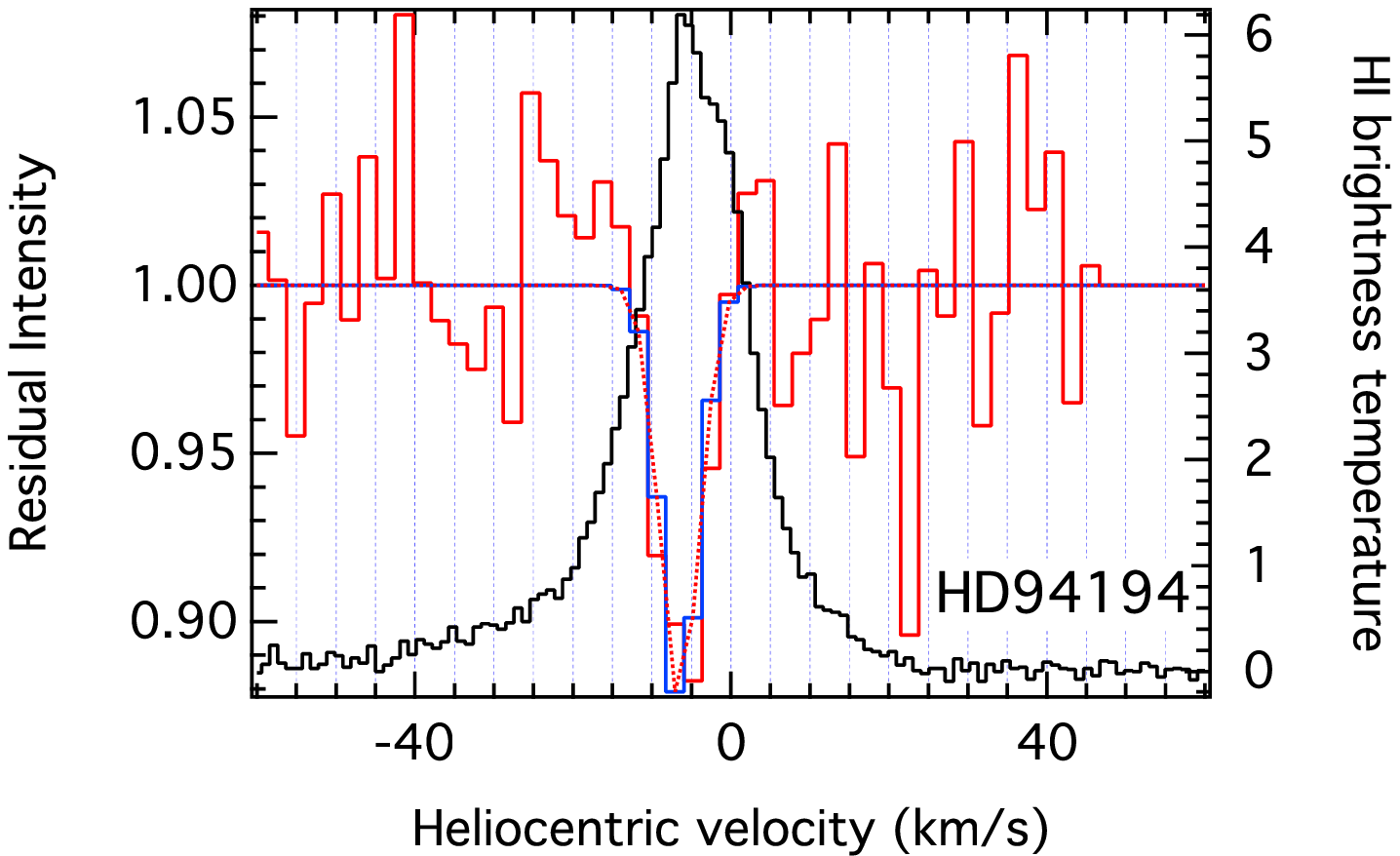}
  	\includegraphics[width=1\linewidth]{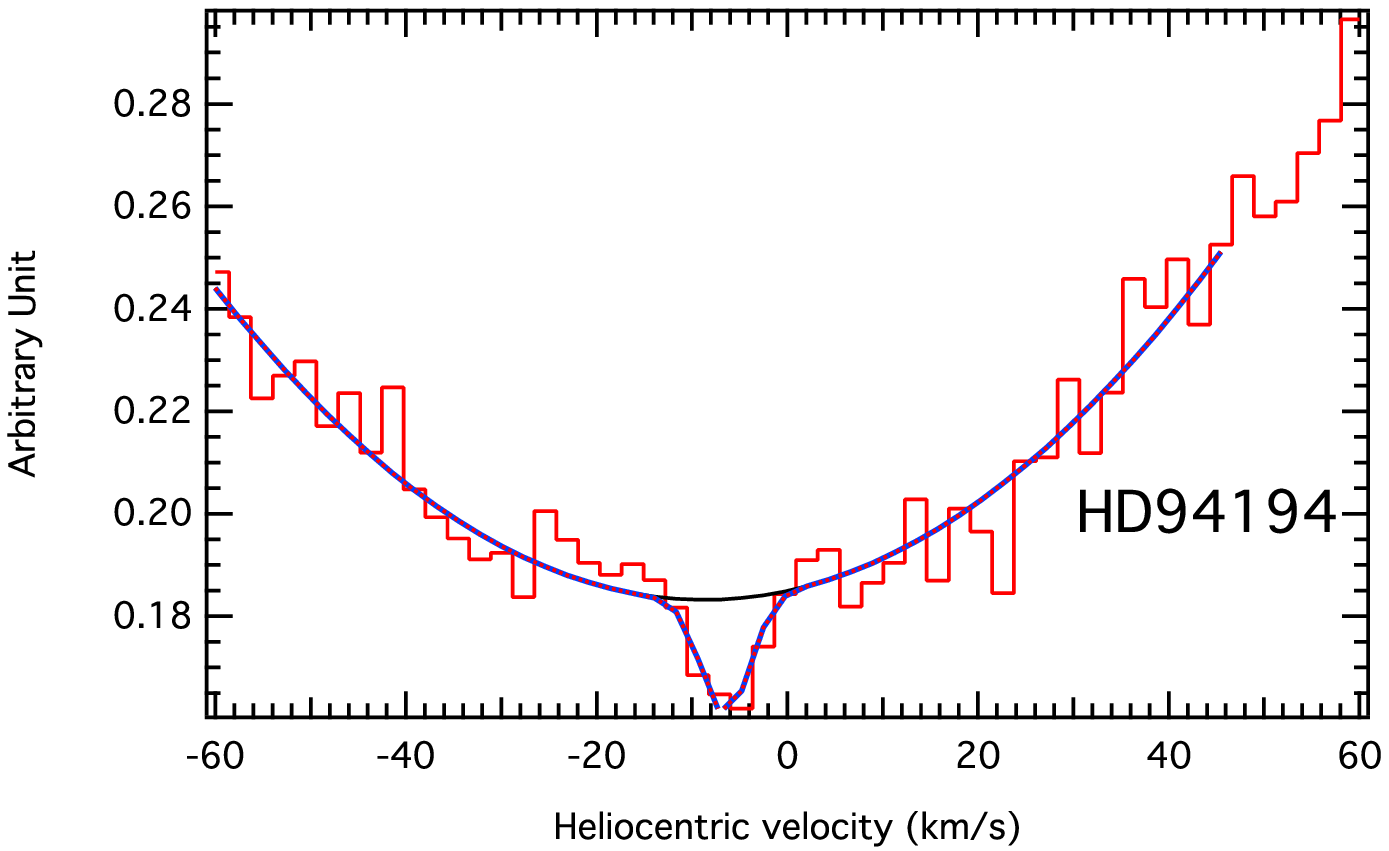}
  	\includegraphics[width=1\linewidth]{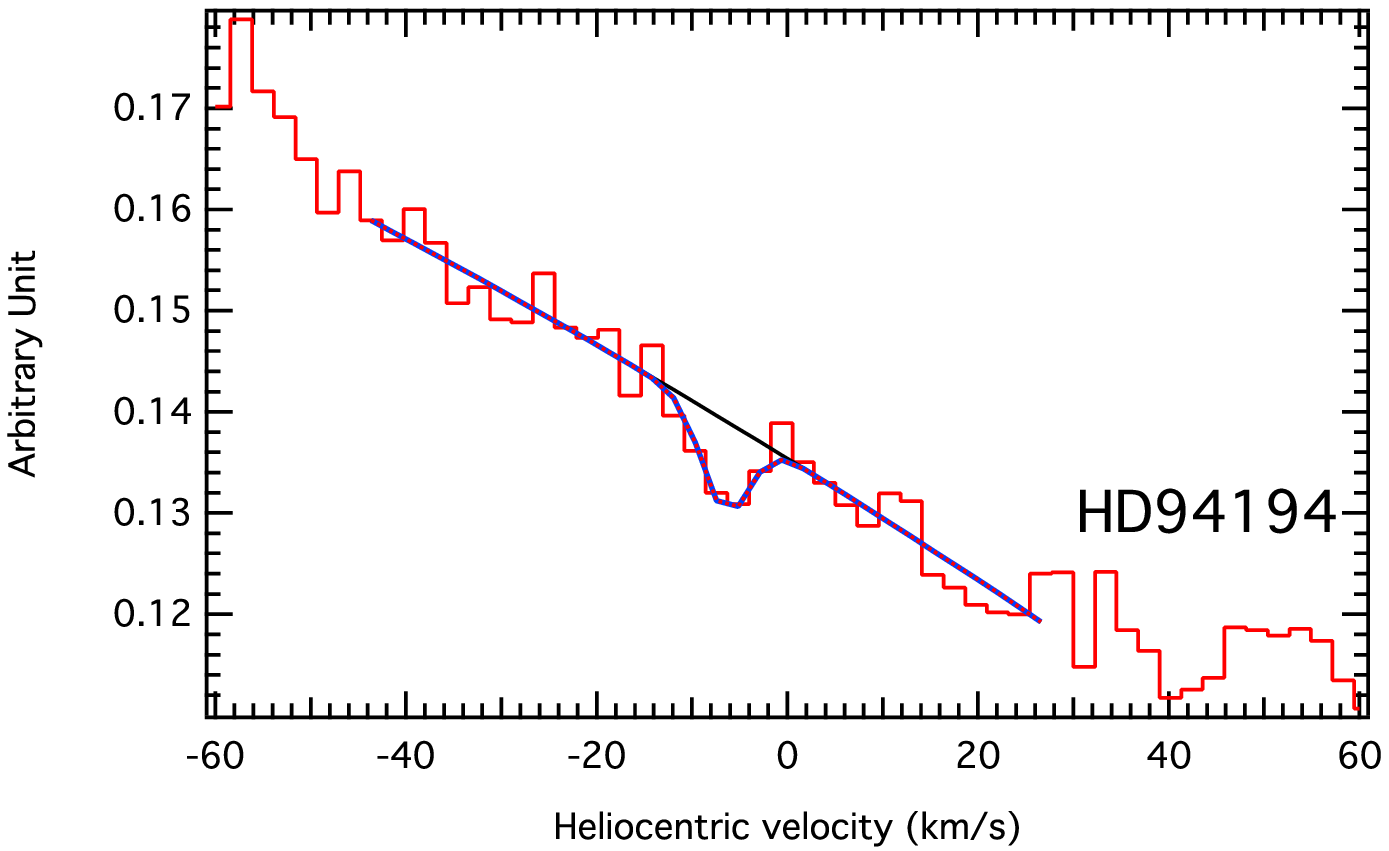}
\end{minipage}\hfill
\begin{minipage}[t]{0.3\linewidth}
\centering
  	\includegraphics[width=1\linewidth]{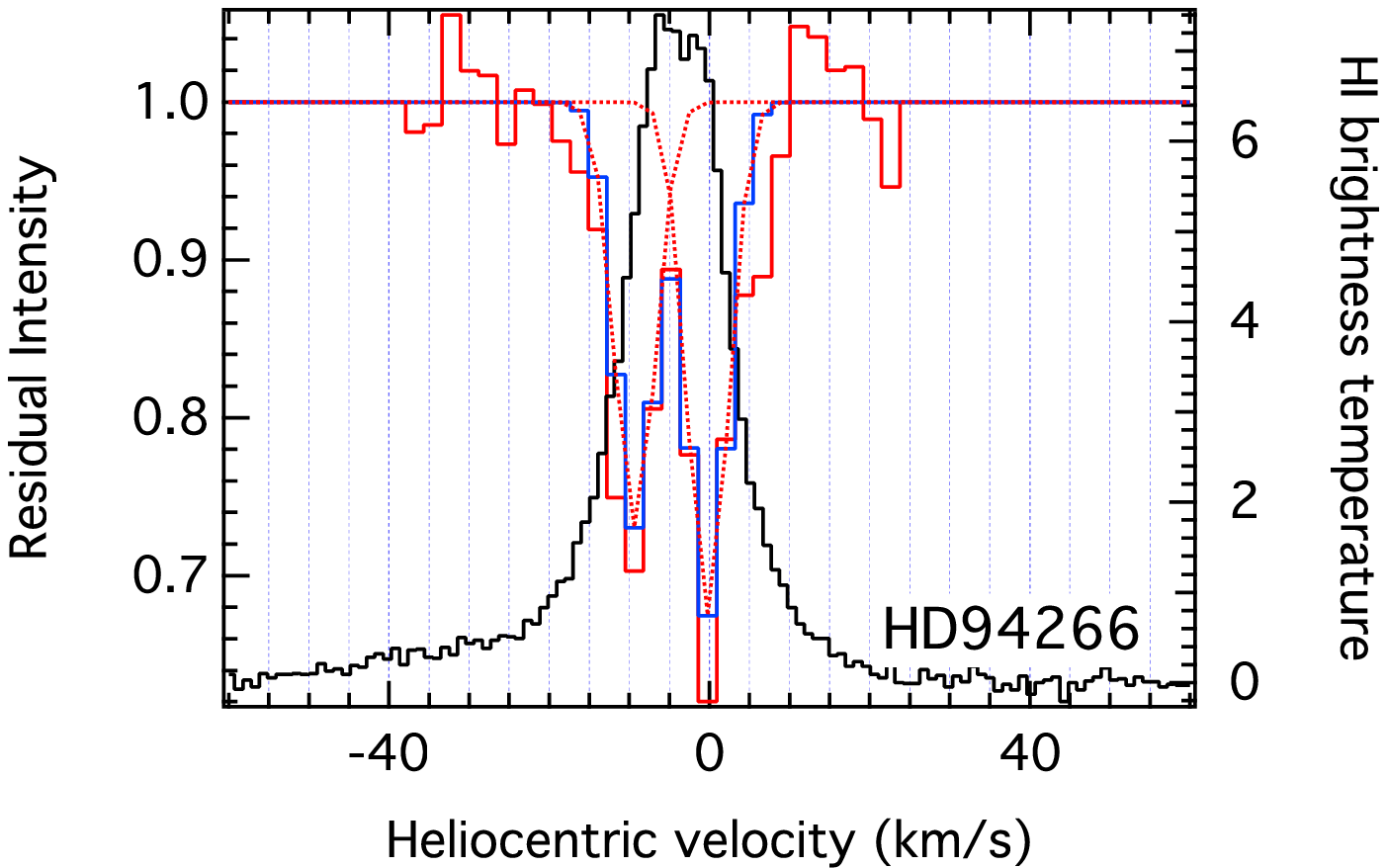}
  	\includegraphics[width=1\linewidth]{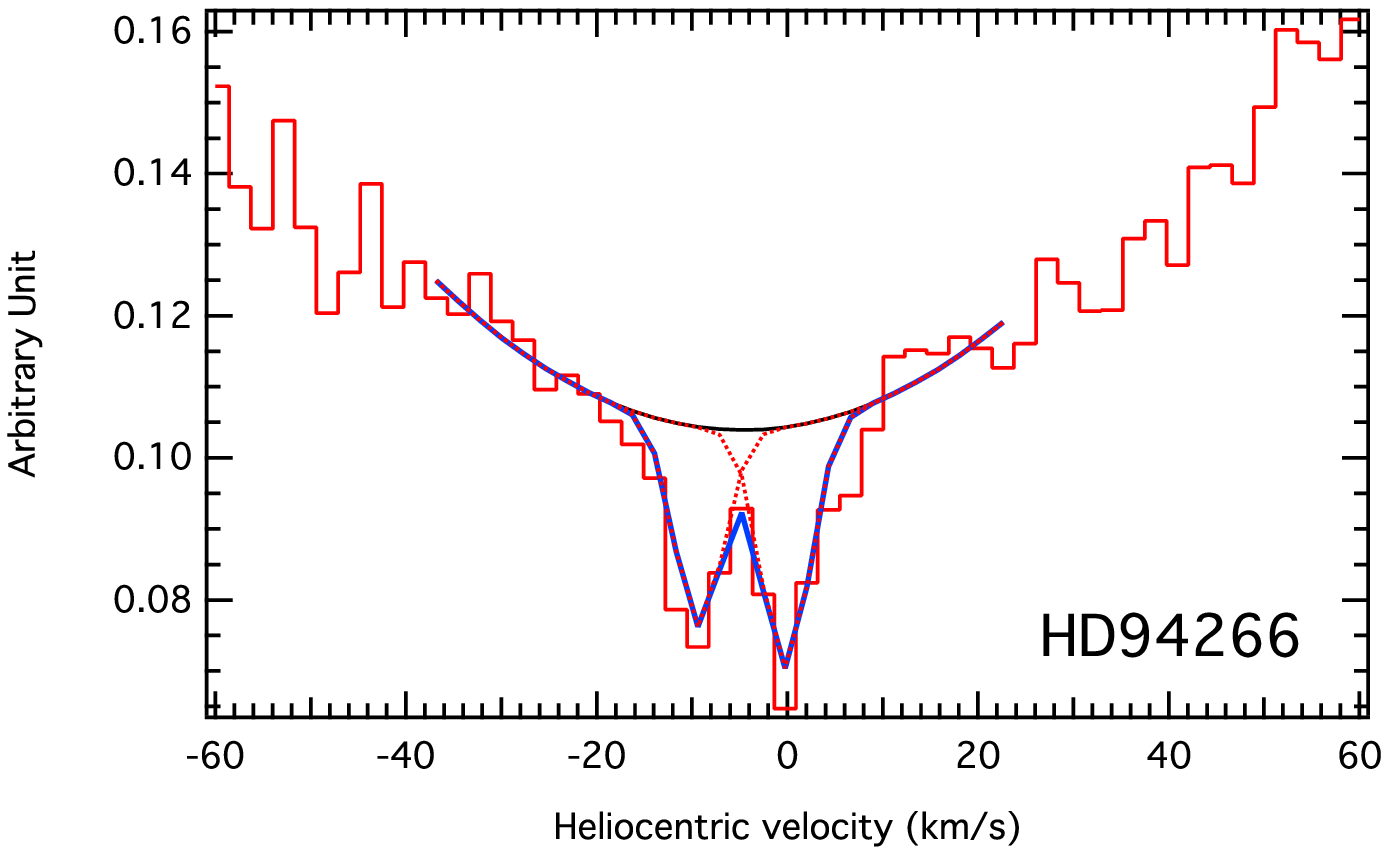}
  	\includegraphics[width=1\linewidth]{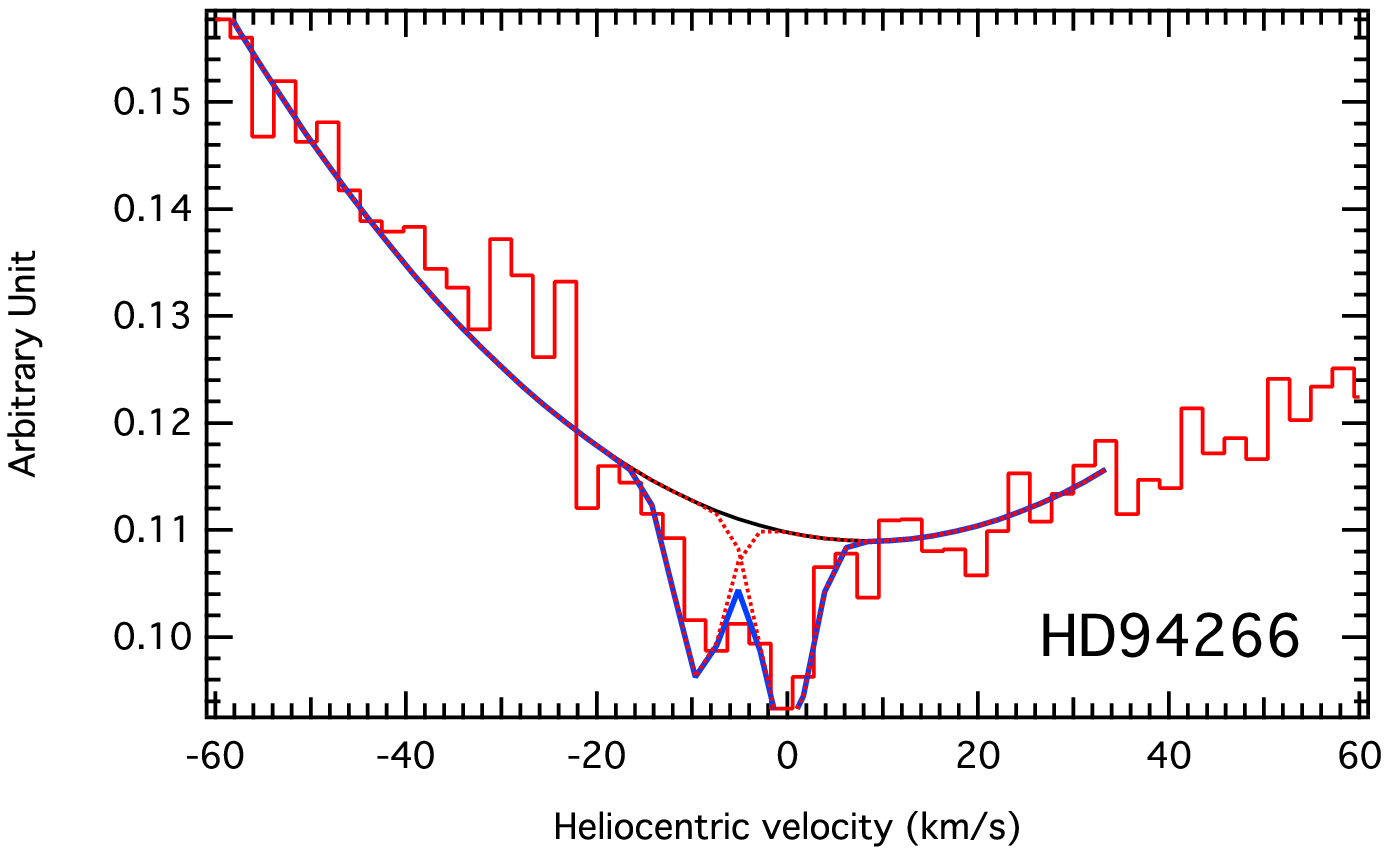}
\end{minipage}\hfill
\begin{minipage}[t]{0.3\linewidth}
\centering
  	\includegraphics[width=1\linewidth]{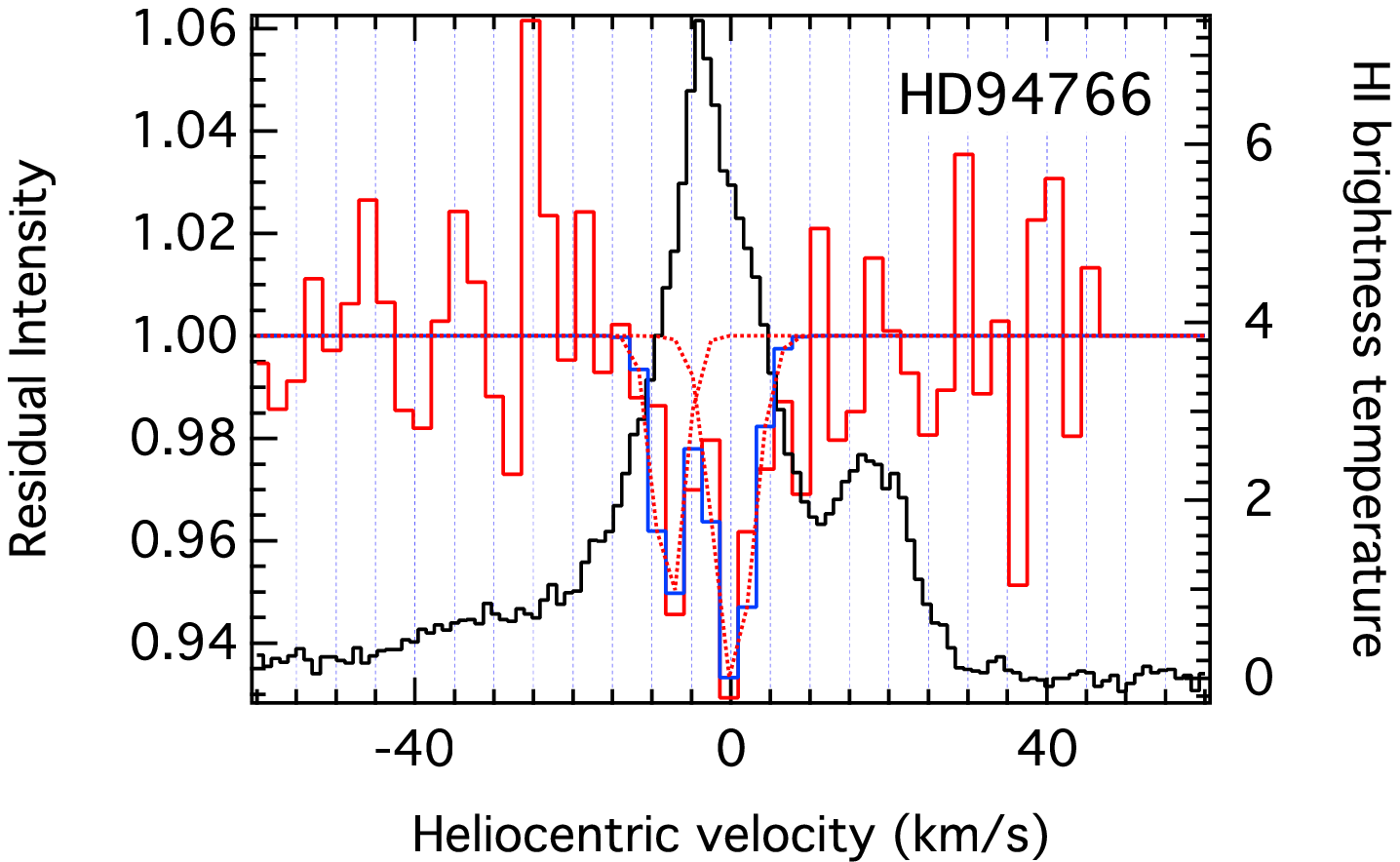}
  	\includegraphics[width=1\linewidth]{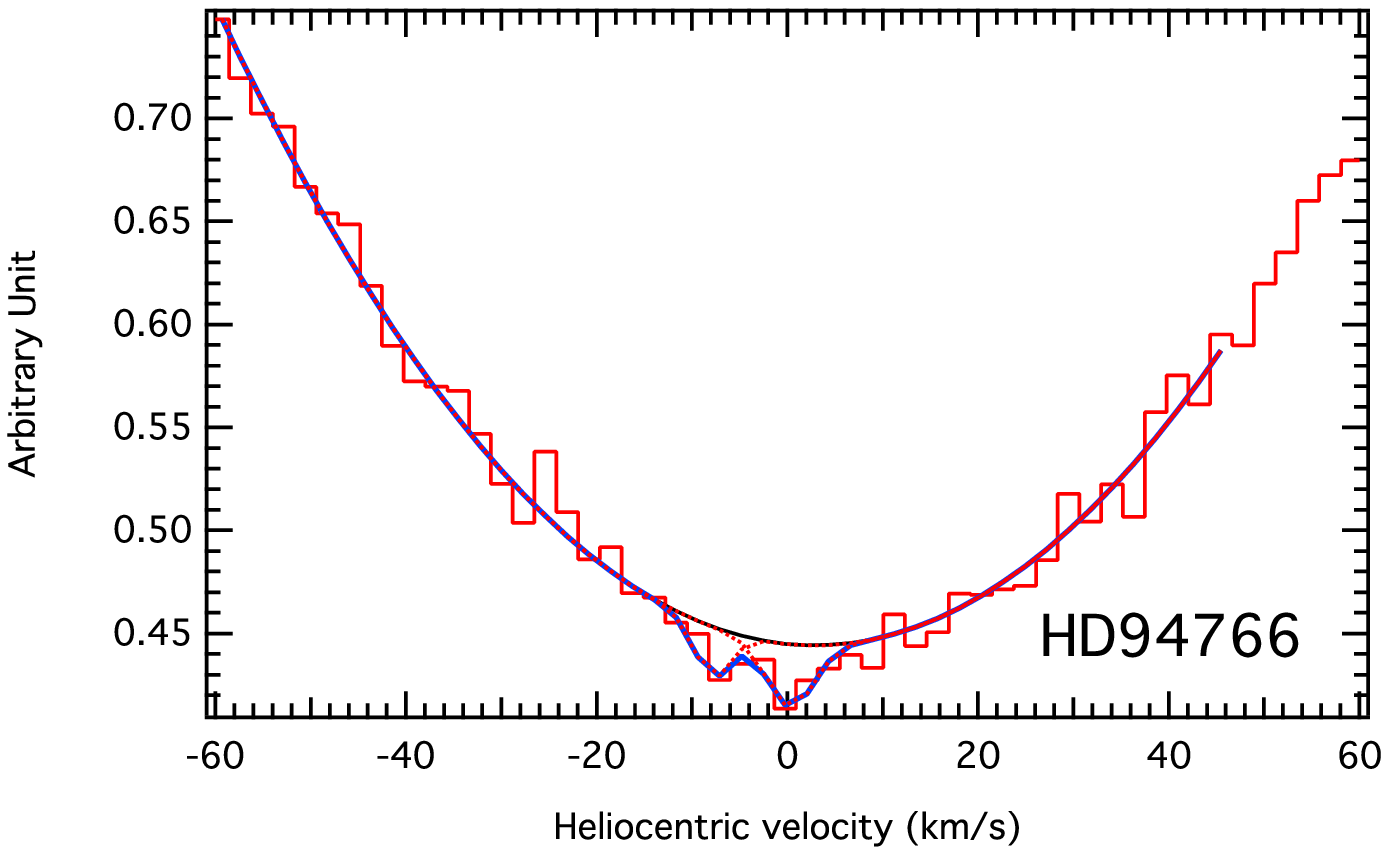}
  	\includegraphics[width=1\linewidth]{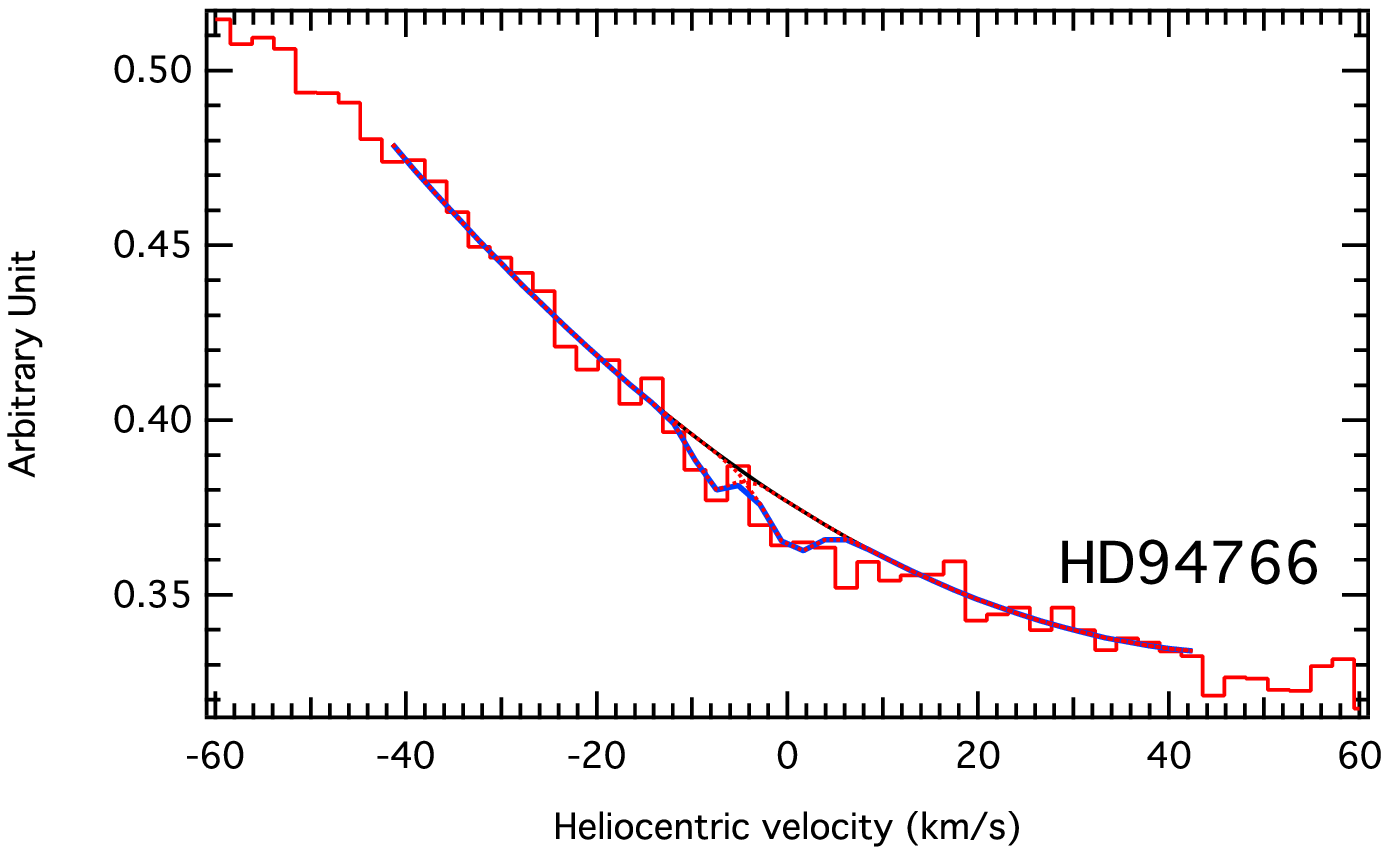}
\end{minipage}
\caption{Same as Fig. \ref{HD94194} but for interstellar CaII {(3934-3968)} absorption for target stars: HD94194, HD94266, and HD94766. From top to bottom: the normalized CaII K, the non-normalized CaII K, and CaII H line.}
\label{HD91494ca}
\end{figure*}

The resulting Doppler velocities, column densities, and the ratios between the NaI and CaII column densities are listed in Table \ref{table1}, as are the coordinates and Hipparcos distances of the target stars. 
The quoted errors are those associated with the profile fitting  alone and do not take into account wavelength calibration uncertainties or uncertainties due to the continuum placement.
The sodium lines are extremely sharp, errors on the central velocity are very small, even smaller than errors associated with the wavelength calibration.
The fitted data and the individual HI emission spectra taken from the LAB Survey for each target direction  are shown in {Fig. \ref{HD94194}, \ref{HD91494ca} and online appendix.}
The Doppler velocities of the detected clouds were compared with the HI emission and conclusions were drawn on the basis of this comparison and  the morphology of the shell.  

For comparison with the HI spectra, the Doppler velocities  were converted from the heliocentric frame (such as listed in Table \ref{table1}) into LSR velocities. The LSR velocities are shown in Fig \ref{FIG:HDSTAR_MAPLSR}.

\section{Results and discussion}

\subsection{Detection vs distance}
For all but two targets, it was possible to detect IS clouds or give a stringent upper limit on the neutral sodium columns. {On the other hand, the NaI clouds have a CaII counterpart for most of the targets. For the remaining stars, the noise makes it impossible to extract a CaII line reliably.}
The exceptions to those detections are HD112097 and HD111893, which turned out to be binaries with a cold companion or to have ambiguous lines, making them unsuitable for classical profile-fitting methods.  
Model-data adjustments at better than one sigma were obtained for one or two clouds only. 
{The} b-values are between 1.3 and 3.2 km.s$^{-1}$ for NaI and between 1.8 and 2.5 km.s$^{-1}$ for CaII. Note that 2.5 corresponds to our upper limit of temperature of 15,000 K, however, inspection of spectra shows that this b-value is appropriate in all cases. 
When simply combined, CaII and NaI b-values do not give reasonable turbulence and temperature values, confirming the classical result that CaII and NaI are not exactly similarly located. 
We did not attempt to discuss further on the physical properties, since our main goal was the cloud location. We simply note that for several stars the b-value corresponds to typical values for cool HI clouds.

The decision as to how much the HI 21 cm radio emission is associated with each NaI and CaII component was made on the basis of of radial velocity coincidence.  
About two-thirds of our recorded stellar spectra possess narrow interstellar NaI absorption lines that coincide with the HI emission velocity interval. 
More precisely, most of the stars with at least one absorbing cloud have a well-defined counterpart to one of the clouds in the HI spectrum in the form of a narrow emission, at the velocity that defines Loop I (based on well-known images and studies).
It was more difficult to clearly identify for the lowest latitudes, where HI  from Loop I  is seen in combination with background components. Still, we can rely on the map of Fig. \ref{FIG:HDSTAR_MAPLSR} drawn for the narrow velocity range [-10,+10] km.s$^{-1}$ (LSR), that follows the Loop I emission from higher to lower latitudes. 
The shell is so well defined that a sodium cloud detected within this velocity interval exactly toward the shell can reasonably be attributed to it.

{Looking at our target coverage and the low-velocity shells from Fig. \ref{HDSTAR_MAP} it is clear that one can distinguish two main features, one located above 70$\degree$  and extending over a broad longitude range, and a second one located below $\simeq$ 70$\degree$ and at longitudes smaller than 260$\degree$. 
The first higher-latitude shell \textbf{(b $\geq$ 70 $\degree$)} extends angularly over about 35$\degree$, but clearly corresponds to a coherent elongated structure. The second cloud could possibly be somewhat disconnected and subsequently start at a different distance from the Sun. 
We note, however, bridges between the two structures seem to be present, as can be seen at longitudes $\simeq$ 70$\degree$ in Fig \ref{HDSTAR_MAP}.  
This is why we distinguished the two regions in the distance analysis and considered separately the two groups of stars that probe those two regions.}

\begin{figure}[!ht] 
\centering\includegraphics[width=\linewidth]{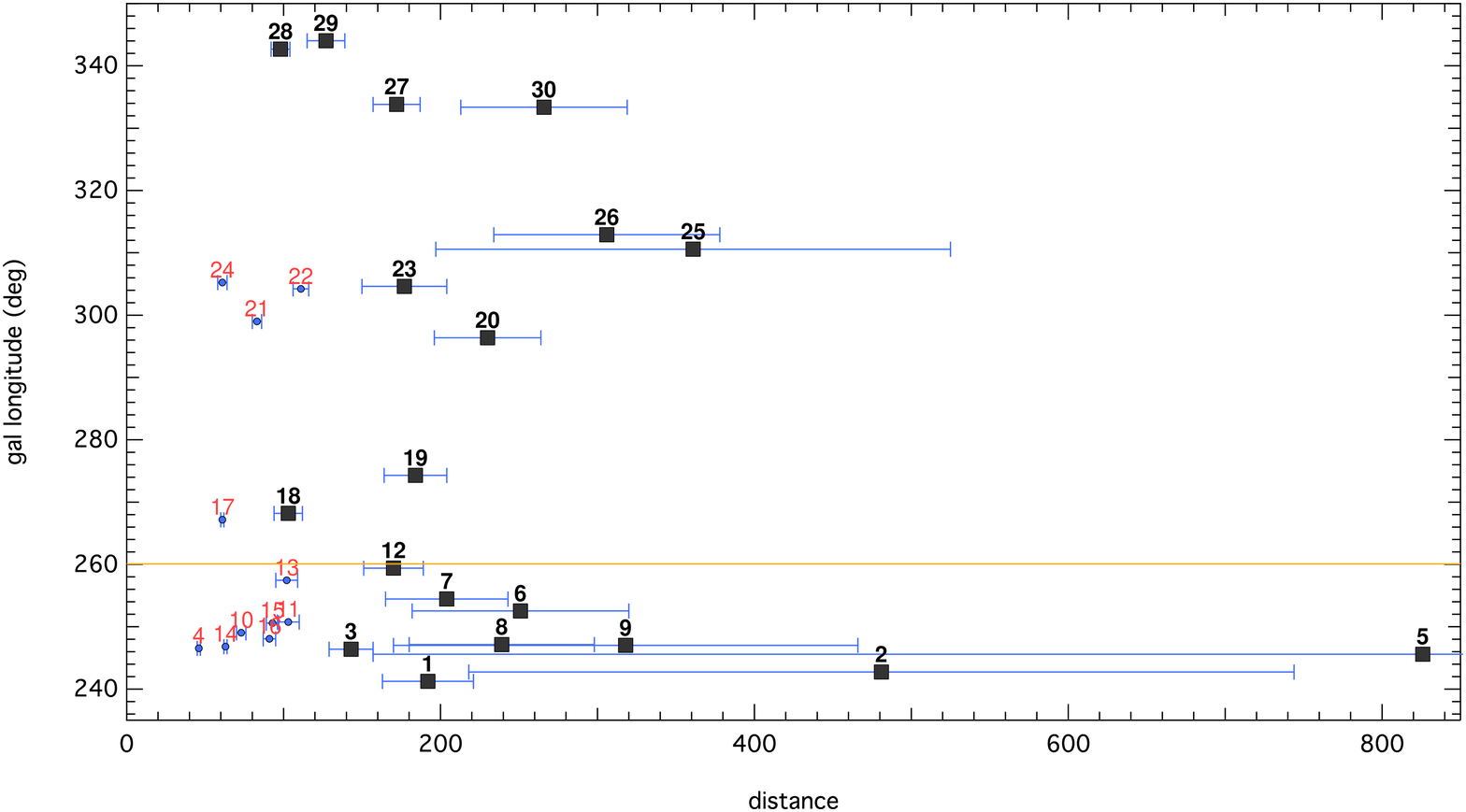}
\centering\includegraphics[width=\linewidth]{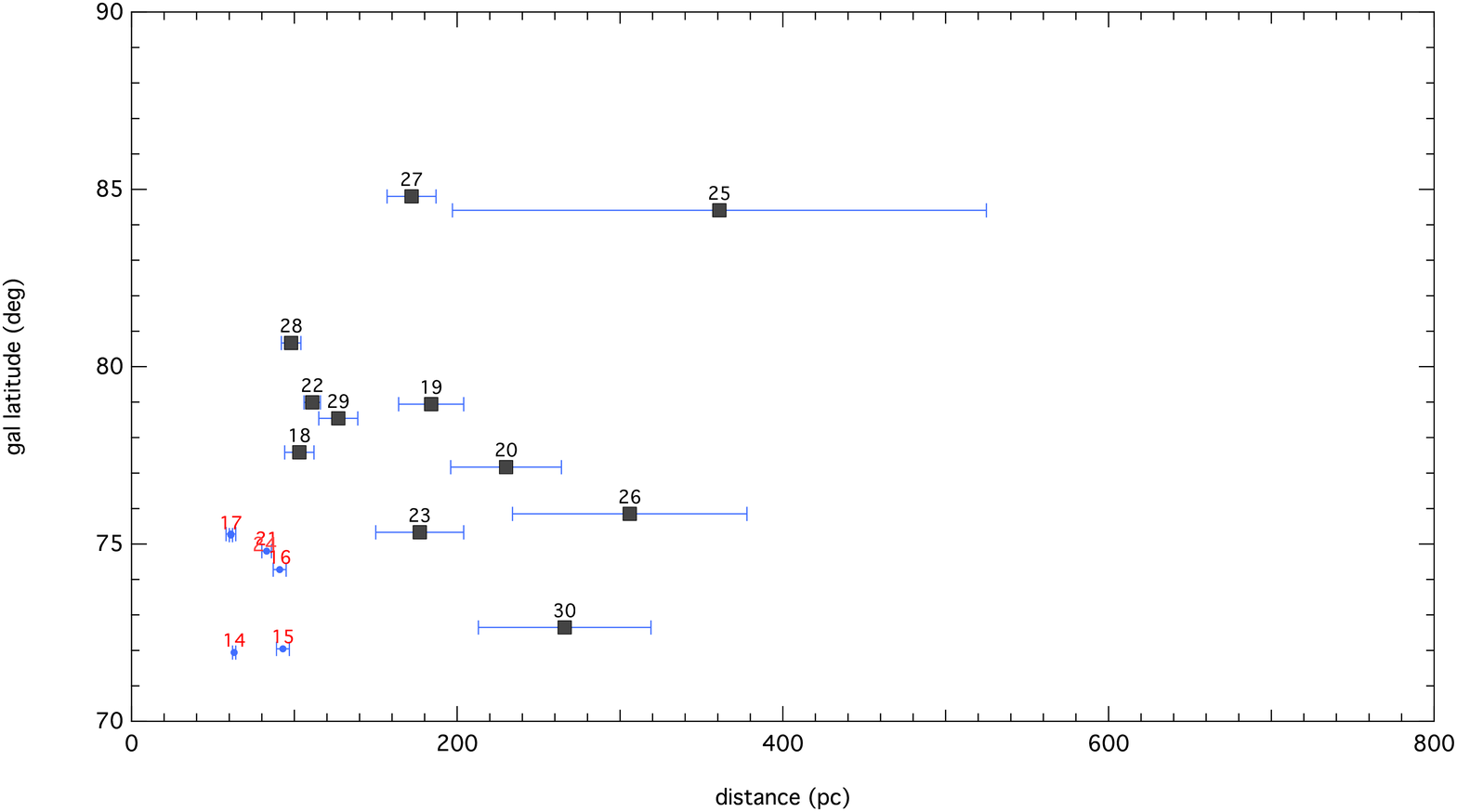}
\centering\includegraphics[width=\linewidth]{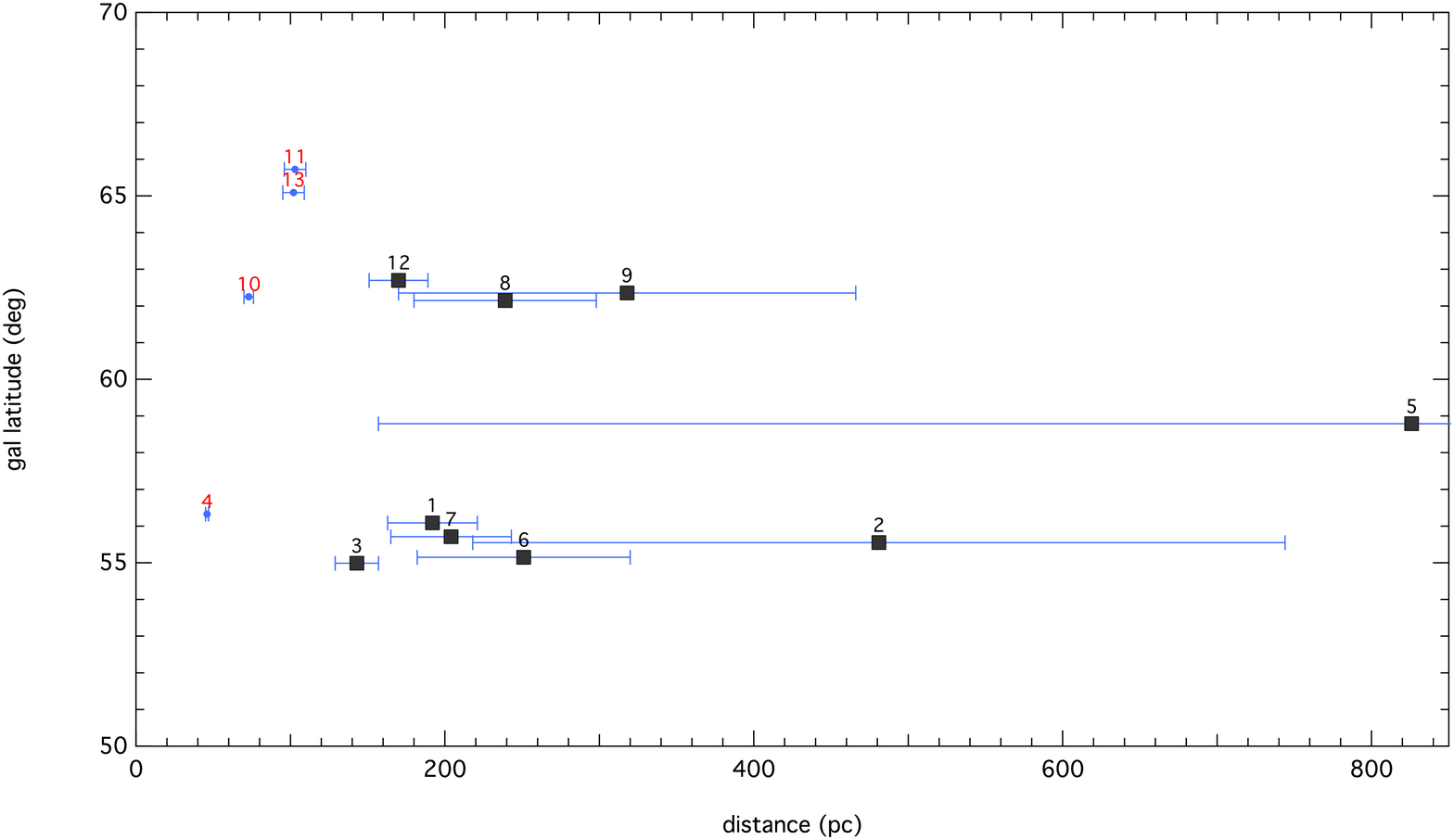}
\caption{{Distribution of the target stars in longitude vs. distance (top) and latitude vs distance (middle for b $\geq$ 70 $\degree$ and bottom for b $\geq$ 70 $\degree$). The stars have been divided into two groups, according their latitudes (see text). In the top graph, the two groups of stars are separated by a yellow line, except for the three targets 14 to 16 that are at low longitudes but seem to be located in the direction of the high-latitude arch.
The red (resp. black) numbers show the target star without (resp. with) interstellar sodium lines at radial velocities identical to the HI loop I radio velocities. The stars with interstellar lines are also shown with rectangular symbols.
}
}
\label{latlong}
\end{figure}

\begin{figure*}[!ht]
\centering
\includegraphics[width=\linewidth,height=68mm]{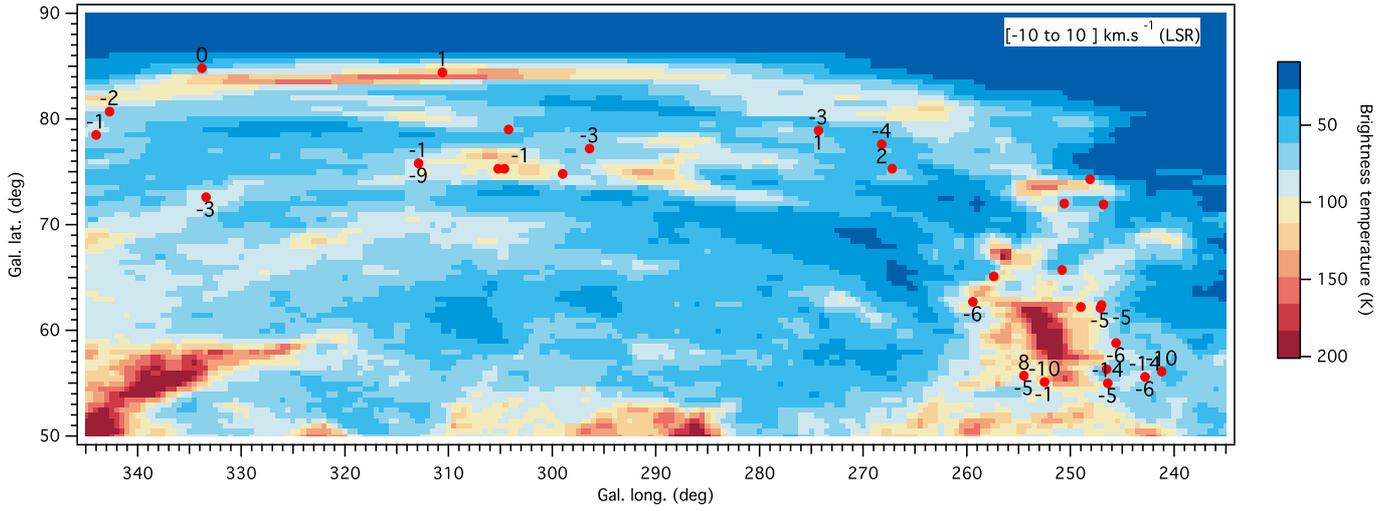}
\caption{Distribution of the detected IS cloud radial velocities in LSR frame from NaI absorptions, {superimposed on  HI emission map -10 to 10 km/s}. When two components are detected, they are indicated above and below the star marker.}
\label{FIG:HDSTAR_MAPLSR}
\end{figure*}

Figure \ref{latlong} illustrates the derivation of the distances {to those Loop I features}. 
{The longitude (resp. latitude) vs. Hipparcos distance intervals} are displayed for all targets, and different colors and rectangular symbols separate targets showing distinct absorptions at  Loop I velocity from those that have no detected absorption lines, whether in sodium or calcium. 
It is immediately clear from Fig. \ref{latlong} that the closest stars belong to the first category, while more distant stars belong to the second one. 

More precisely, most of the stellar spectra that do not have interstellar NaI absorption lines are located  at a distance d $\lesssim$ 100 pc: HD95382 (46 pc), HD98747 (73 pc), HD100518 (103 pc), HD101470 (102 pc), HD102660 (63 pc), HD103152 (93 pc), HD103877 (91 pc), HD106661 (61 pc), and HD111164 (83 pc). 
The closest star that possesses interstellar NaI lines is HD115403 (star 28) at a distance of 98 $\pm$ 6 pc.
The most distant star that does not have any firmly detected line (NaI and CaII) is HD100518 (star 11) at a distance of 103 $\pm$ 7 pc.
However, HD101470 (star 13) at 102 $\pm$ 7 pc also gives a similar result. 

{Bearing in mind that the higher latitude filamentary structures and the lower longitude-latitude cloud are separate entities, we concluded that: (i) the shortest limit of about 98 $\pm$ 6  pc strictly applies to the highest latitude arcs that culminate at \textit{b}=+85$\degree$ and +75$\degree$; (ii) the structure around (\textit{l},\textit{b})= (250$\degree$, +55$\degree$) has fewer constraints, and could be as distant as 143 $\pm$ 14 pc (or as close as 100 parsecs).} 

{When} using the total set of stars or equivalently assuming that all the arches probed here start at the same distance from the Sun, and considering that the detection of either NaI or CaII defines a cloud, this distance can be bracketed to a narrow interval: d(Loop I) = 100 $\pm$ 4 pc, from being smaller than (98 + 6) and larger than (103 - 7) due to the two targets HD115403 and HD100518. We note that there are no contradictory results among the stars, which suggests that the distance toward the various portions of the arches does not vary much. Also, the variability of the column densities reflects well either the star location with respect to the densest part of the HI filament or the distance (i.e., small columns correspond to close stars, which are embedded in the front part of the cloud or to stars offset from the brightest HI regions).

\begin{figure*}[!ht]
\centering
\includegraphics[width=\linewidth,height=68mm]{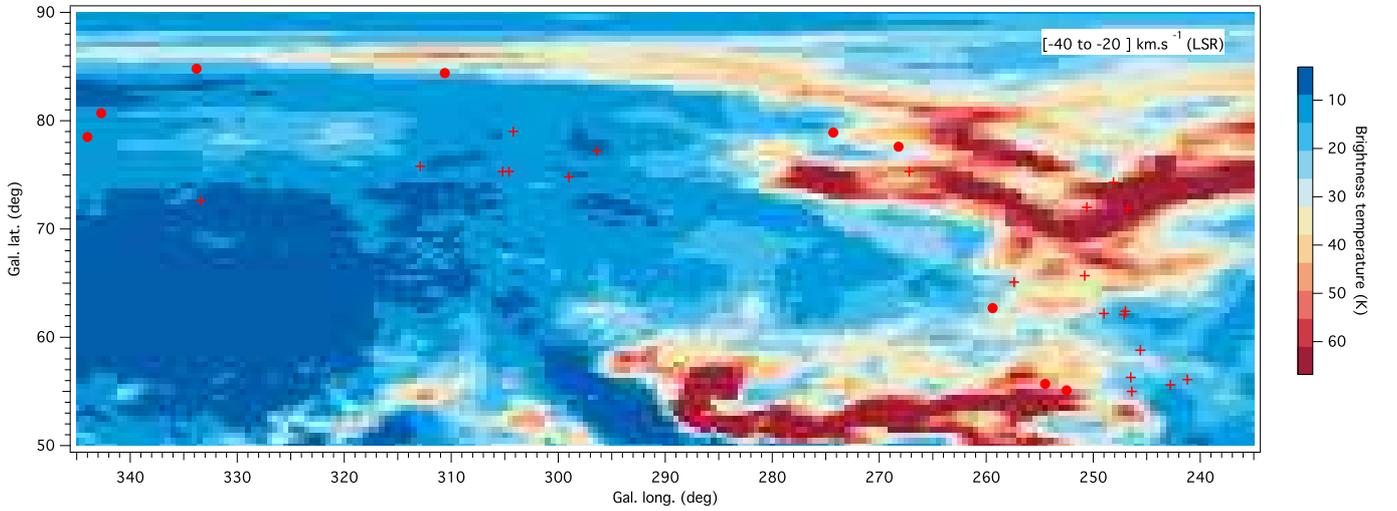}
\caption{Target locations with respect to the second, negative velocity structure, {as seen in the HI map of the gas in the velocity range -40 to -20 km/s}. This structure is not detected in any spectrum. Targets that constrain the best the minimum distance are shown as filled circles. }
\label{secondloop}
\end{figure*}

\begin{figure*}[!ht]
\centering
\includegraphics[height=90mm]{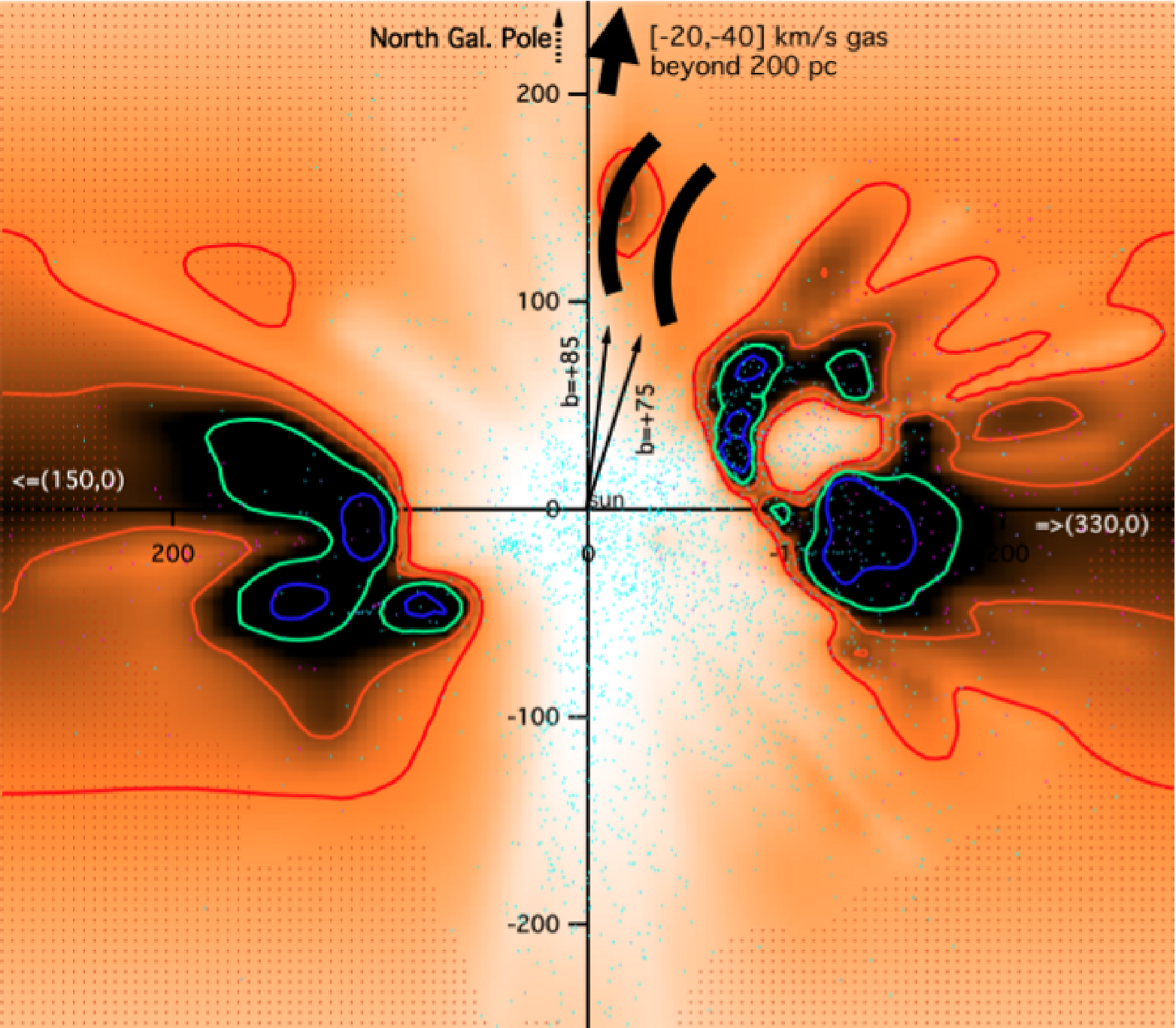}
\caption{Locations of the HI arches detected in absorption w.r.t. the local ISM global structure, {drawn as thick black curve lines}. The map represents a planar cut  through the 3D opacity distribution  derived by Vergely et al, (2010) from extinction data. The Sun is at the center (0,0 coordinates) and the north pole is directed to the top of the figure. The plane is oriented along the 150-330 $\degree$ longitude direction. Units are parsecs. Black represents high-opacity regions, white non-reddening medium. The spatial resolution of the map is 15 pc (see Vergely et al, for explanations).  Iso-opacity contours have been added to emphasize the dense areas. Turquoise and green small dots are the stars close to the Plane included in the inverted dataset {and their distribution gives an idea of the constraints on the gas location}. Pale dots indicate regions where no inversion was made due to insufficient data. {The -1 below the plane on the right is the first digit of the -100 pc label (X counted positively toward the anticenter).} We have represented the two arches detected in this plane, i.e., toward \textit{b}=+85 and +75 $\degree$. The first arch falls close to the dust cloud found by inversion of extinction data. The lower velocity gas is indicated by arrows pointing to longer distances. The cavity towards \textit{l},\textit{b}=(330$\degree$,+20$\degree$) was very likely blown by stars from the Upper-Scorpius OB association. Interestingly, the HI gas here located falls in the continuity of the masses of dust, whose boundaries defining the local {\it chimney} are inclined w.r.t. to the polar axis. }
\label{cut_scheme}
\end{figure*}


\subsection{The low-velocity arcs}

Looking at HI data and our velocities in more detail, it appears that a low LSR velocity component in the [-6,0] km.s$^{-1}$ range is found in all regions, as shown in Fig \ref{FIG:HDSTAR_MAPLSR}.
Another component within the [-14,-9] km.s$^{-1}$ range is also found for some stars more distant than about 140 pc (e.g., star 3), but only at low latitude and in a small area. 
It is very likely not related to the northern shell, and we will no longer consider this cloud. 
Thus, the distance of 100 $\pm$4 pc quoted above, {or the two distance limits of 98 and 143 parsecs} for the two separate latitude ranges, do indeed correspond to the low LSR velocity gas (note that Doppler velocities indicated on the map are in the LSR, converted from the heliocentric velocities of Table 1). 
Calcium line shifts are identical to the sodium values, within uncertainties due to calibration and noise level, and confirm this pattern. Finally, for all stars that are distant enough to possess absorption lines, there is a very coherent set of LSR radial velocities (see Fig. \ref{FIG:HDSTAR_MAPLSR}), with LSR Doppler varying from a value close to zero at the highest latitude, to -5,-6 km.s$^{-1}$ around (\textit{l},\textit{b})= (350$\degree$, +55$\degree$).  

These absorption data definitely demonstrate that Loop I HI high-latitude, low-velocity shells are very nearby features, in agreement with stellar light polarization data. {Indeed, we note that our distance estimates are entirely compatible with inferences from stellar light polarization data. 
We used the catalog of polarization data of \cite{heiles00} and extracted all targets located within the sky fraction of Fig \ref{HDSTAR_MAP}. 
They show a global increase of the polarized light percentage at about 100-150 parsecs. 
Using a few stars located as closely as possible to our targets as well as combining the low-velocity HI column and the polarization fraction, we obtained reasonable polarization-HI ratios, or equivalent polarization-reddening ratios, similar to those reported by} \cite{santos11} {at lower latitudes. 
This also agrees with the general direction of the polarization angle, suggesting that the magnetic field follows the high-latitude arch.}

Furthermore,  the detected clouds are the gaseous counterpart of the dust cloud located by inversion in the opacity maps of Vergely et al, (2010), whose closest part begins at $\simeq$ 120 parsecs (see  Fig. \ref{cut_scheme}). 
The small difference in the distance may be attributed to the scarcity of extinction data, parallax uncertainties, {and map resolution (the minimum cloud size in those maps is 15 pc)}. 
The whole radio structure associated to the continuum and HI  is thus found to be consistently located with about 100 parsecs.  Simple geometric considerations show that such a pattern implies a very low-expansion velocity of the shell, assuming it is centered as close as 140 pc (see section 1).

\subsection{The negative velocity structure seen in HI}

Those new measurements also shed light on another structure characterized by significantly more negative Doppler velocities, fainter than the main low-velocity arch seen in Fig \ref{HDSTAR_MAP} {and located mainly} at latitudes {above 85$\degree$}. 
This structure, visible in the [-40, -20] km.s$^{-1}$ (LSR) HI map shown in Fig. \ref{secondloop}, is located slightly above the low-velocity loop, corresponds in the HI spectra to a broader velocity distribution, and is not so conspicuous. 
To our knowledge, it has not been discussed independently from Loop I. 
{In the}  [-40, -20] km.s$^{-1}$ interval, it differs  from Loop I in the first quadrant at low latitudes (in particular, it does not end above the NPS but at more positive longitudes). 
About half of the individual HI spectra in the fitted interstellar line figures (Fig. \ref{HD94194}, \ref{HD91494ca}, and in the online appendix)
very clearly reveal this negative velocity structure and the narrow emission at $\simeq$ zero {LSR} Doppler shift.
But this negative velocity structure is also present in all spectra, although less detached.  
What is remarkable here is that there are NO detected absorptions corresponding to this arch, even for the most distant target stars. 
Despite being fainter and broader, it should be detected in absorption for a number of our target stars that have a smooth and flat continuum, assuming that the neutral sodium and the ionized calcium have similar relative abundances as in the main Loop. 
Figure \ref{secondloop} identifies the nine targets that correspond to this criterion and their distance. 
Constraints on distances based on those targets can be estimated from the individual spectra {(see fitted interstellar line figures)} by comparing HI and NaI or HI and CaII. 
At least four of those targets imply a minimum distance to this second HI structure of $\simeq$ 200 parsecs, with no upper limit owing to the absence of any detection.
 As a consequence, it is likely that this gas is completely detached from the front shell. 


\subsection{On the shell geometry}

The minimum distance to the {high-latitude} shell of $\simeq$ 100 parsecs can be used in conjunction with the shell coordinates to discuss its geometry and its origin. 
{As discussed in more details by \cite{santos11}, there are three main classes of models aiming at explaining the Local Bubble-Loop I structure. A common point is the presence of one or more spherically expanding shells centered close to the Plane at \textit{l}$\simeq$ 330$\degree$. According to the first class of models, both the Local cavity (LC) and the Loop I cavity possess expanding shells, which are currently colliding, with an interaction region that forms a dense ring (\cite {EggerAschenbach}). Here, the closest northern shell should be related to the LC expansion front. The second class considers that the Loop I cavity X-ray emission and the HI shells correspond to a recent explosive event, having pushed away and reheated a much older cavity. The third model, with a globally similar scenario, explains the radio continuum polarization data with two colliding shells, both generated within about the same region as in the Sco-Cen association. Recently, \cite{reiscorradi08}, \cite{santos11} and \cite{Reisetal} have questioned the existence of the interaction ring, which is associated to the first model only (see also \cite{frisch11}}). 

According to our results and in agreement with the stellar polarization data, there is no detection of gas closer than about 100 pc, even toward our lowest latitude targets. {We emphasize that constraints on the gas distance from absorption are stronger than when deduced from polarization, because polarization  depends on the homogeneity of the grain orientations, which introduces uncertainties into their interpretation.} Our results imply that either the shell starts at our threshold distance, i.e., 100 pc, or that it is detected only when the line of sight is tangential to it, where the HI {and NaI columns reach maximum values}. This would also agree with the null velocity. But in the case of spherical shells, and according to the most simple geometrical rules, the shell center must be either at low b and rather distant (at about 270 pc if  \textit{b}=17$\degree$, i.e., excluding the Sco-Cen area), or, if nearby, at a significant distance from the Plane (for a distance to the Sun of about 140 pc \textit{b}=40 $\degree$, i.e., 80 pc above the Plane). Such an offset from the Plane resembles the solution found by Wolleben (2007) for the radio continuum Loop S2; here, however, the center is significantly more distant from the Plane. On the other hand, while the tangential directions to the two S1 and S2 loops derived by Wolleben at the highest latitude correspond well to the directions of the +85 and +75 $\degree$ arcs seen in HI, we find it difficult to reconcile such loops with our threshold distance. Finally, we note that all the nearby clouds detected within about 50 pc are warm and ionized, without any very narrow absorption such as those we find here, except for the Leo cloud, which is in a direction opposite to the present shells. As a consequence, we believe that the absence of gas detection closer than 100 pc at {high-latitudes (above 75$\degree$}) can hardly be explained by tangential effects. Instead, we believe that the cold HI is actually present only beyond about 100 pc.

In Fig. \ref{cut_scheme}, we have drawn the high-latitude sodium/HI clouds superimposed on the opacity distribution derived by Vergely et al, (2010), based on inverted extinctions.  
The figure shows the dust distribution in a vertical plane  containing the Sun and oriented along the \textit{l}=330$\degree$-150$\degree$ axis. 
The map reveals the so-called Local Bubble, which is devoid of dense, sodium-bearing gas, and the ISM concentrations, which define the limits of this \textit{chimney} across the disk. Interestingly, the shell arcs seem to start along the extrapolated inclined boundary of the {\it chimney}. This picture suggests that the northern shells are produced by the same events that formed the inclined {\it wall } of gas and dust that shows up in the maps towards the galactic center hemisphere, {favoring models of the second or third types, according to the \cite{santos11} classification. But it also disfavors a model according to which the very extended area shining in 3/4 keV X-rays corresponds to an extremely hot bubble expanding on both sides of the Galactic Plane. {We believe that} future, more detailed 3D maps should help to better understand the global structure and determine what type of exploding events have shaped the local distribution of gas and dust.

\begin{acknowledgements}
We deeply thank the members of the Pic-du-Midi NARVAL team for their efforts to ensure high-quality data. We also thank S\'{e}verine Raimond, who participated in the target selection.
\end{acknowledgements}


\Online
\begin{appendix}

\textbf{Online Appendix}: NaI D2/D1 and CaII K,H absorption data and models (continuation of Figures 1 and 2 from the article)
\begin{figure*}[!h]
\begin{minipage}[t]{0.3\linewidth}
\centering
  	\includegraphics[width=1\linewidth]{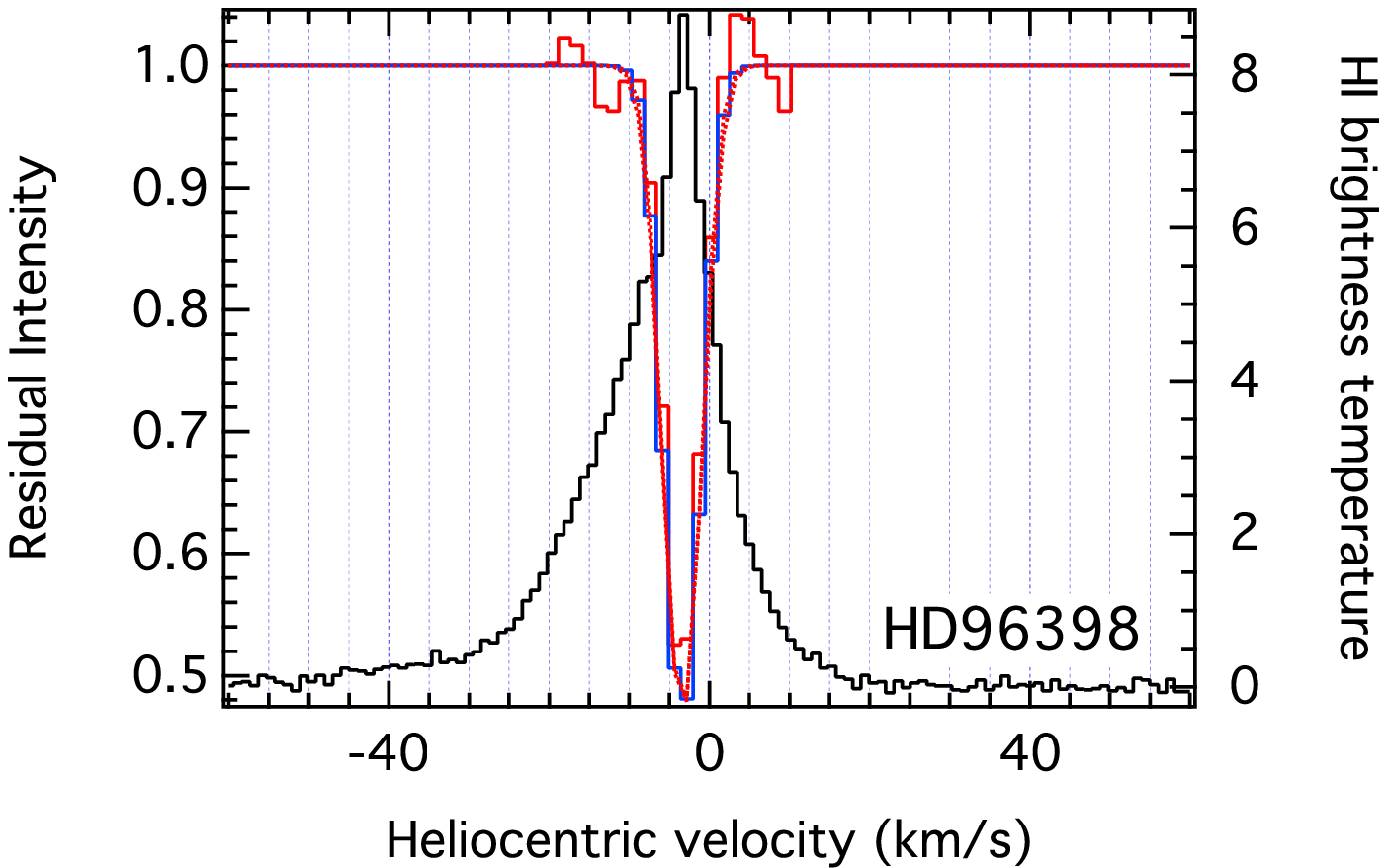}
  	\includegraphics[width=1\linewidth]{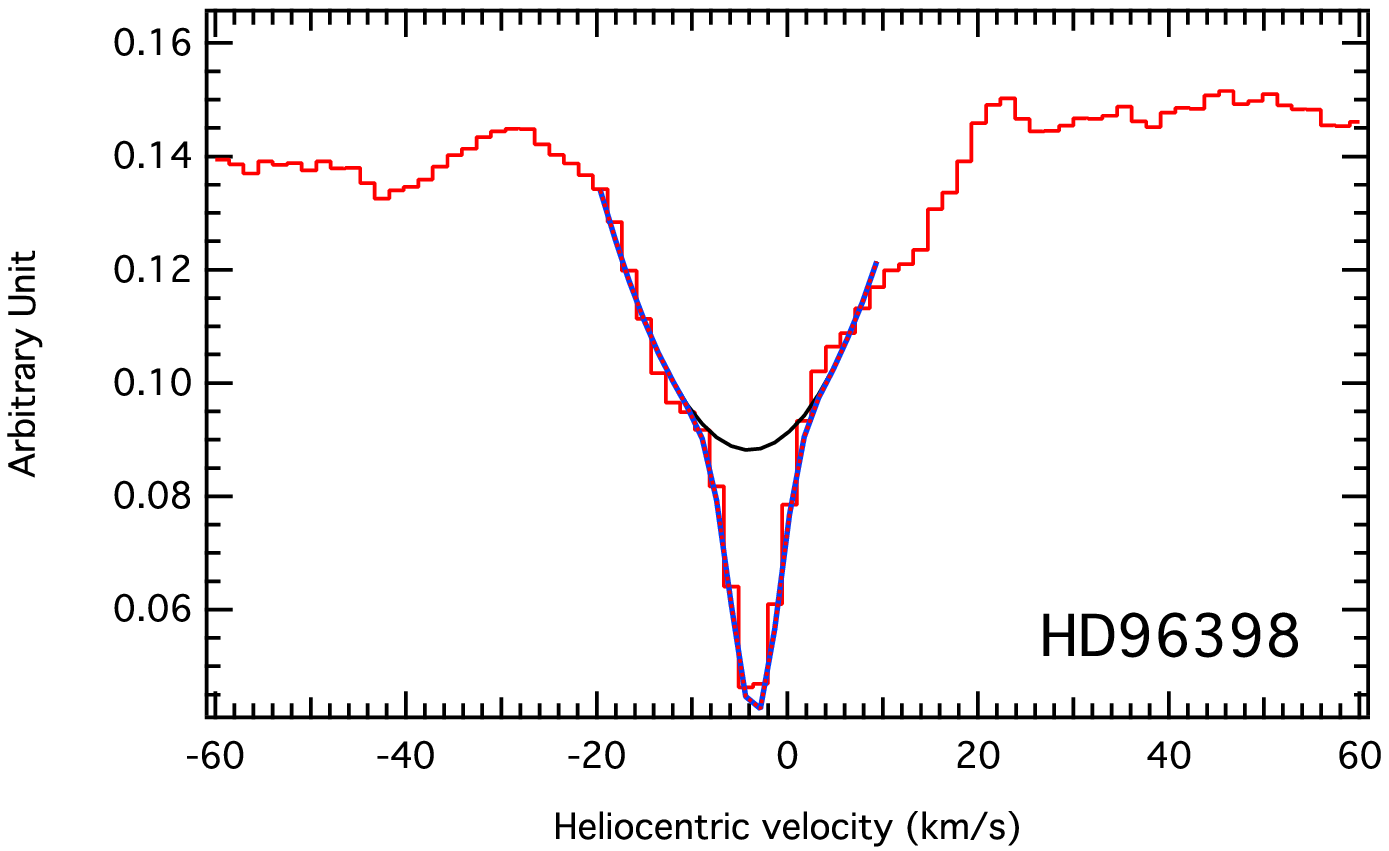}
  	\includegraphics[width=1\linewidth]{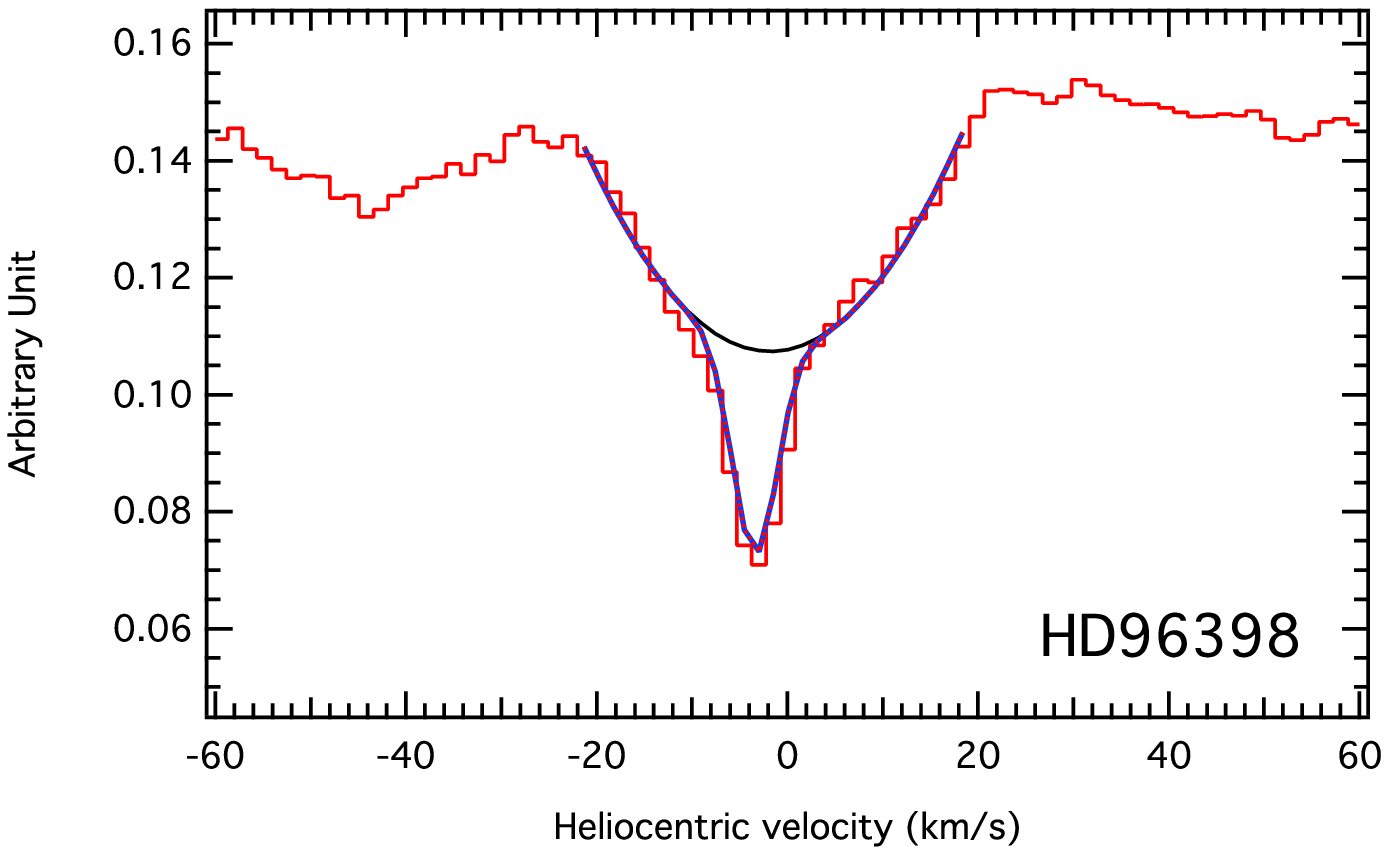}
\end{minipage}\hfill
\begin{minipage}[t]{0.3\linewidth}
\centering
  	\includegraphics[width=1\linewidth]{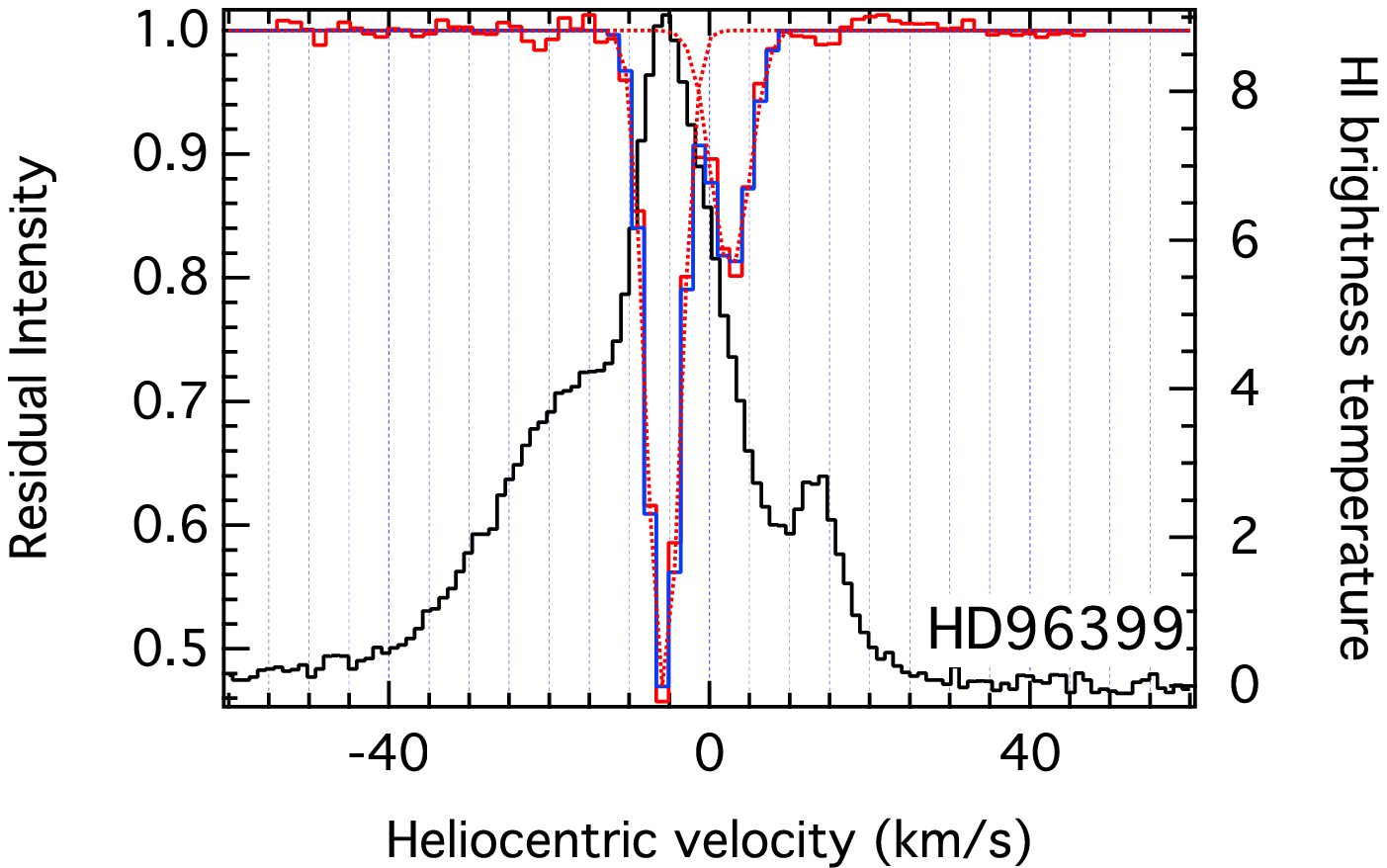}
  	\includegraphics[width=1\linewidth]{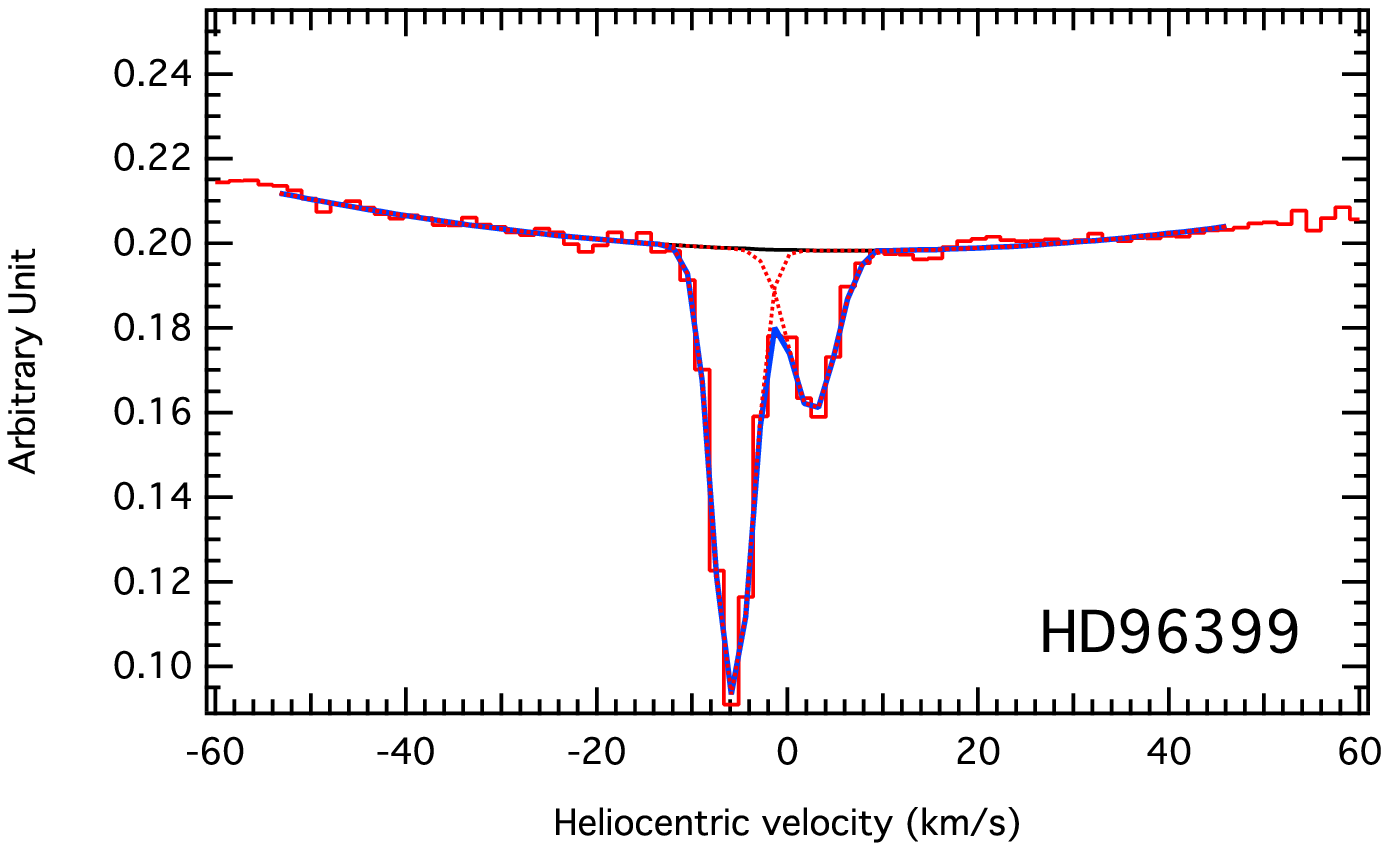}
  	\includegraphics[width=1\linewidth]{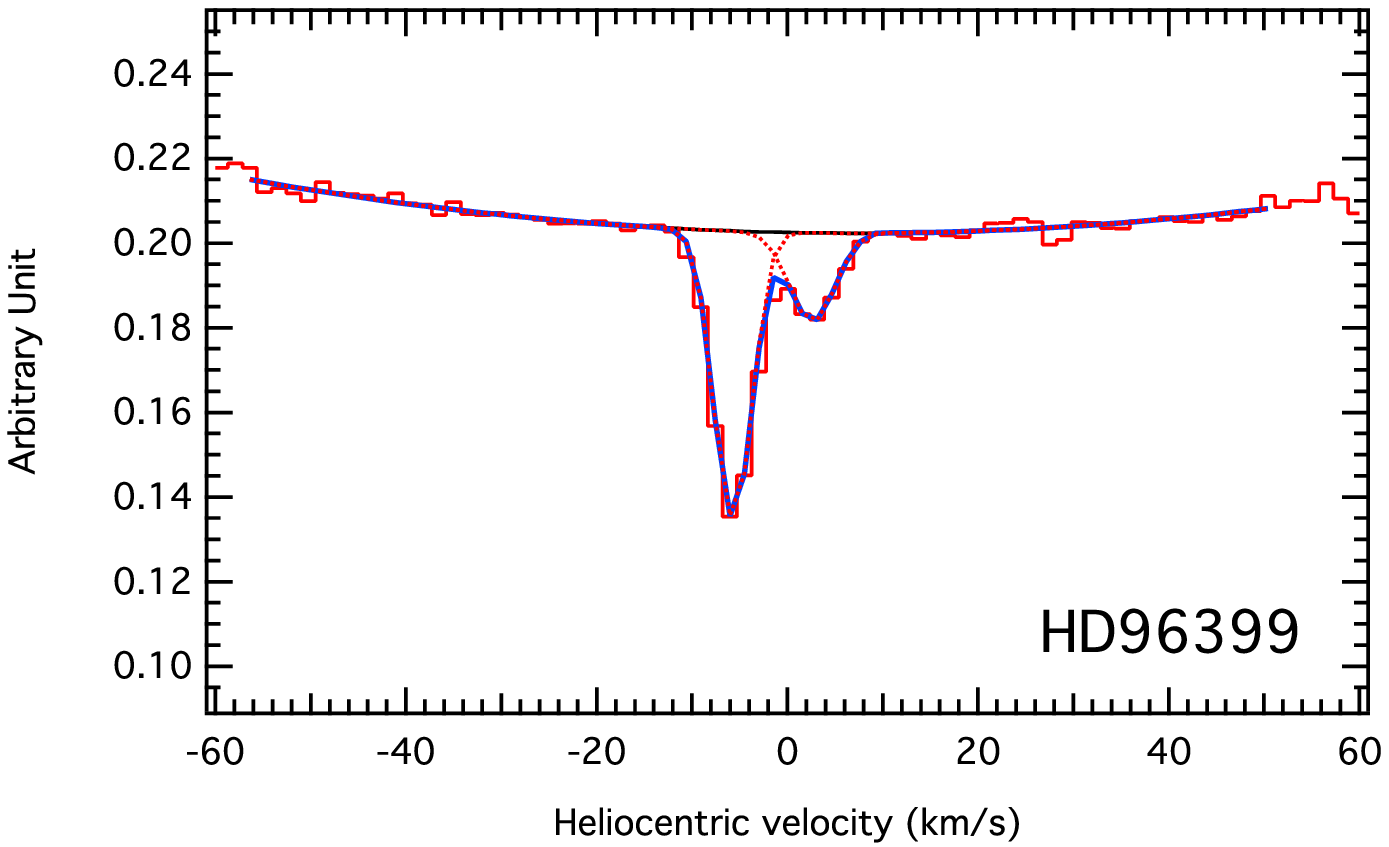}
\end{minipage}\hfill
\begin{minipage}[t]{0.3\linewidth}
\centering
  	\includegraphics[width=1\linewidth]{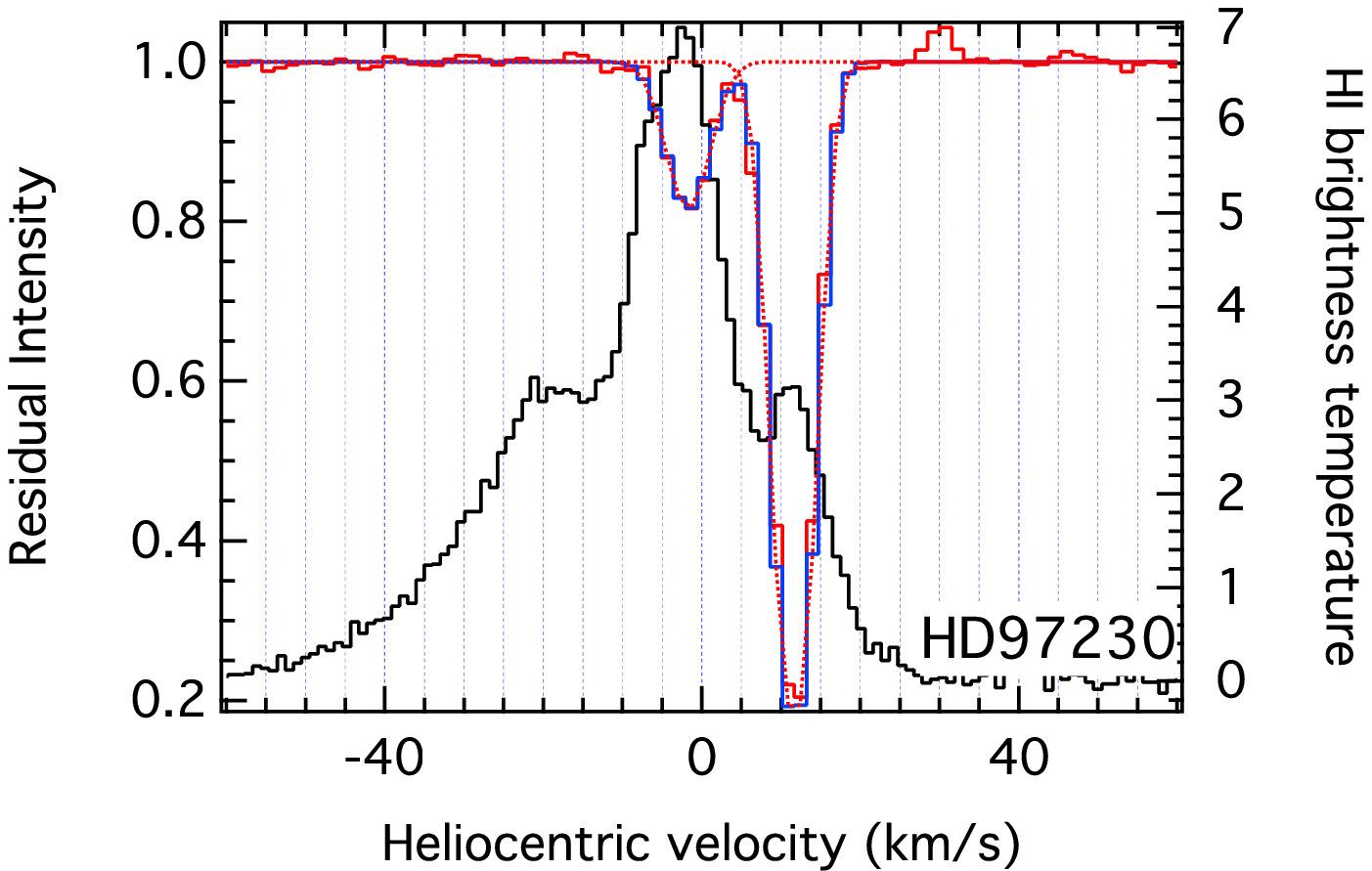}
  	\includegraphics[width=1\linewidth]{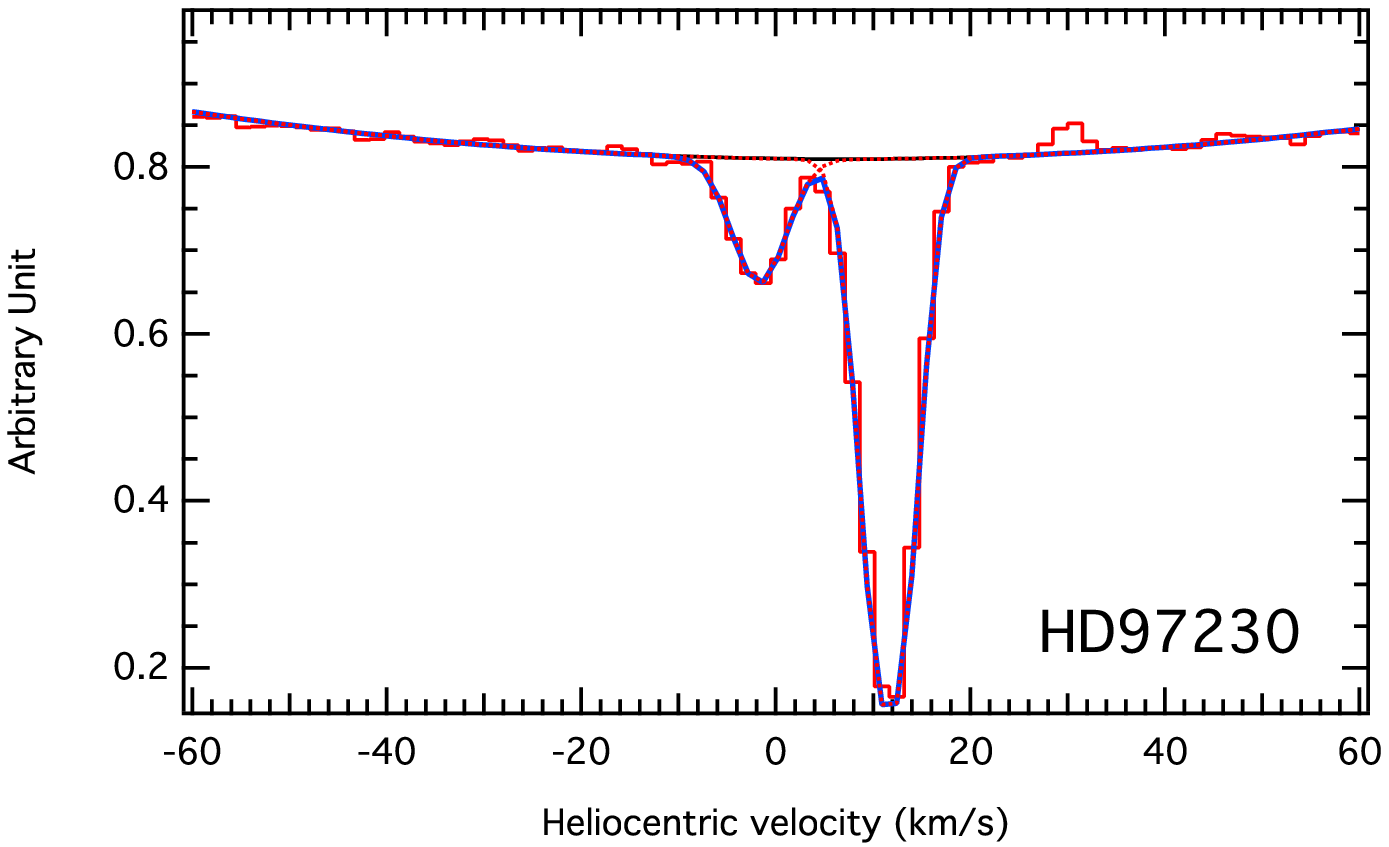}
  	\includegraphics[width=1\linewidth]{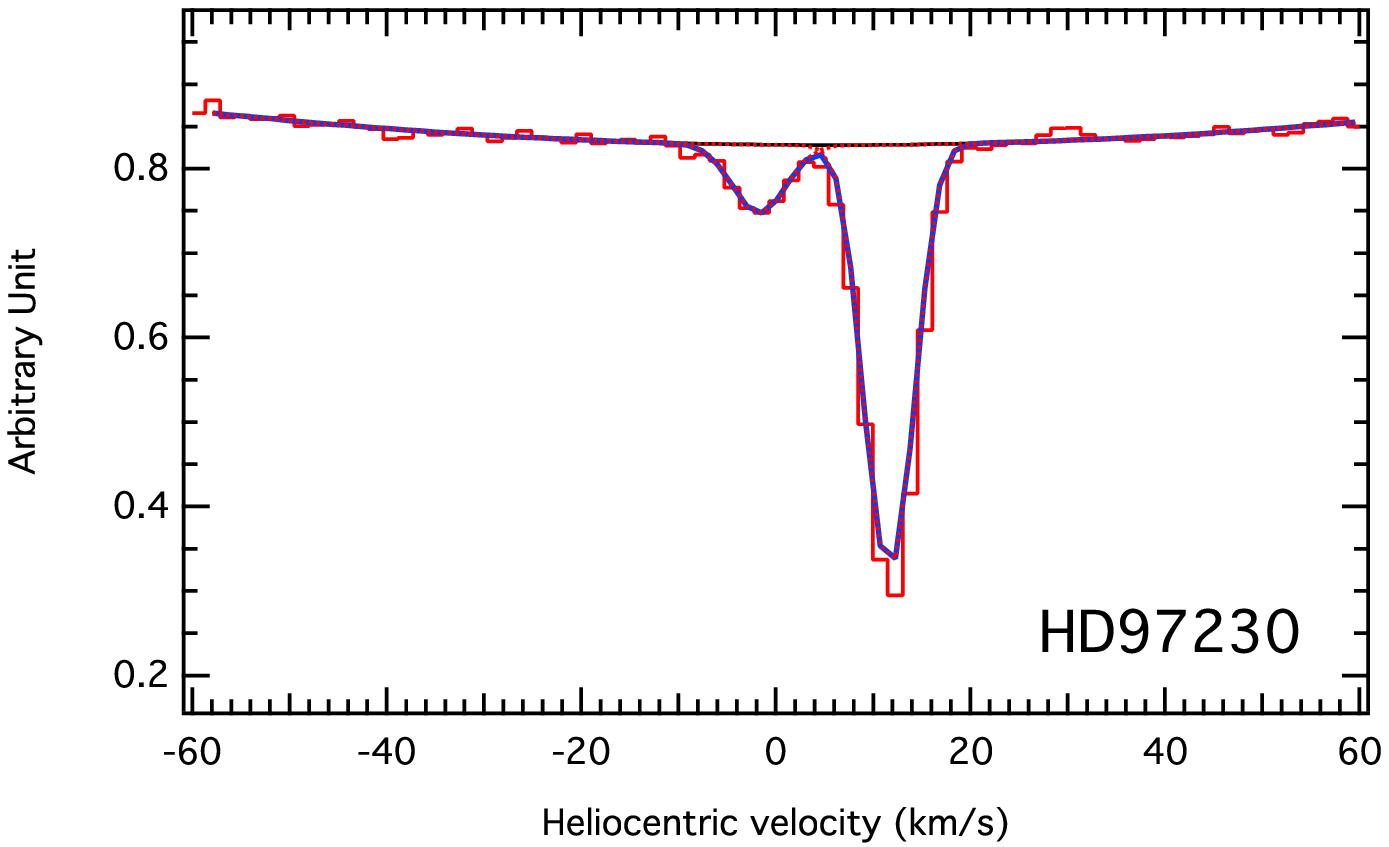}
\end{minipage}
\caption{Interstellar 5889  \AA\  NaI-D2 and 5895 \AA\  NaI-D1  absorption lines for the stars HD96398, HD96399, and HD97230. The first row shows the normalized interstellar NaI-D2 line profile (red histogram line) superimposed on the HI emission spectrum (black histogram line). The second and the third rows show the unnormalized interstellar NaI D2 and D1, respectively line profiles (red histogram line) superimposed on the fit model (blue line) and the individual model components (dotted red line). Data and models for the first three targets HD94194, HD94266, and HD94766 are shown in the article.}
\label{HD96398} 
\end{figure*}

\begin{figure*}[!h]
\begin{minipage}[t]{0.3\linewidth}
\centering
  	\includegraphics[width=1\linewidth]{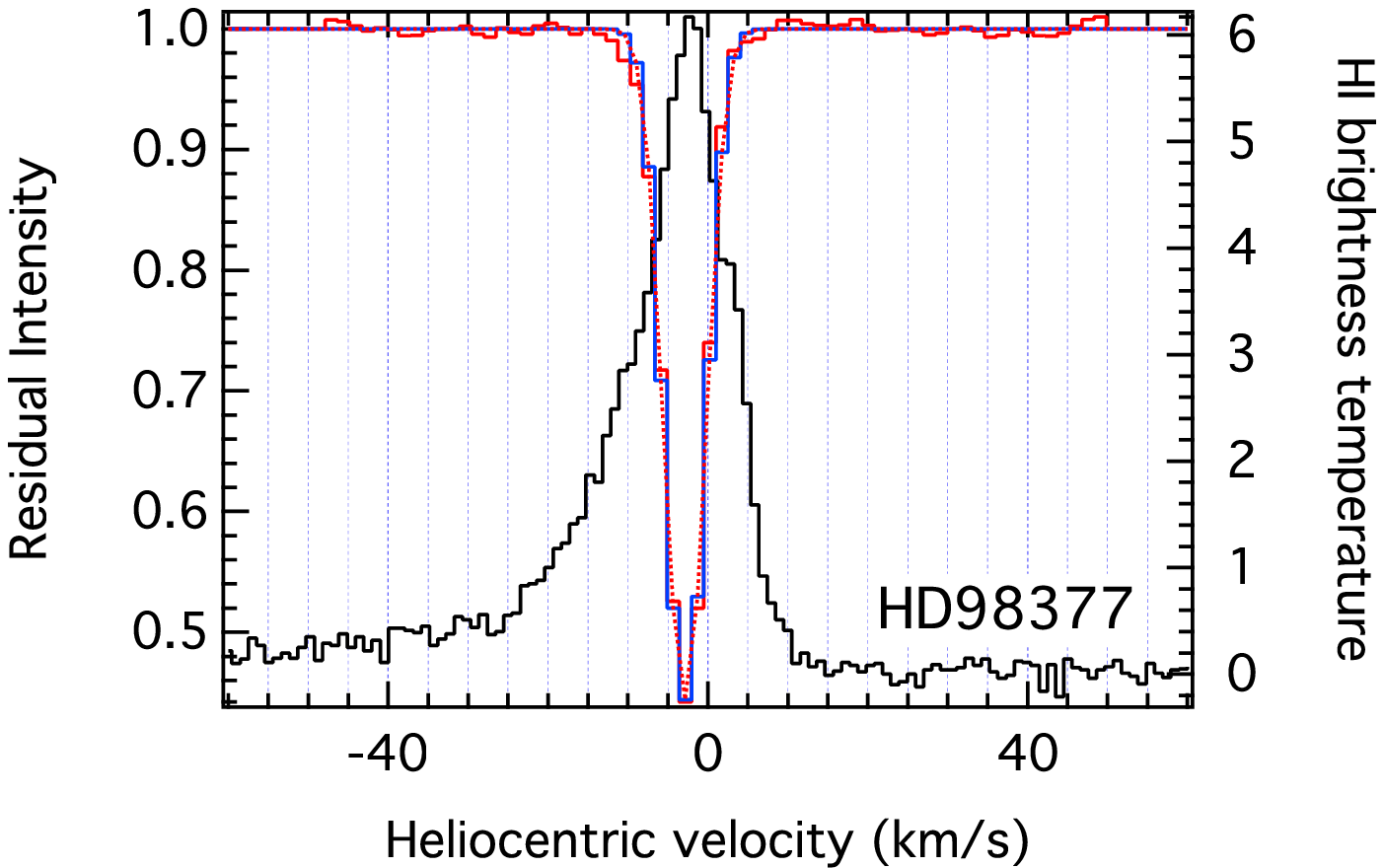}
  	\includegraphics[width=1\linewidth]{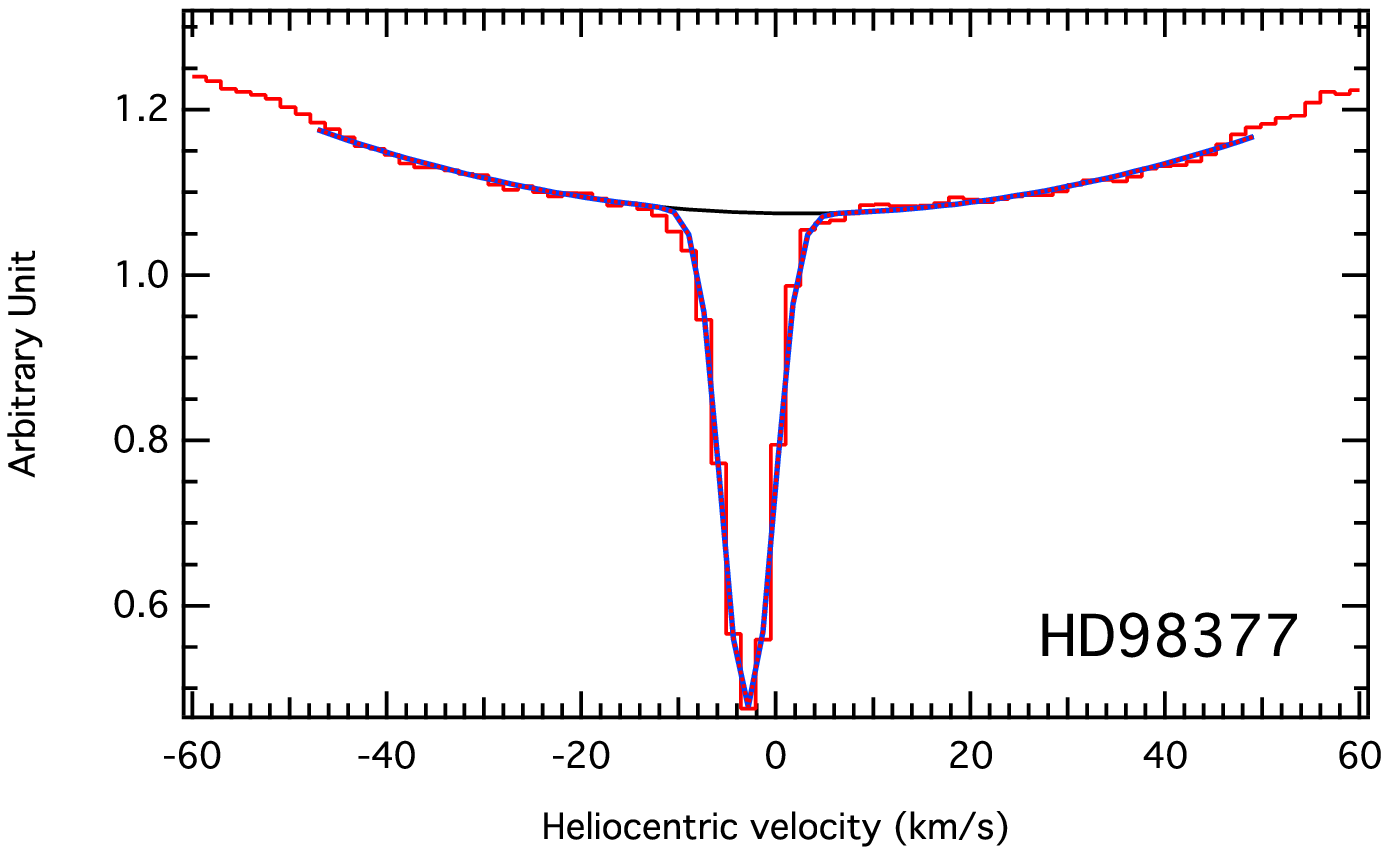}
  	\includegraphics[width=1\linewidth]{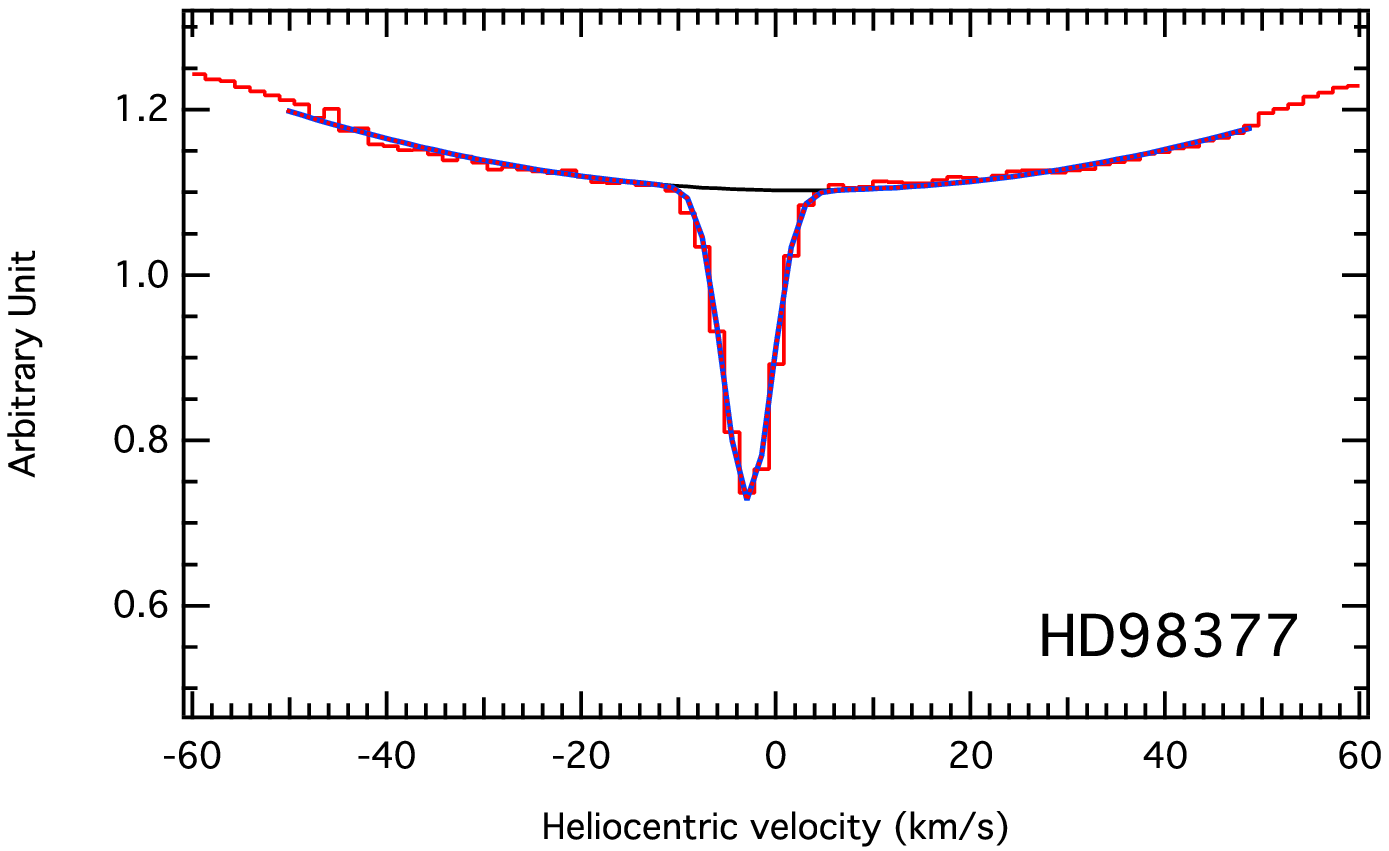}
\end{minipage}\hfill
\begin{minipage}[t]{0.3\linewidth}
\centering
  	\includegraphics[width=1\linewidth]{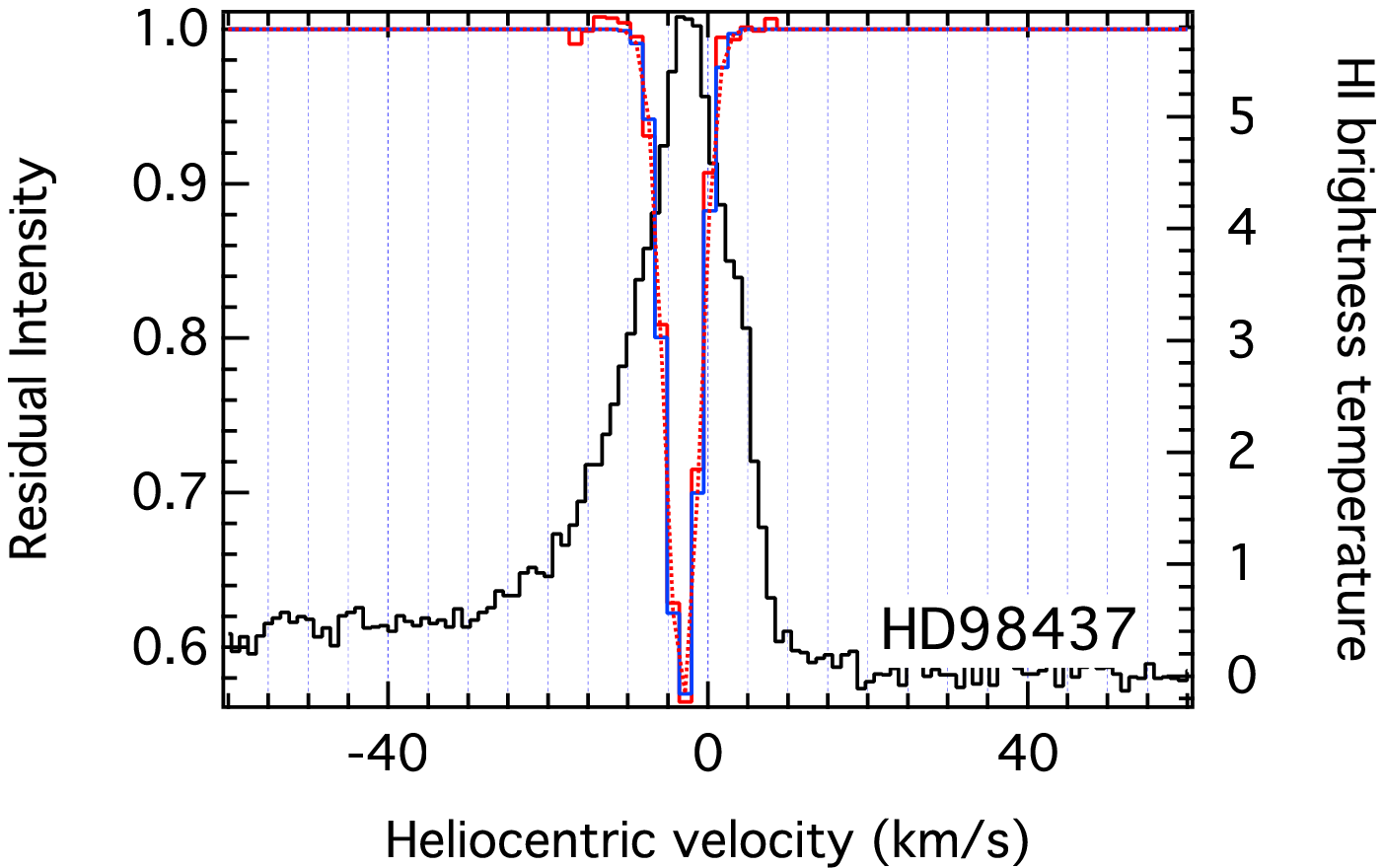}
  	\includegraphics[width=1\linewidth]{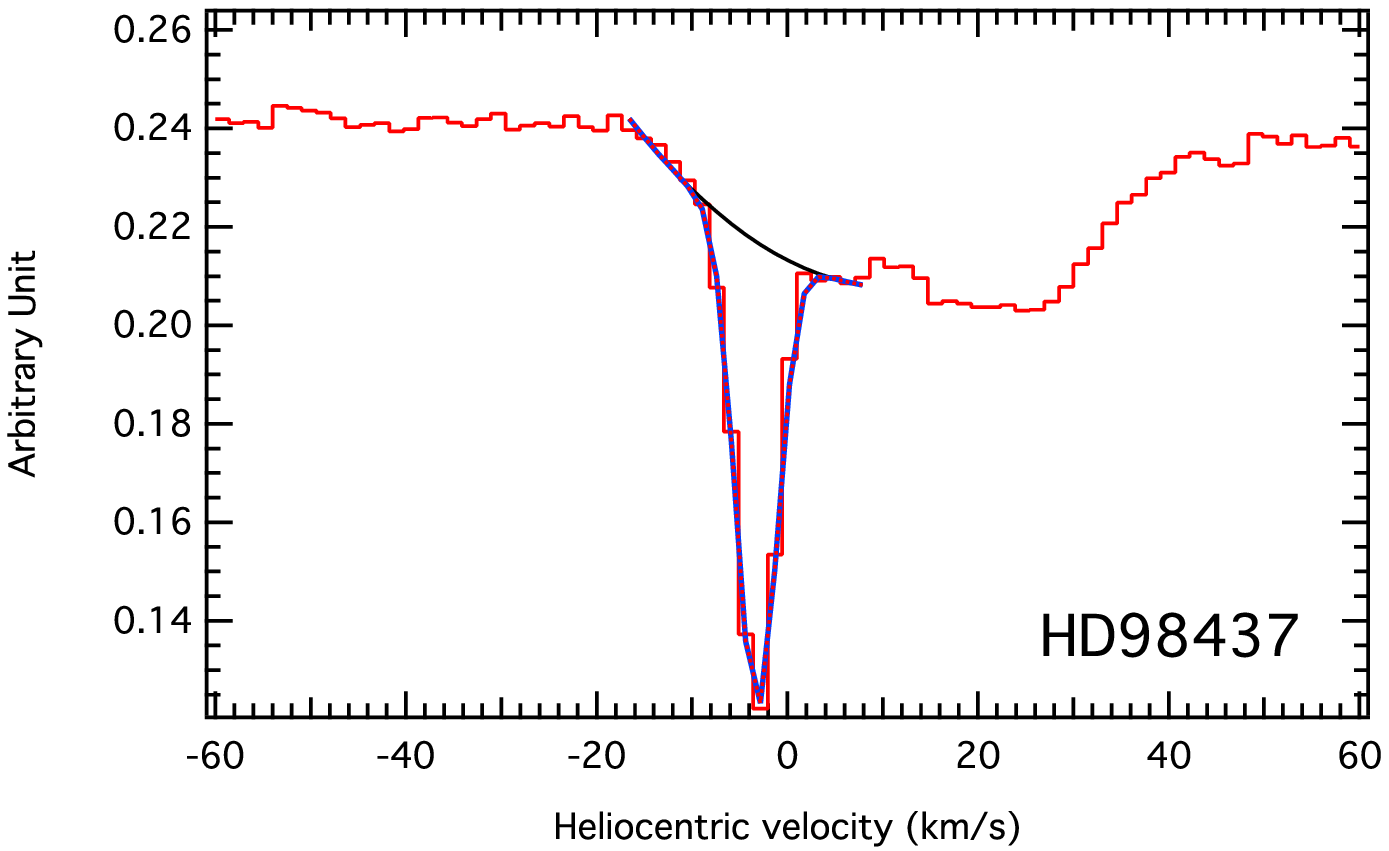}
  	\includegraphics[width=1\linewidth]{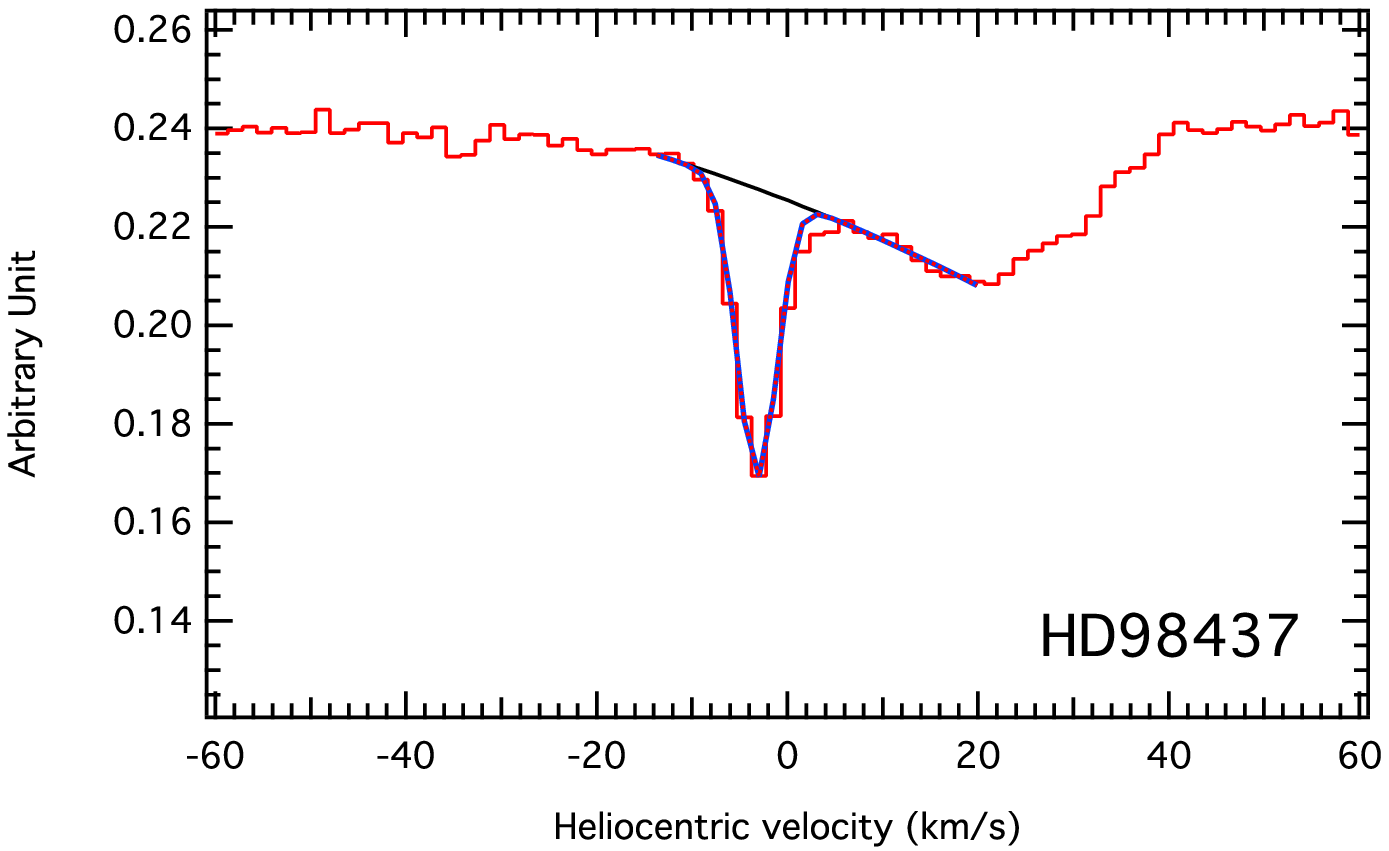}
\end{minipage}\hfill
\begin{minipage}[t]{0.3\linewidth}
\centering
  	\includegraphics[width=1\linewidth]{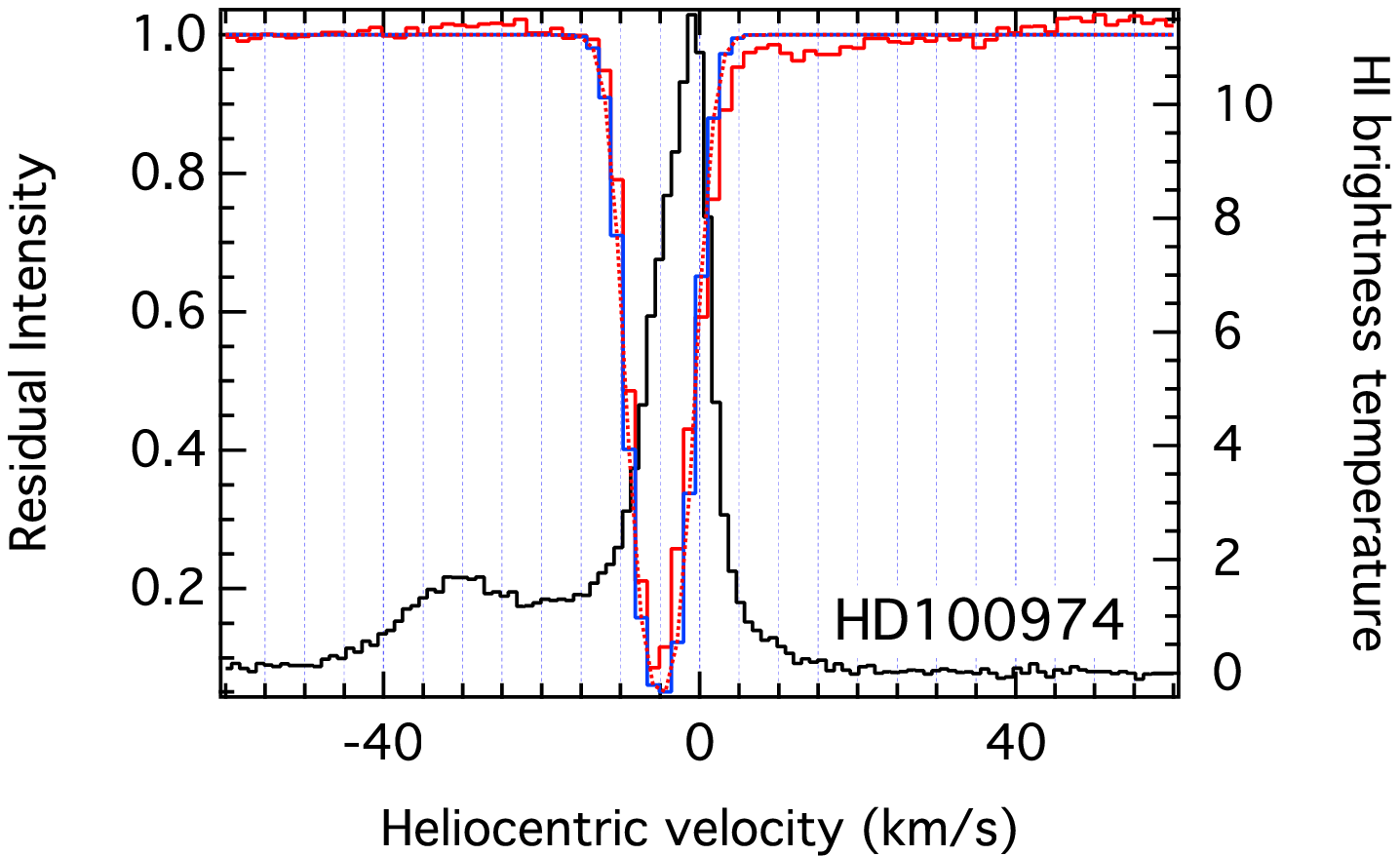}
  	\includegraphics[width=1\linewidth]{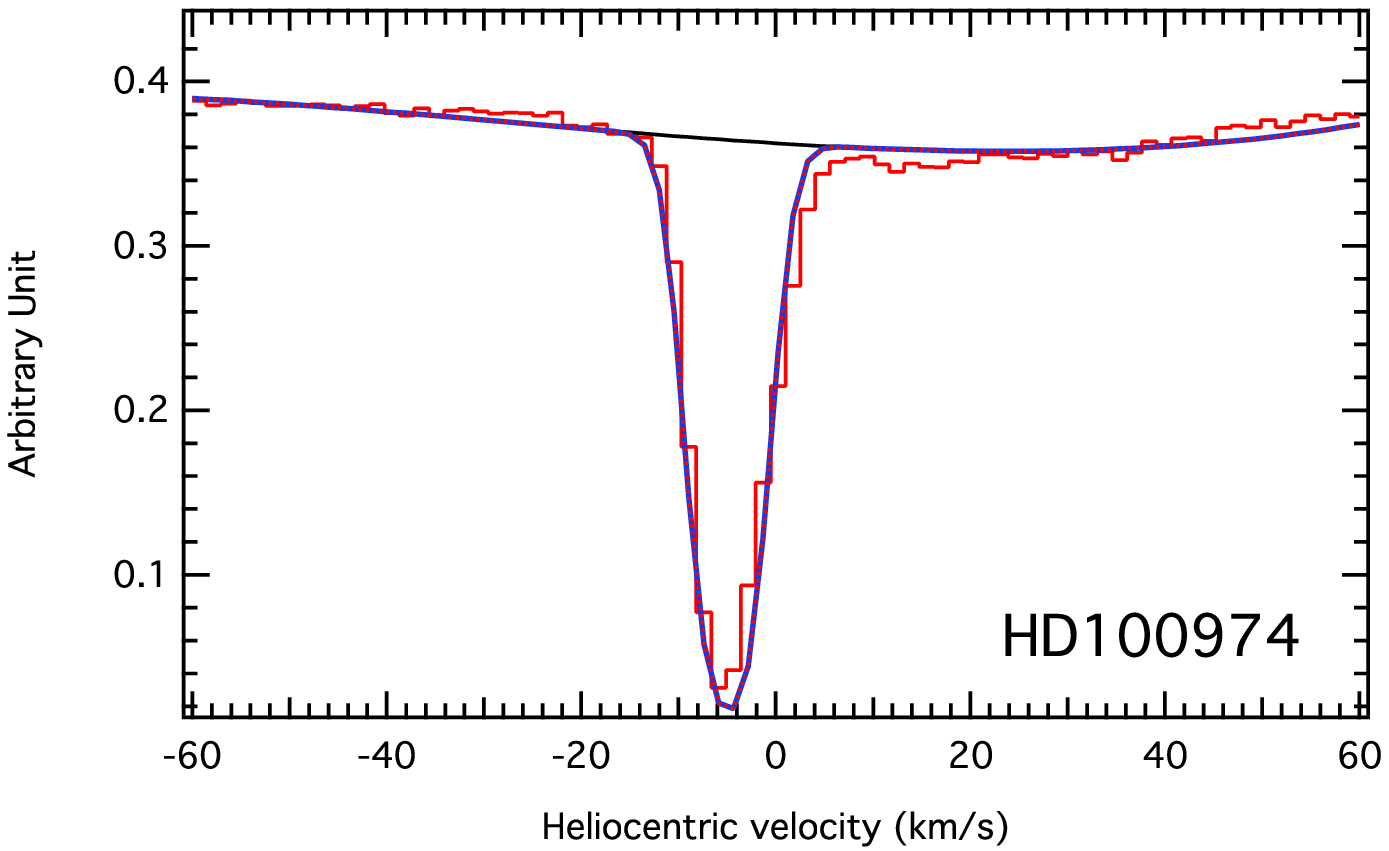}
  	\includegraphics[width=1\linewidth]{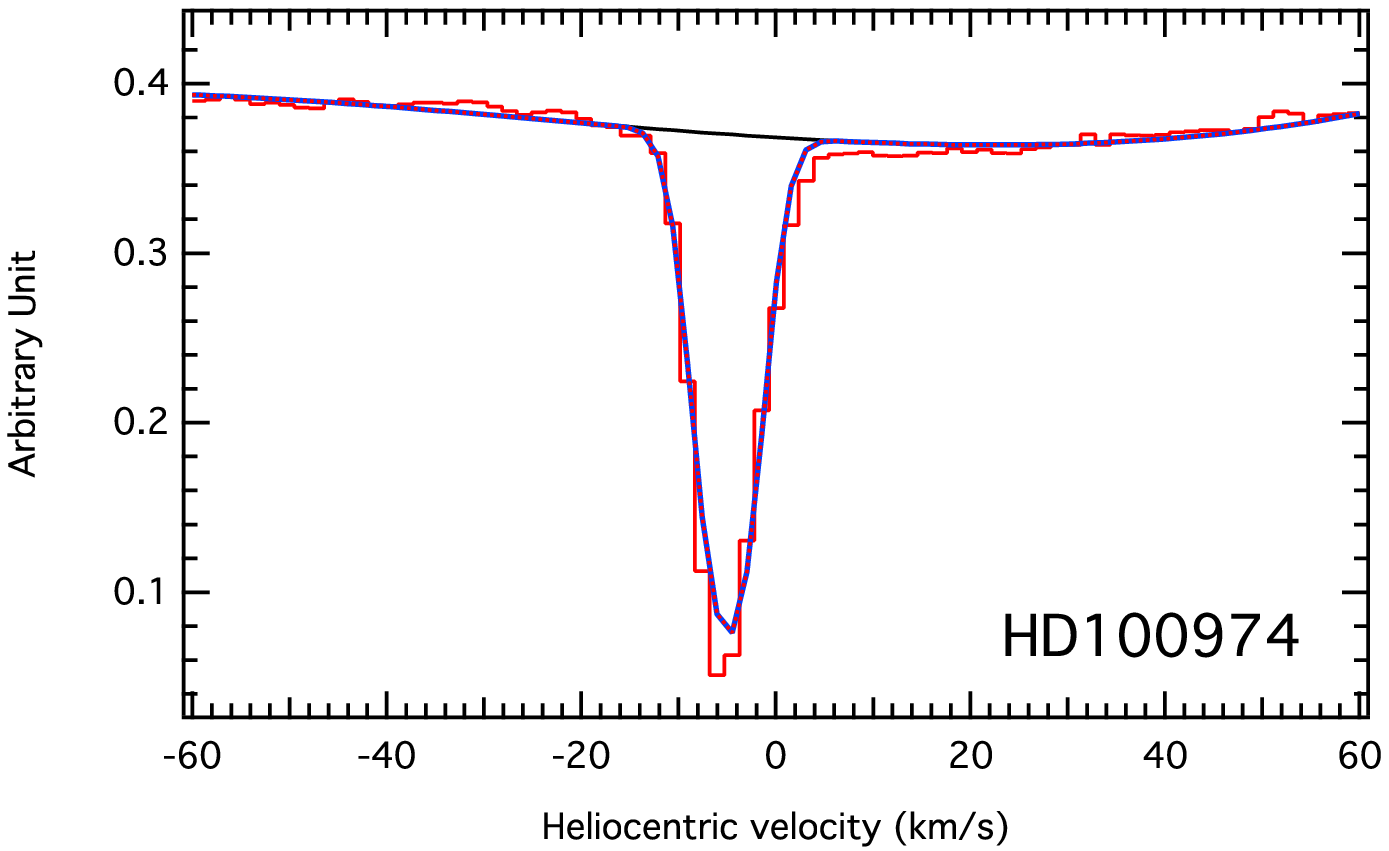}
\end{minipage}
\caption{Same as Fig. \ref{HD94194} {(in the article)}  but for HD98377, HD98437, and HD100974}
\label{HD98377}
\end{figure*}

\begin{figure*}[!h]
\begin{minipage}[t]{0.3\linewidth}
\centering
  	\includegraphics[width=1\linewidth]{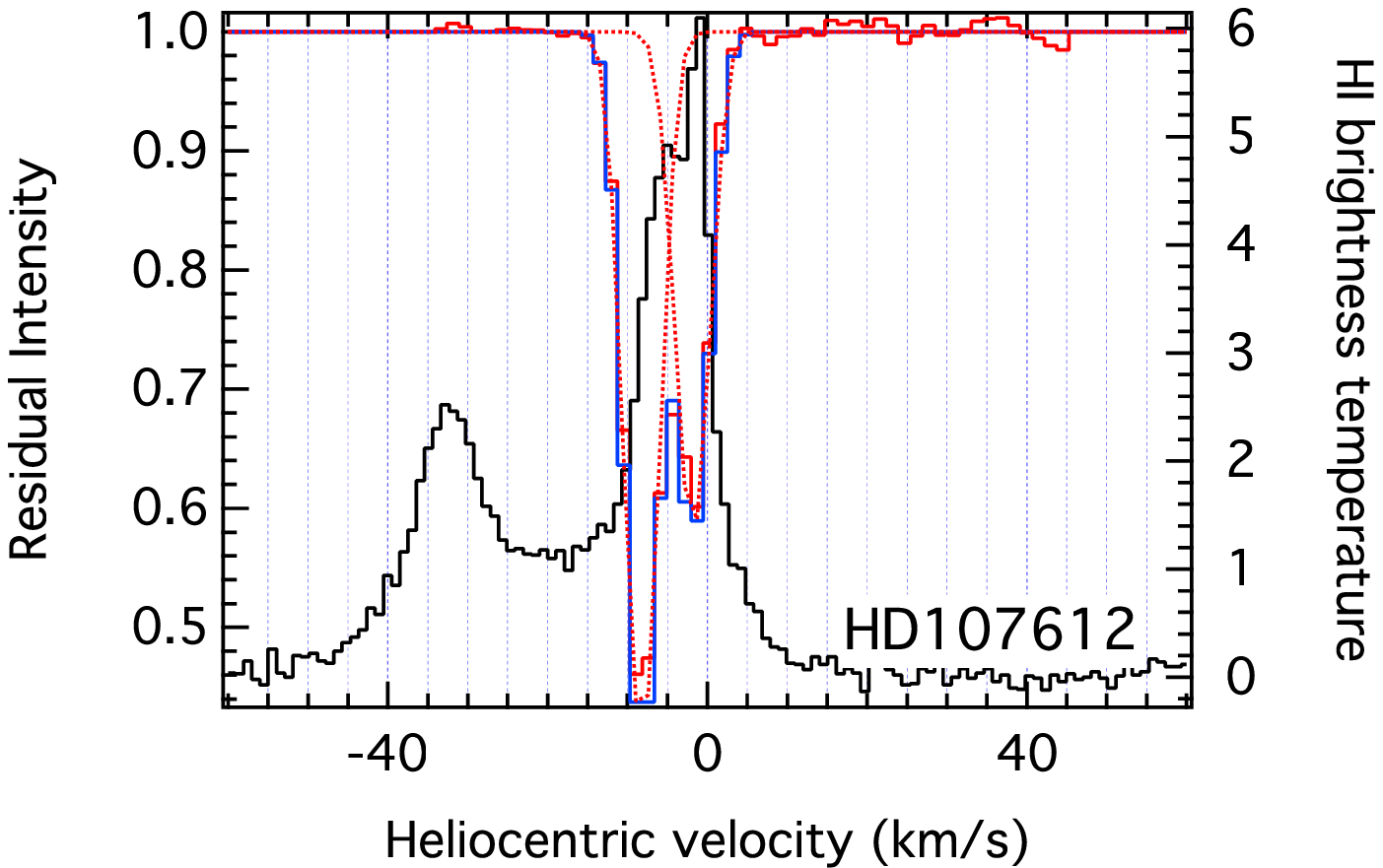}
  	\includegraphics[width=1\linewidth]{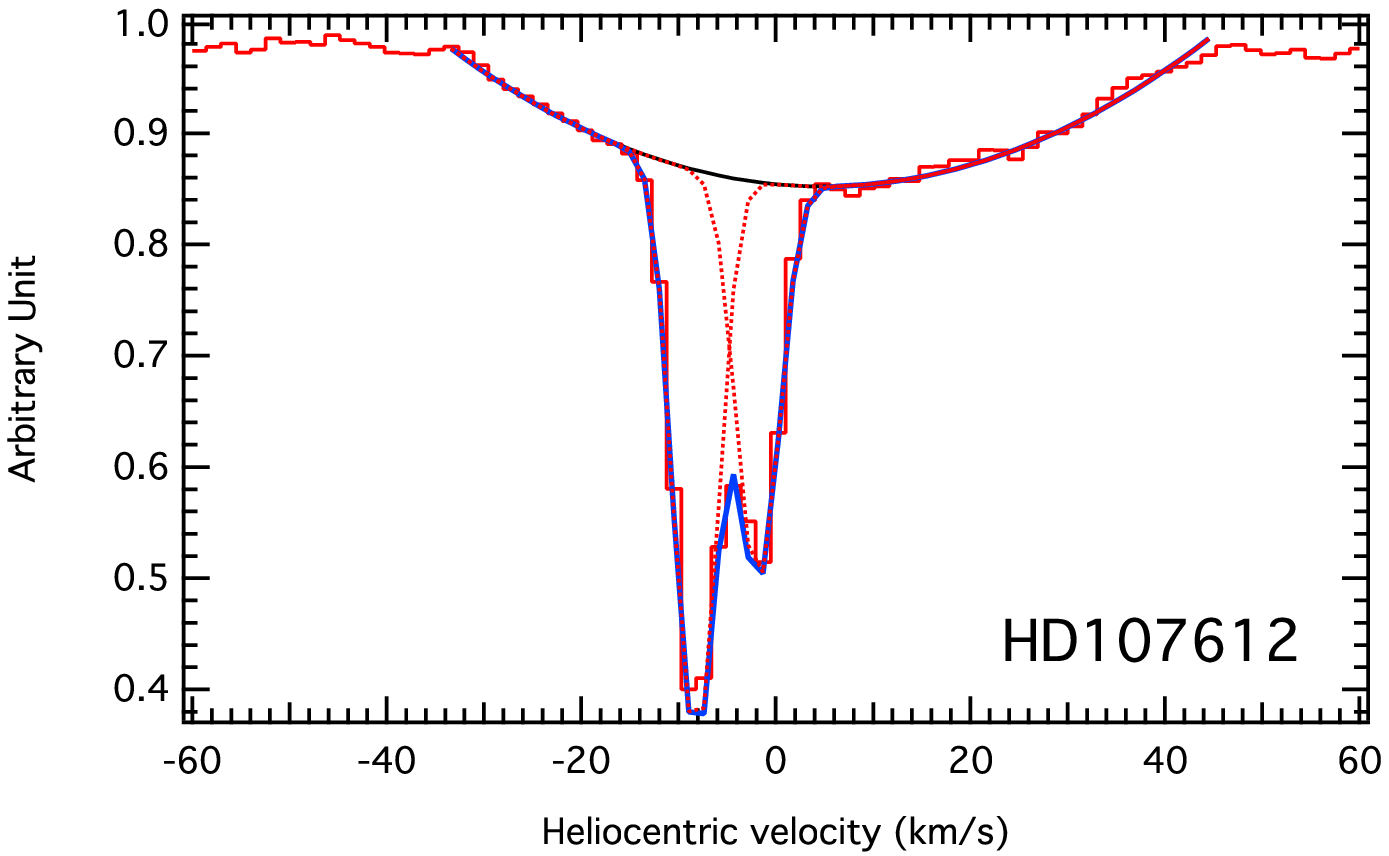}
  	\includegraphics[width=1\linewidth]{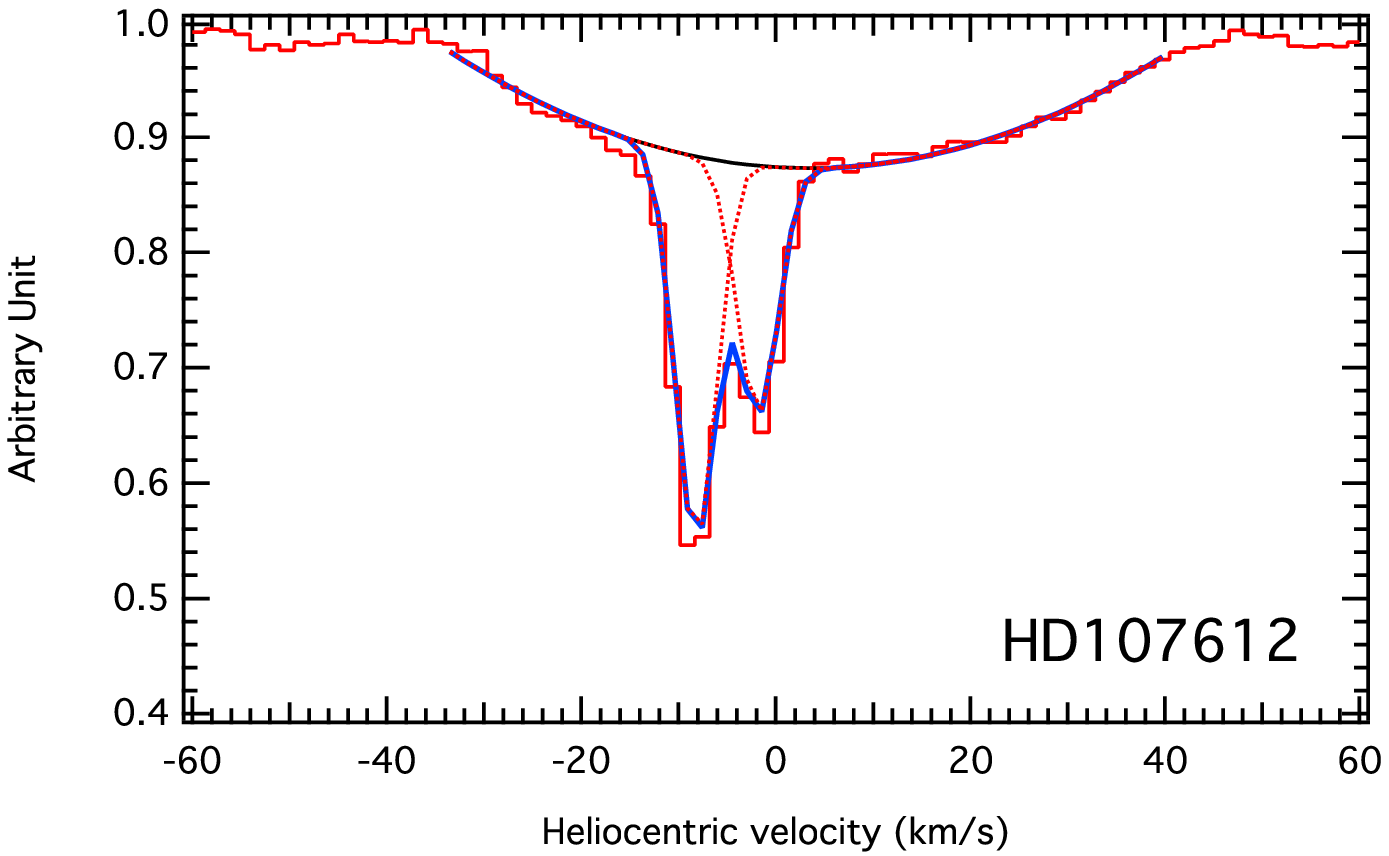}
\end{minipage}\hfill
\begin{minipage}[t]{0.3\linewidth}
\centering
  	\includegraphics[width=1\linewidth]{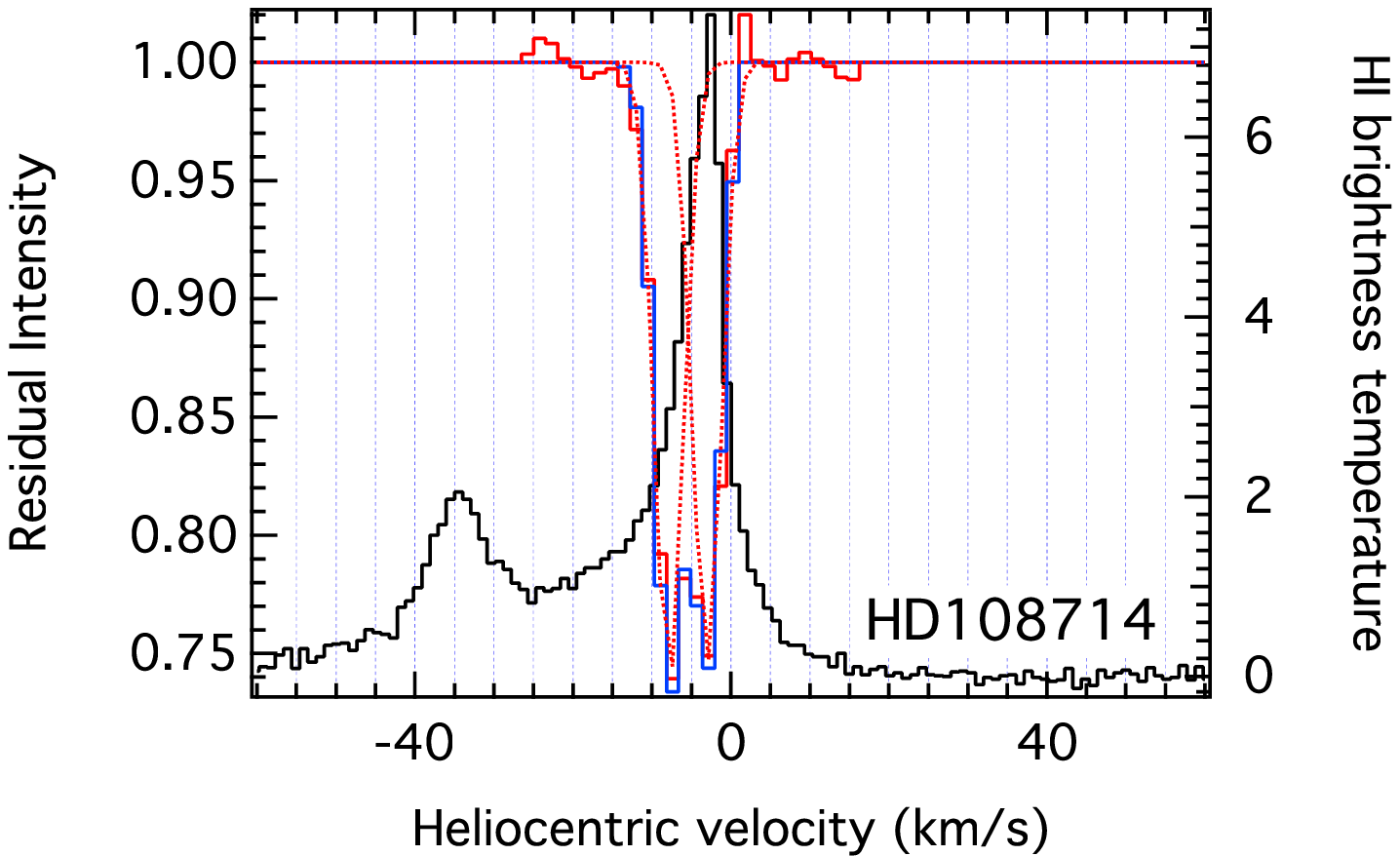}
  	\includegraphics[width=1\linewidth]{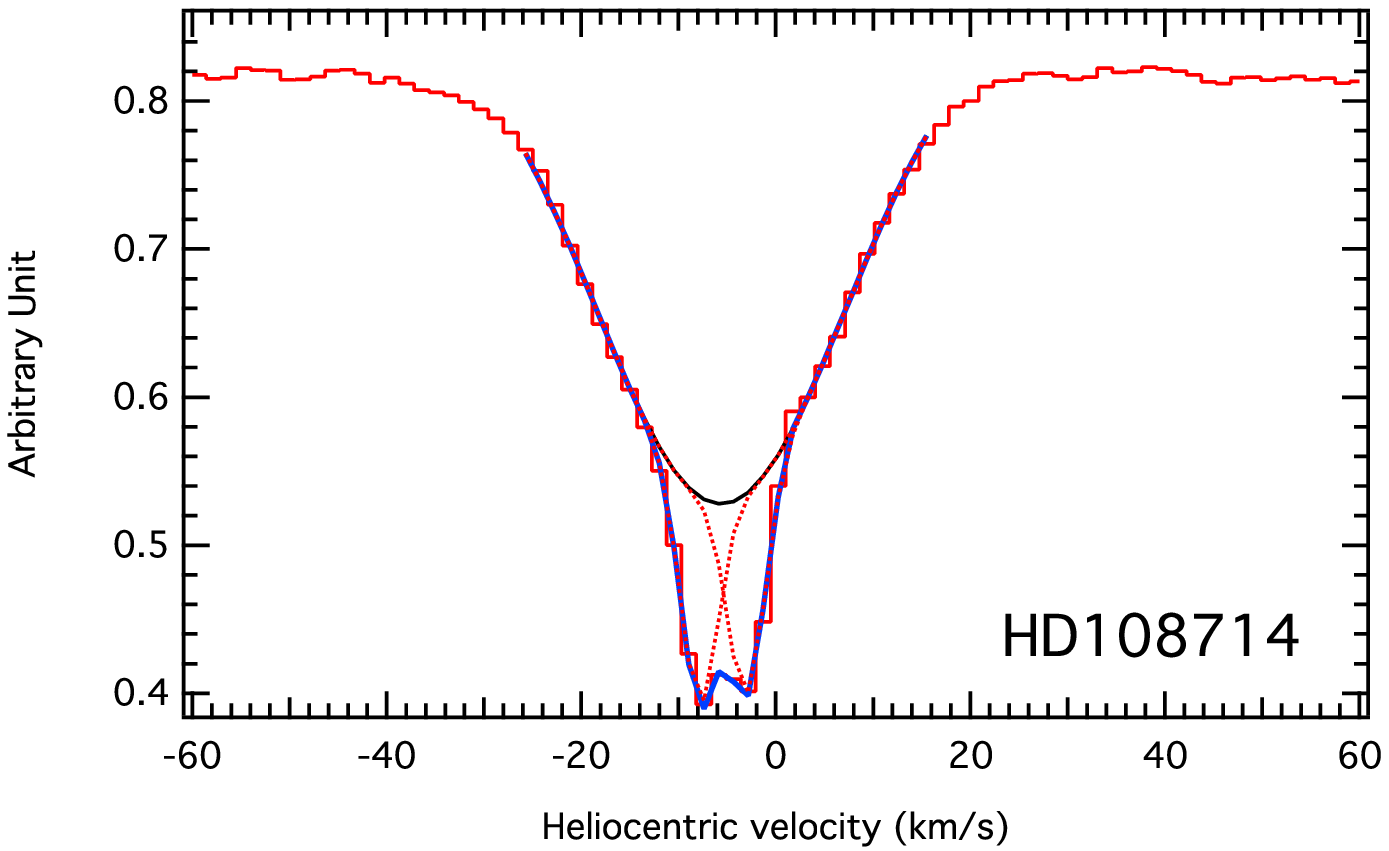}
  	\includegraphics[width=1\linewidth]{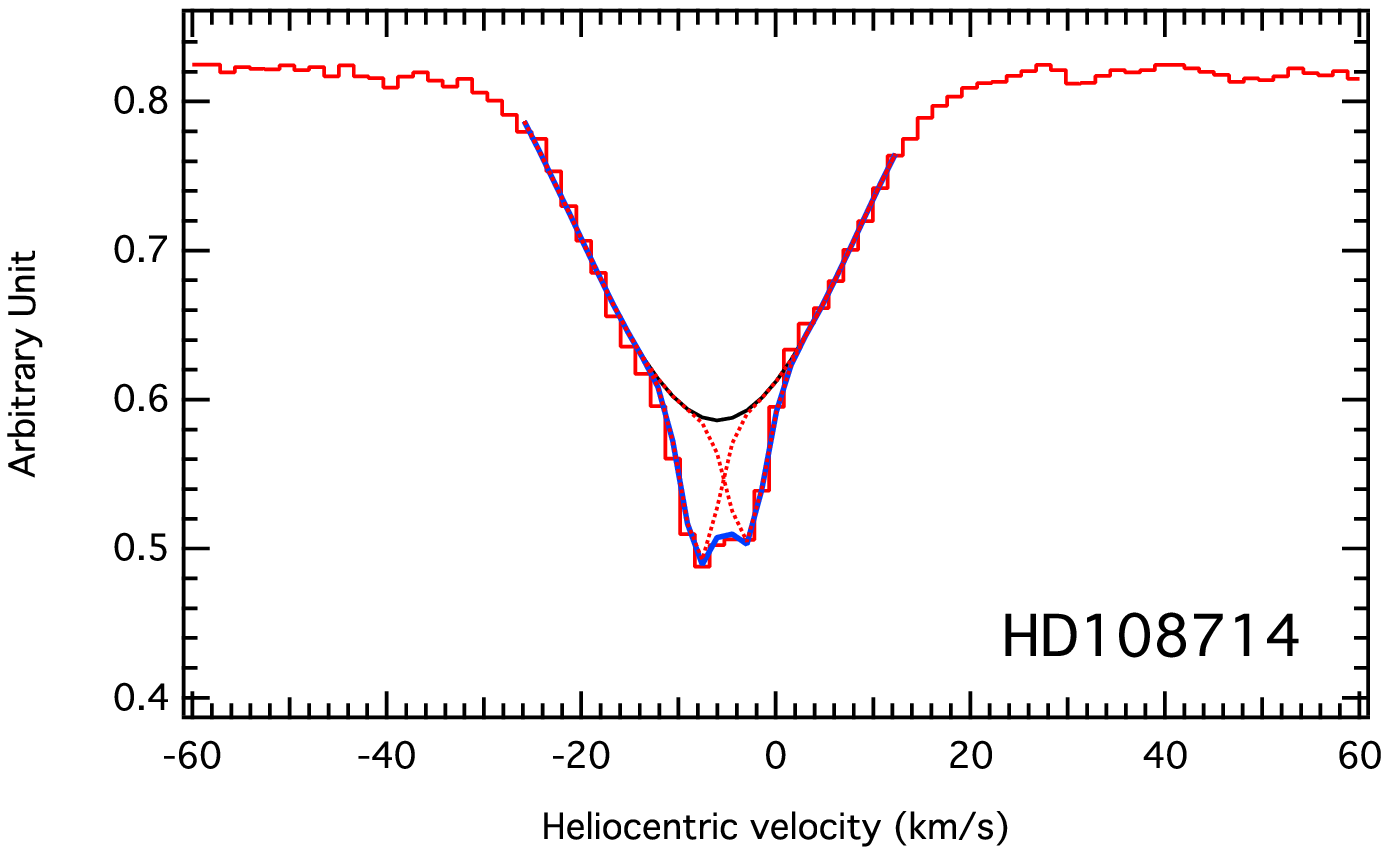}
\end{minipage}\hfill
\begin{minipage}[t]{0.3\linewidth}
\centering
  	\includegraphics[width=1\linewidth]{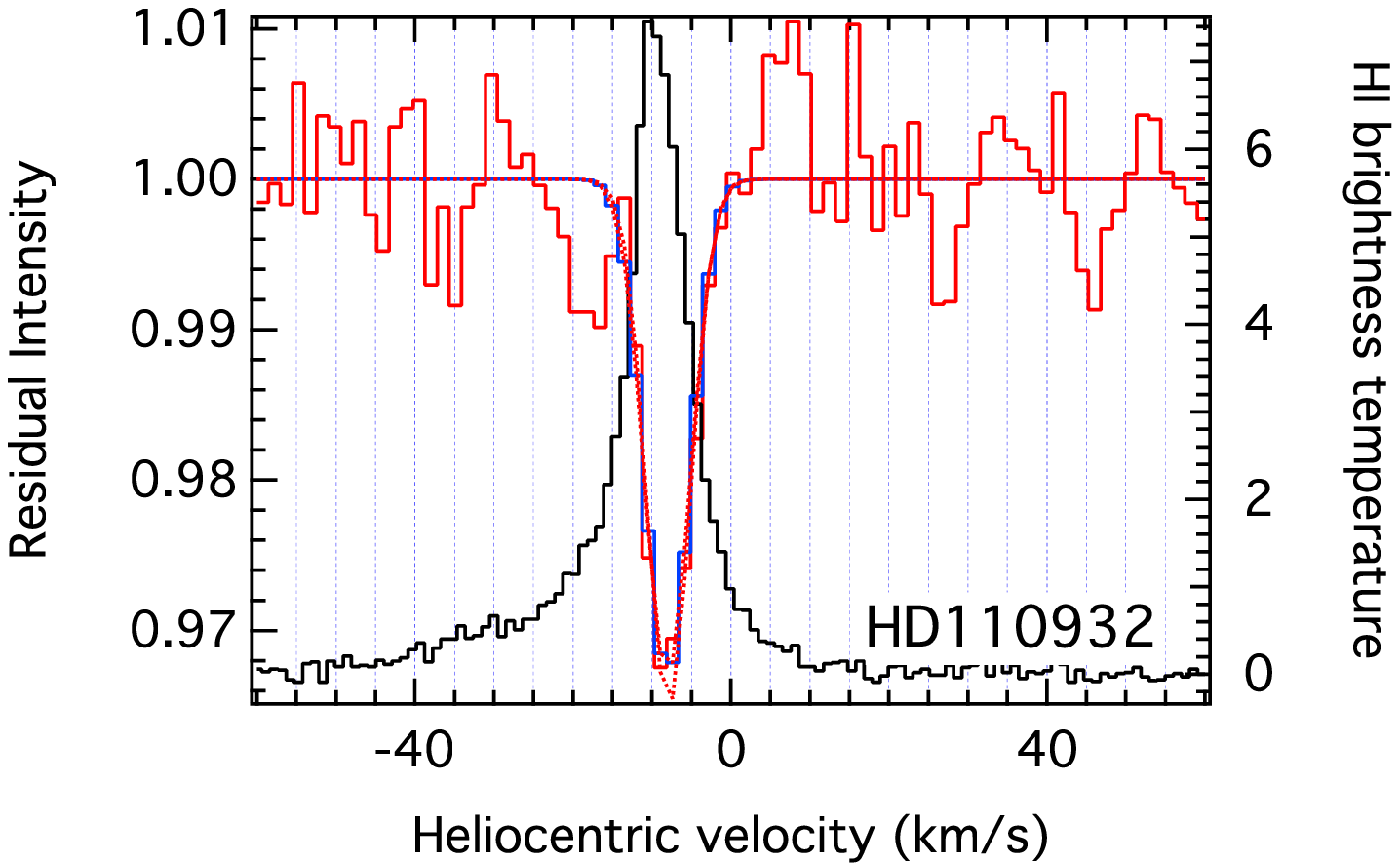}
  	\includegraphics[width=1\linewidth]{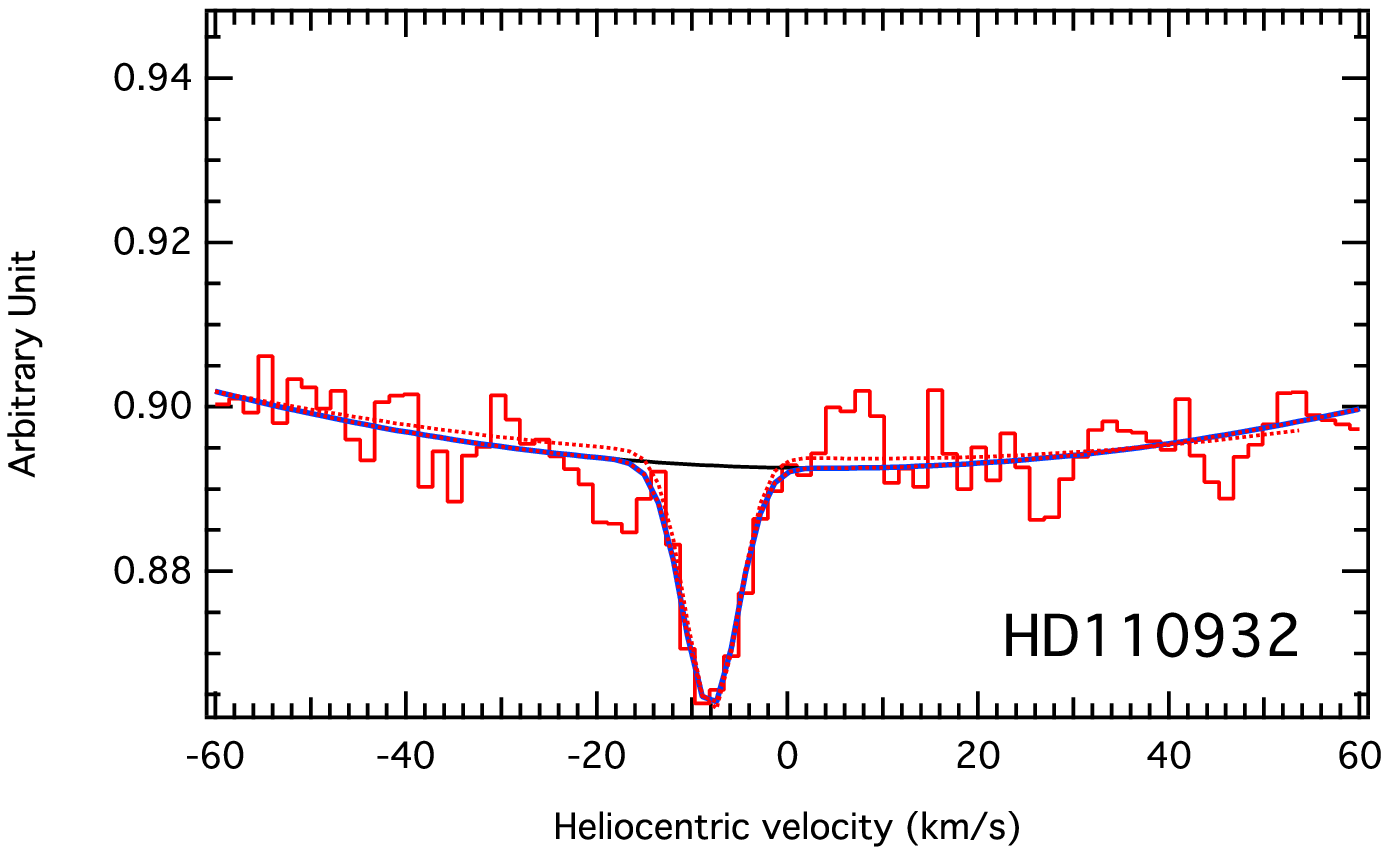}
  	\includegraphics[width=1\linewidth]{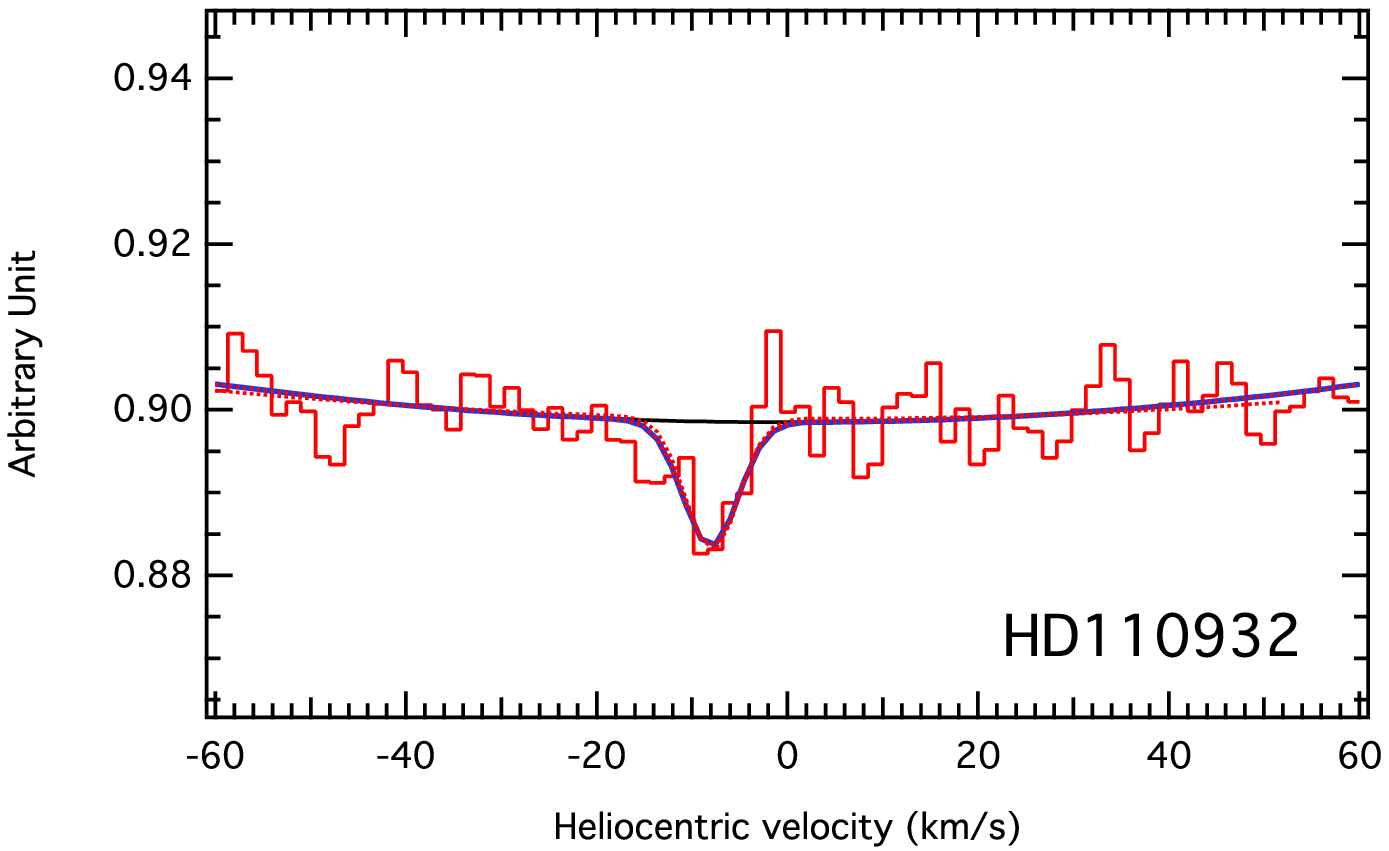}
\end{minipage}
\label{HD107612}
\caption{Same as Fig. \ref{HD94194} {(in the article)}  but for HD107612, HD108714, and HD110932}
\end{figure*}

\begin{figure*}[!h]
\begin{minipage}[t]{0.3\linewidth}
\centering
  	\includegraphics[width=1\linewidth]{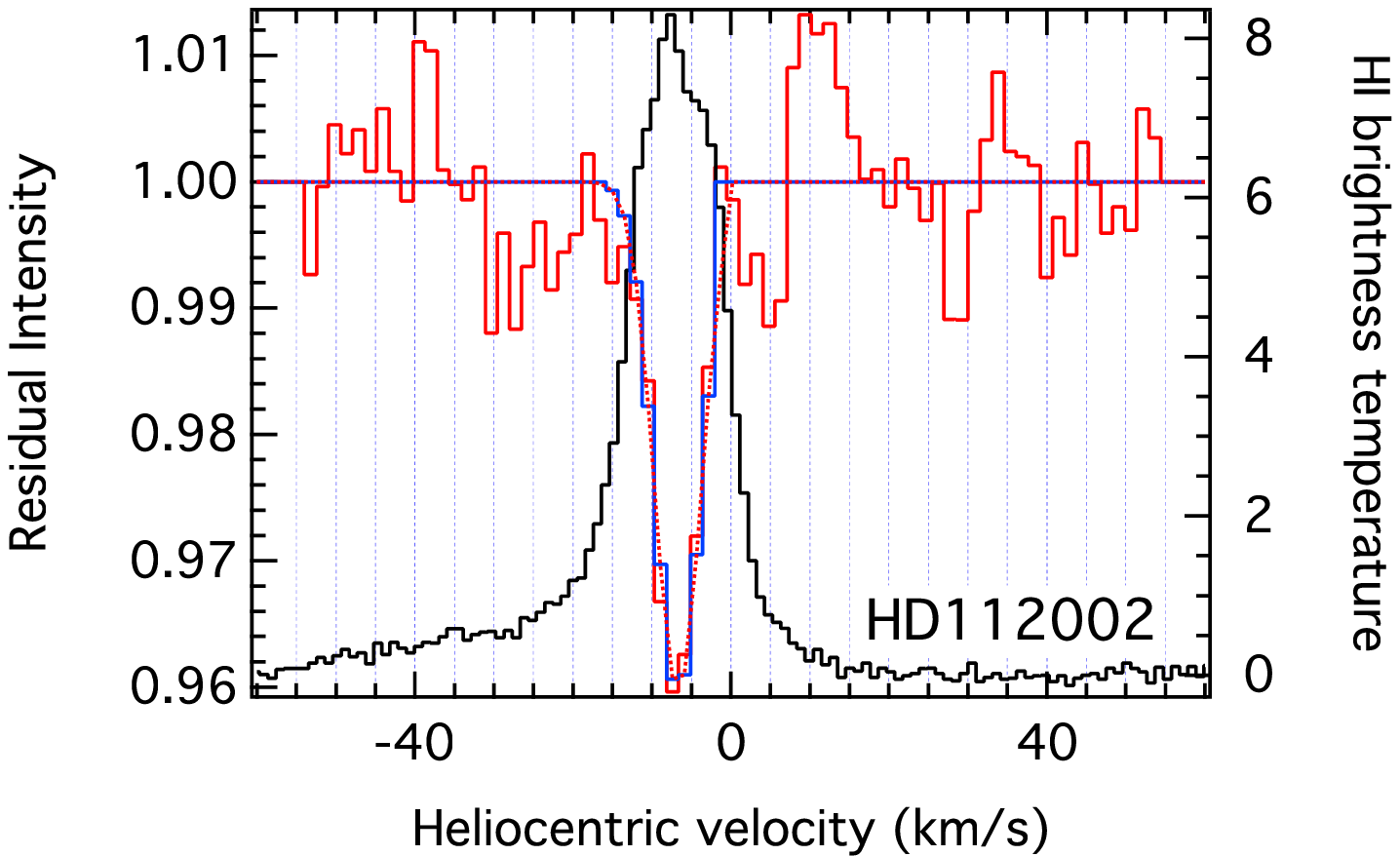}
  	\includegraphics[width=1\linewidth]{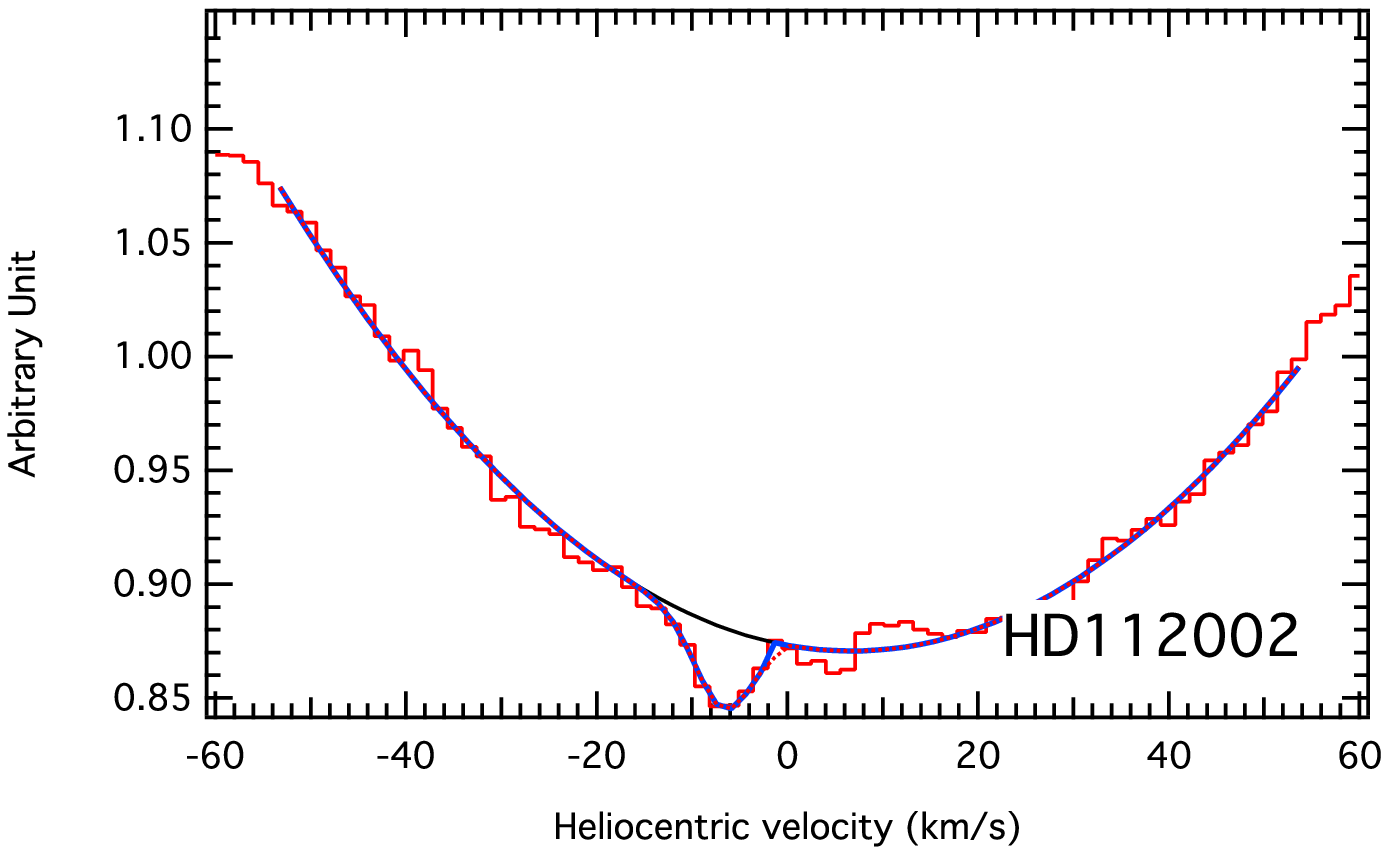}
  	\includegraphics[width=1\linewidth]{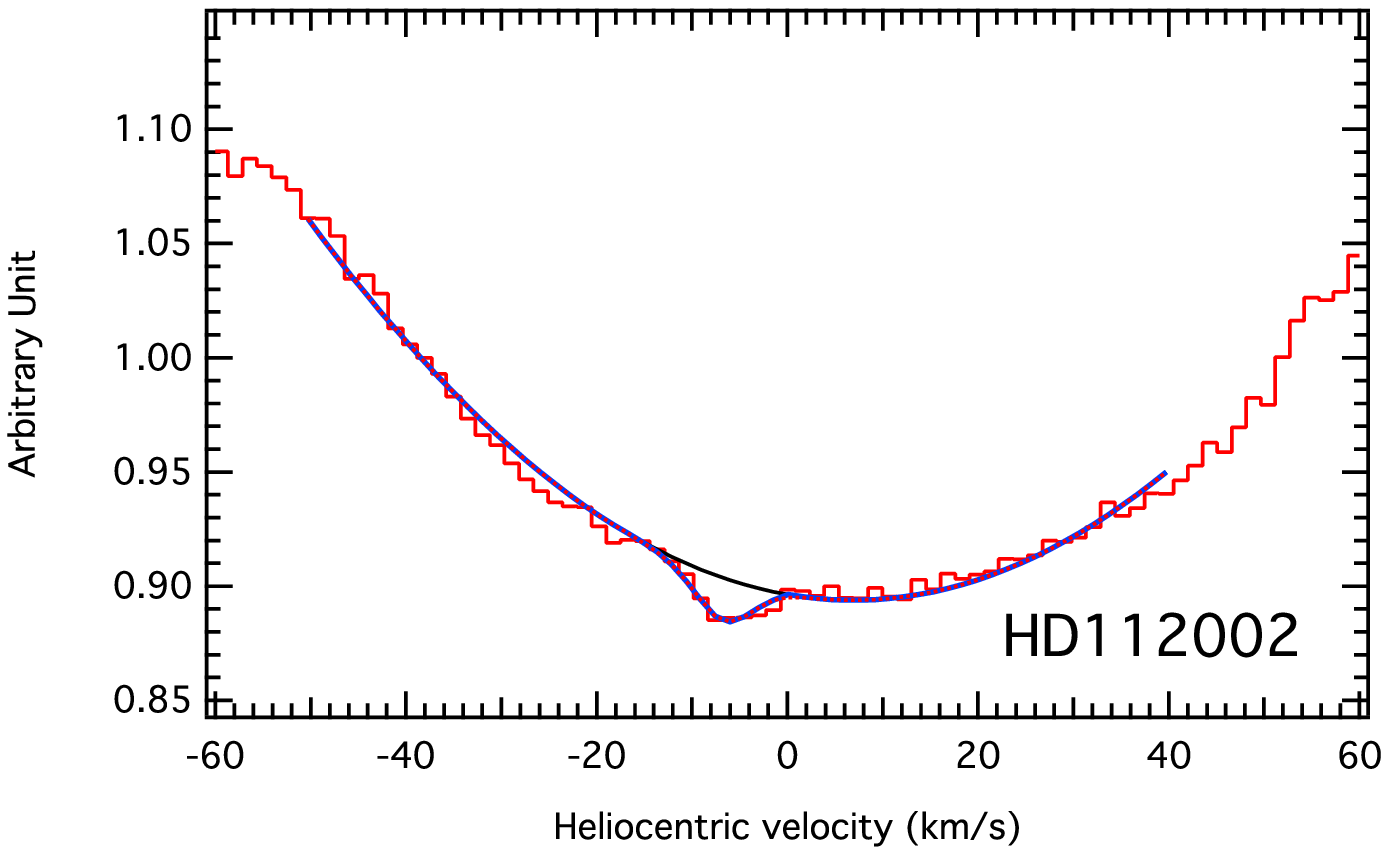}
\end{minipage}\hfill
\begin{minipage}[t]{0.3\linewidth}
\centering
  	\includegraphics[width=1\linewidth]{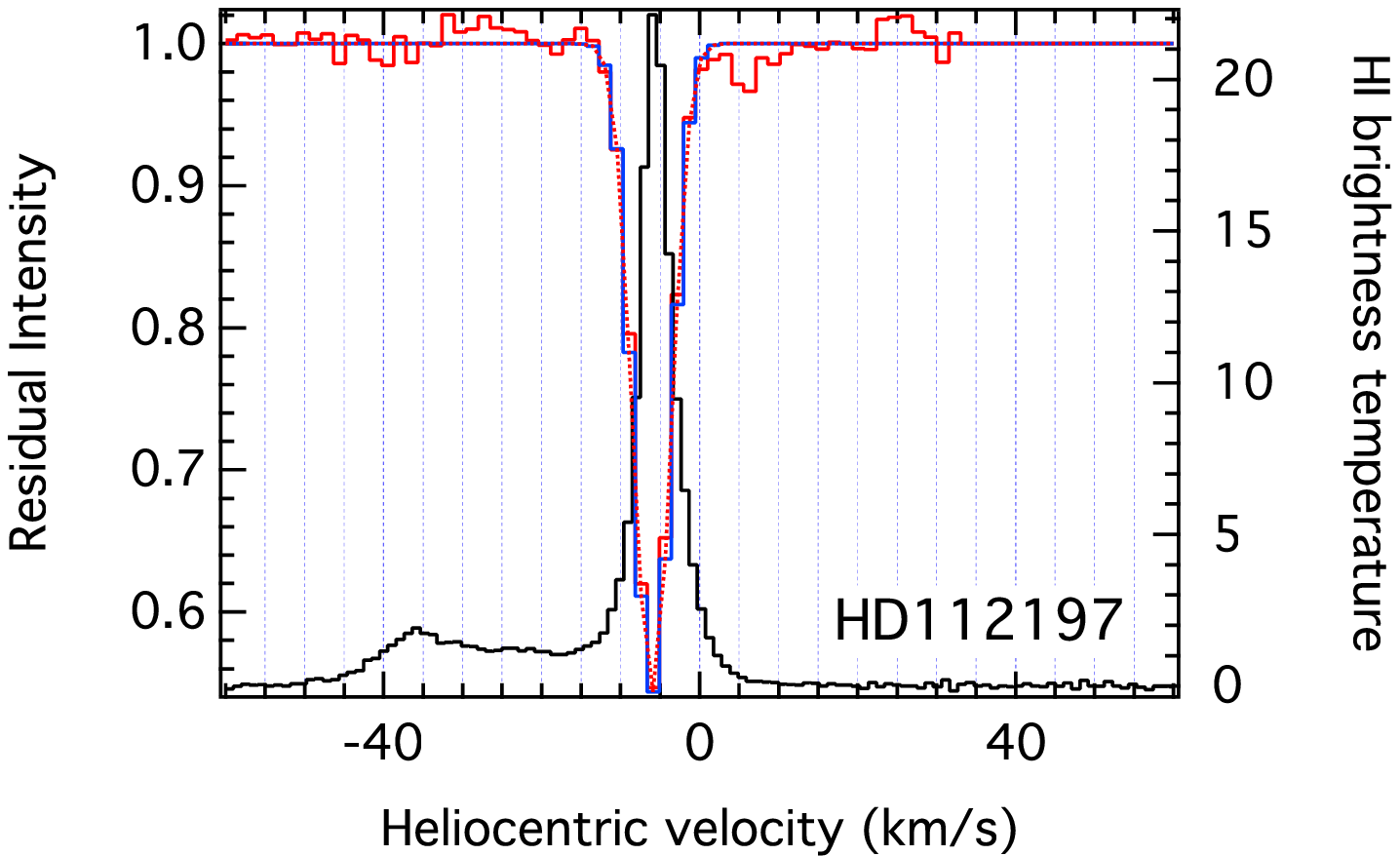}
  	\includegraphics[width=1\linewidth]{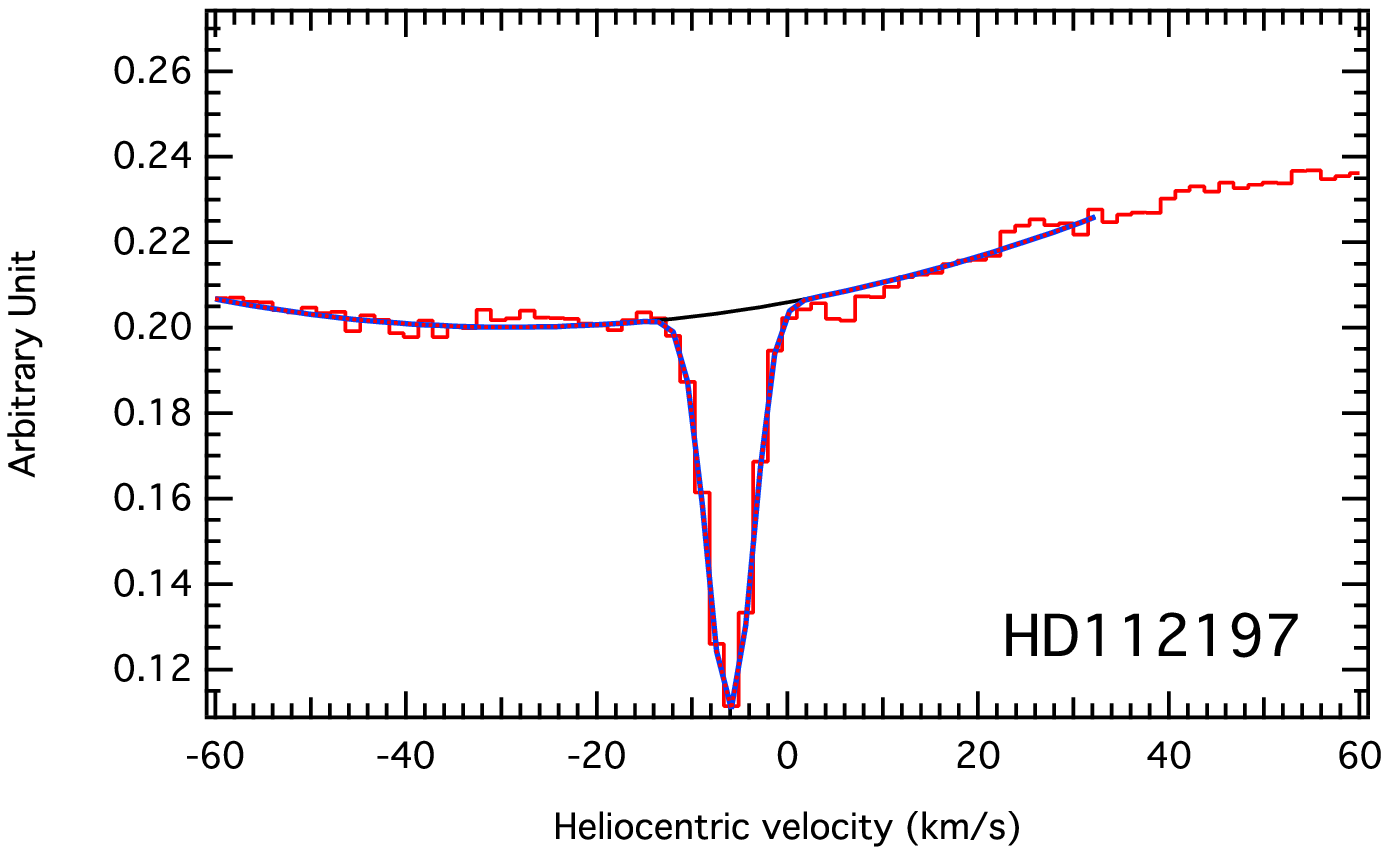}
  	\includegraphics[width=1\linewidth]{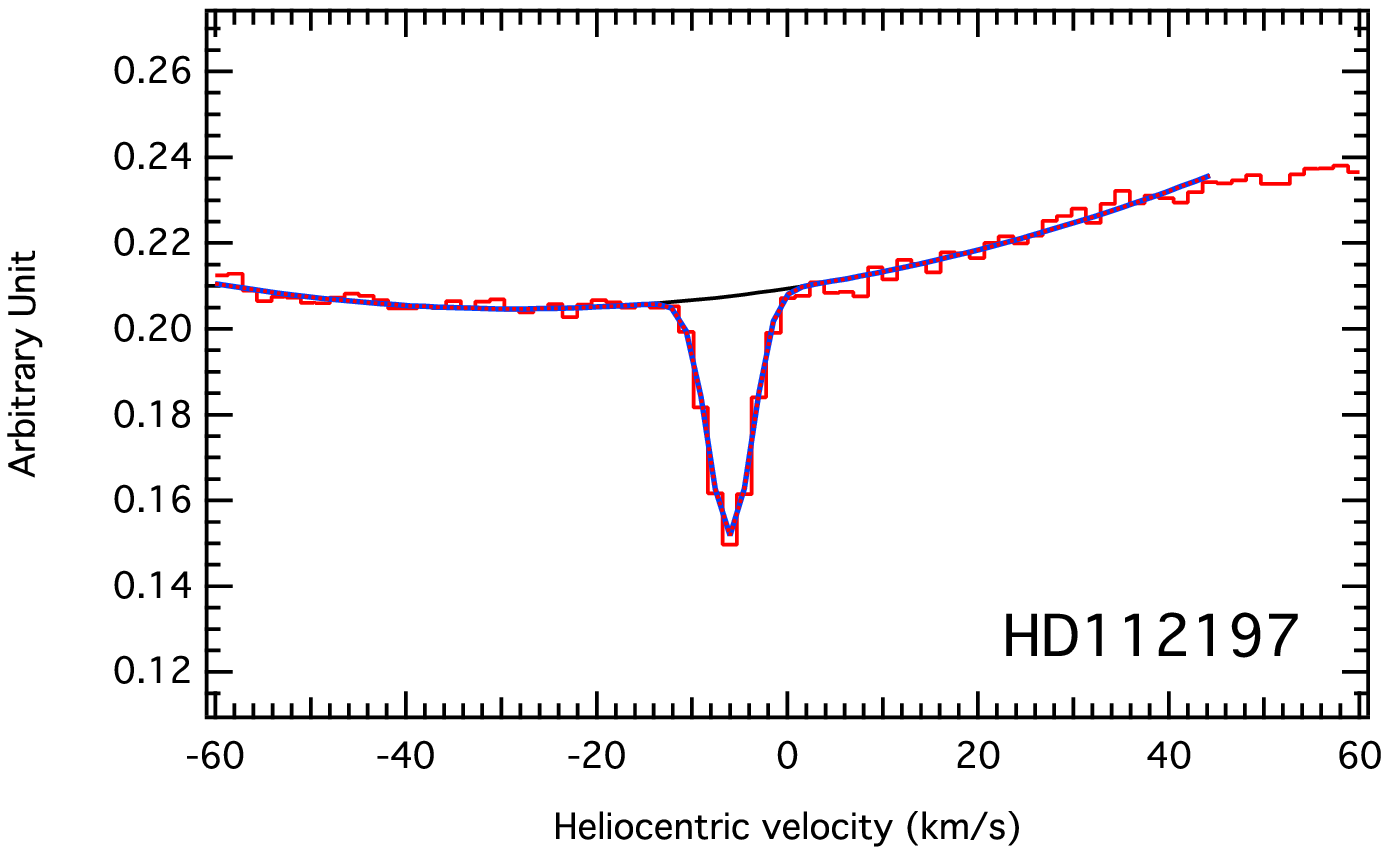}
\end{minipage}\hfill
\begin{minipage}[t]{0.3\linewidth}
\centering
  	\includegraphics[width=1\linewidth]{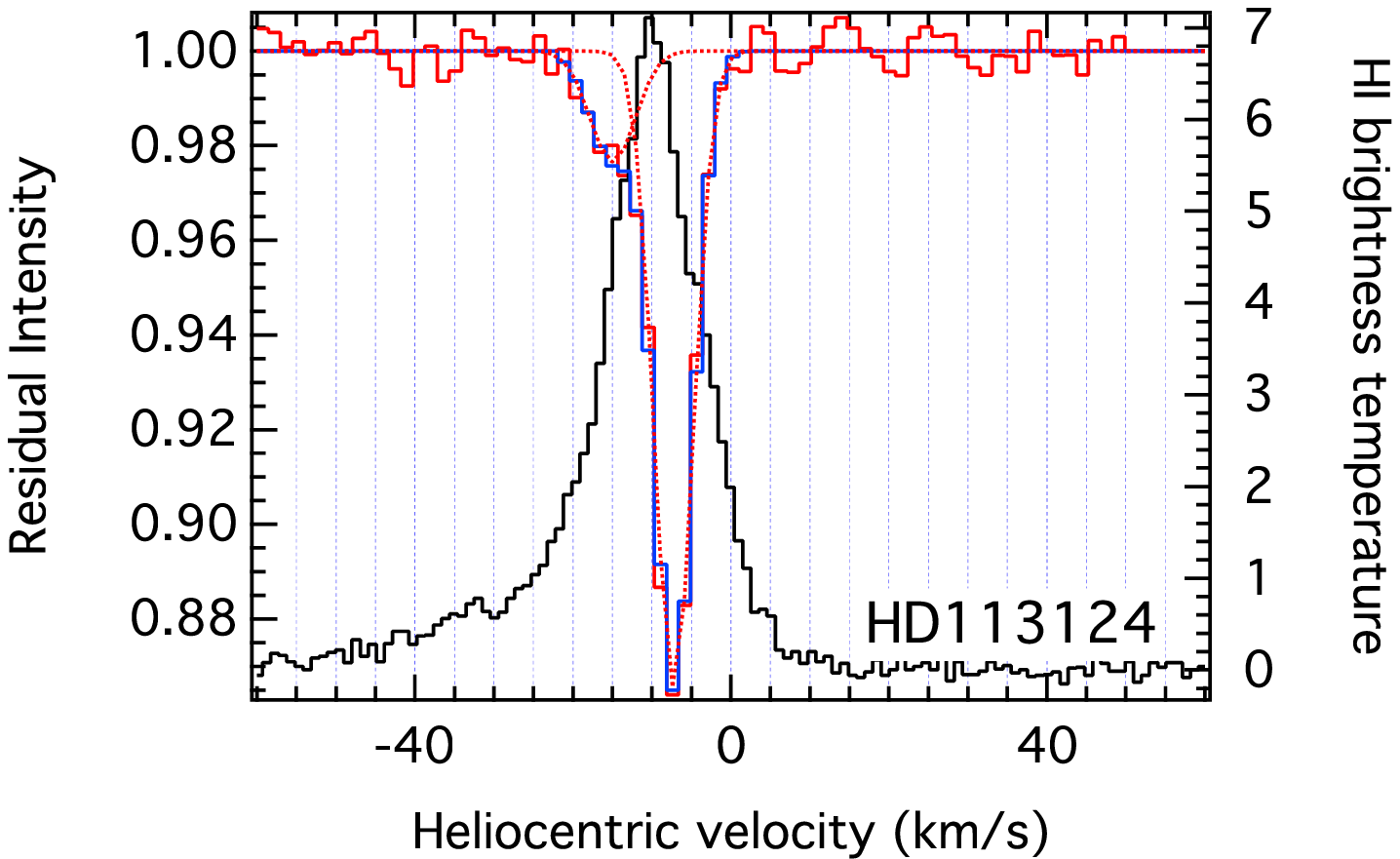}
  	\includegraphics[width=1\linewidth]{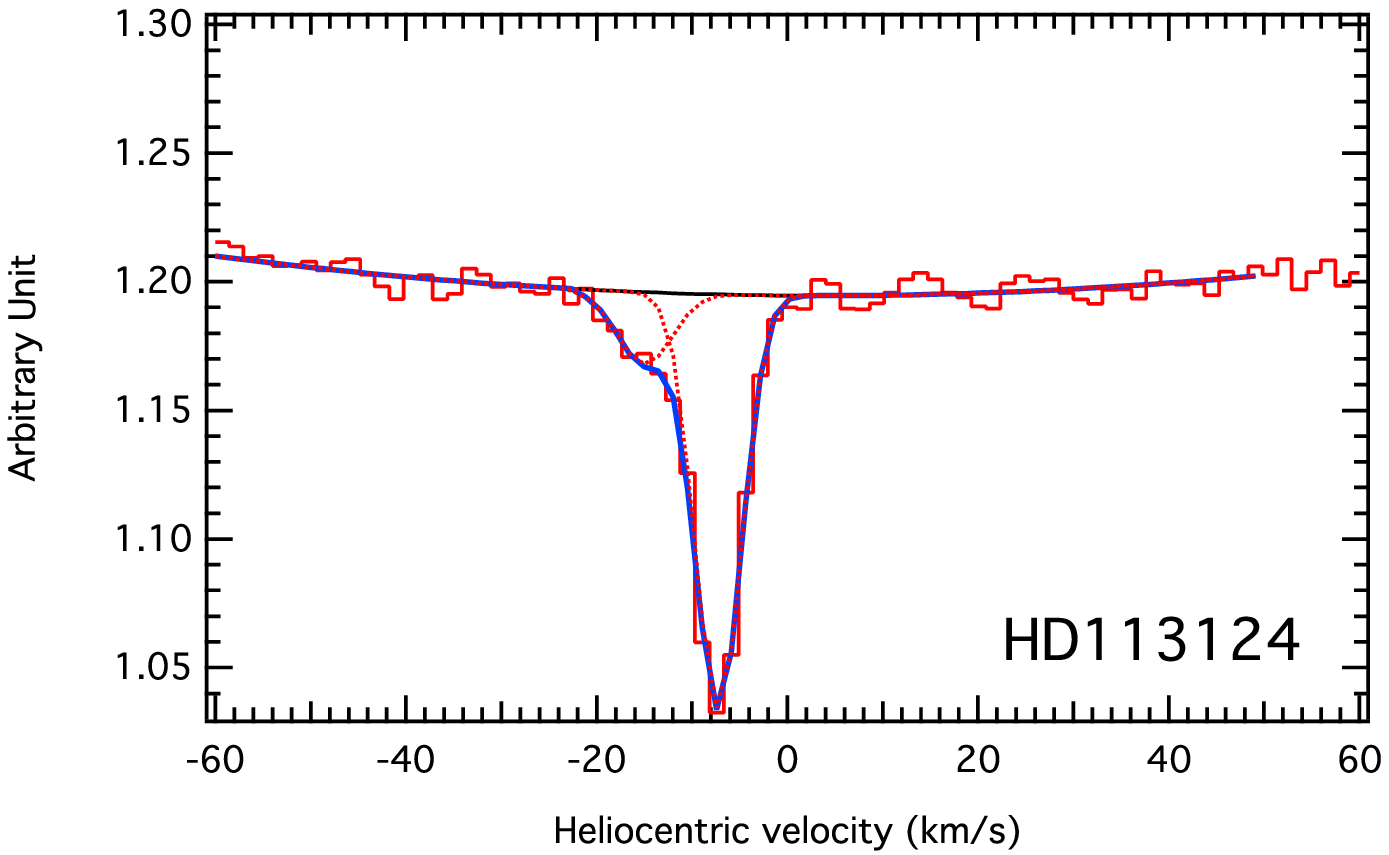}
  	\includegraphics[width=1\linewidth]{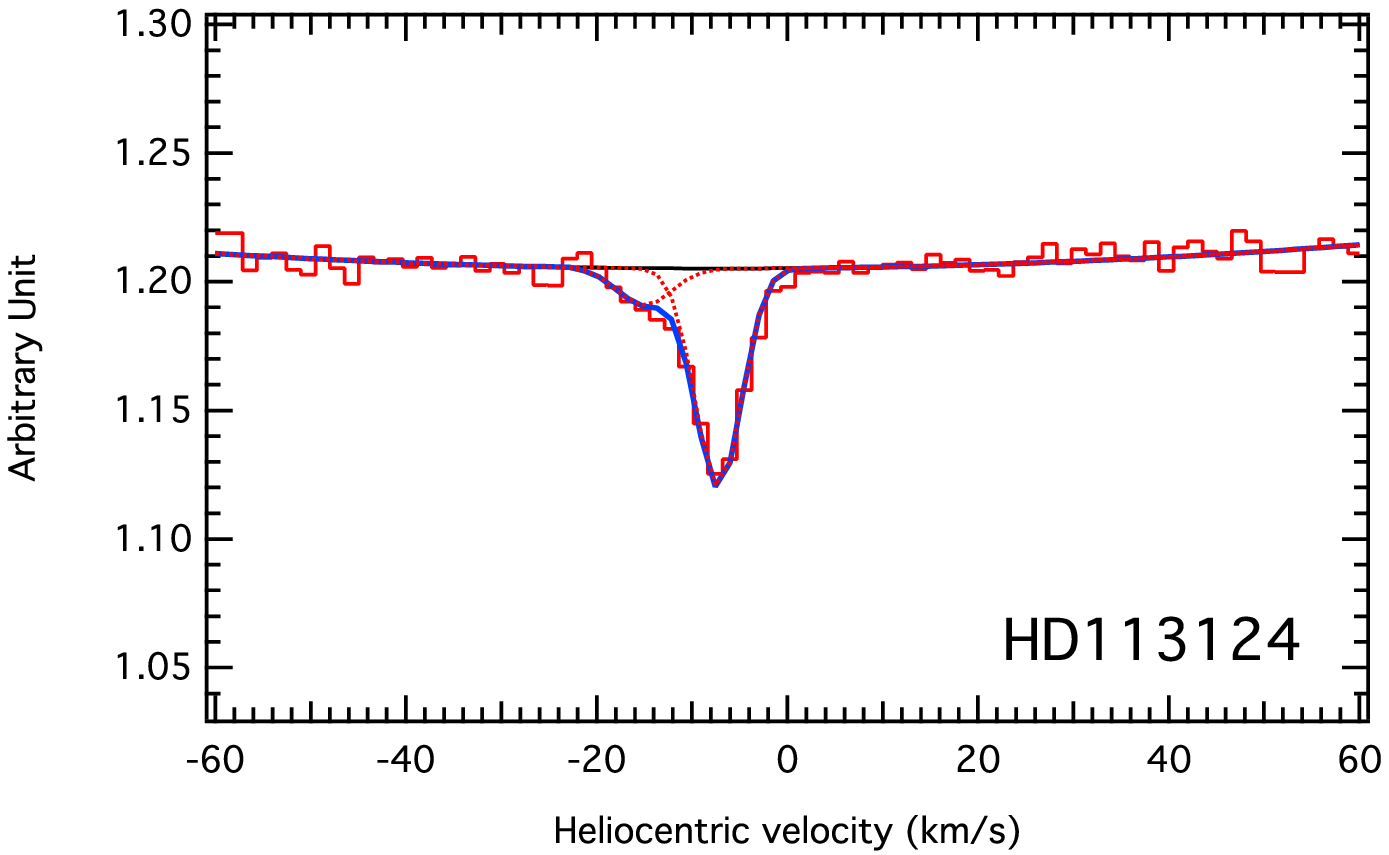}
\end{minipage}\hfill
\caption{Same as Fig. \ref{HD94194} {(in the article)}  but for HD112002, HD112197, and HD113124}
\end{figure*}

\begin{figure*}[!h]
\begin{minipage}[t]{0.24\linewidth}
\centering
	\includegraphics[width=1\linewidth]{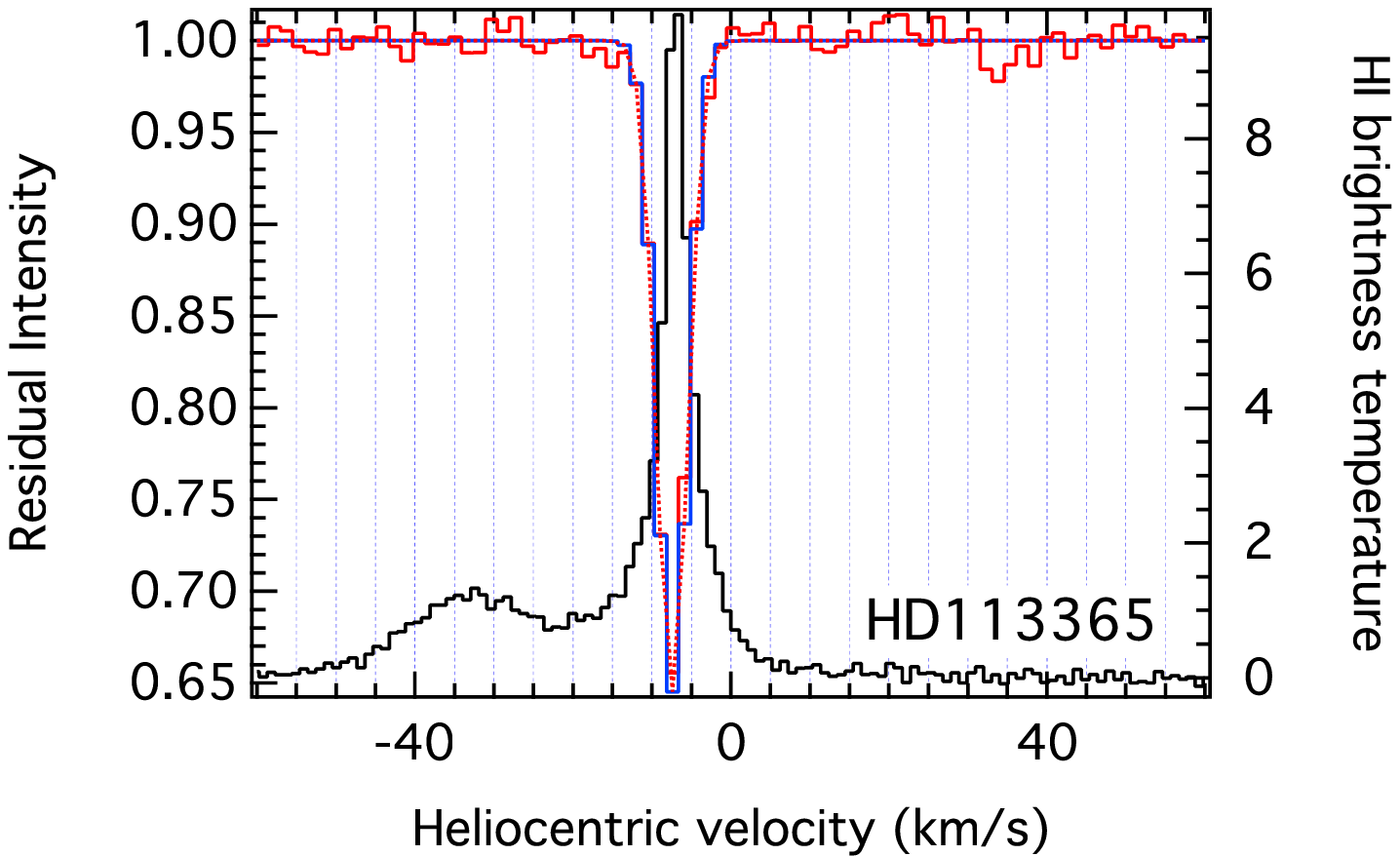}
  	\includegraphics[width=1\linewidth]{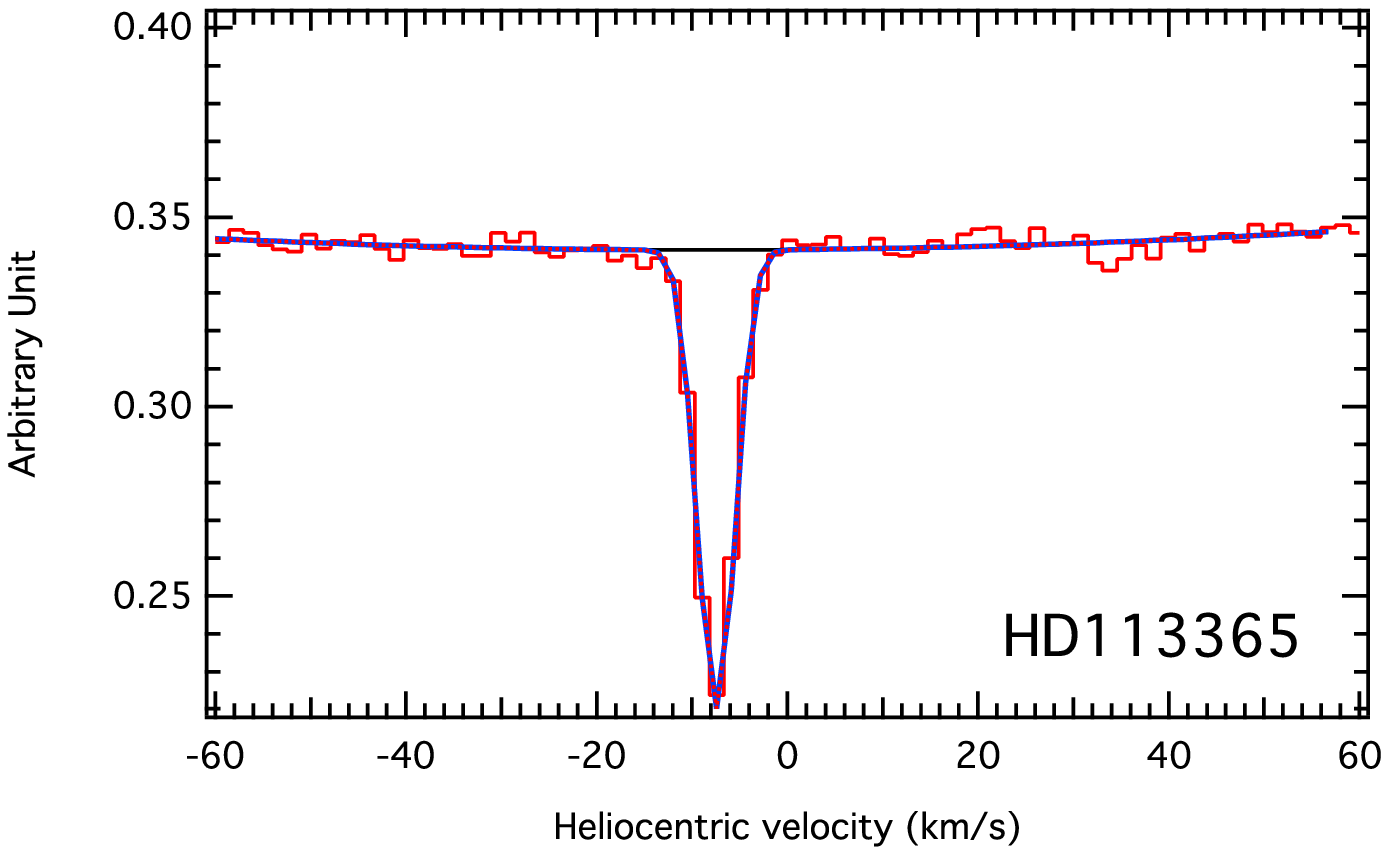}
  	\includegraphics[width=1\linewidth]{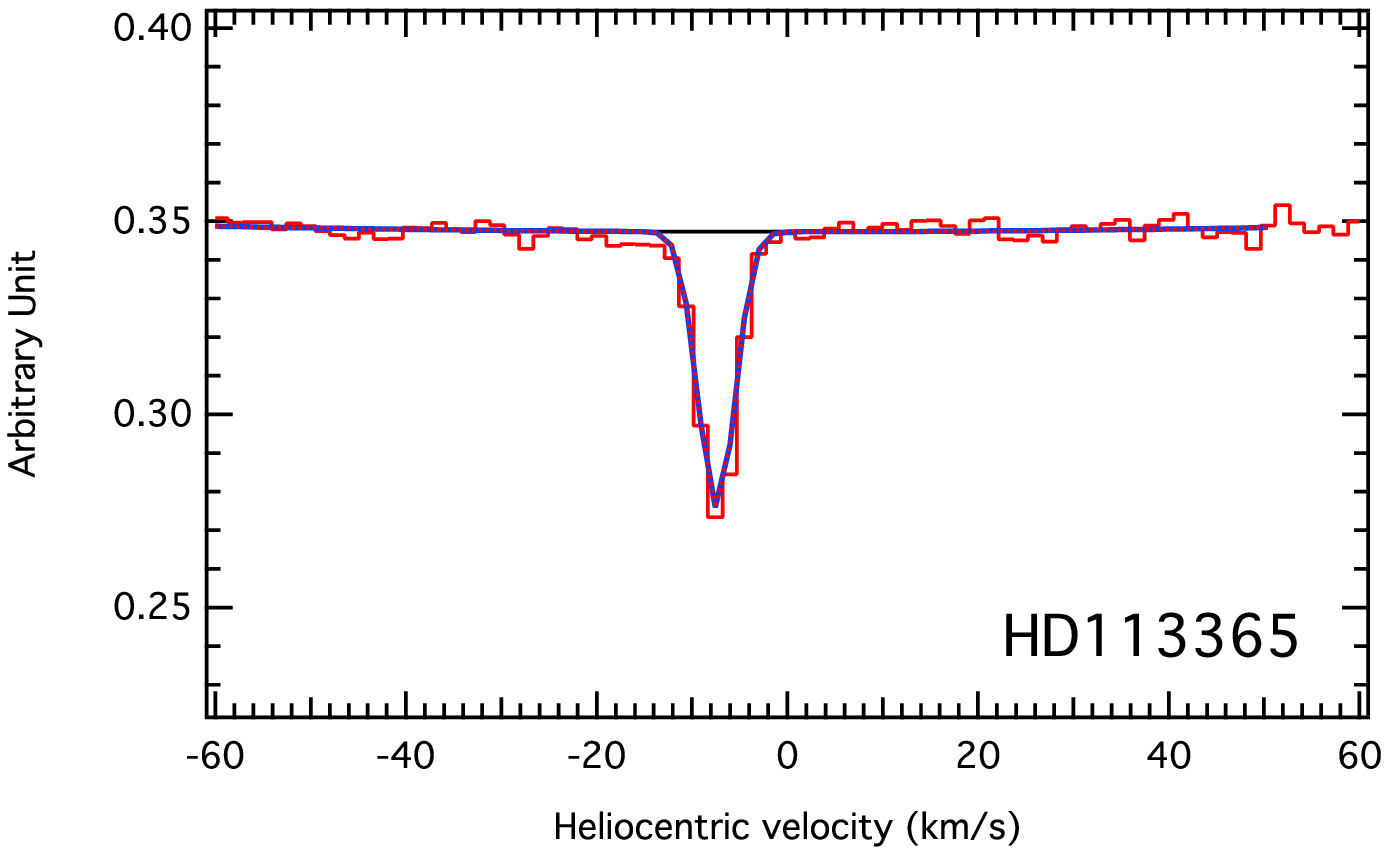}
\end{minipage}\hfill
\begin{minipage}[t]{0.24\linewidth}
\centering
  	\includegraphics[width=1\linewidth]{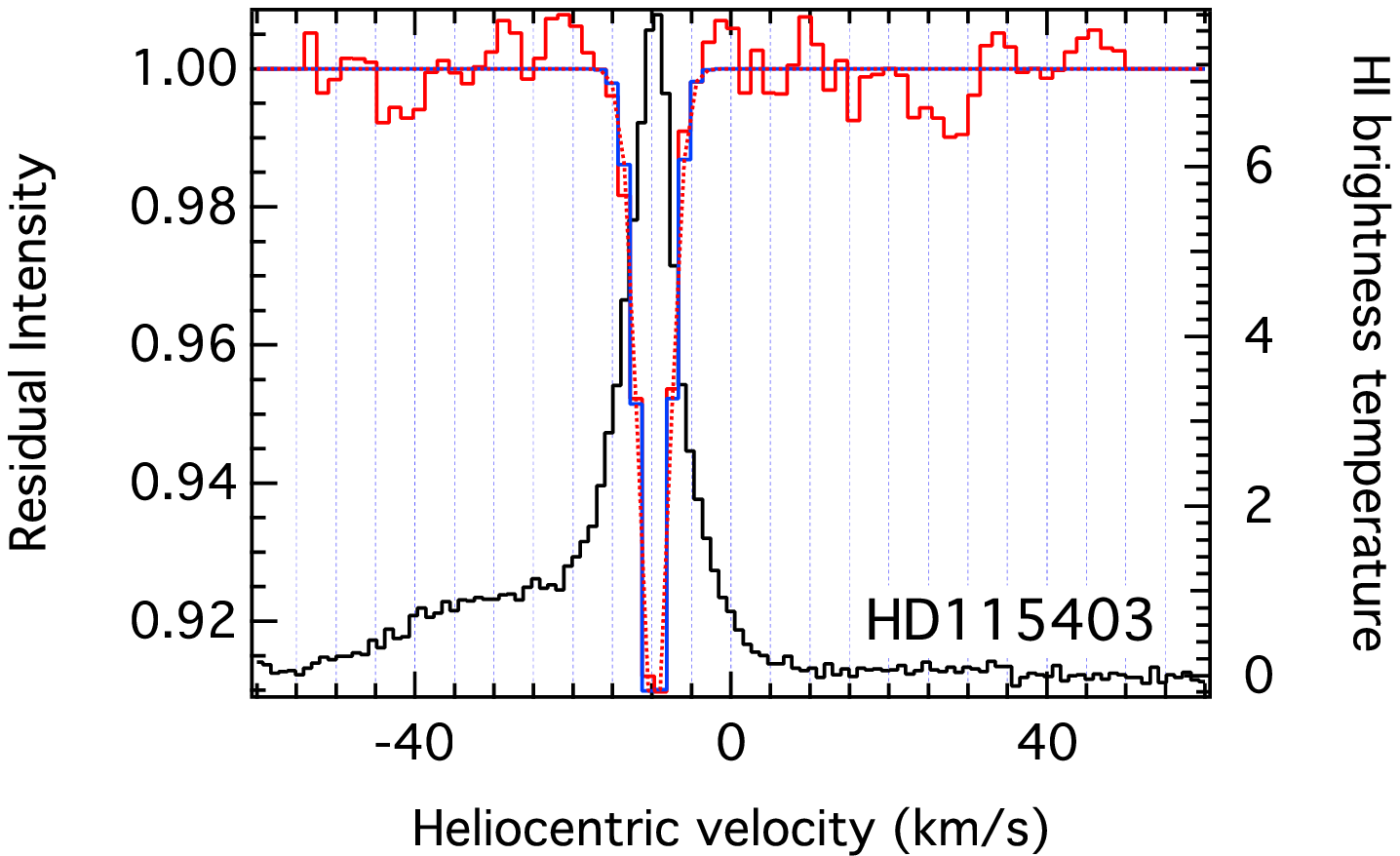}
  	\includegraphics[width=1\linewidth]{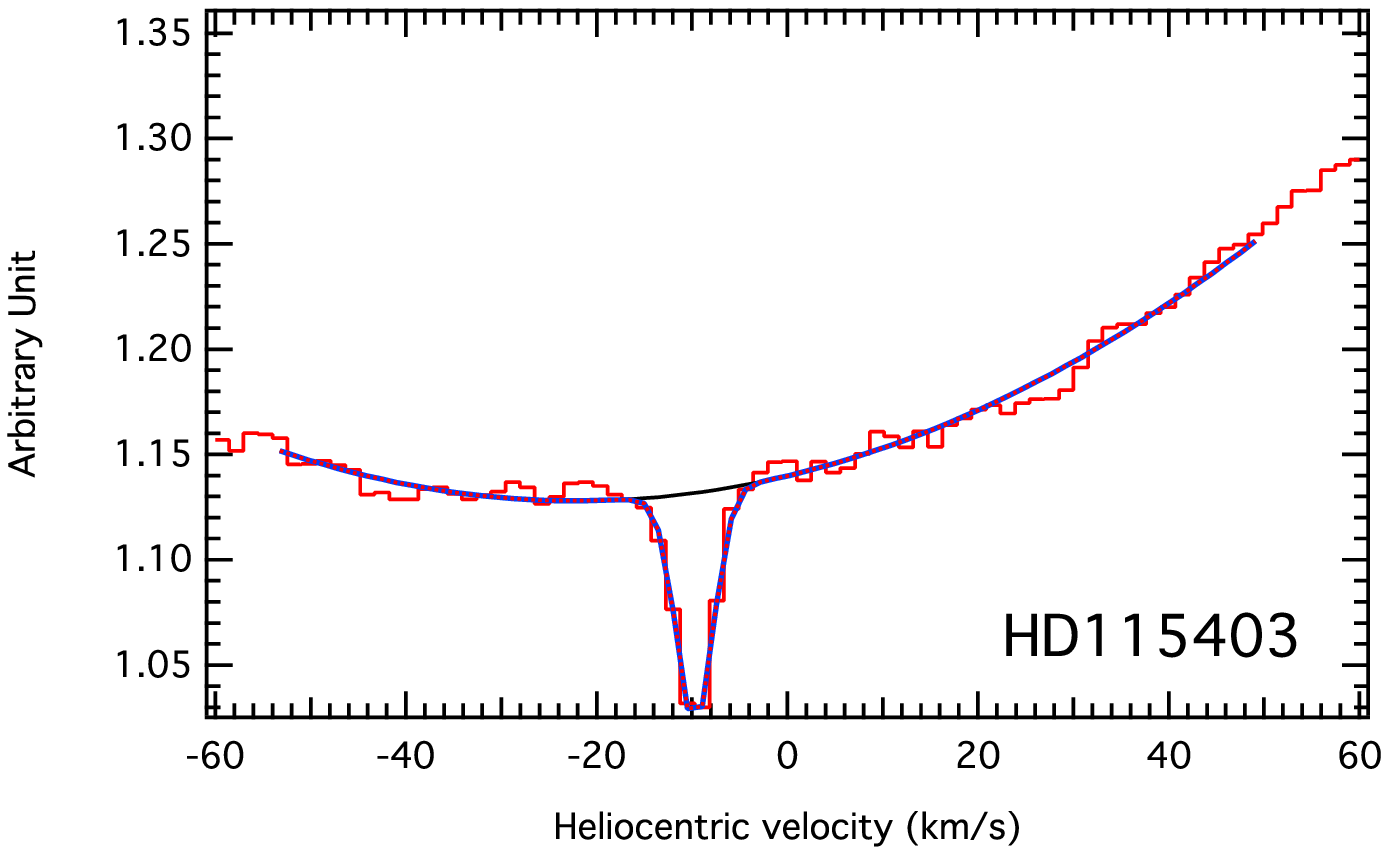}
  	\includegraphics[width=1\linewidth]{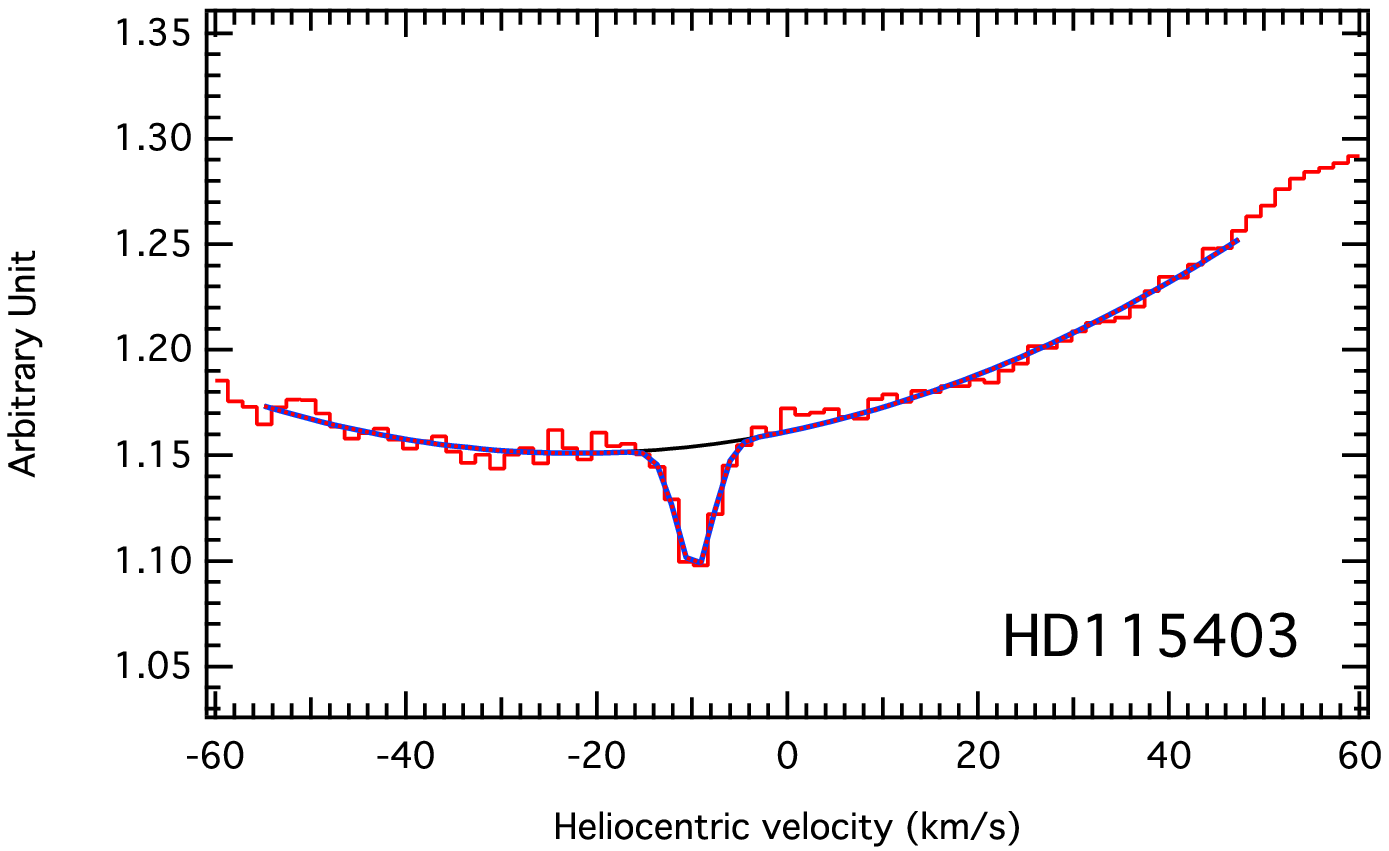}
\end{minipage}
\begin{minipage}[t]{0.24\linewidth}
\centering
  	\includegraphics[width=1\linewidth]{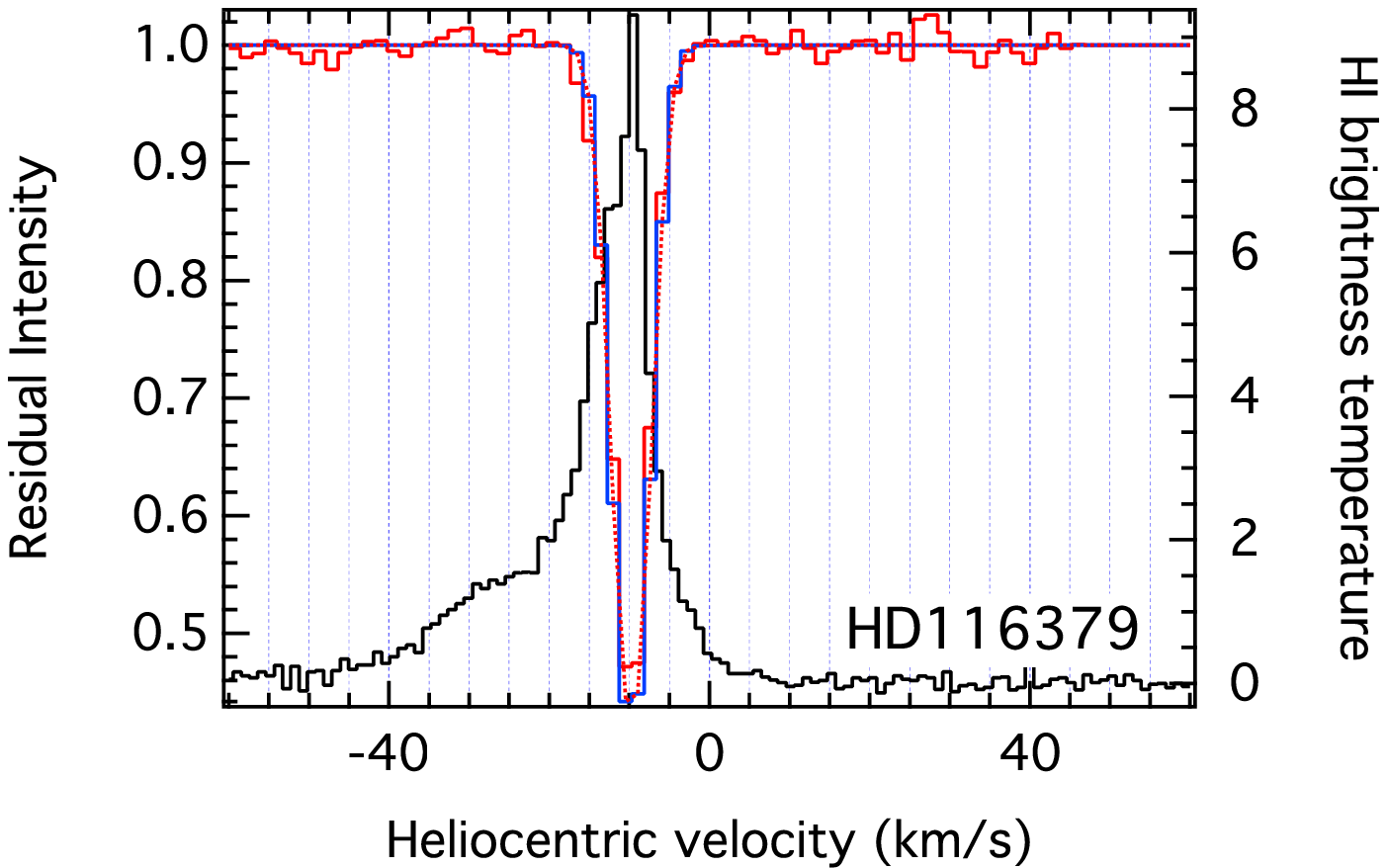}
  	\includegraphics[width=1\linewidth]{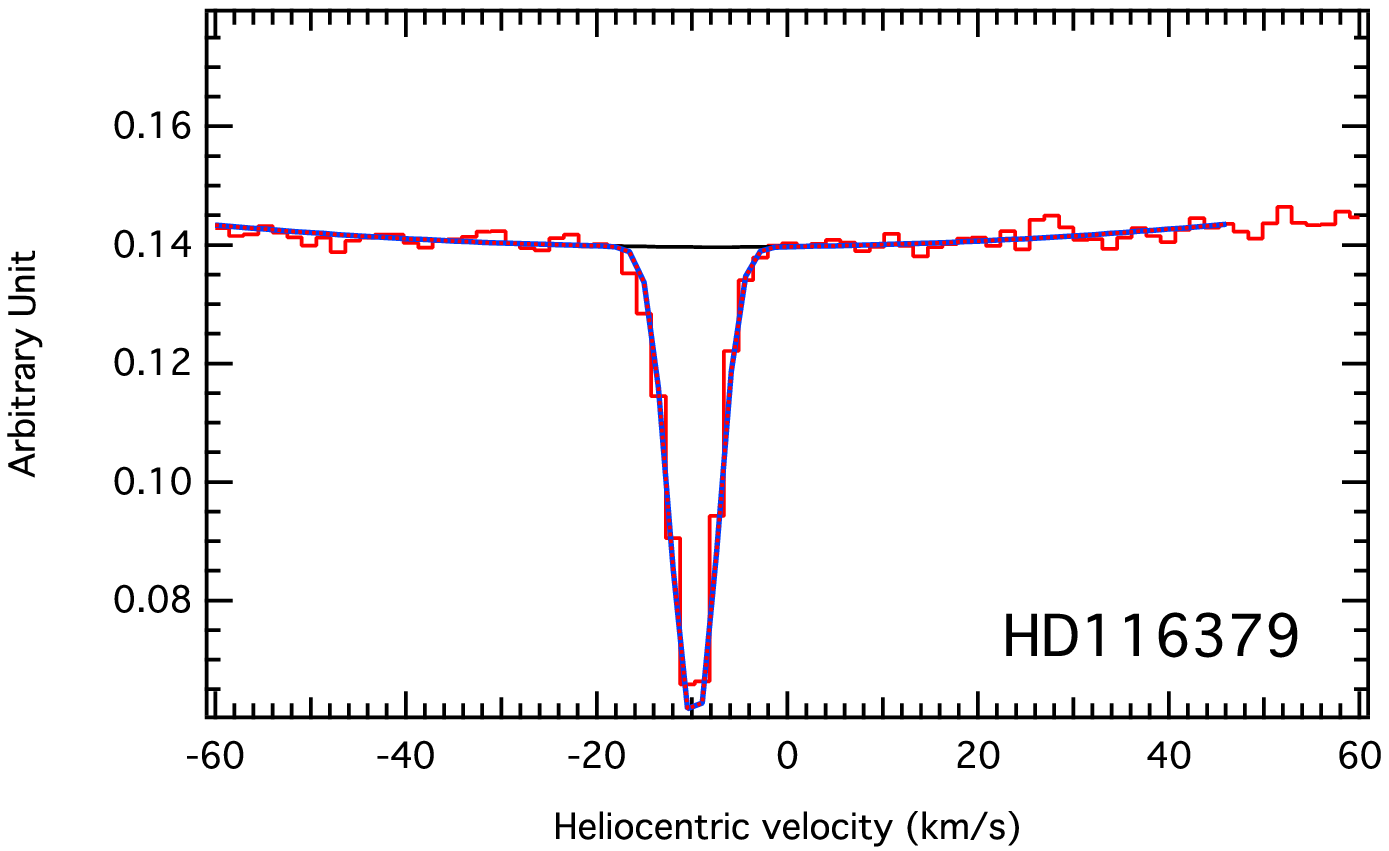}
  	\includegraphics[width=1\linewidth]{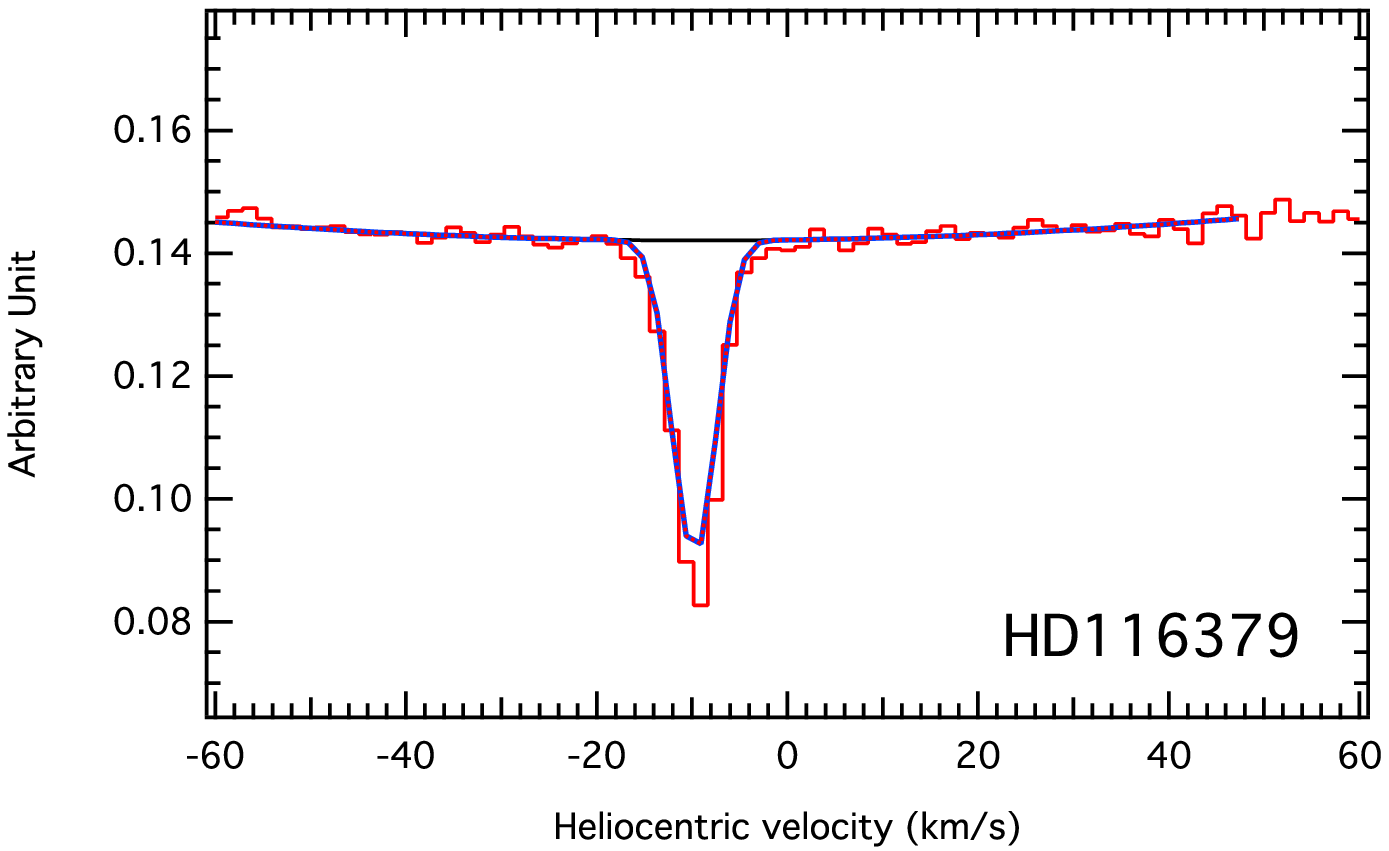}
\end{minipage}\hfill
\begin{minipage}[t]{0.24\linewidth}
\centering
  	\includegraphics[width=1\linewidth]{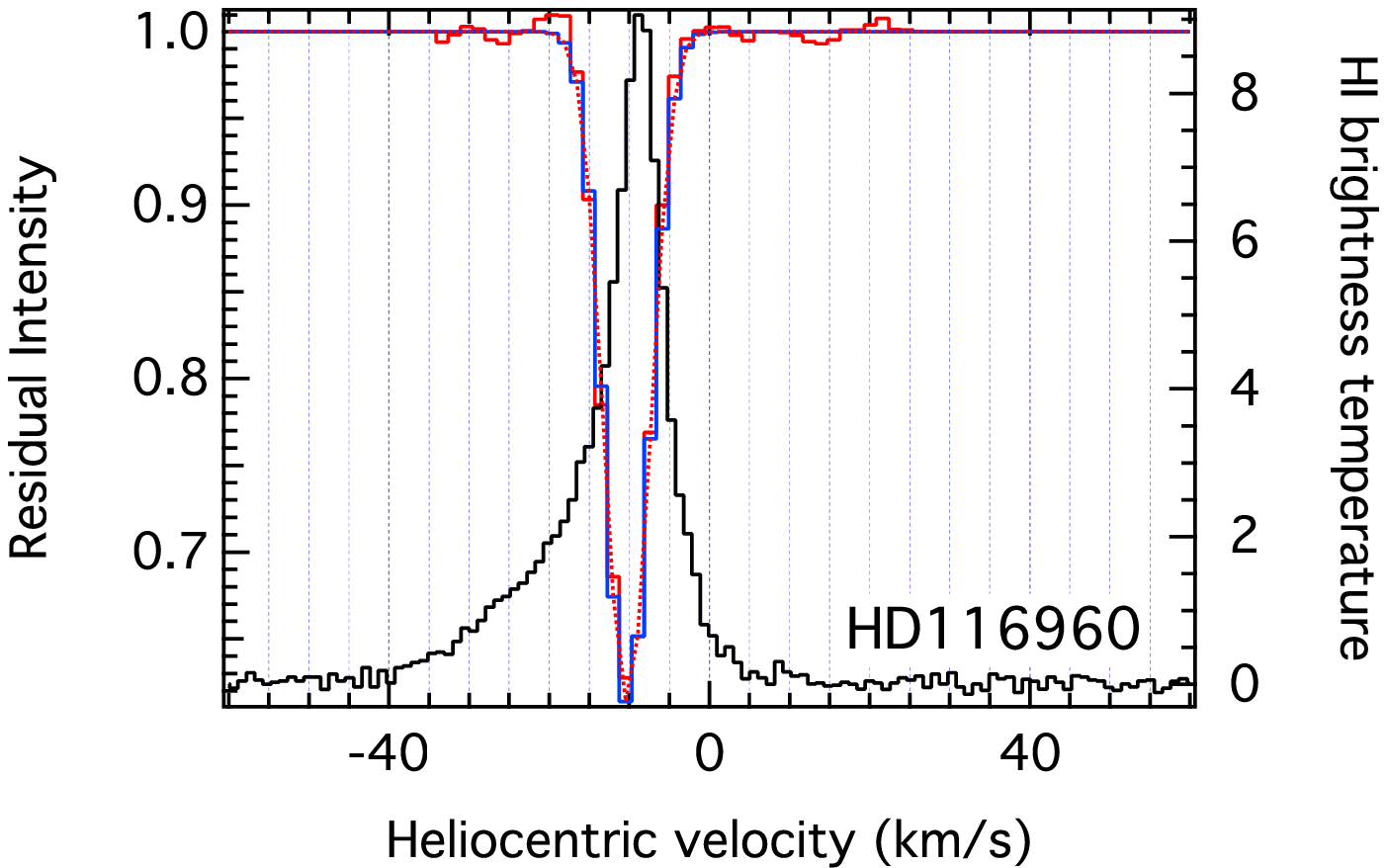}
  	\includegraphics[width=1\linewidth]{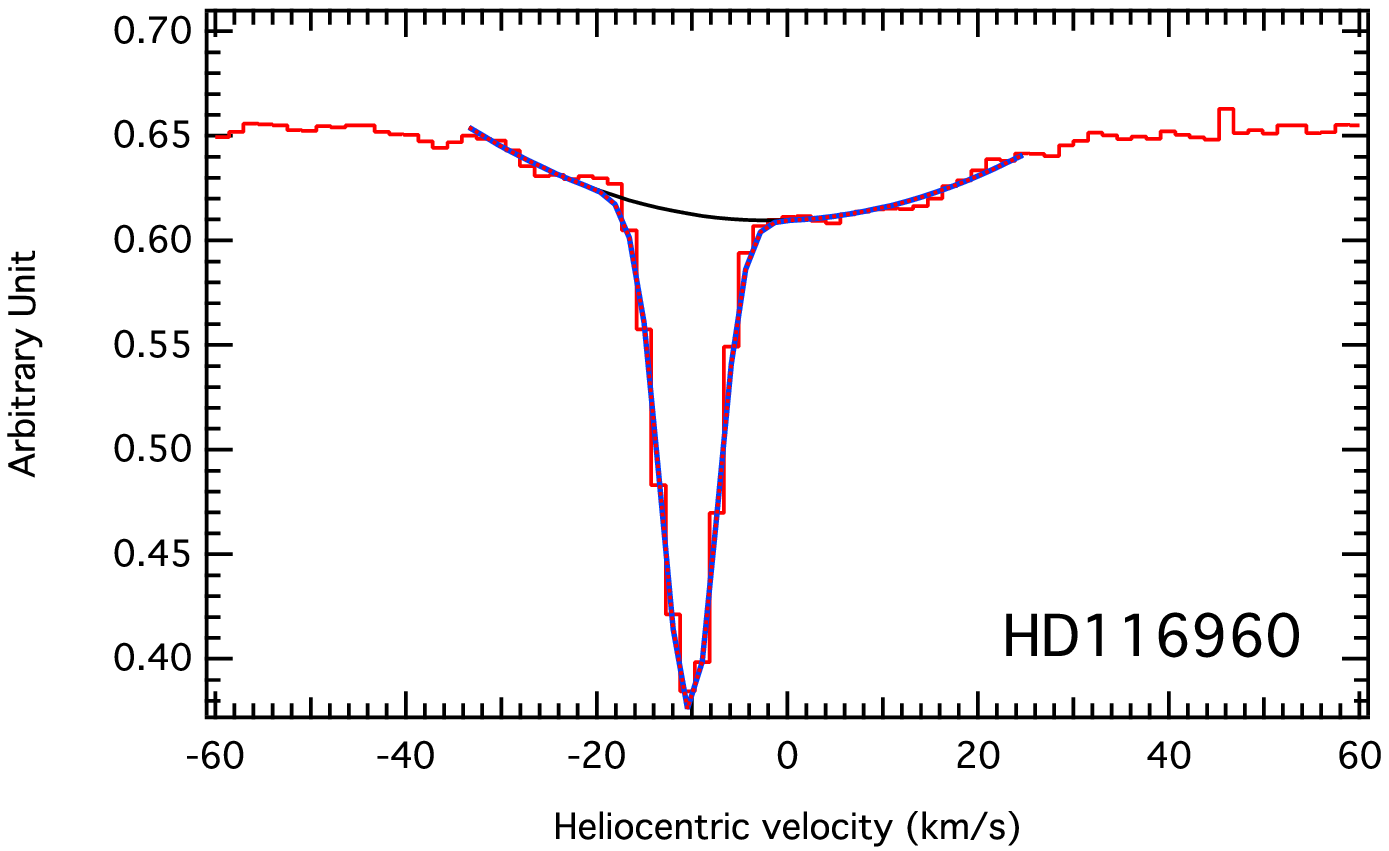}
  	\includegraphics[width=1\linewidth]{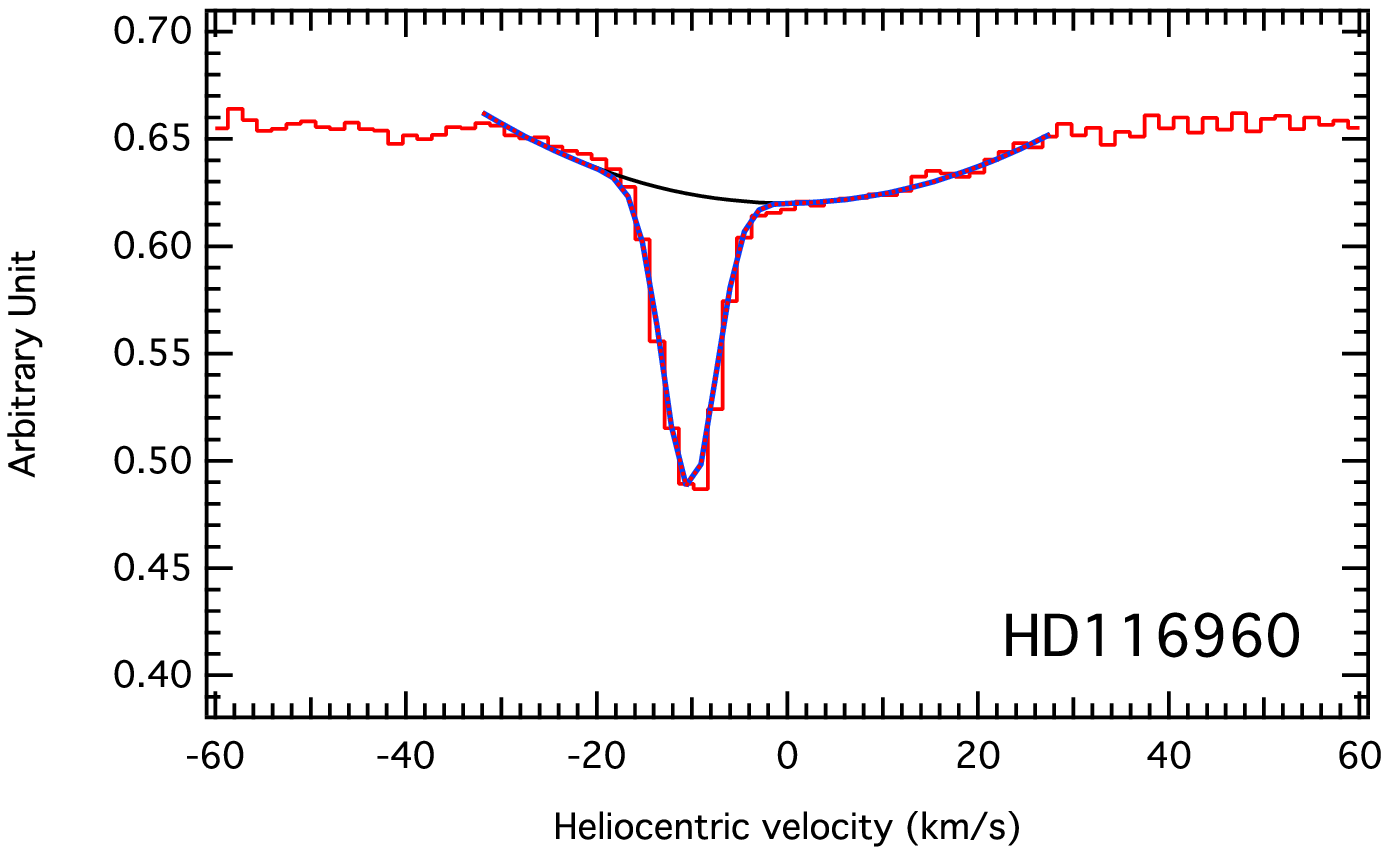}
\end{minipage}\hfill
\caption{Same as Fig. \ref{HD94194} {(in the article)}  but for HD113365, HD115403, HD116379, and HD116960}
\end{figure*}


\begin{figure*}[!h]
\begin{minipage}[t]{0.3\linewidth}
\centering
  	\includegraphics[width=1\linewidth]{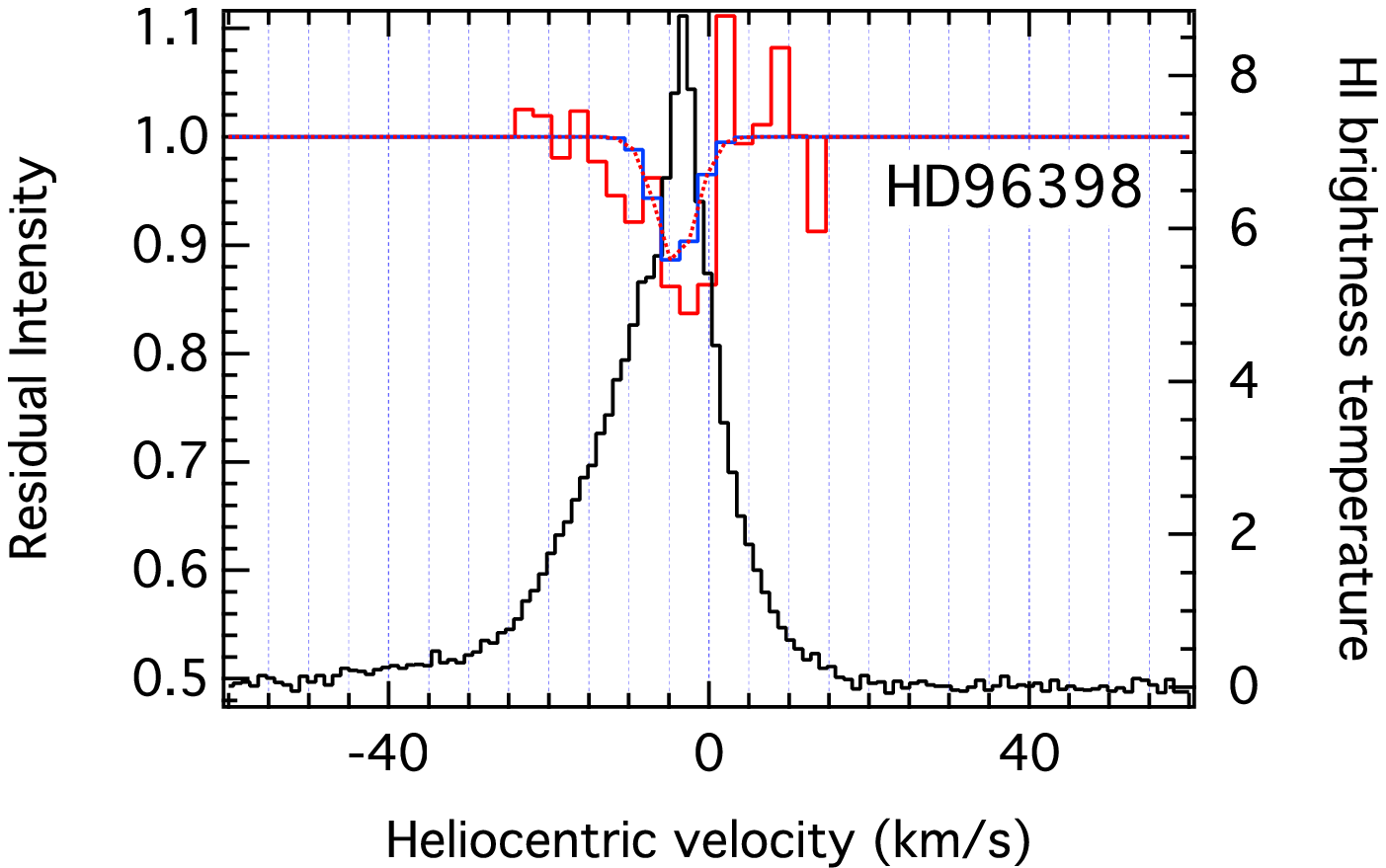}
  	\includegraphics[width=1\linewidth]{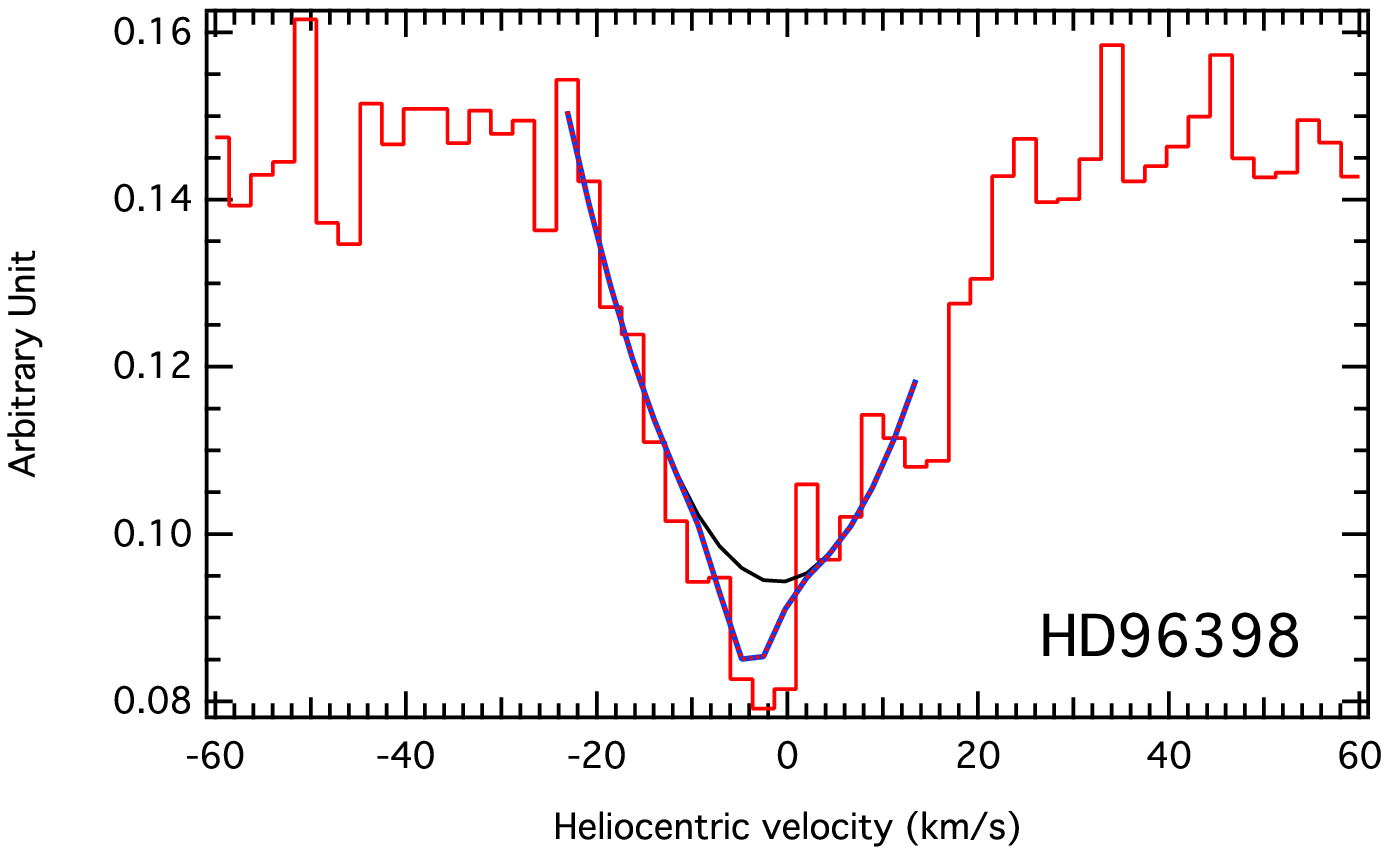}
  	\includegraphics[width=1\linewidth]{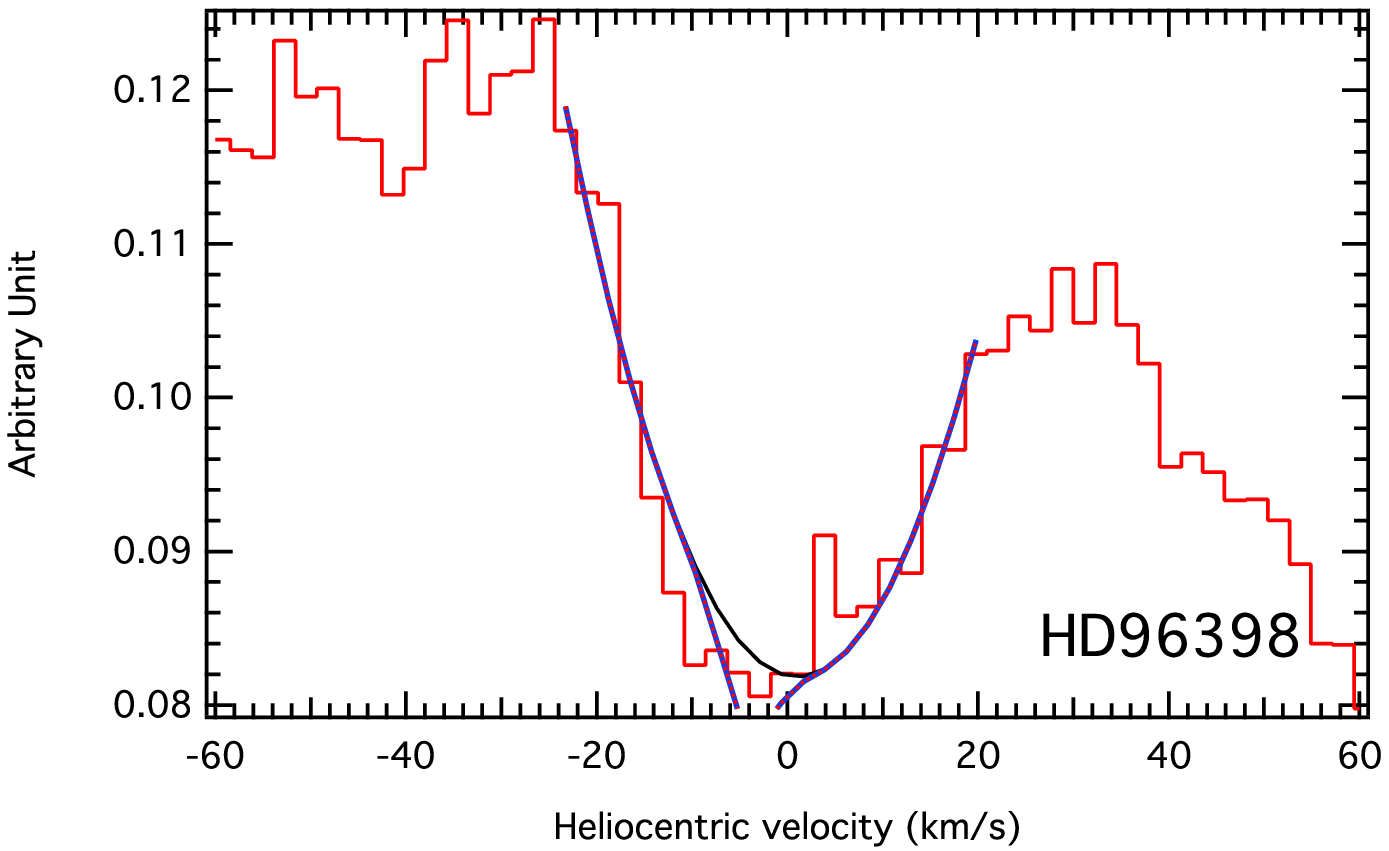}
\end{minipage}\hfill
\begin{minipage}[t]{0.3\linewidth}
\centering
  	\includegraphics[width=1\linewidth]{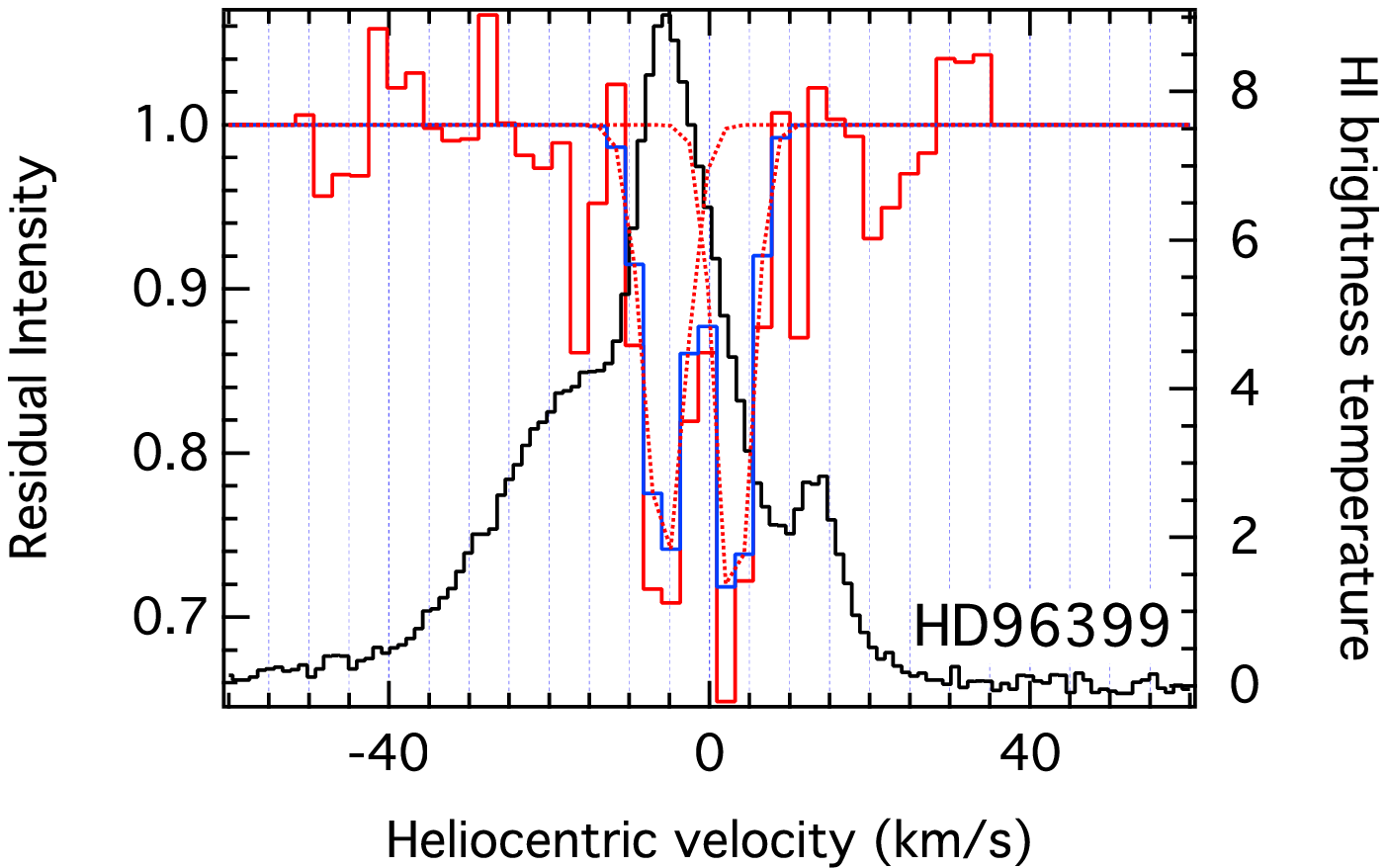}
  	\includegraphics[width=1\linewidth]{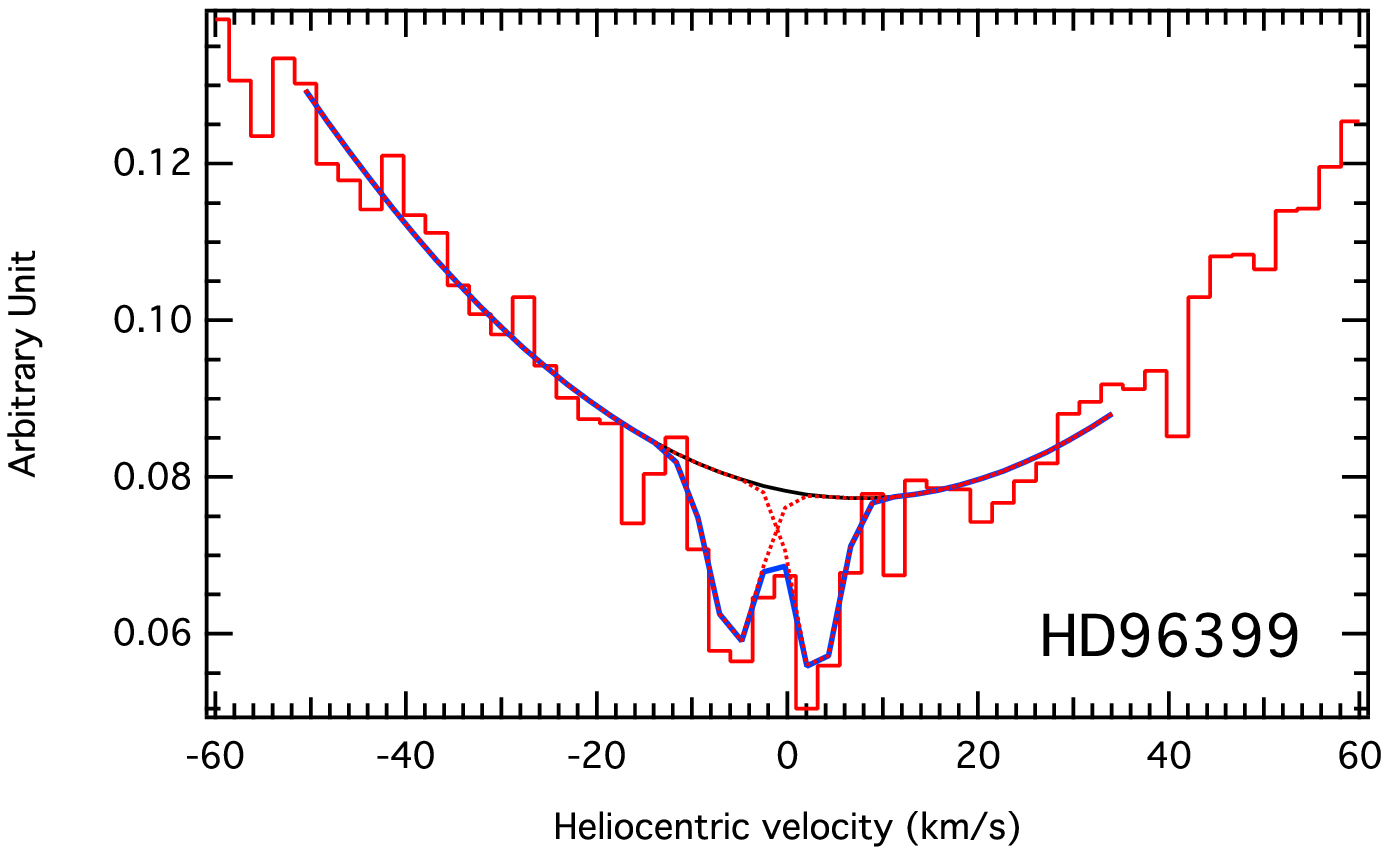}
  	\includegraphics[width=1\linewidth]{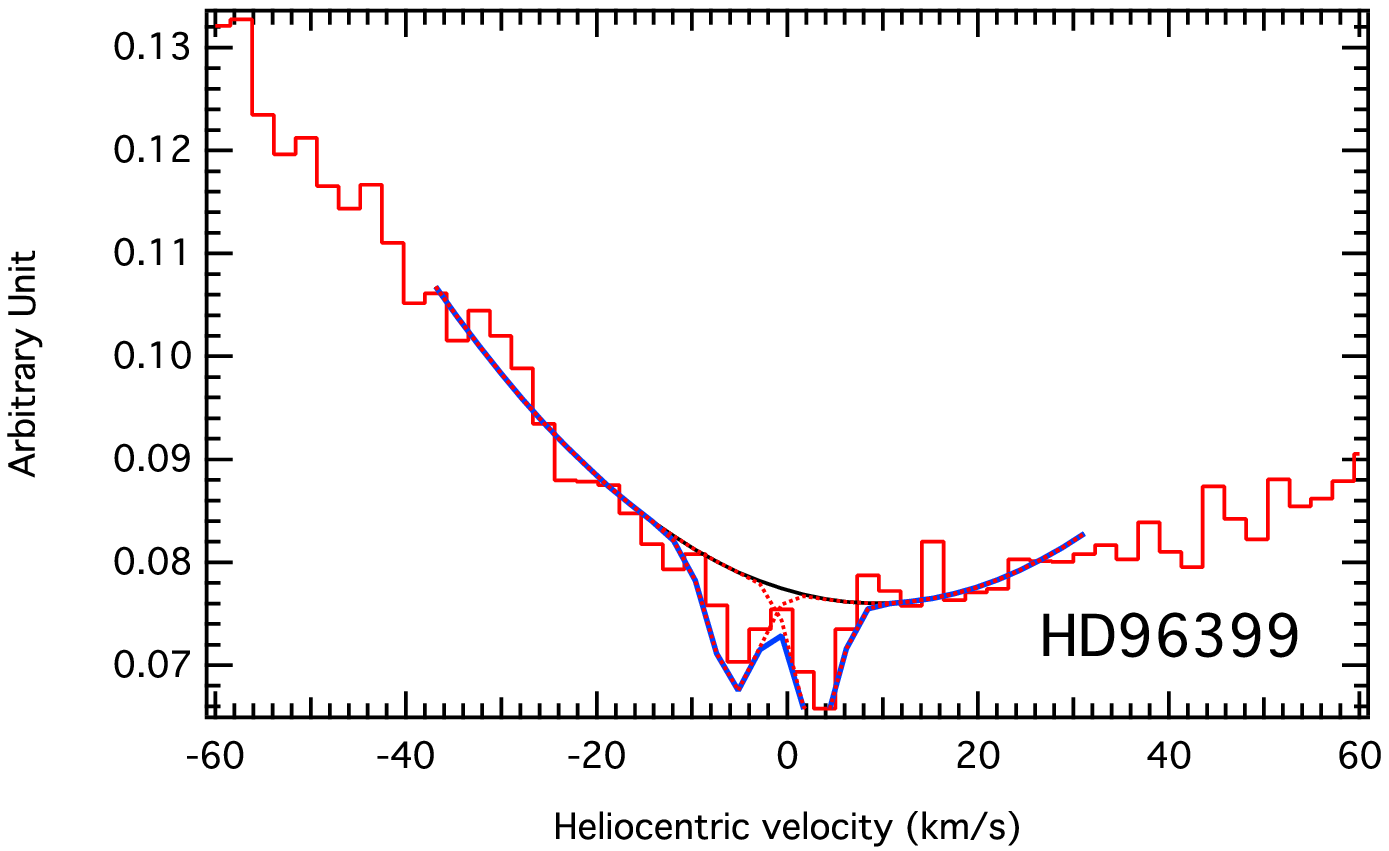}
\end{minipage}\hfill
\begin{minipage}[t]{0.3\linewidth}
\centering
  	\includegraphics[width=1\linewidth]{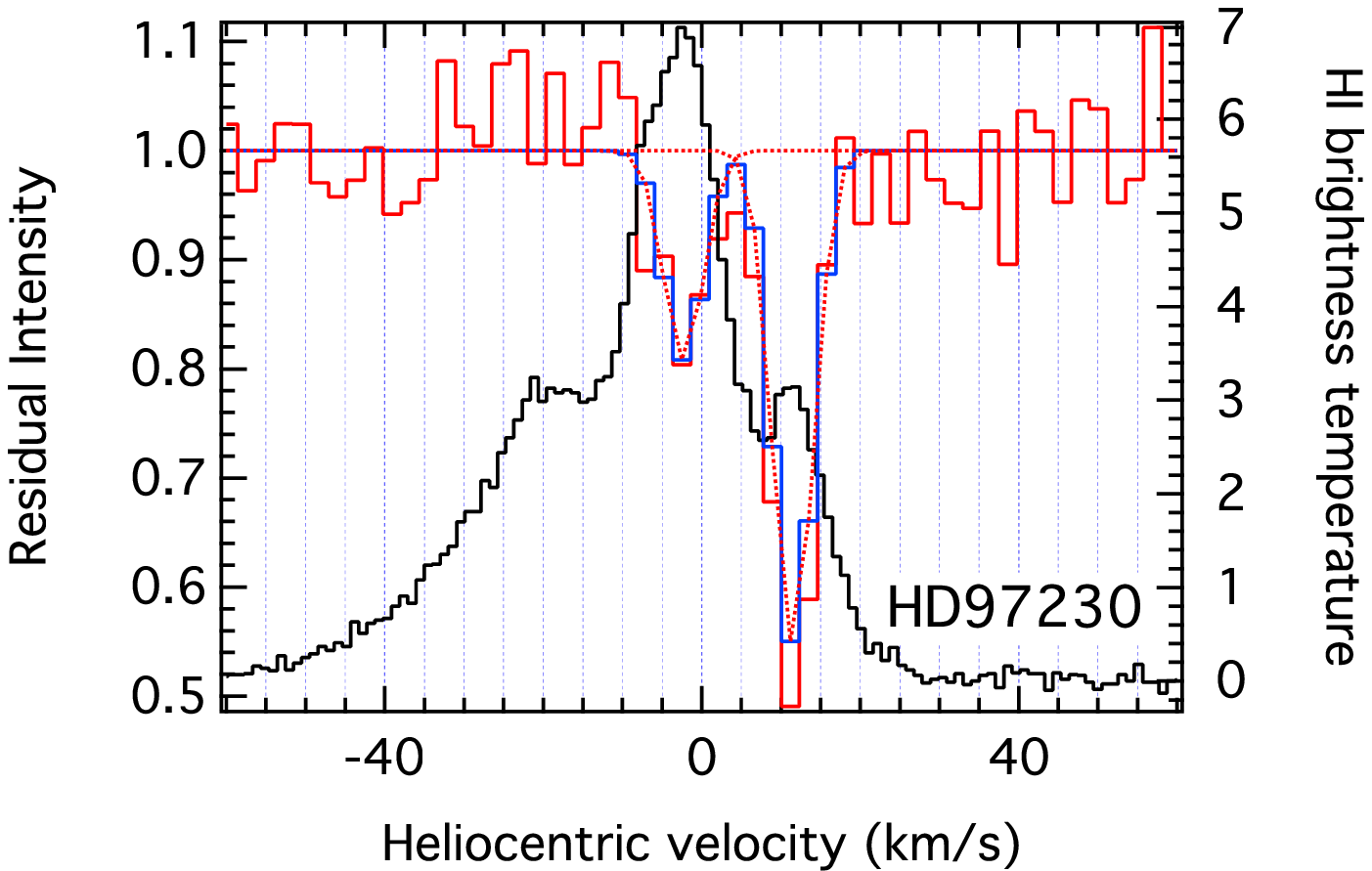}
  	\includegraphics[width=1\linewidth]{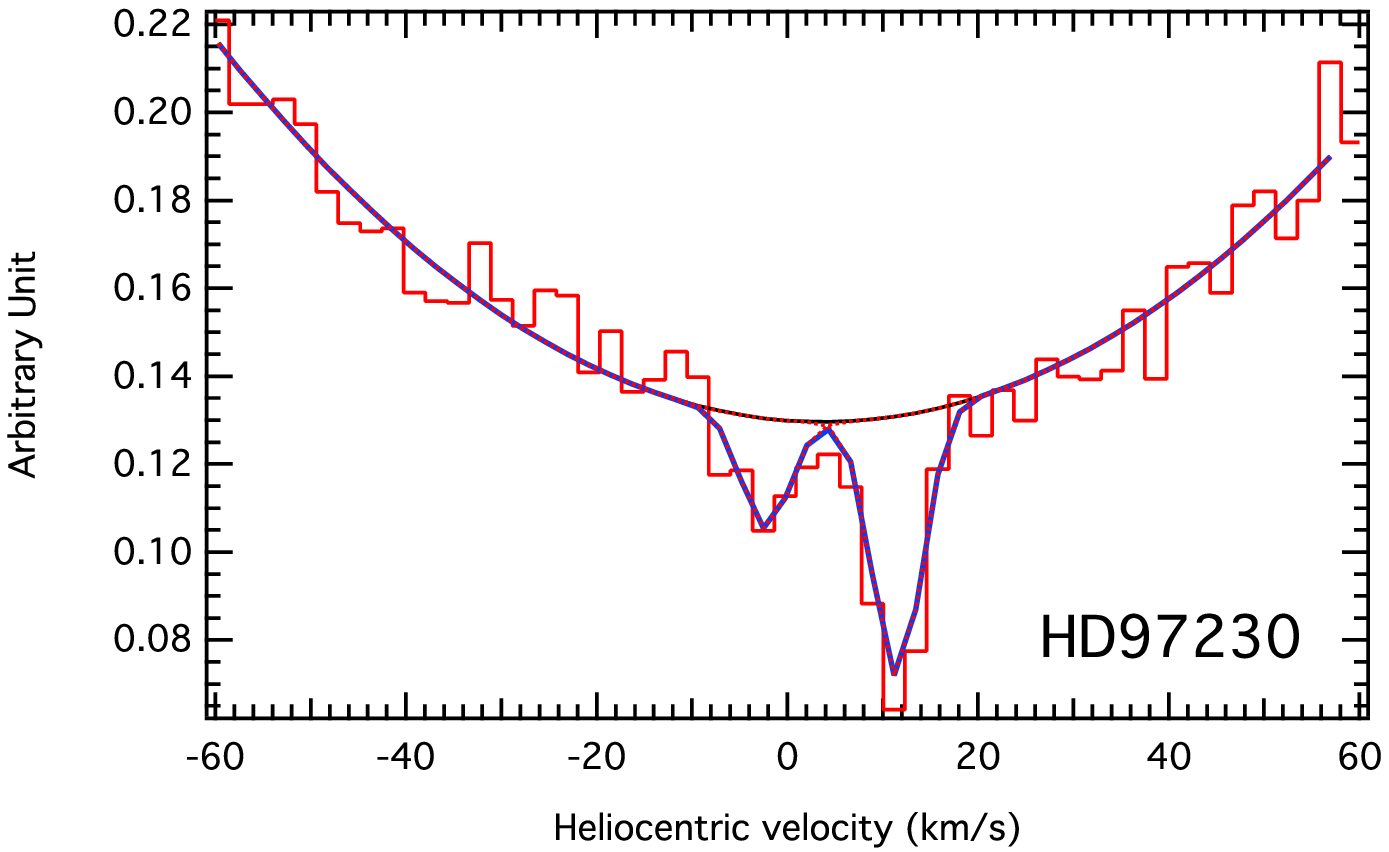}
  	\includegraphics[width=1\linewidth]{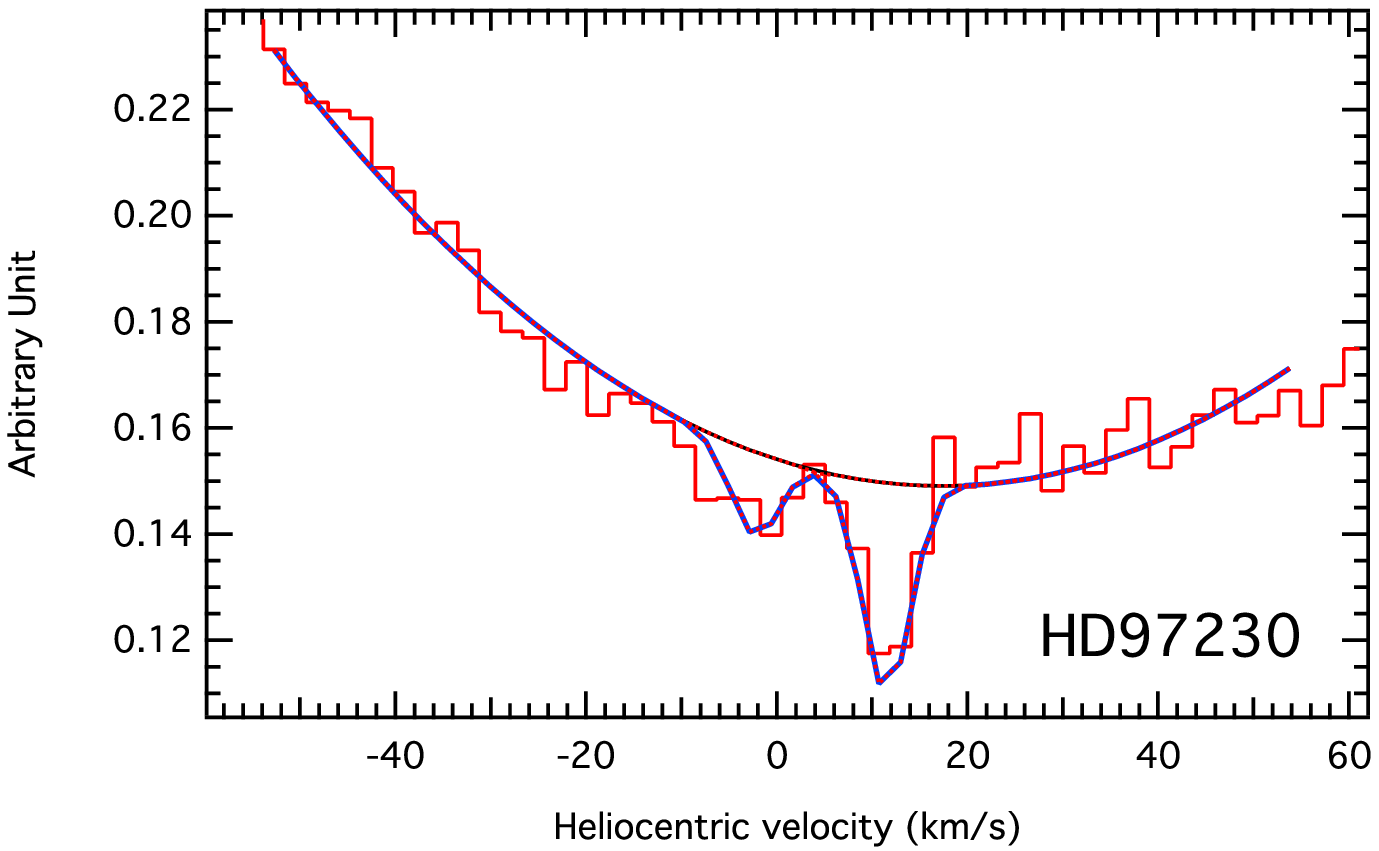}
\end{minipage}
\caption{Same as Fig. \ref{HD94194} {(in the article)}  but for interstellar CaII absorption for target stars: HD96398, HD96399, and HD97230}
\end{figure*}

\begin{figure*}
\begin{minipage}[t]{0.3\linewidth}
\centering
  	\includegraphics[width=1\linewidth]{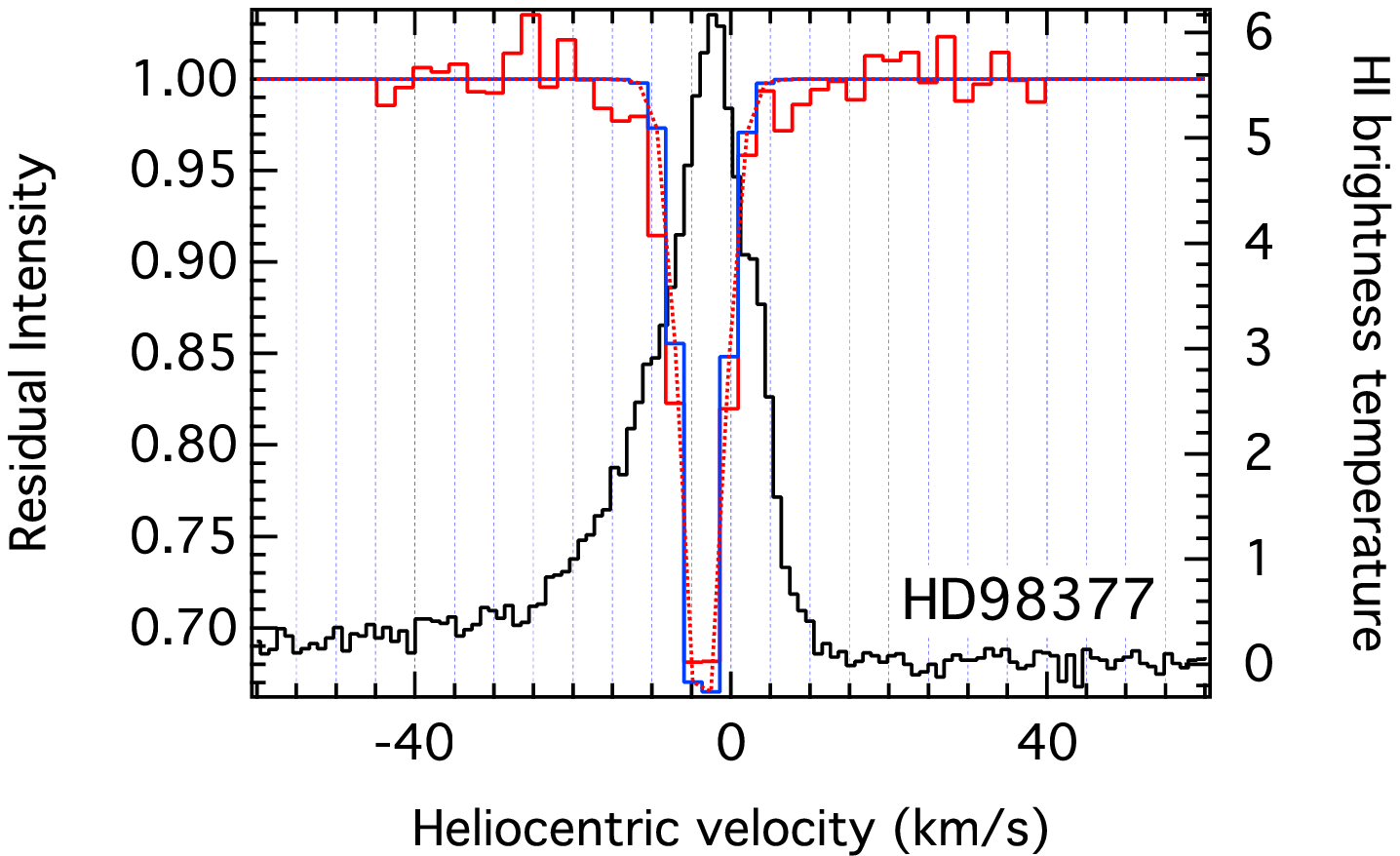}
  	\includegraphics[width=1\linewidth]{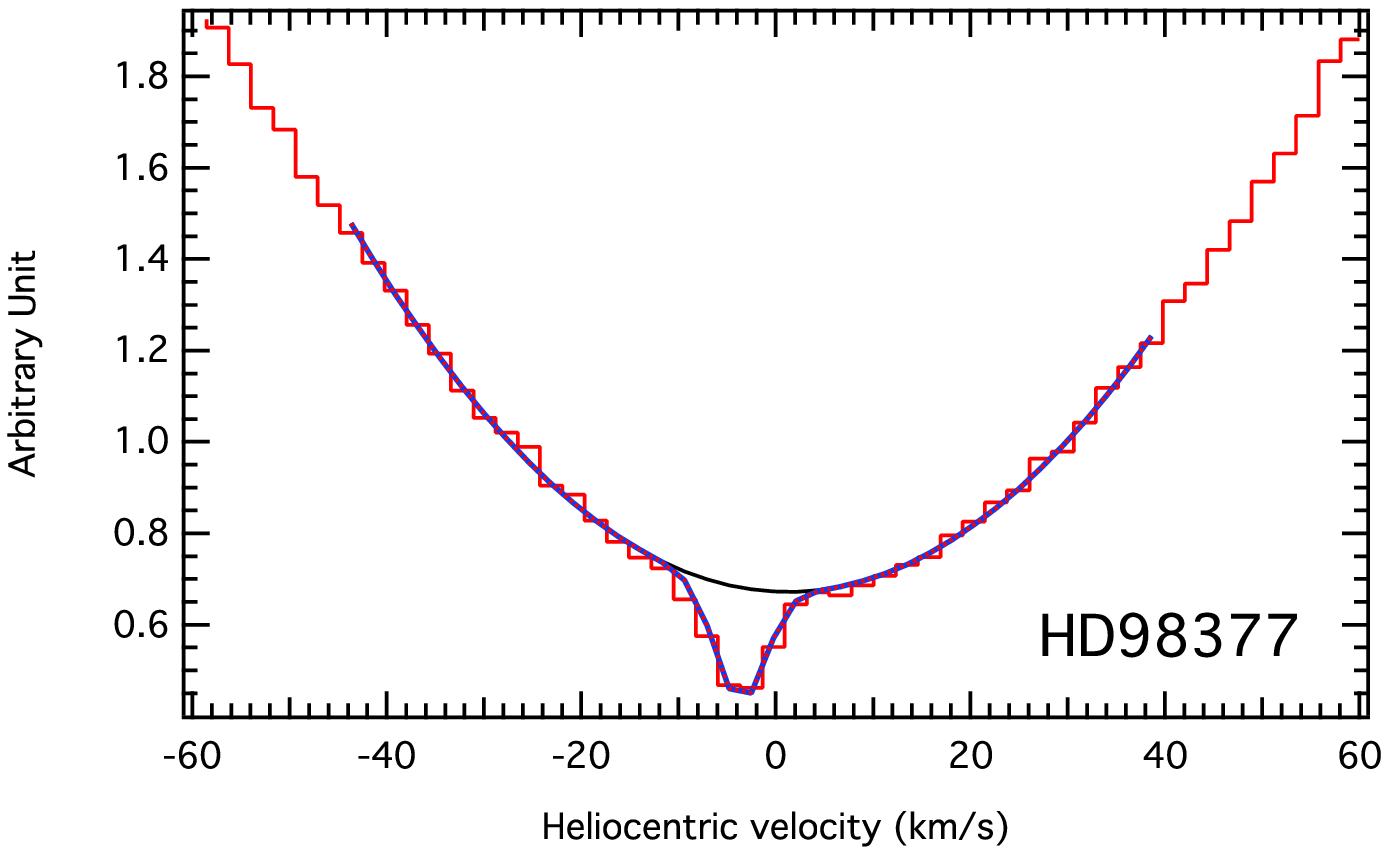}
  	\includegraphics[width=1\linewidth]{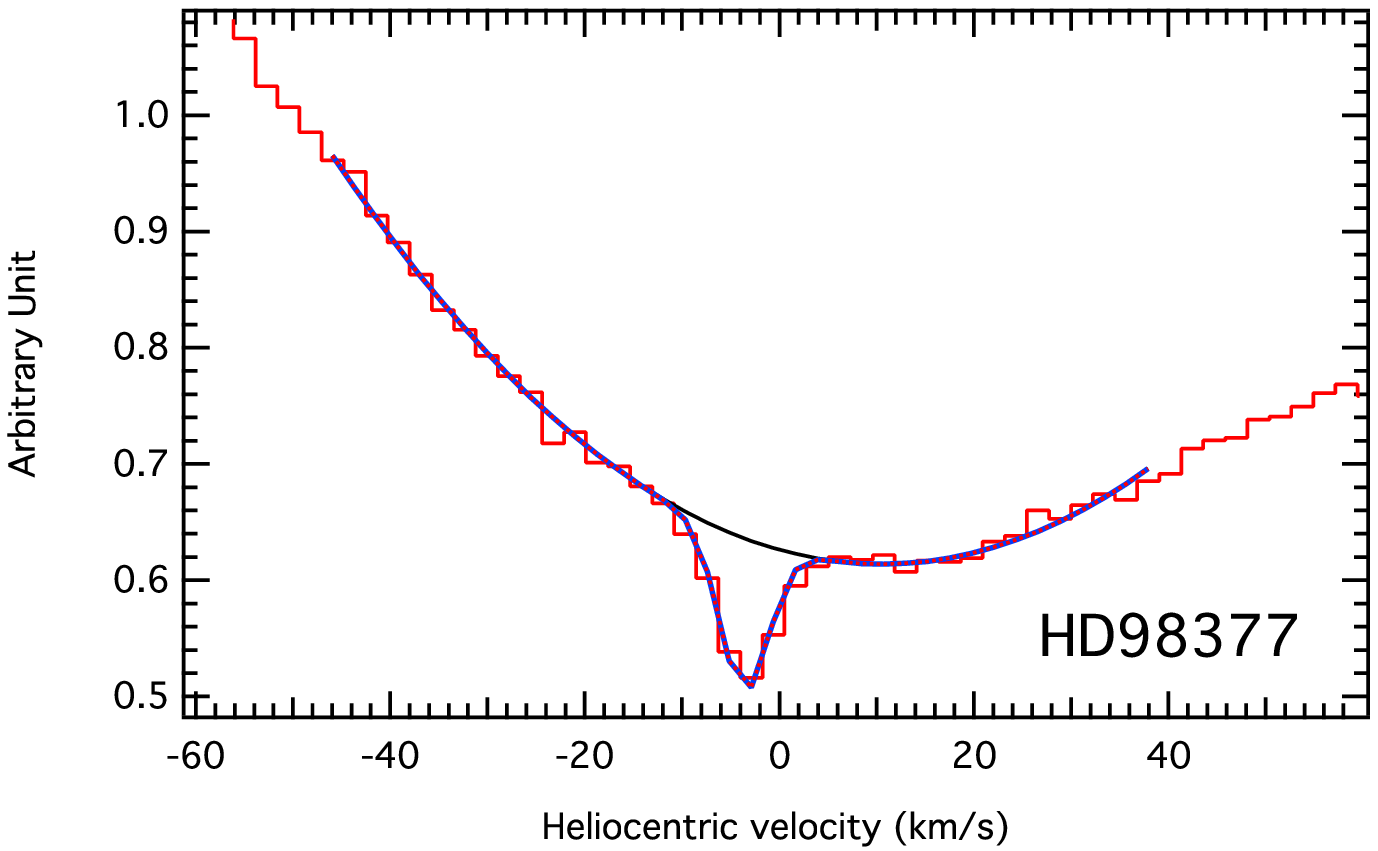}
\end{minipage}\hfill
\begin{minipage}[t]{0.3\linewidth}
\centering
  	\includegraphics[width=1\linewidth]{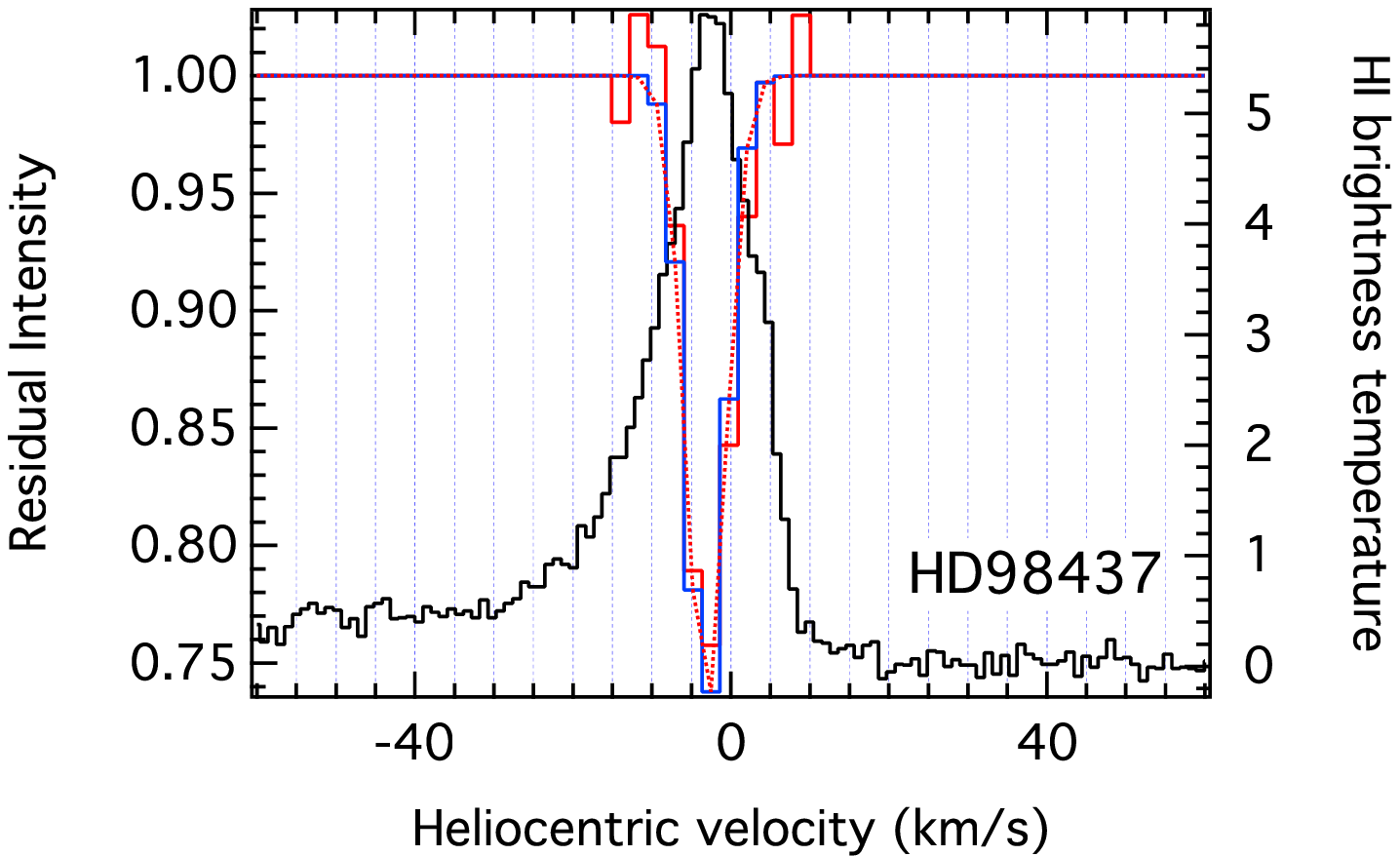}
  	\includegraphics[width=1\linewidth]{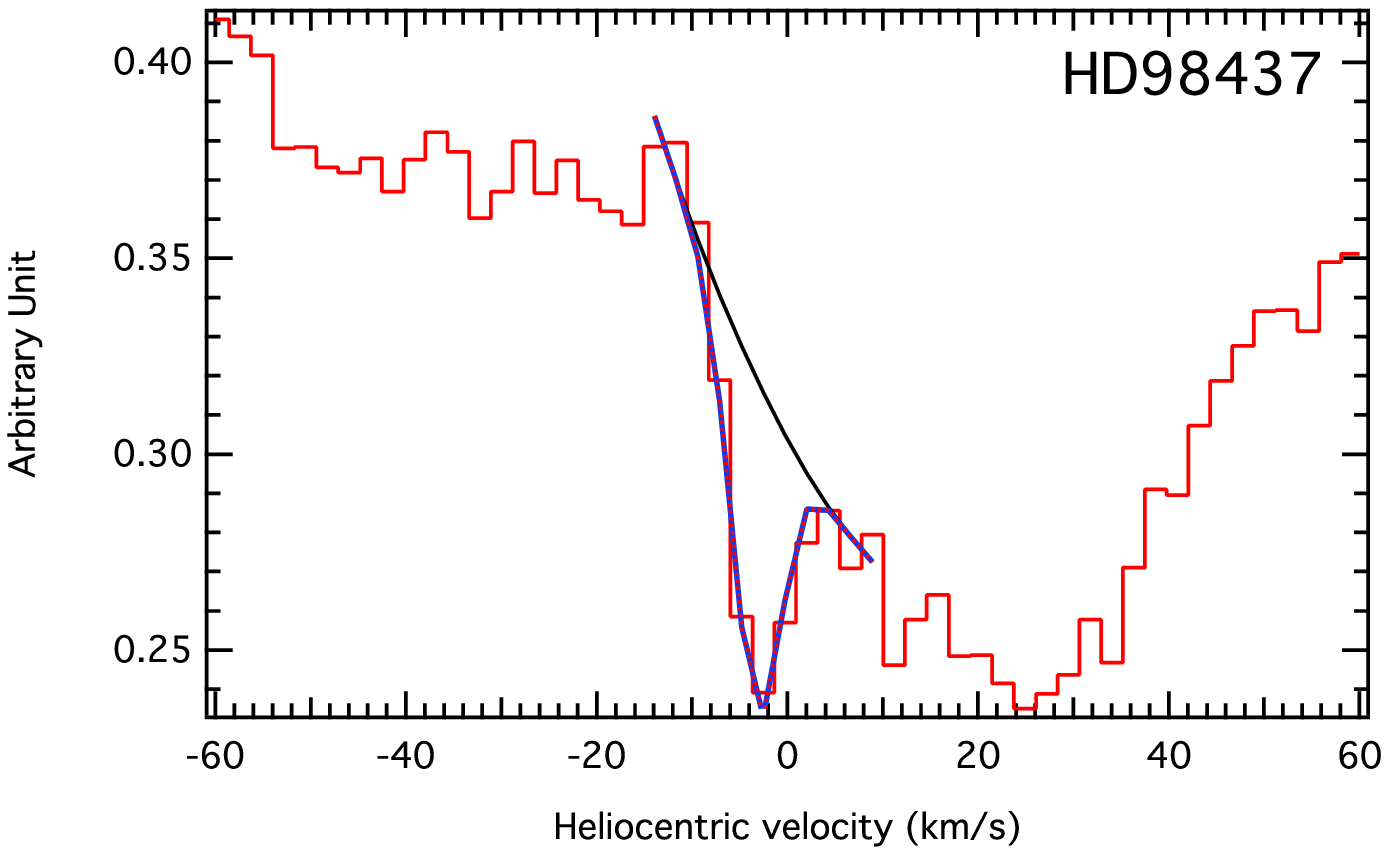}
  	\includegraphics[width=1\linewidth]{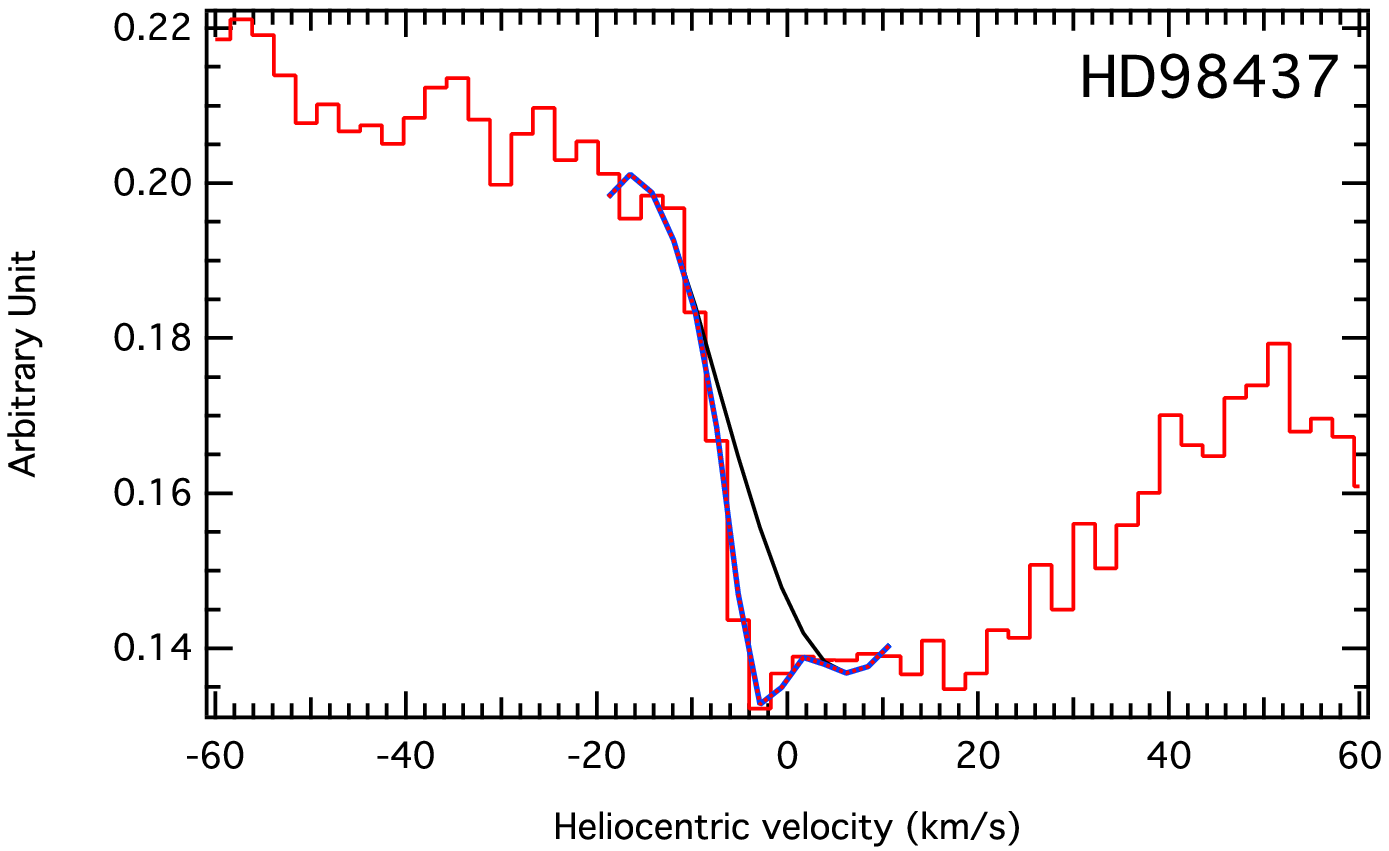}
\end{minipage}\hfill
\begin{minipage}[t]{0.3\linewidth}
\centering
  	\includegraphics[width=1\linewidth]{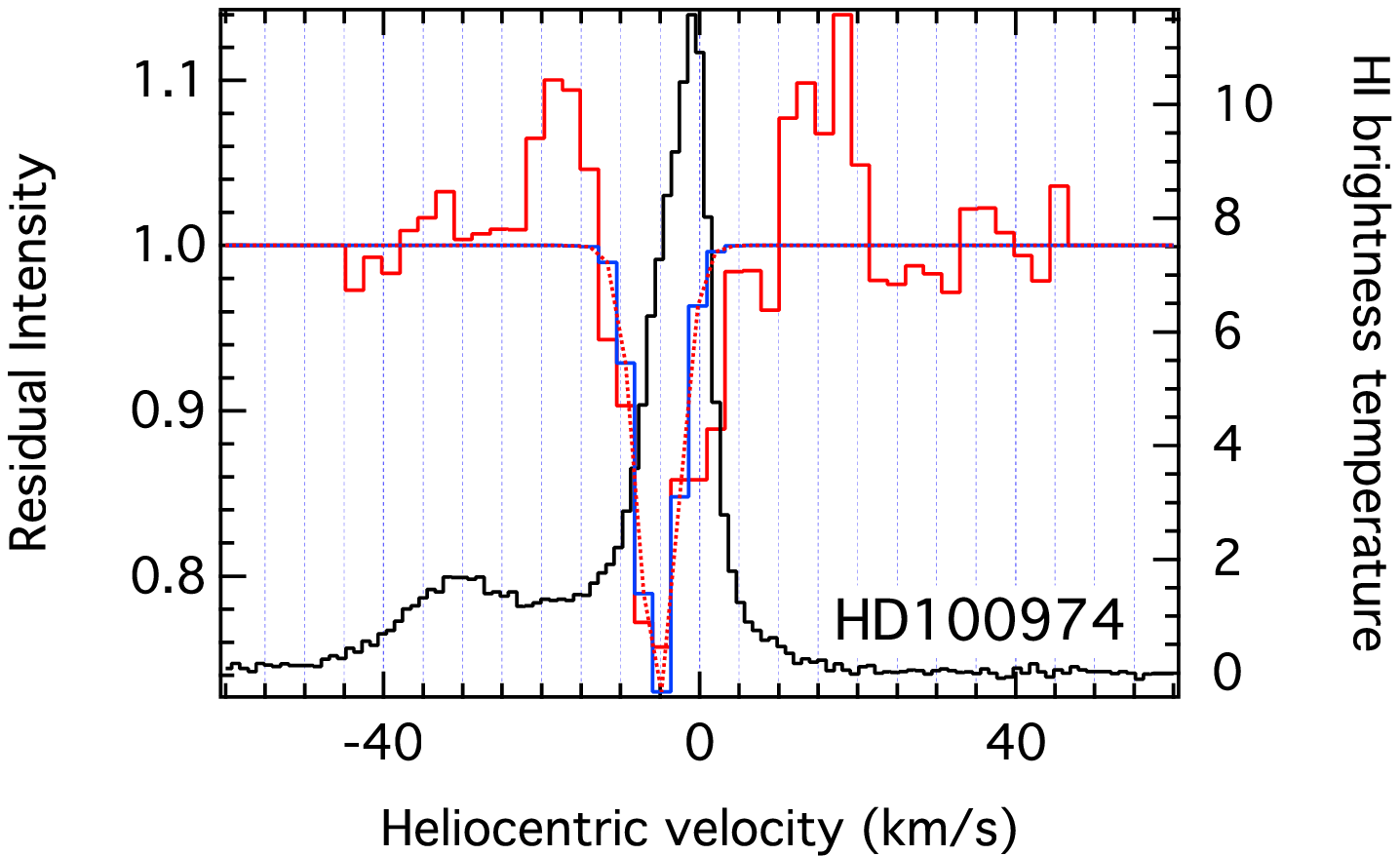}
  	\includegraphics[width=1\linewidth]{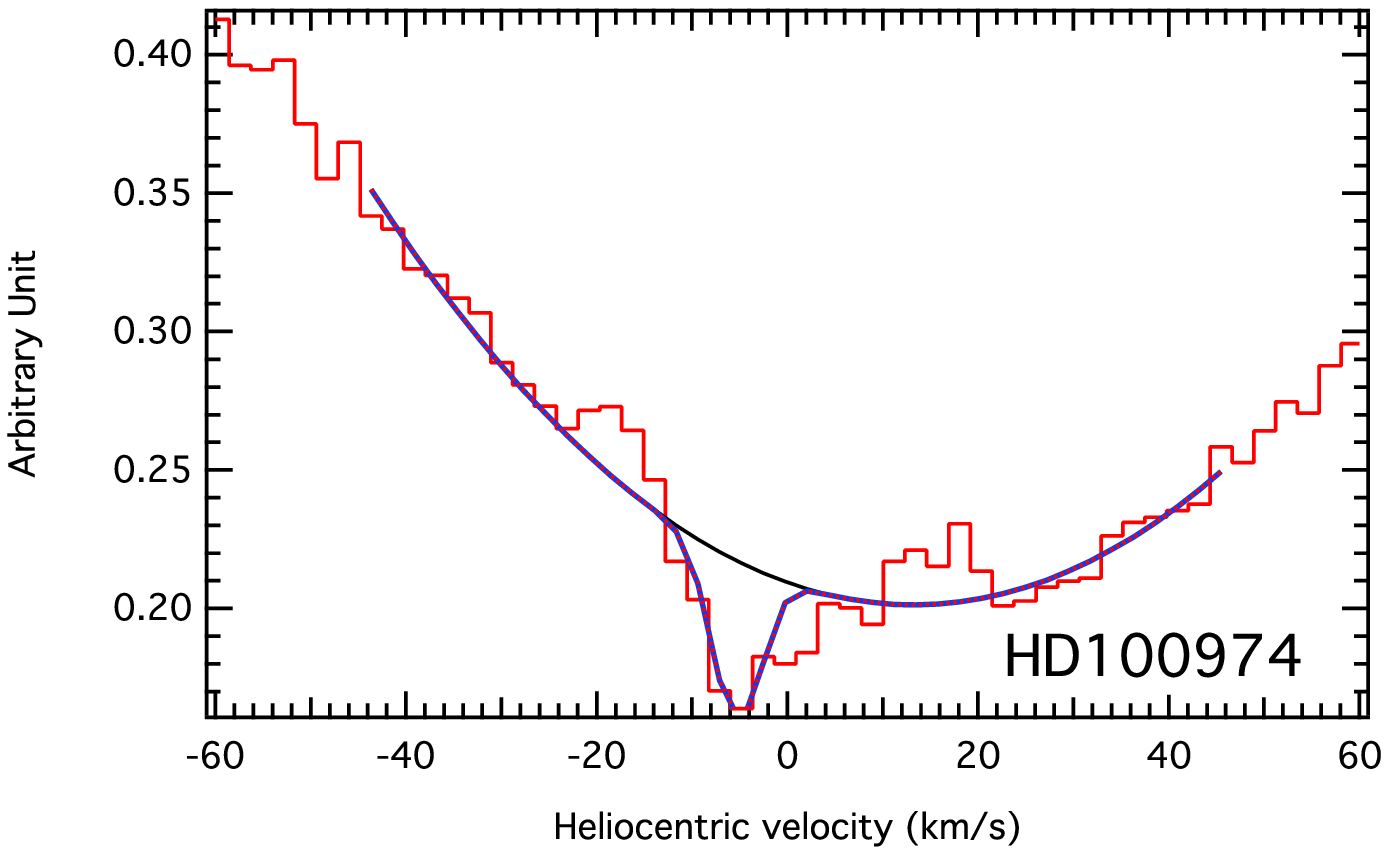}
  	\includegraphics[width=1\linewidth]{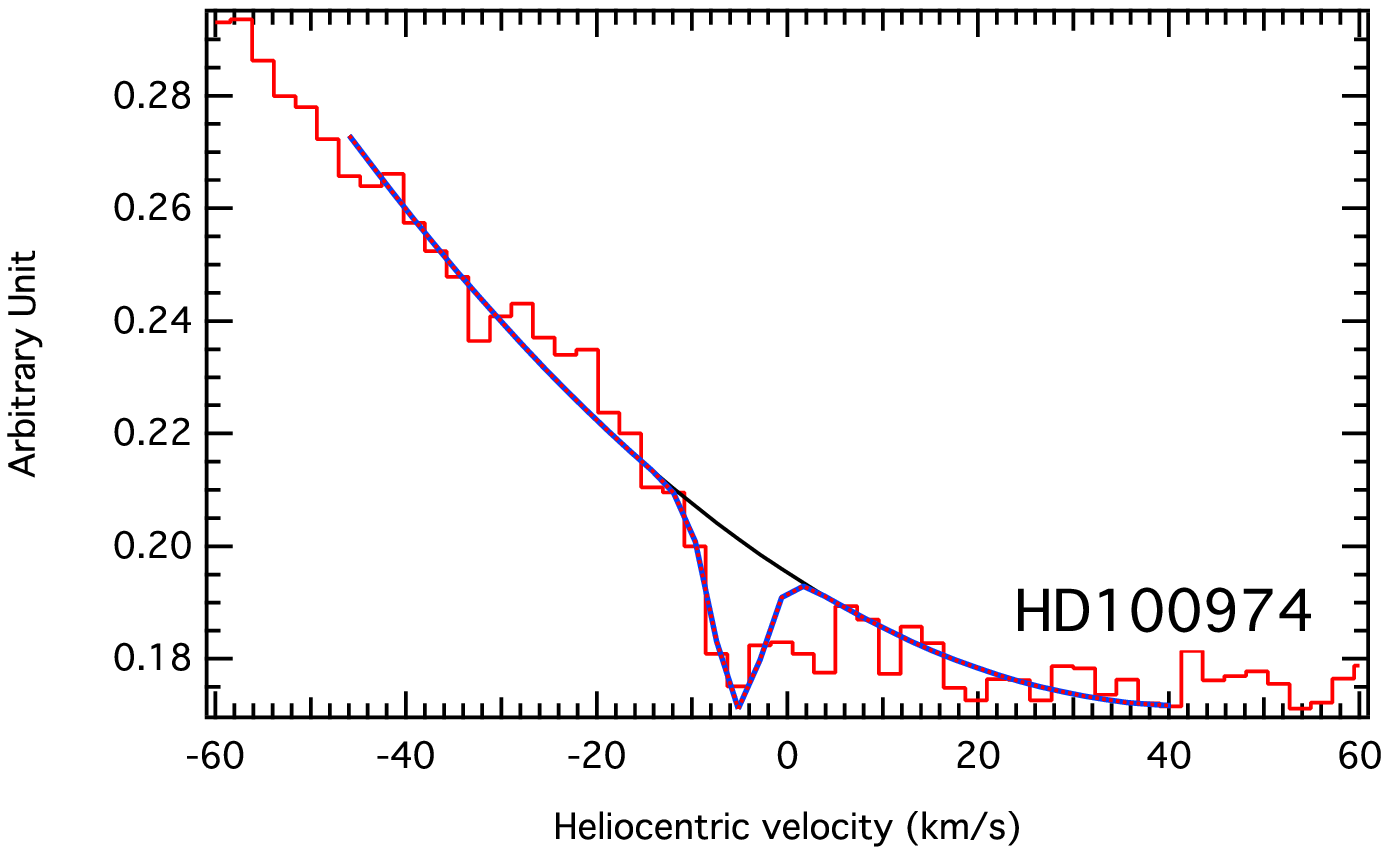}
\end{minipage}
\caption{Same as Fig. \ref{HD94194} {(in the article)}  but for interstellar CaII absorption for target stars: HD98377, HD98437, and HD100974}
\end{figure*}

\begin{figure*}
\begin{minipage}[t]{0.3\linewidth}
\centering
  	\includegraphics[width=1\linewidth]{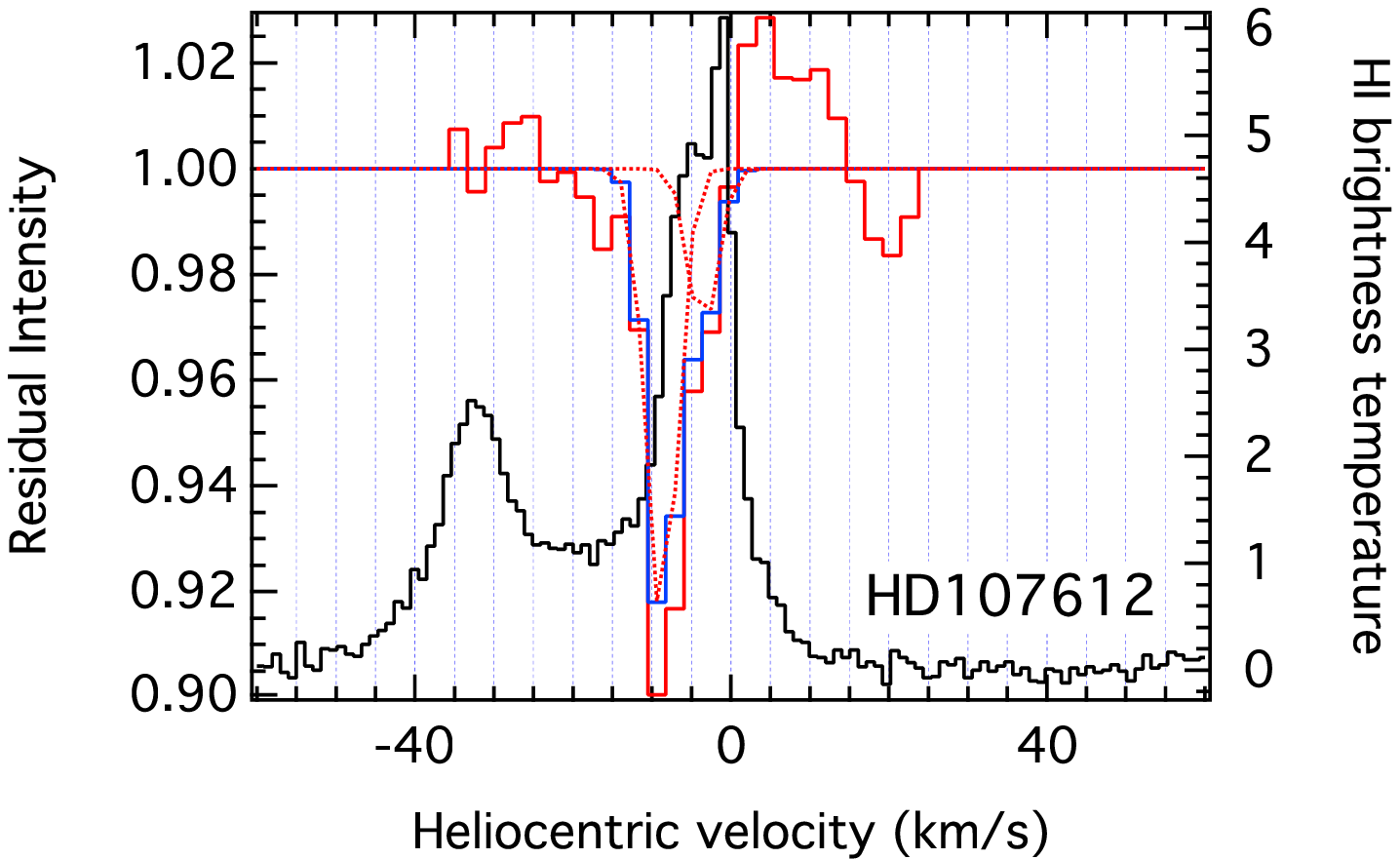}
  	\includegraphics[width=1\linewidth]{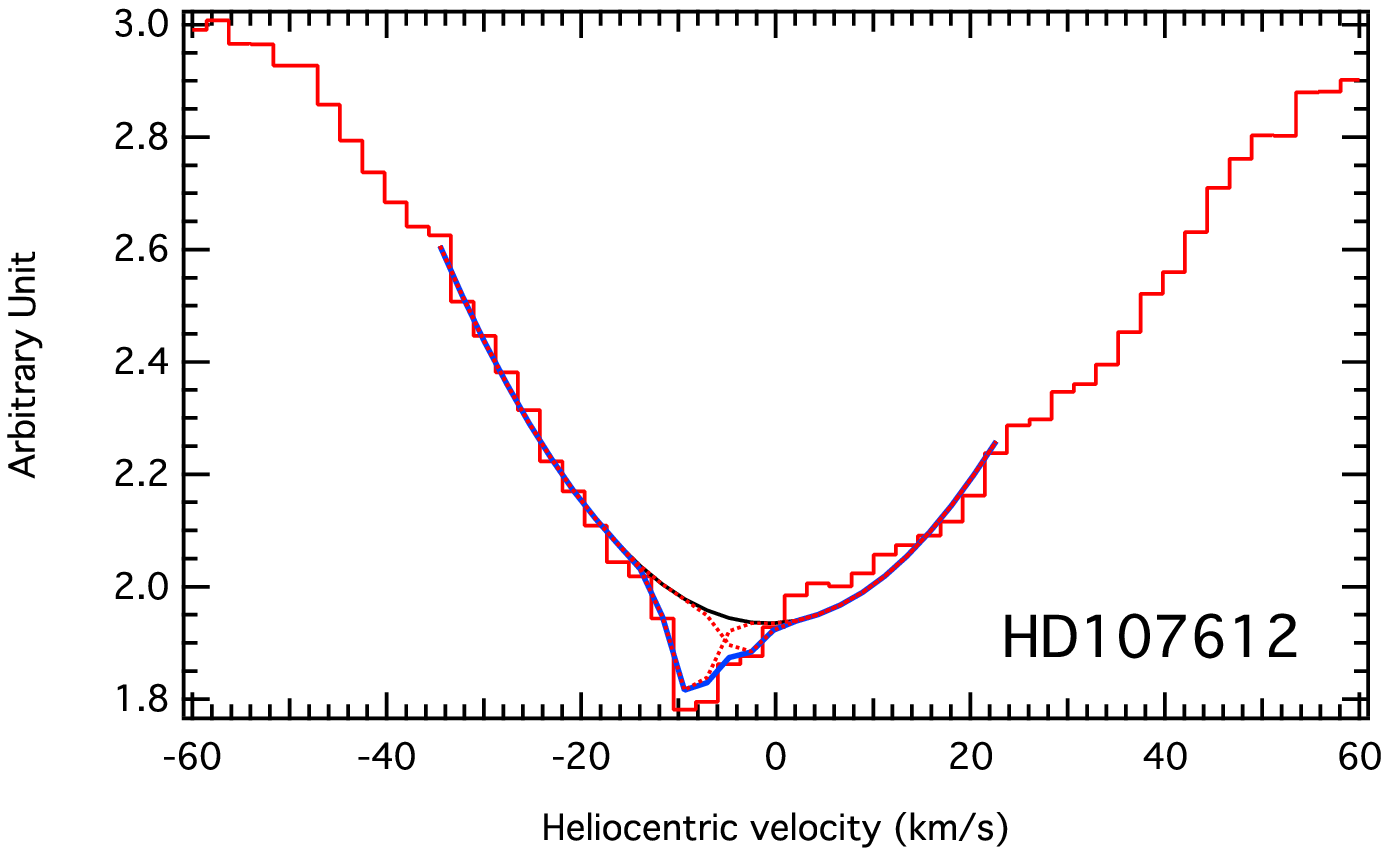}
  	\includegraphics[width=1\linewidth]{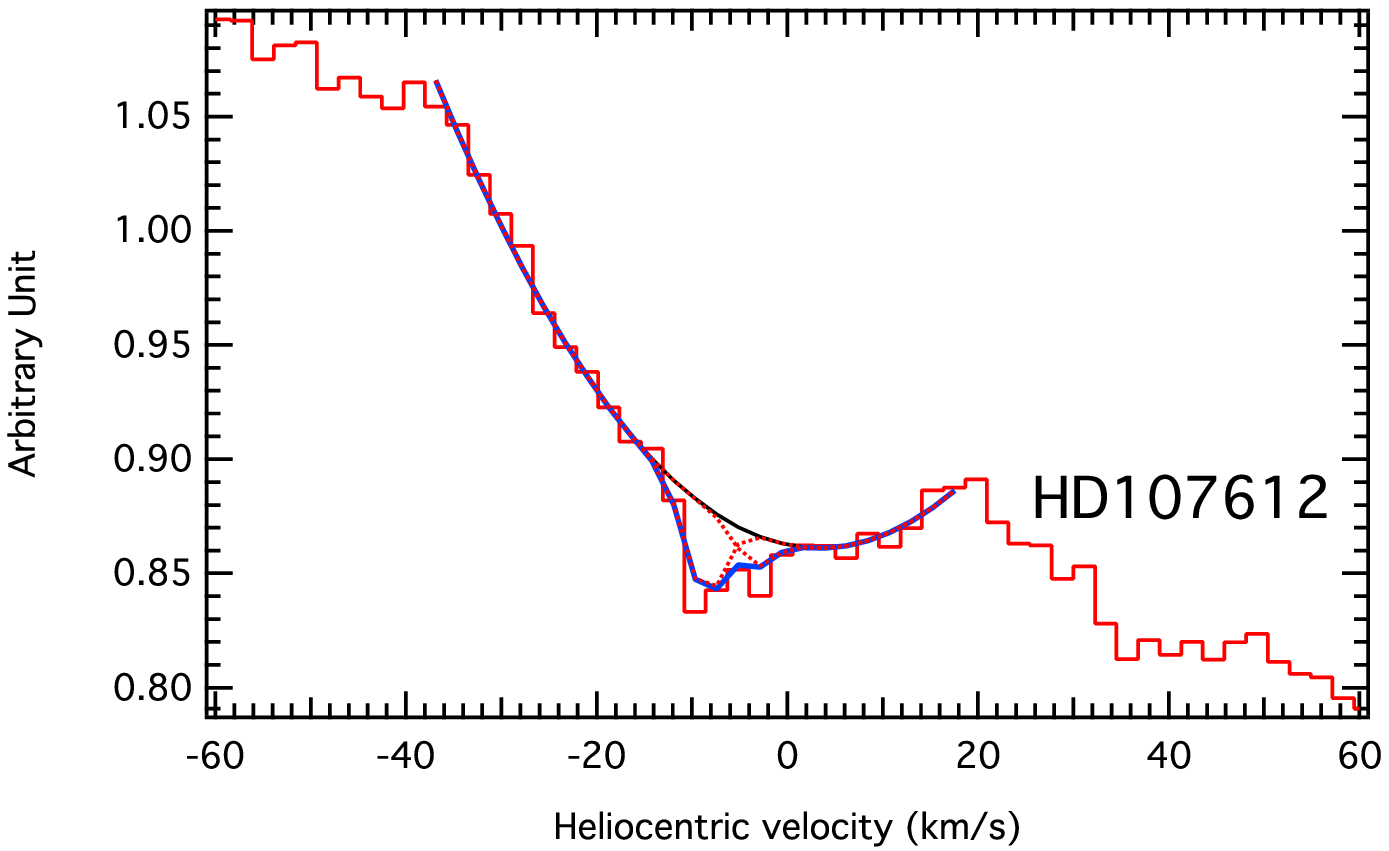}
\end{minipage}\hfill
\begin{minipage}[t]{0.3\linewidth}
\centering
  	\includegraphics[width=1\linewidth]{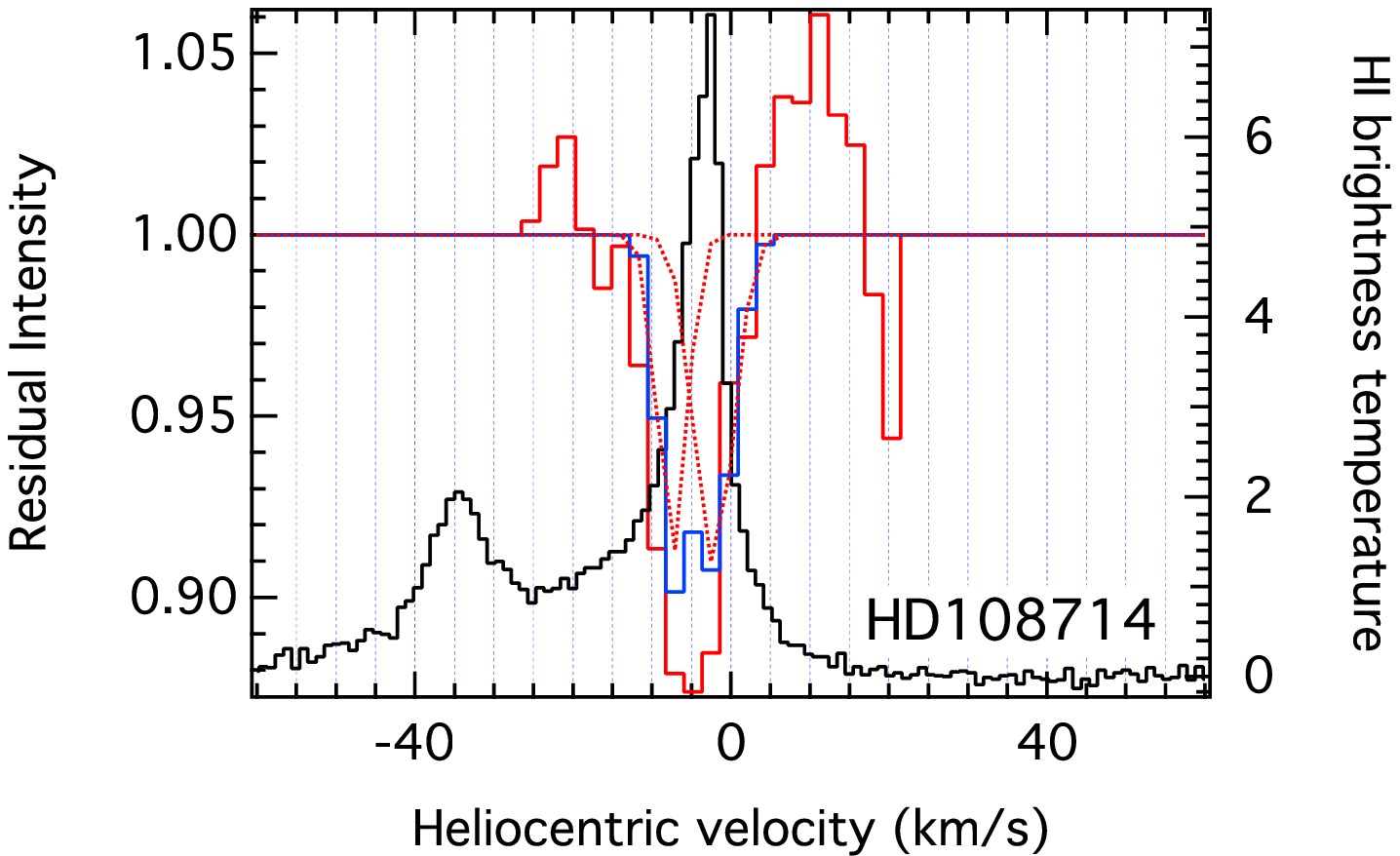}
  	\includegraphics[width=1\linewidth]{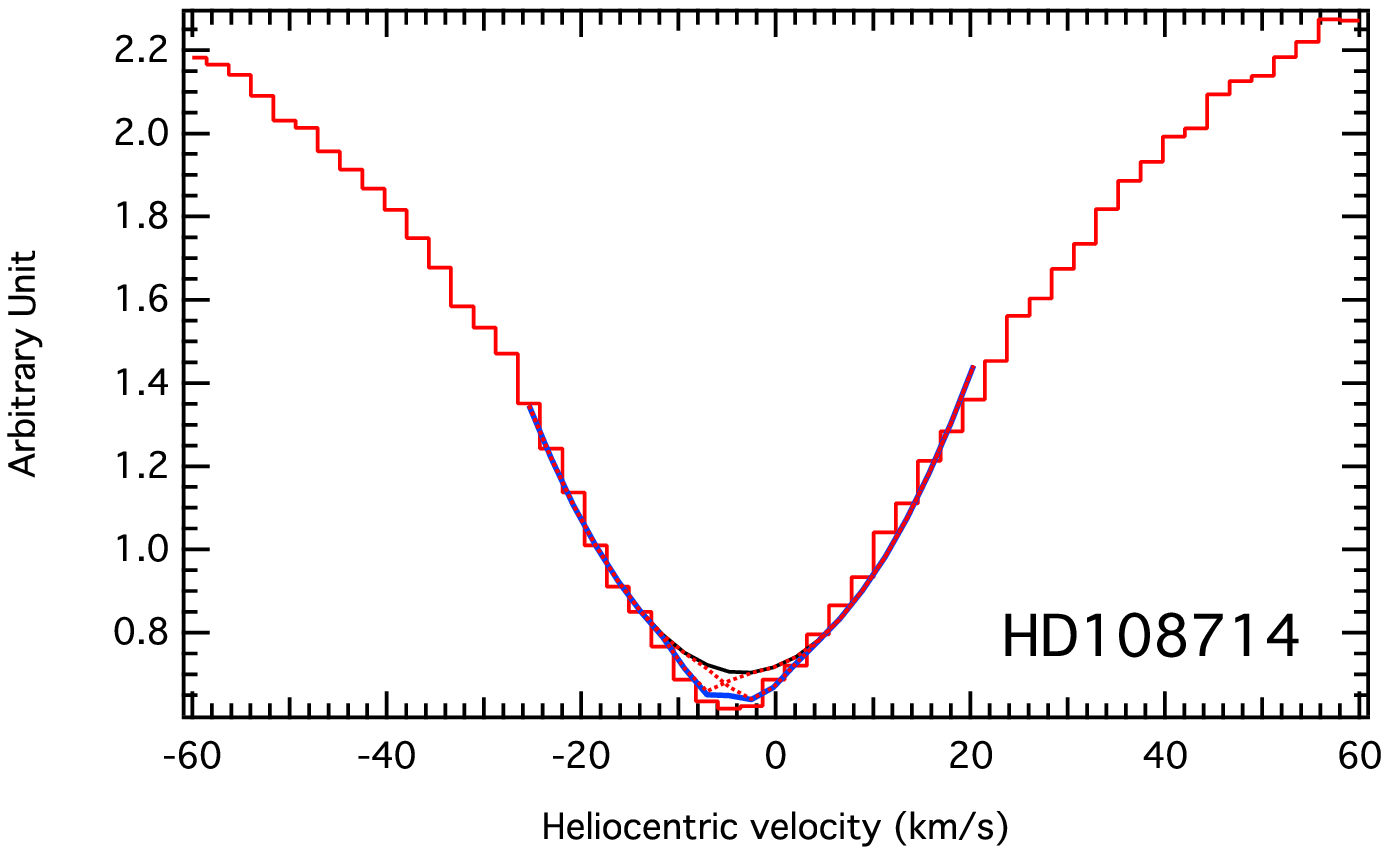}
  	\includegraphics[width=1\linewidth]{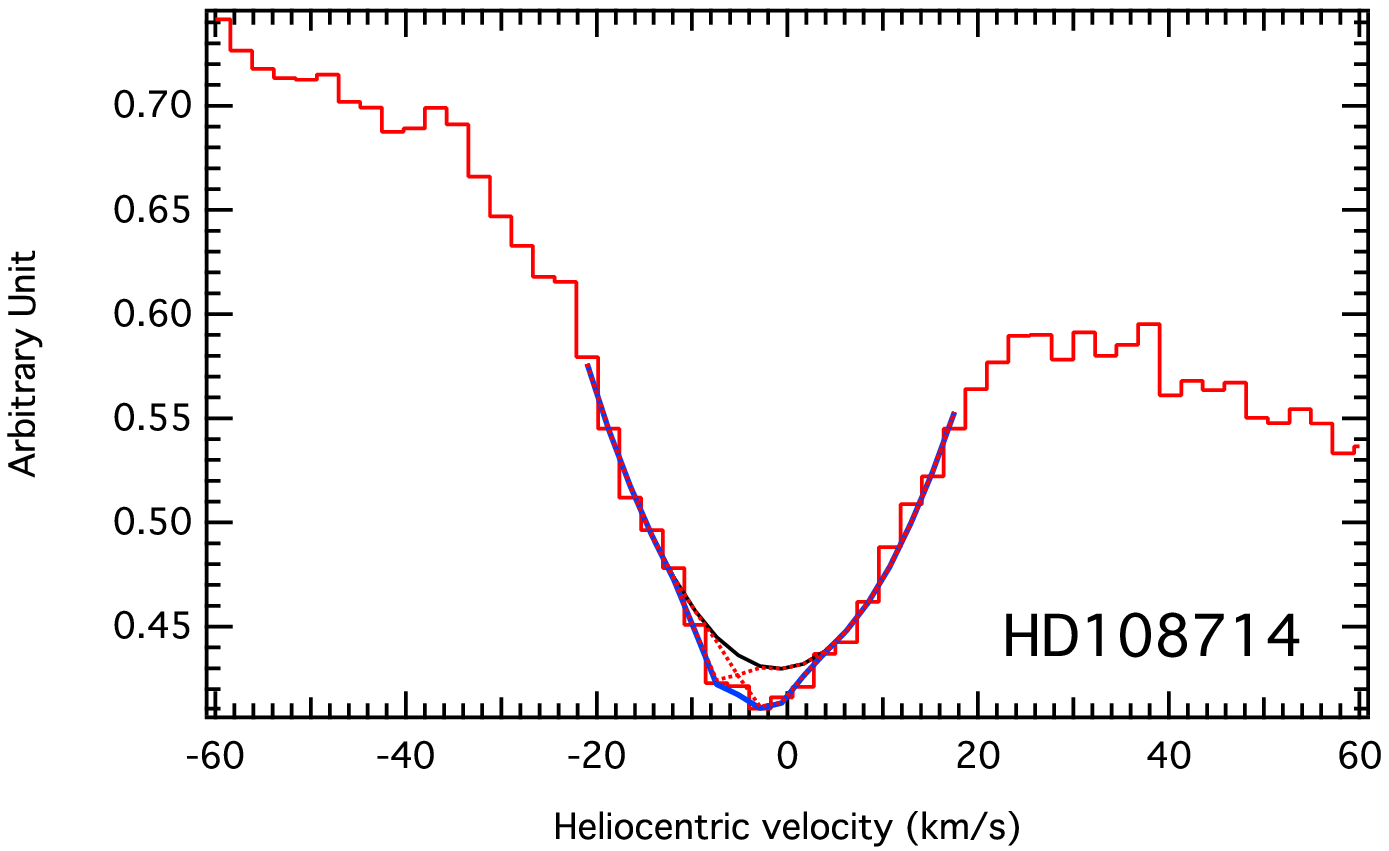}
\end{minipage}\hfill
\begin{minipage}[t]{0.3\linewidth}
\centering
  	\includegraphics[width=1\linewidth]{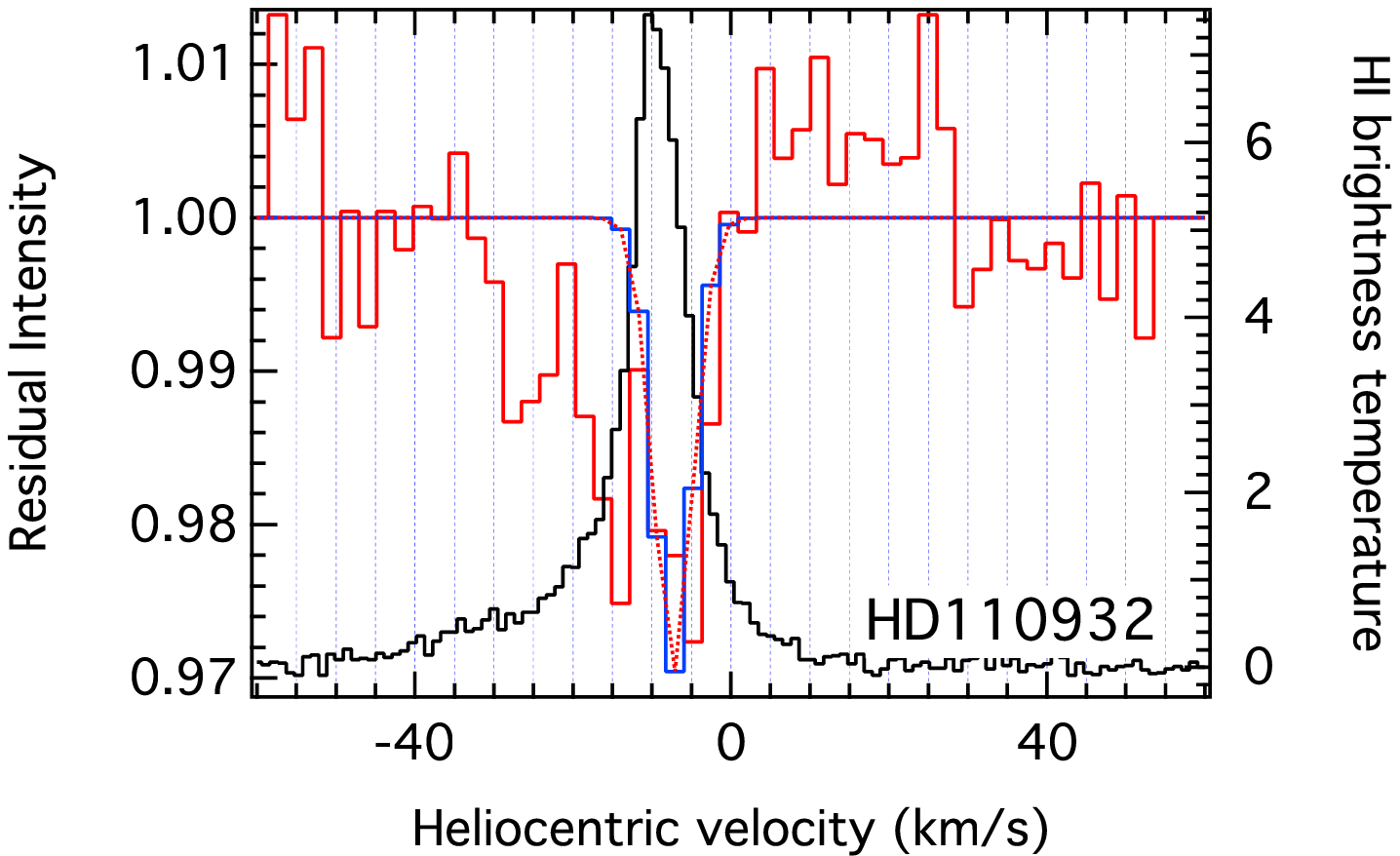}
  	\includegraphics[width=1\linewidth]{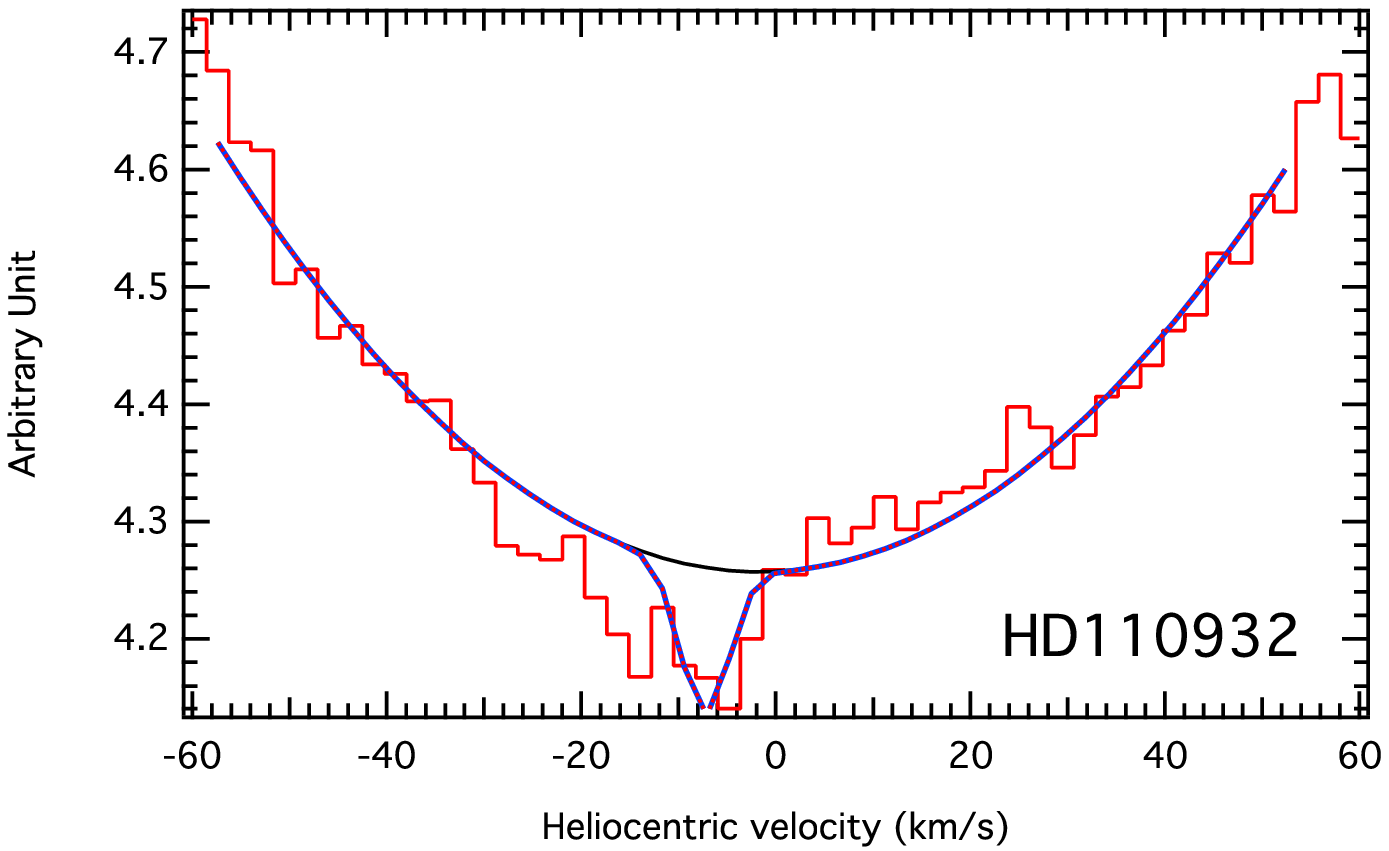}
  	\includegraphics[width=1\linewidth]{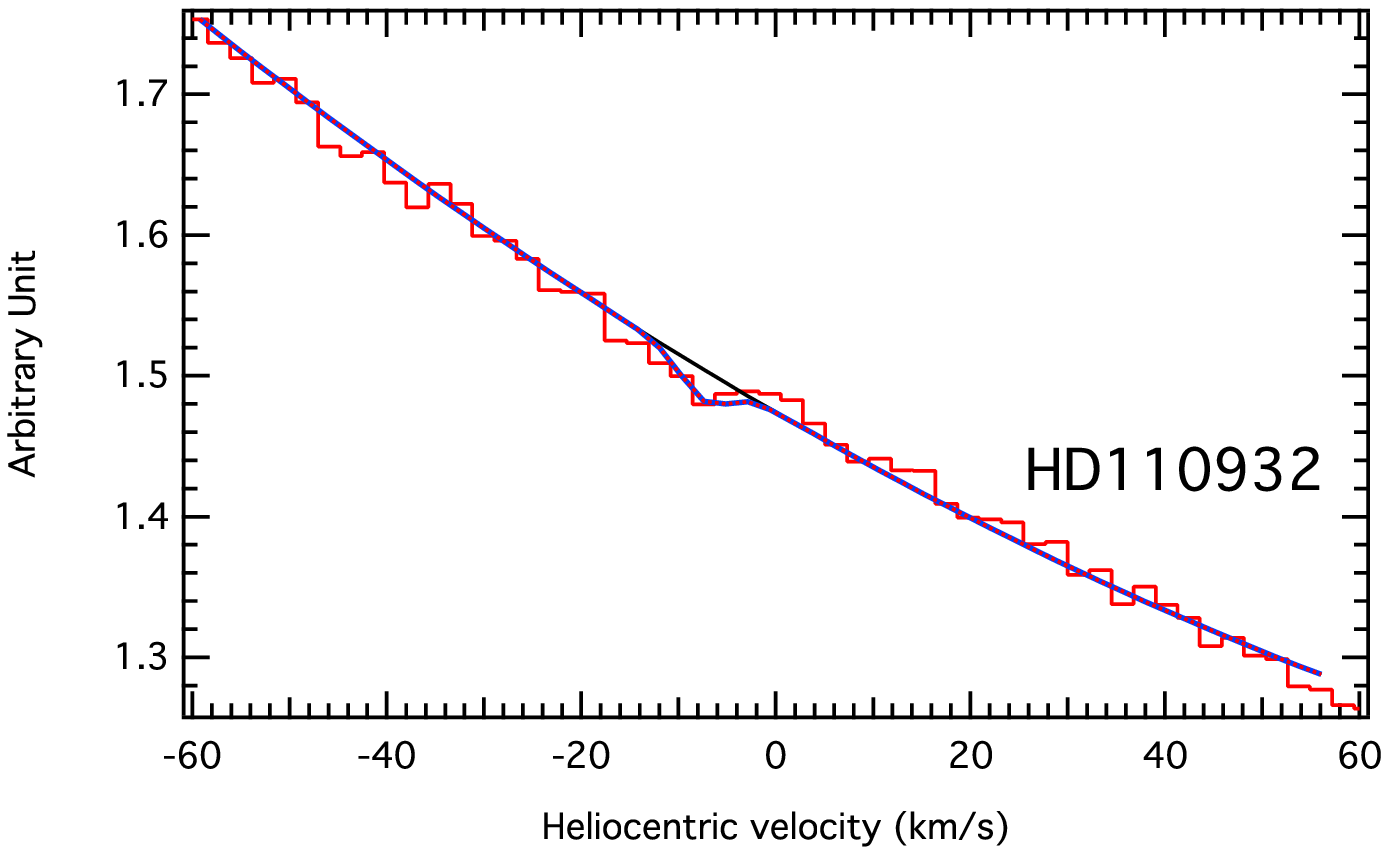}
\end{minipage}
\caption{Same as Fig. \ref{HD94194} {(in the article)}  but for interstellar CaII absorption for target stars: HD107612, HD108714, and HD110932}
\end{figure*}

\begin{figure*}[!h]
\begin{minipage}[t]{0.3\linewidth}
\centering
  	\includegraphics[width=1\linewidth]{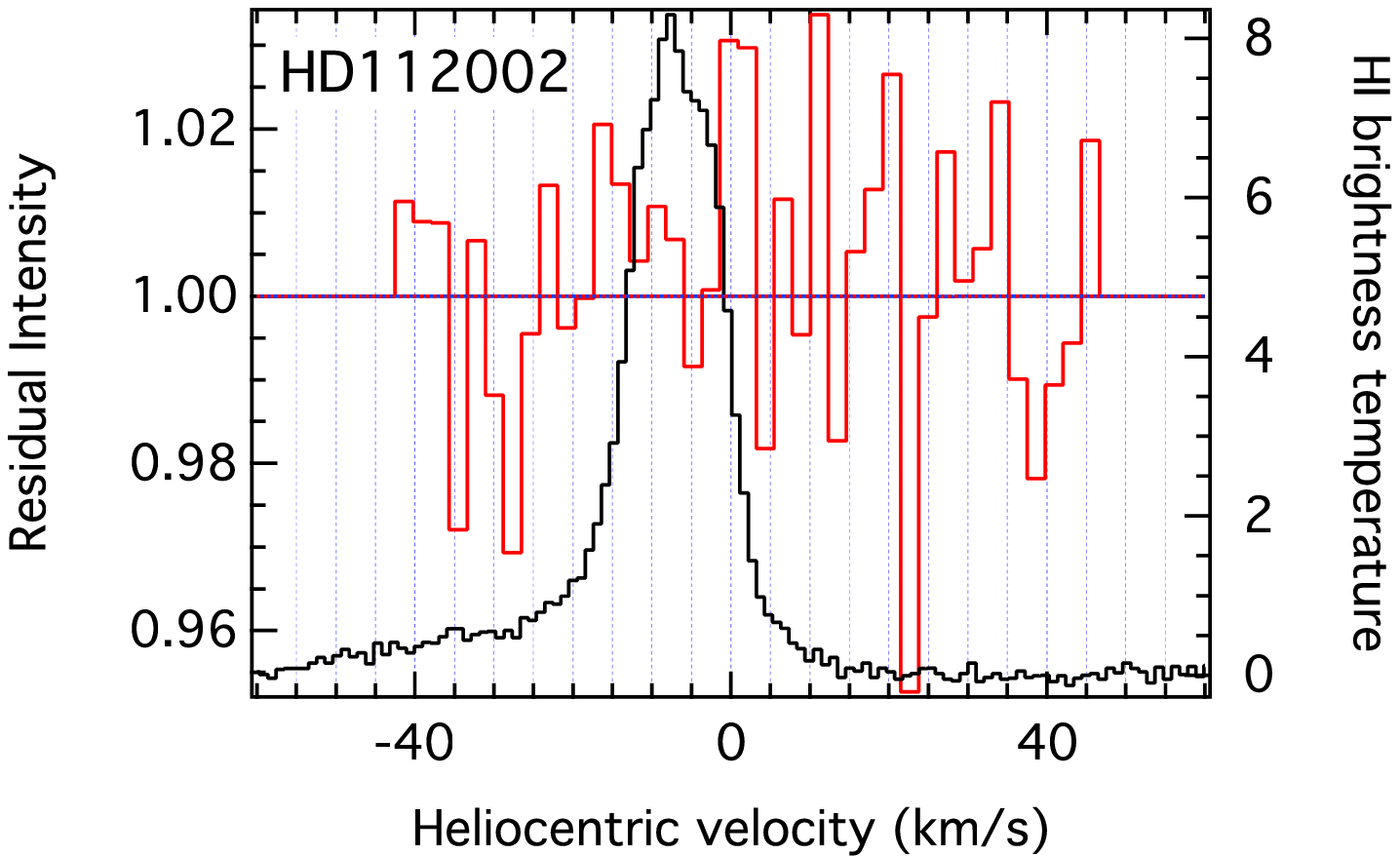}
  	\includegraphics[width=1\linewidth]{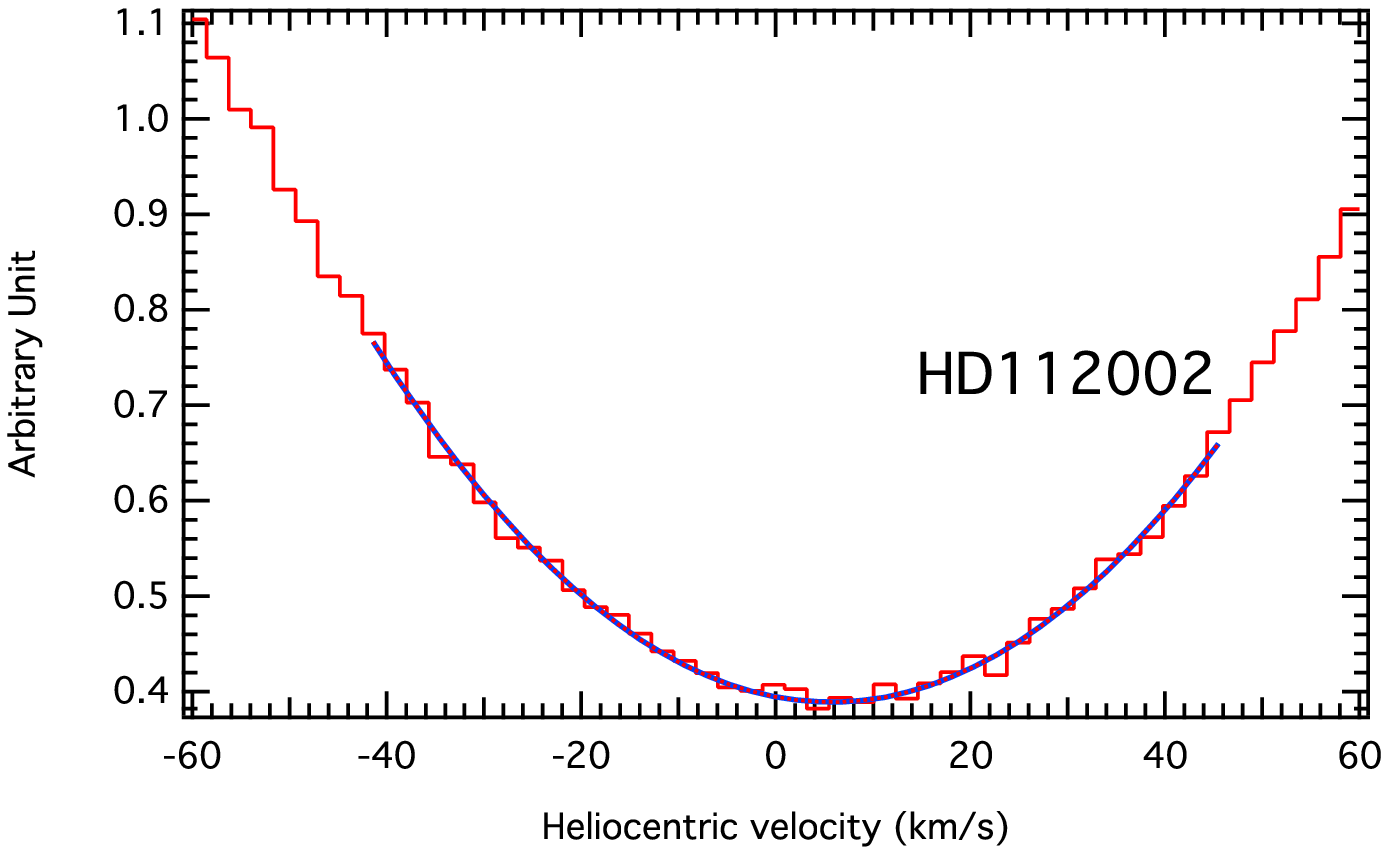}
  	\includegraphics[width=1\linewidth]{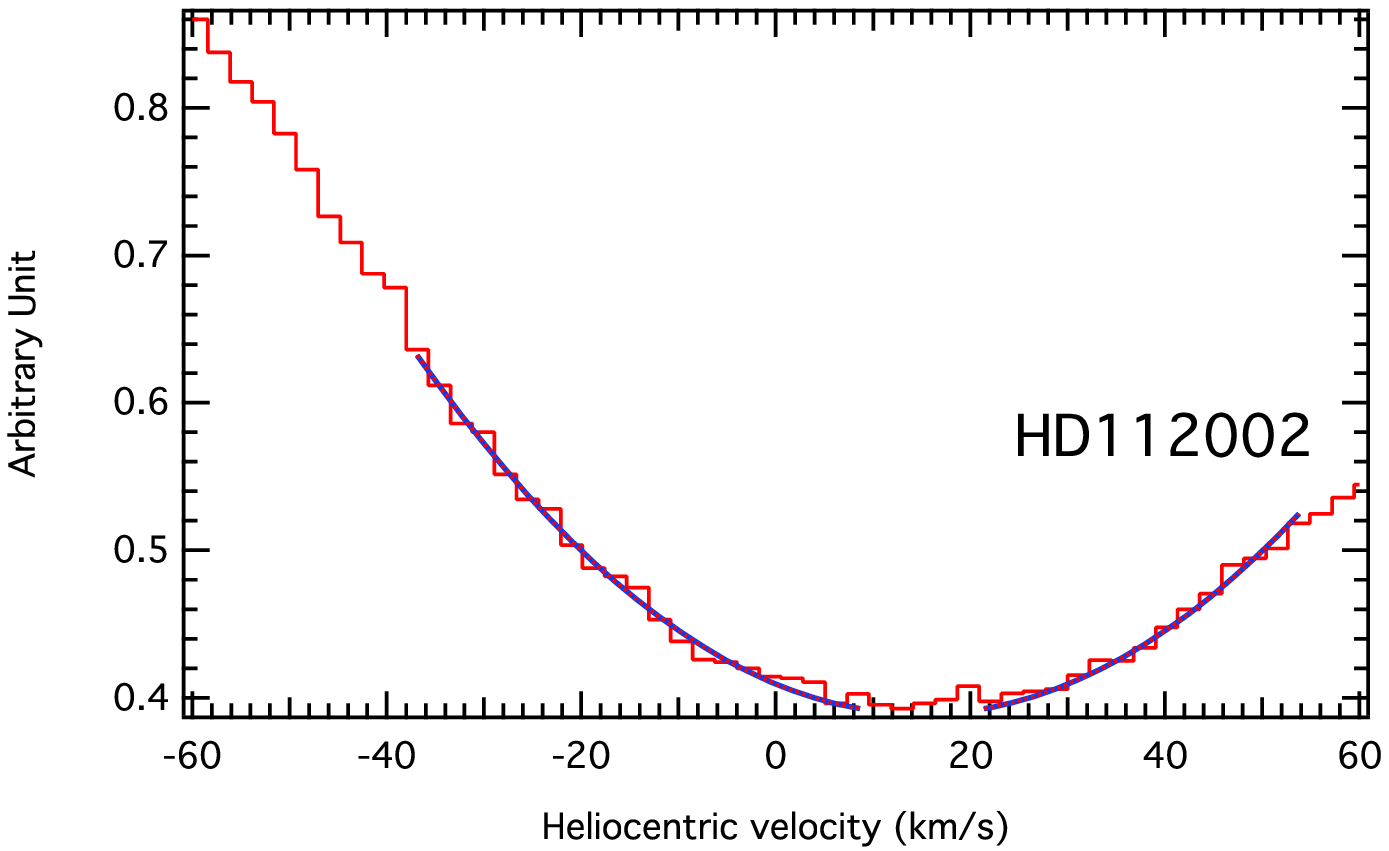}
\end{minipage}\hfill
\begin{minipage}[t]{0.3\linewidth}
\centering
  	\includegraphics[width=1\linewidth]{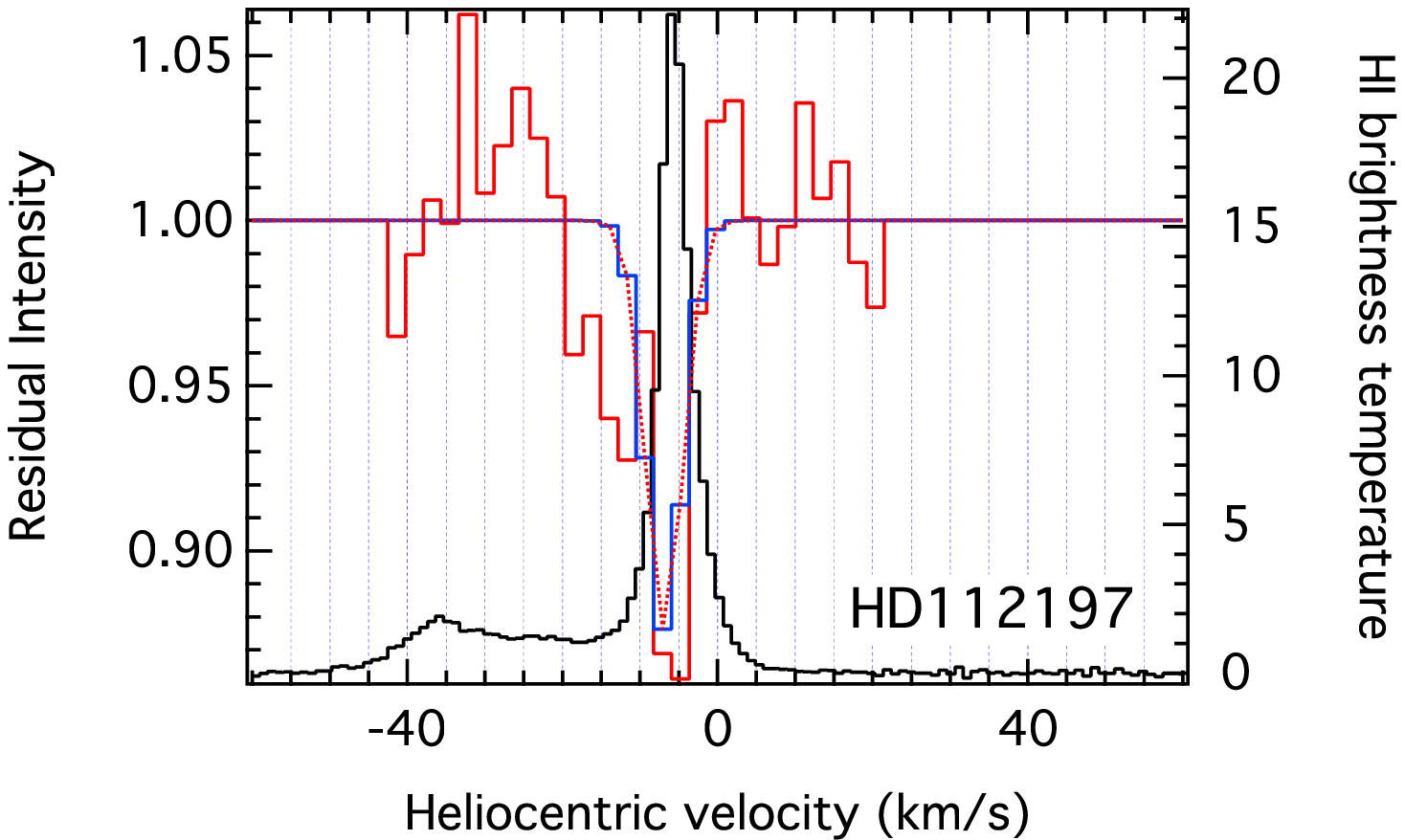}
  	\includegraphics[width=1\linewidth]{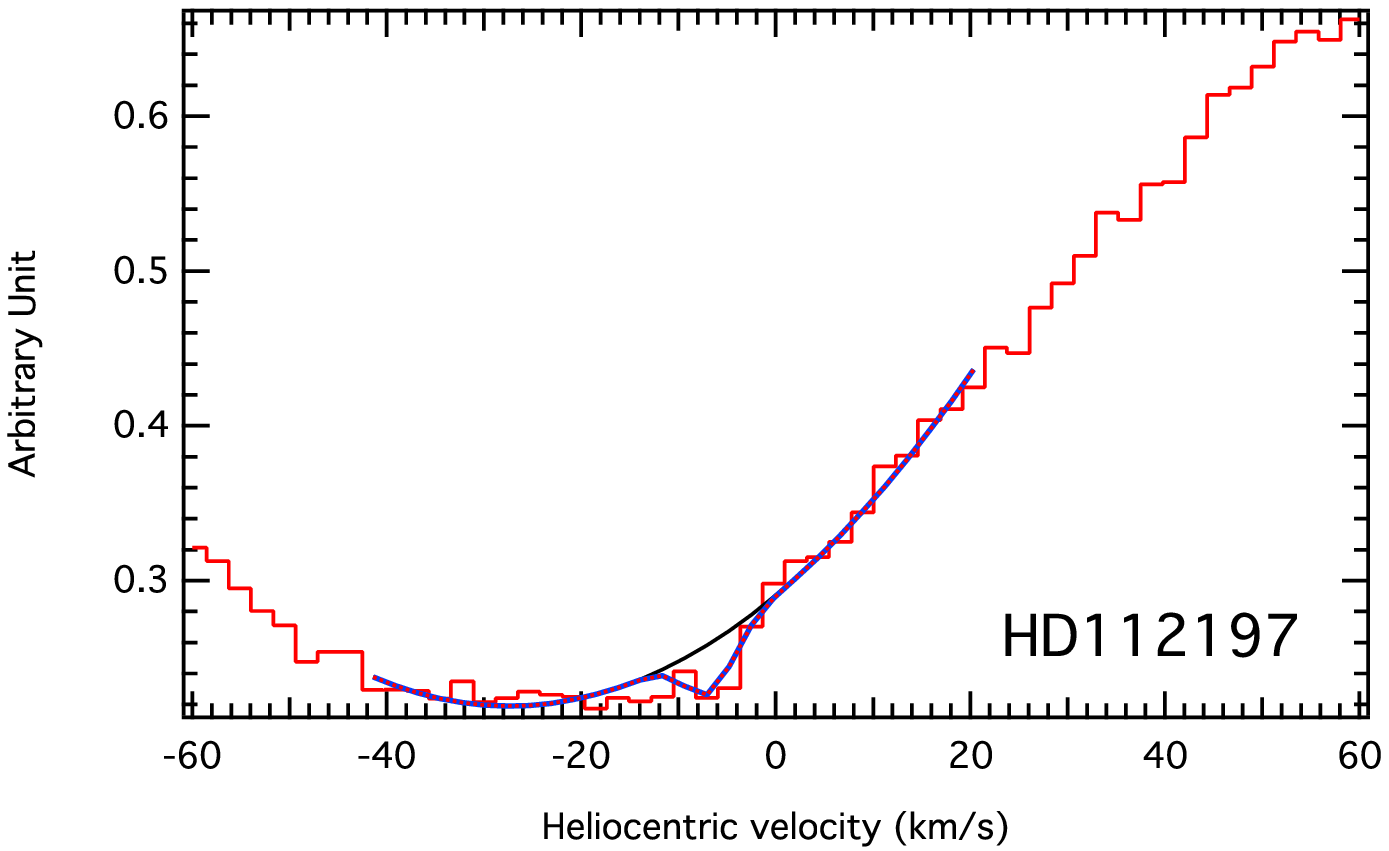}
  	\includegraphics[width=1\linewidth]{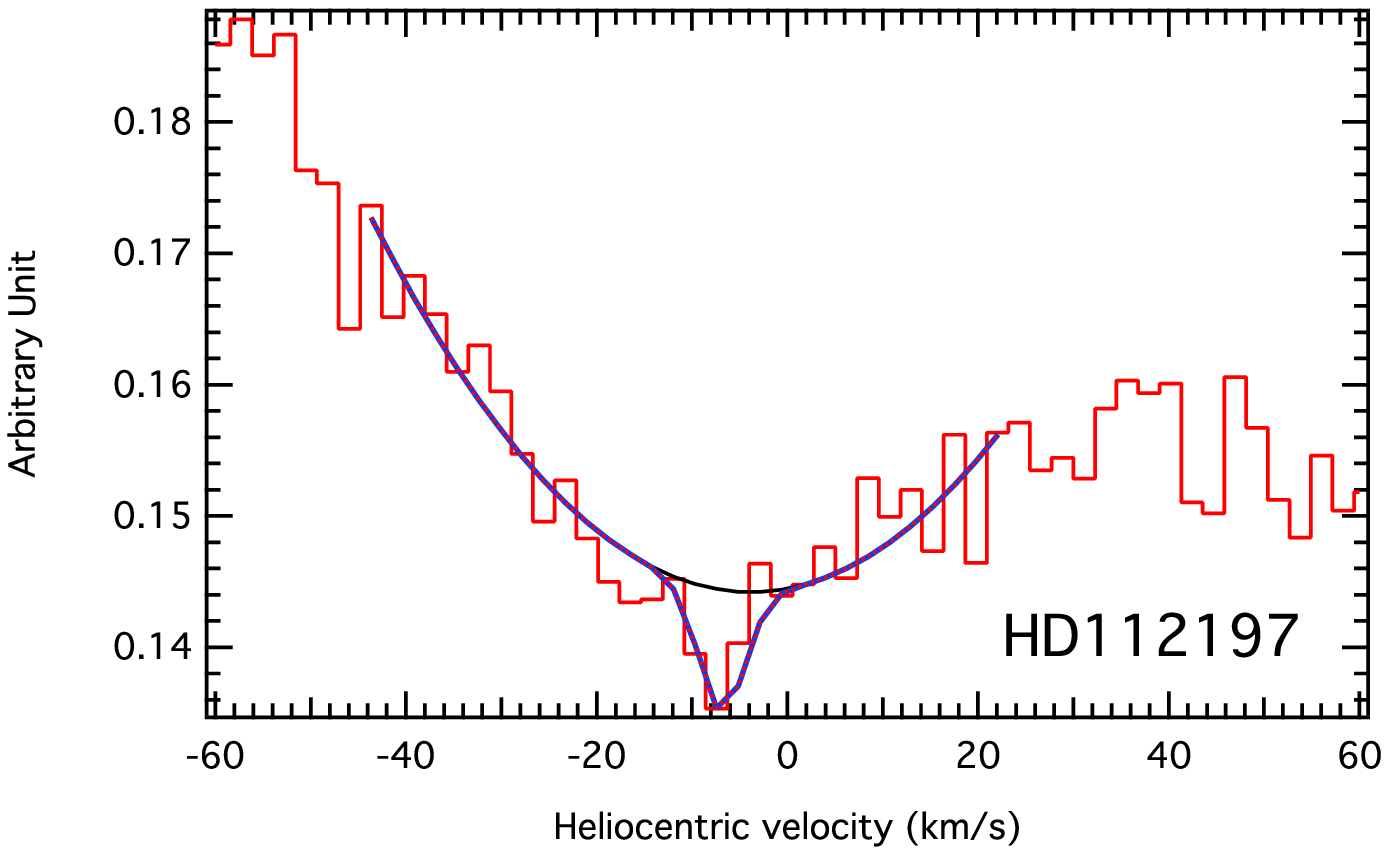}
\end{minipage}\hfill
\begin{minipage}[t]{0.3\linewidth}
\centering
  	\includegraphics[width=1\linewidth]{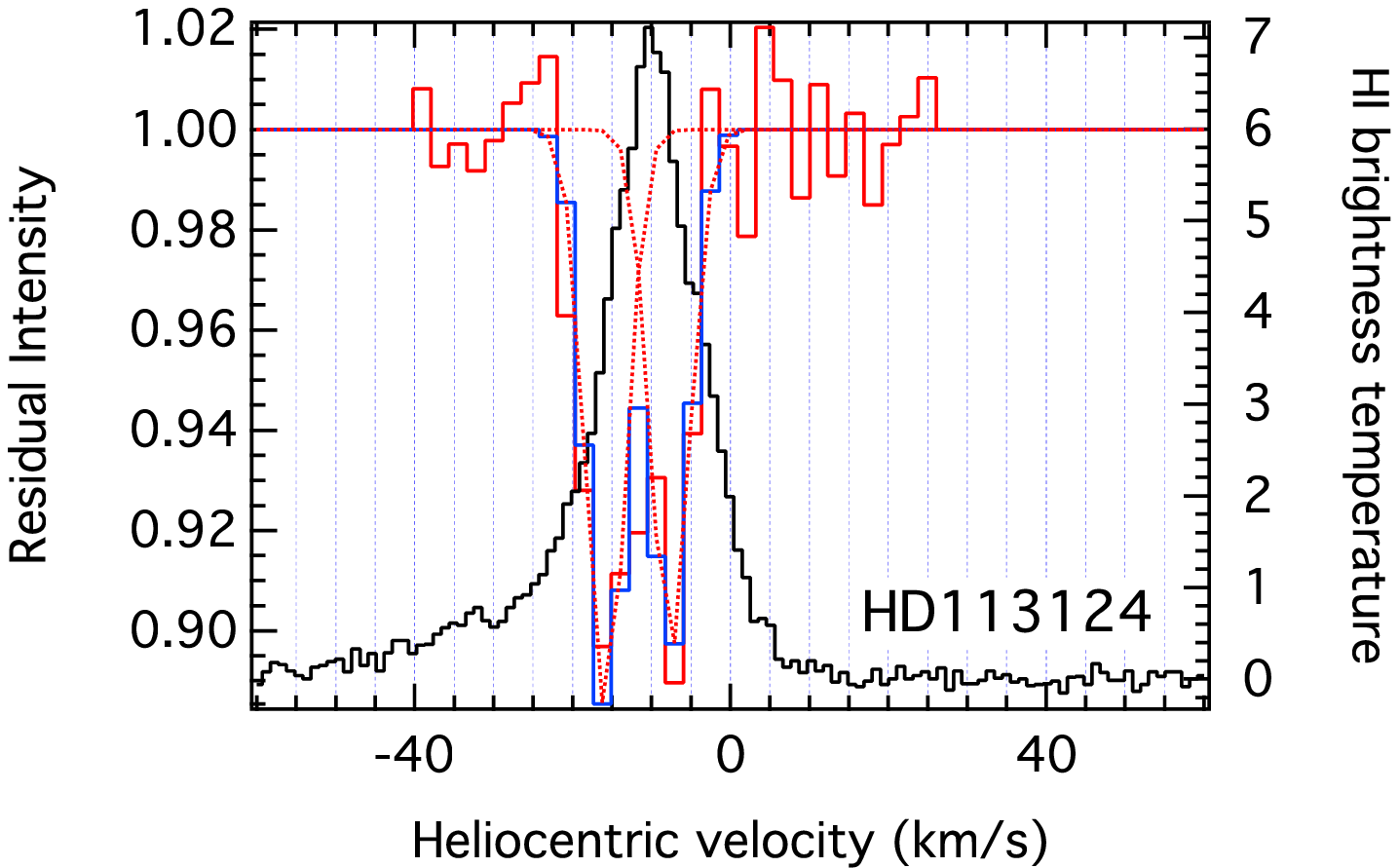}
  	\includegraphics[width=1\linewidth]{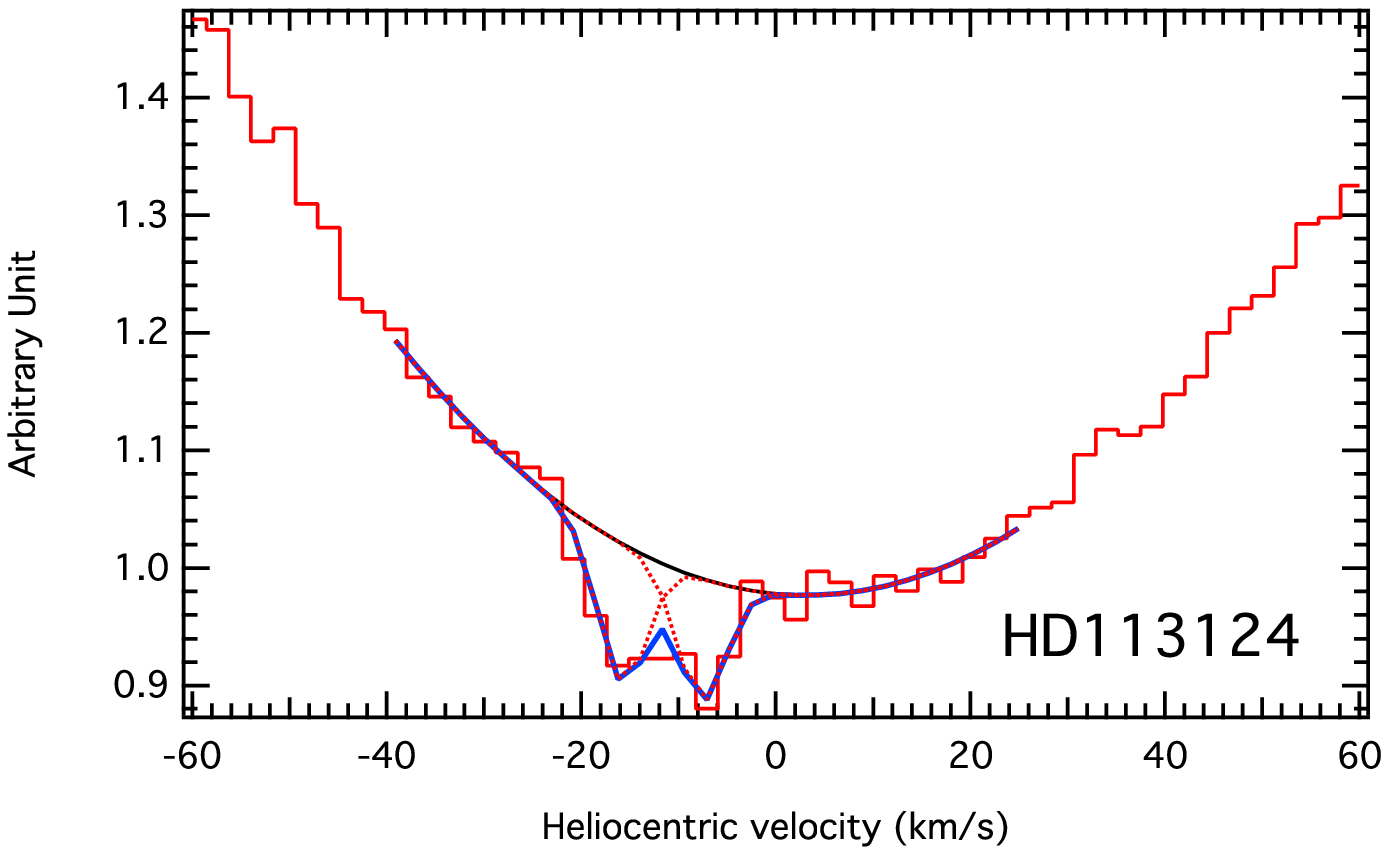}
  	\includegraphics[width=1\linewidth]{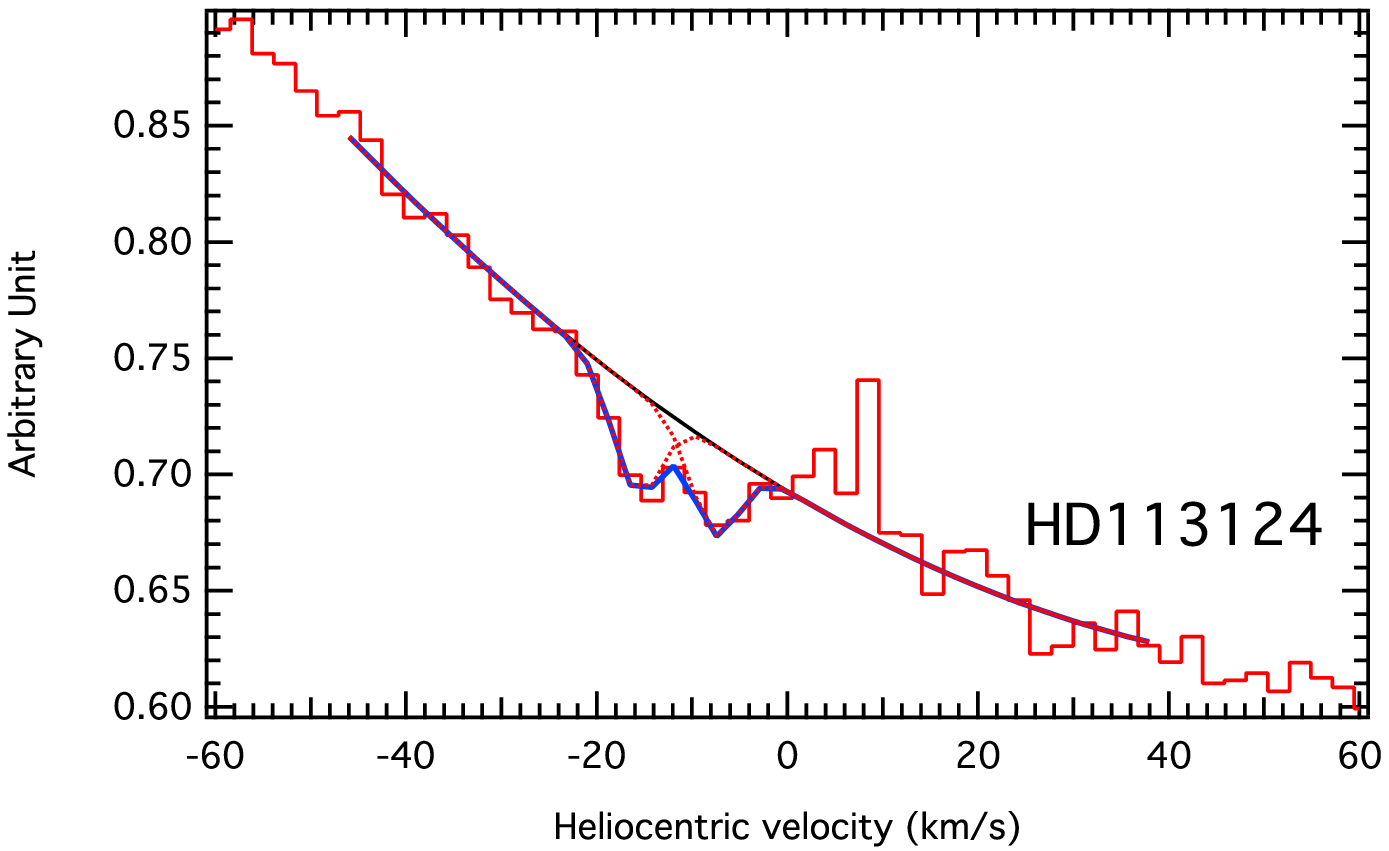}
\end{minipage}\hfill
\caption{Same as Fig. \ref{HD94194} {(in the article)}  but for interstellar CaII absorption for target stars: HD112002, HD112197, and HD113124}
\end{figure*}

\begin{figure*}[!h]
\begin{minipage}[t]{0.24\linewidth}
\centering
	\includegraphics[width=1\linewidth]{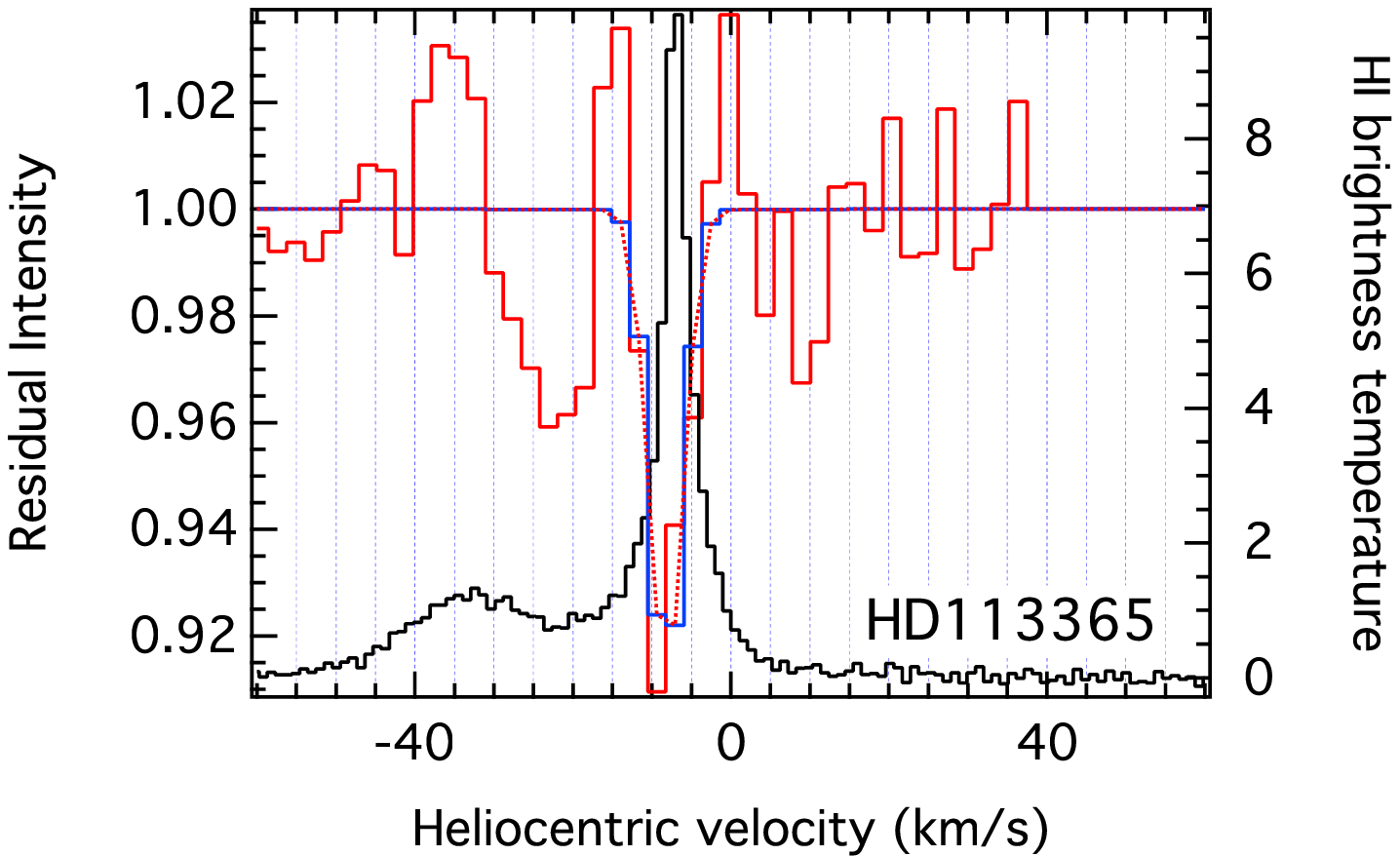}
  	\includegraphics[width=1\linewidth]{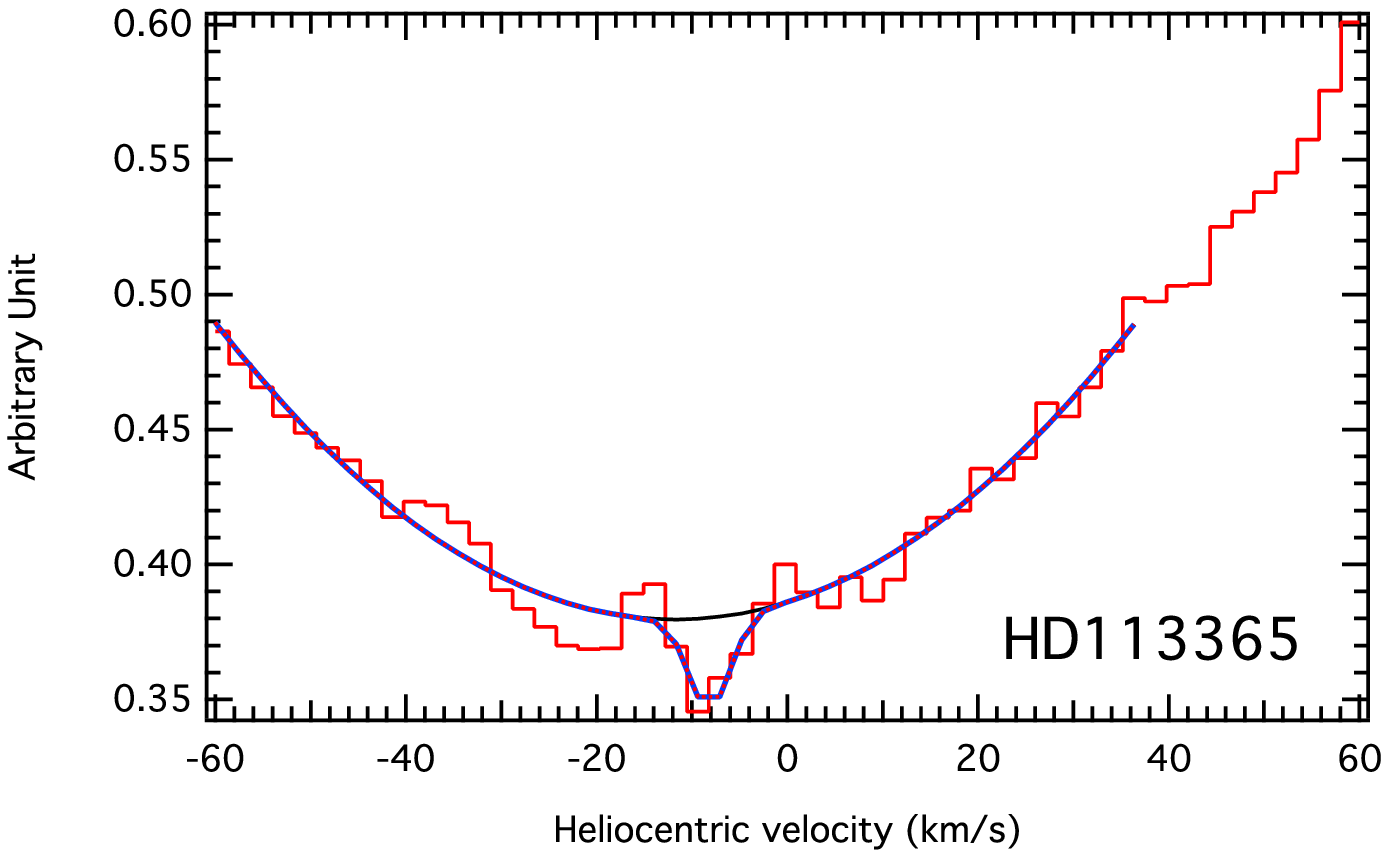}
  	\includegraphics[width=1\linewidth]{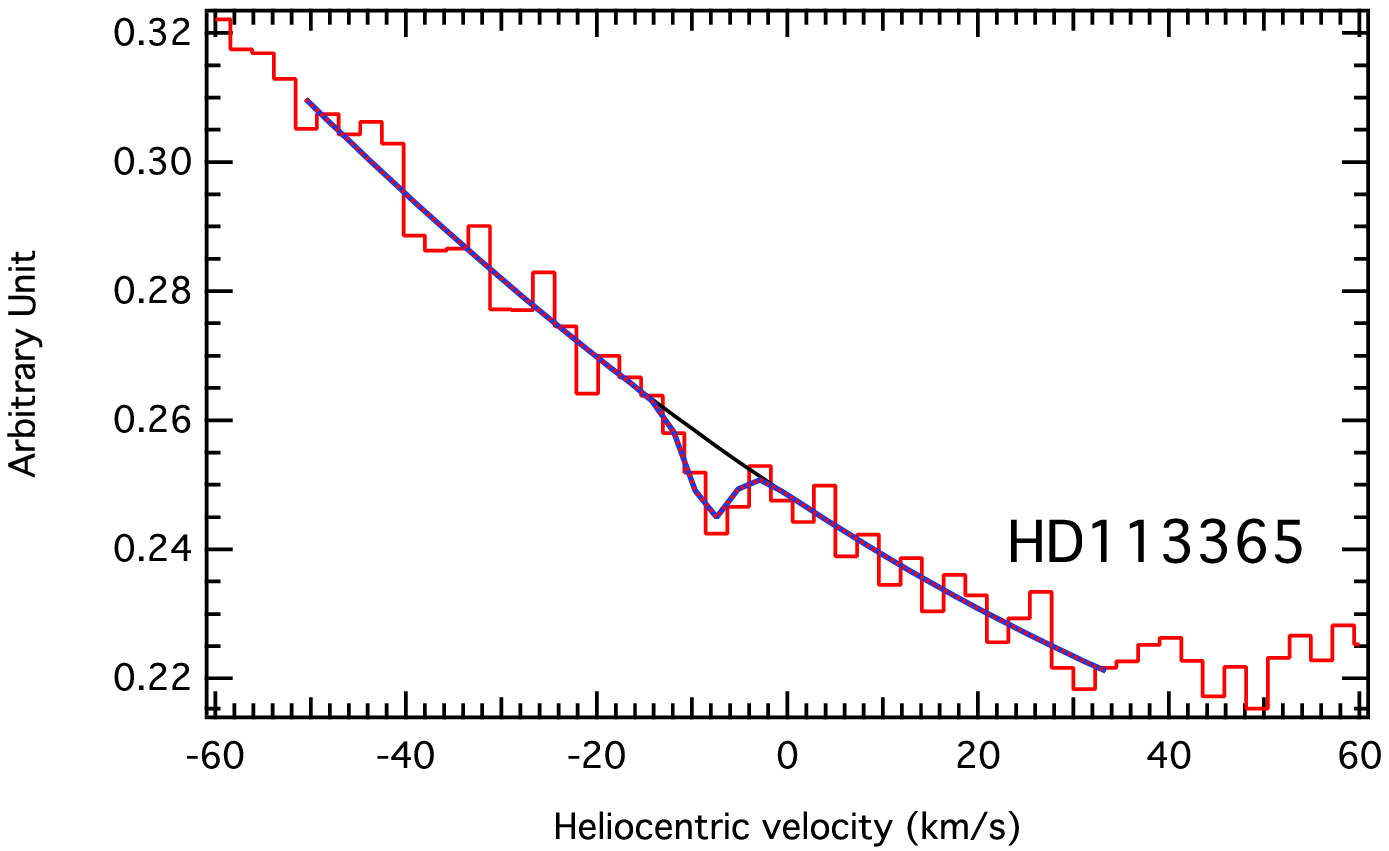}
\end{minipage}\hfill
\begin{minipage}[t]{0.24\linewidth}
\centering
  	\includegraphics[width=1\linewidth]{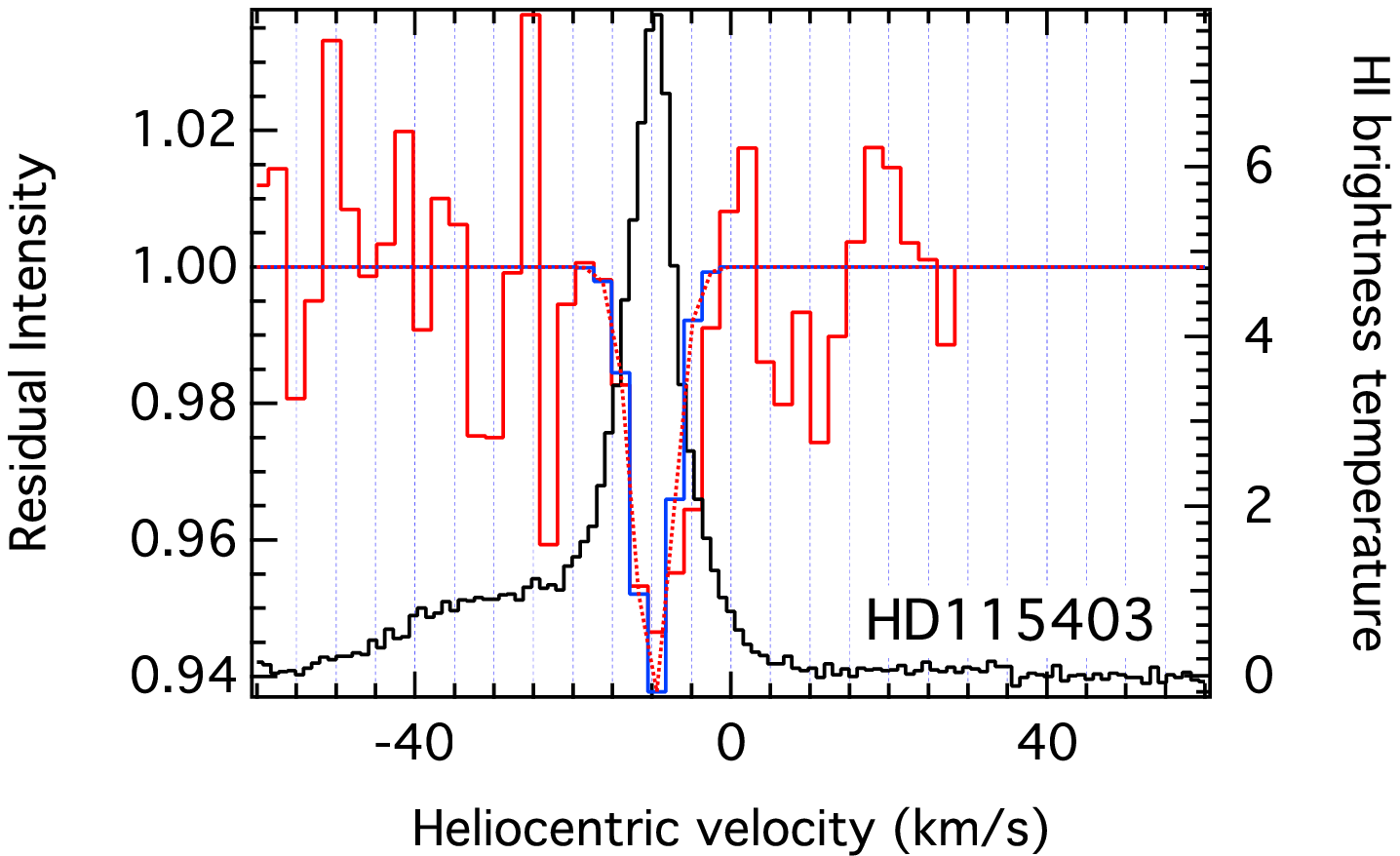}
  	\includegraphics[width=1\linewidth]{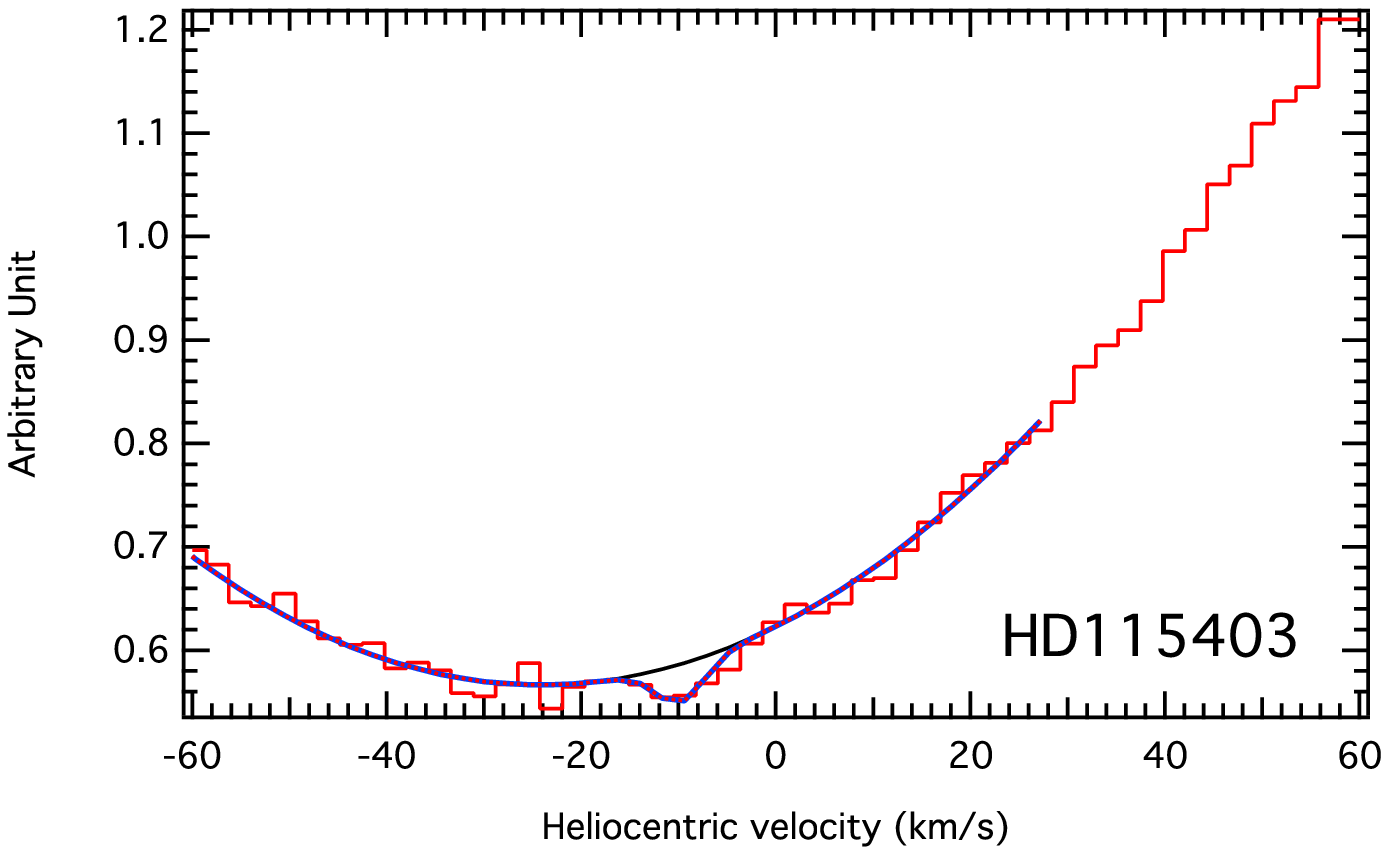}
  	\includegraphics[width=1\linewidth]{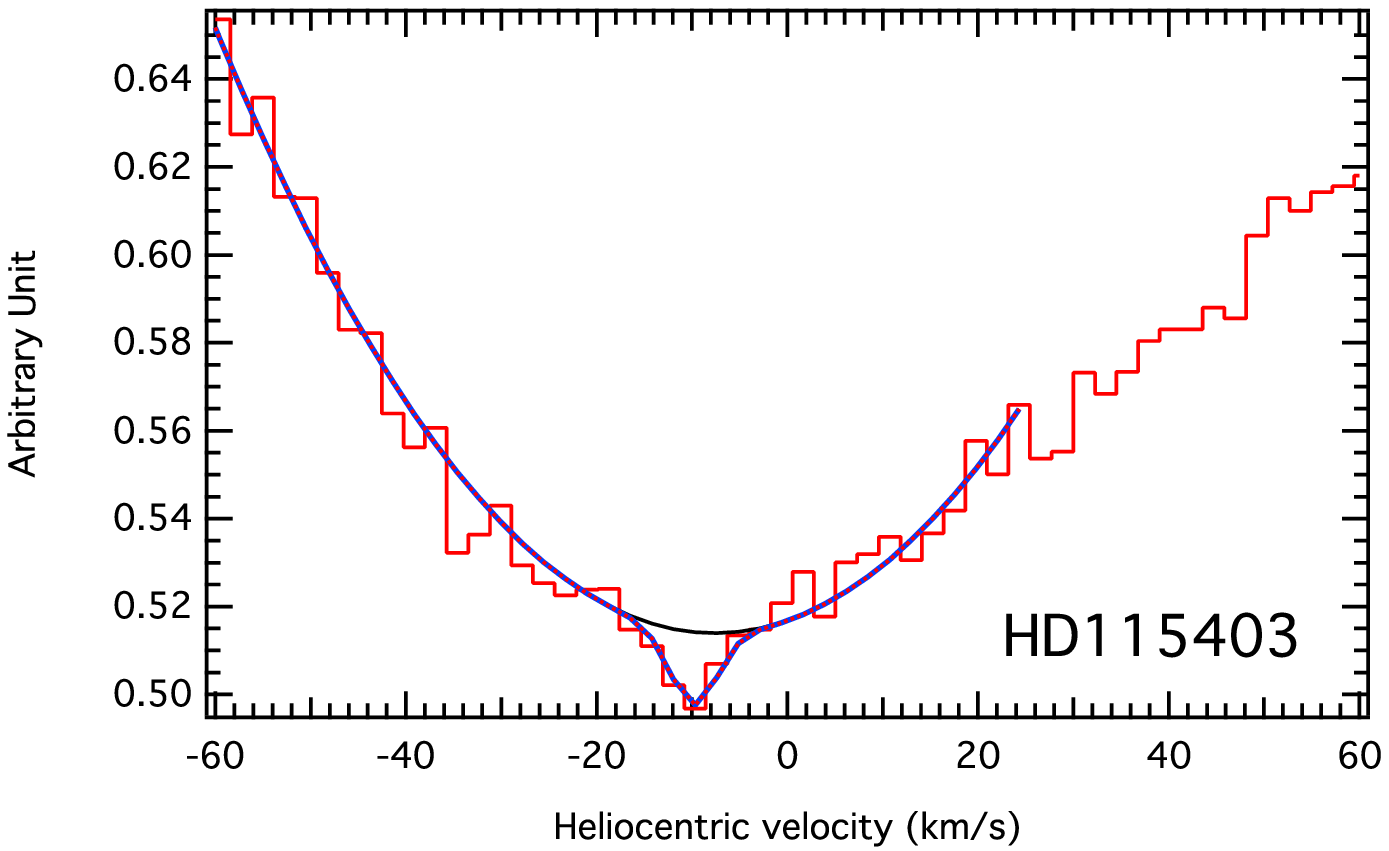}
\end{minipage}
\begin{minipage}[t]{0.24\linewidth}
\centering
  	\includegraphics[width=1\linewidth]{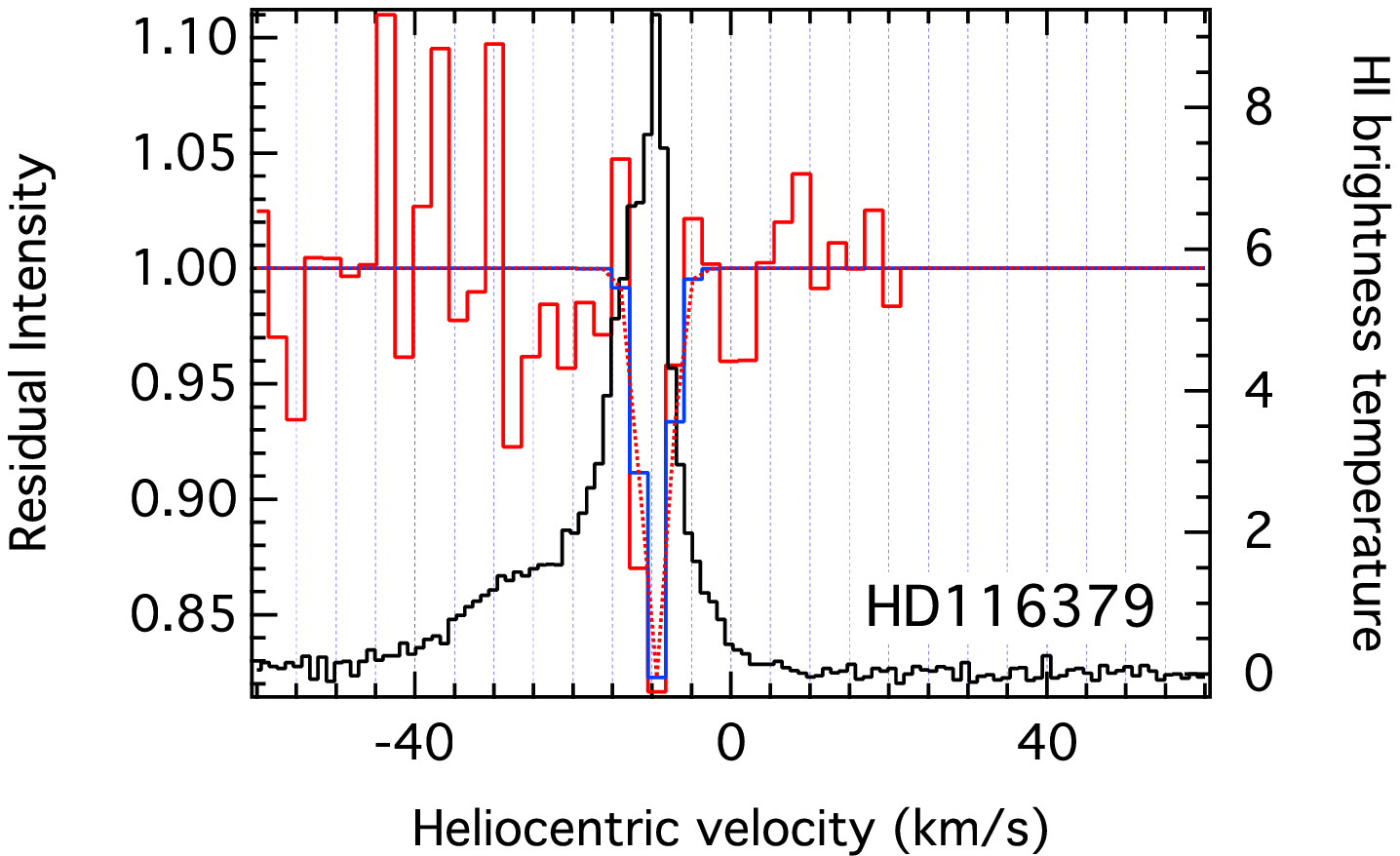}
  	\includegraphics[width=1\linewidth]{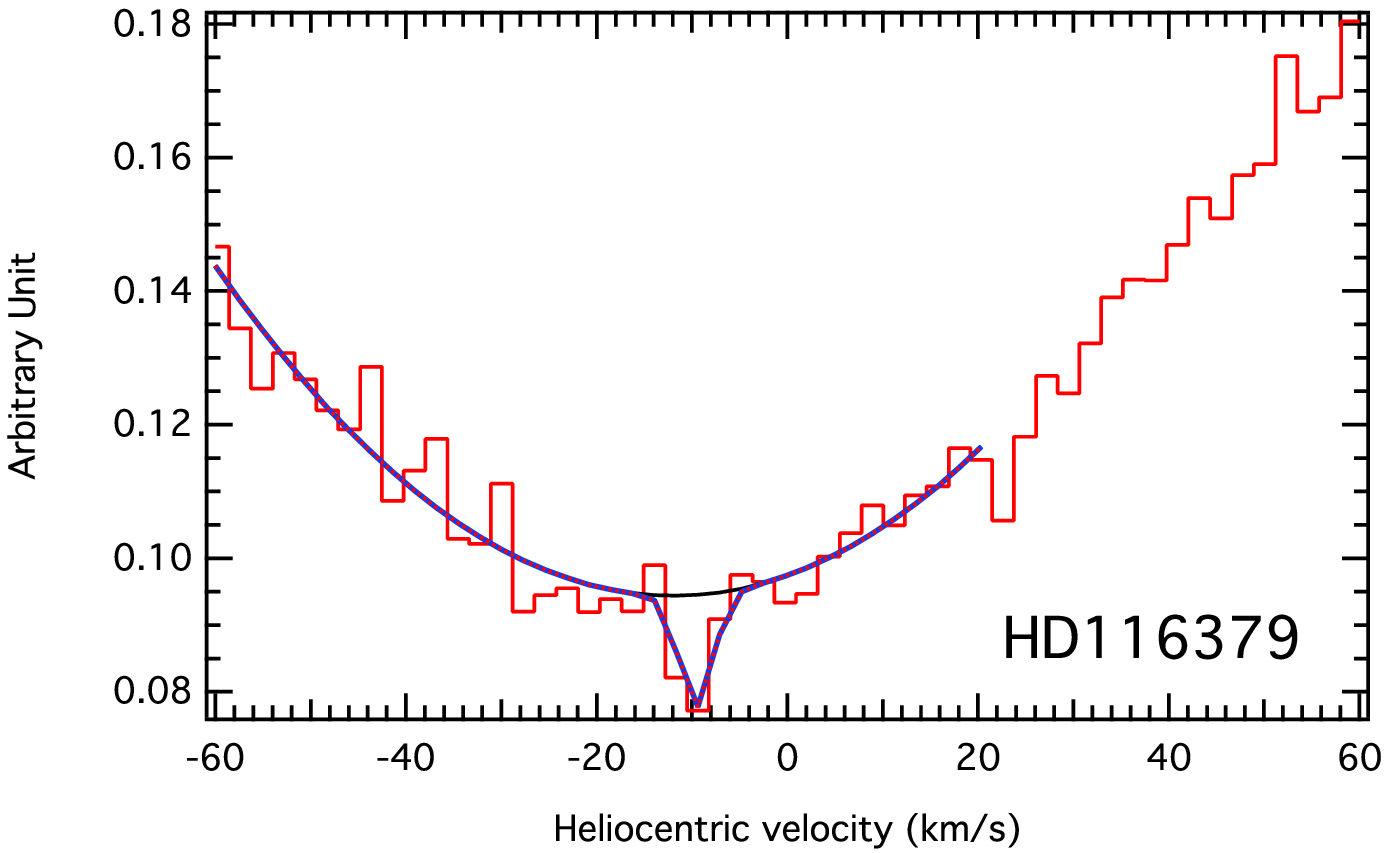}
  	\includegraphics[width=1\linewidth]{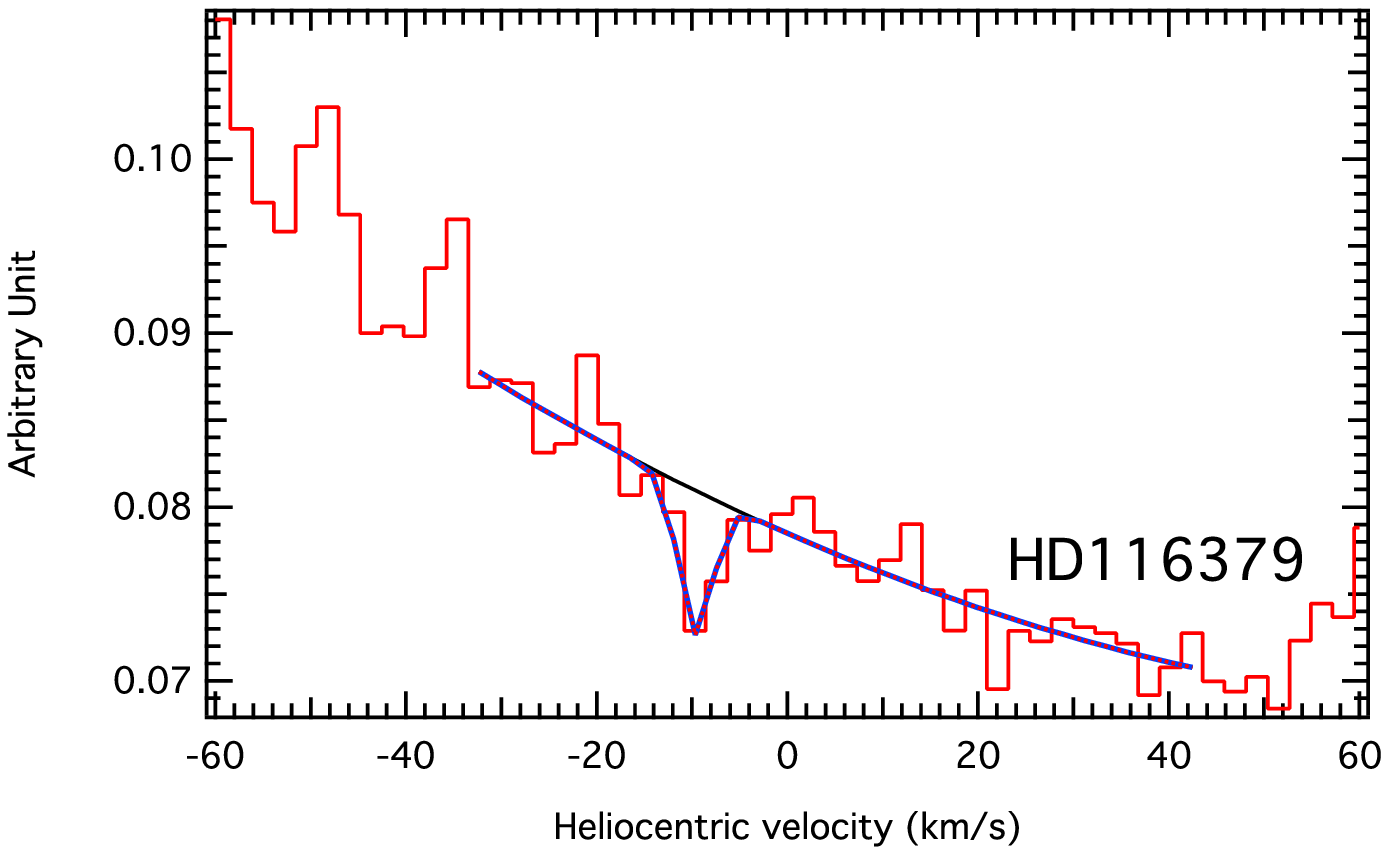}
\end{minipage}\hfill
\begin{minipage}[t]{0.24\linewidth}
\centering
  	\includegraphics[width=1\linewidth]{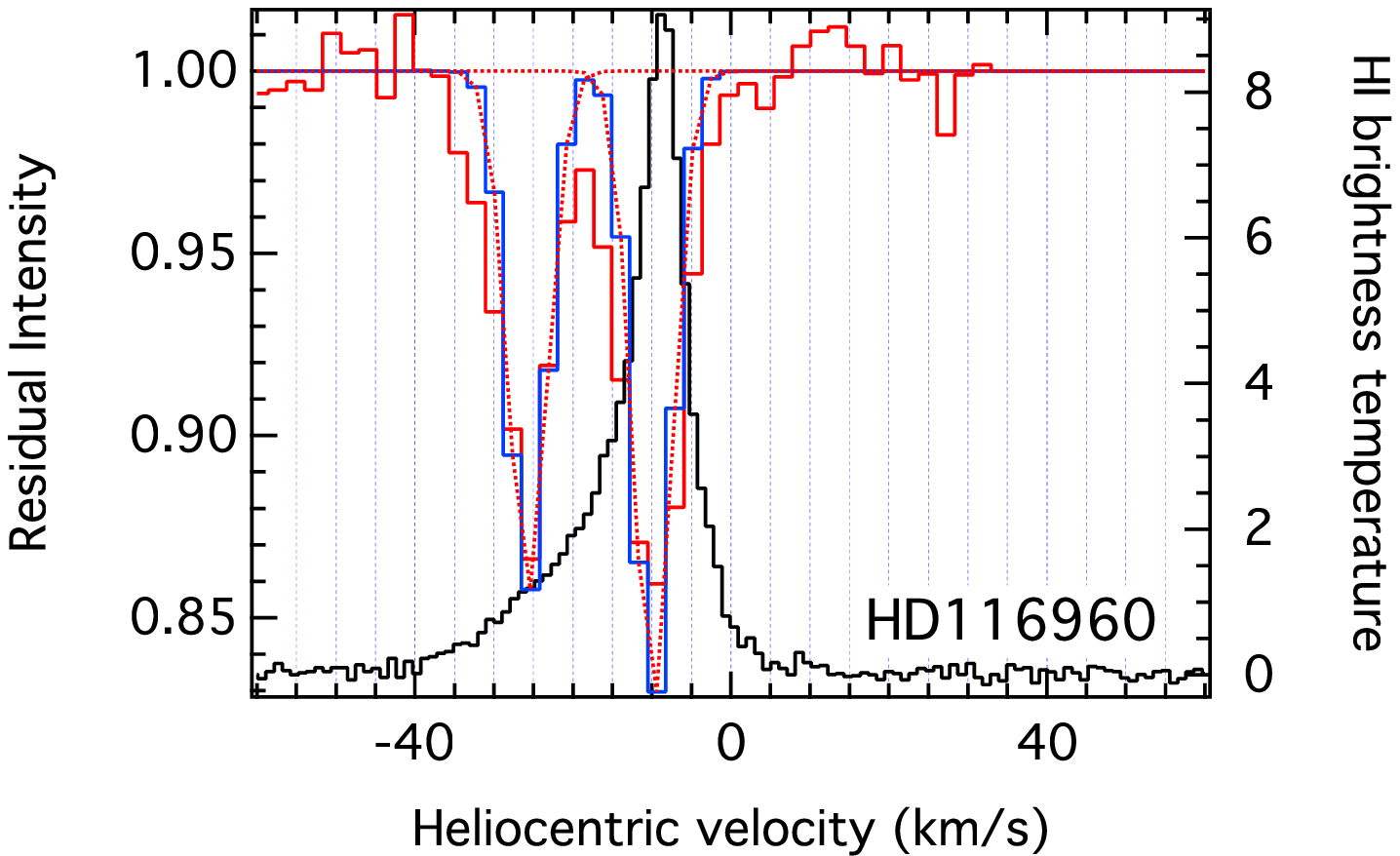}
  	\includegraphics[width=1\linewidth]{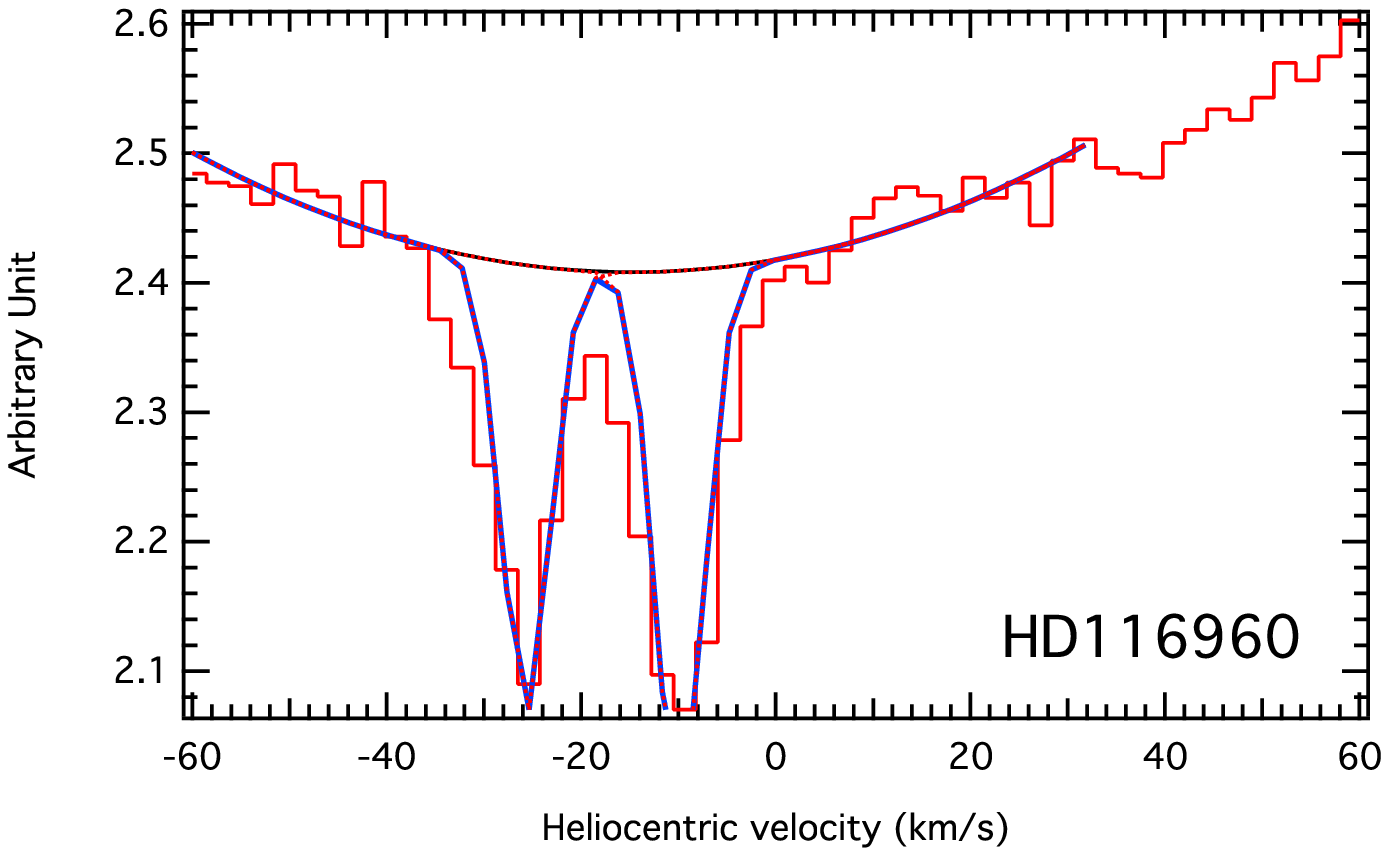}
  	\includegraphics[width=1\linewidth]{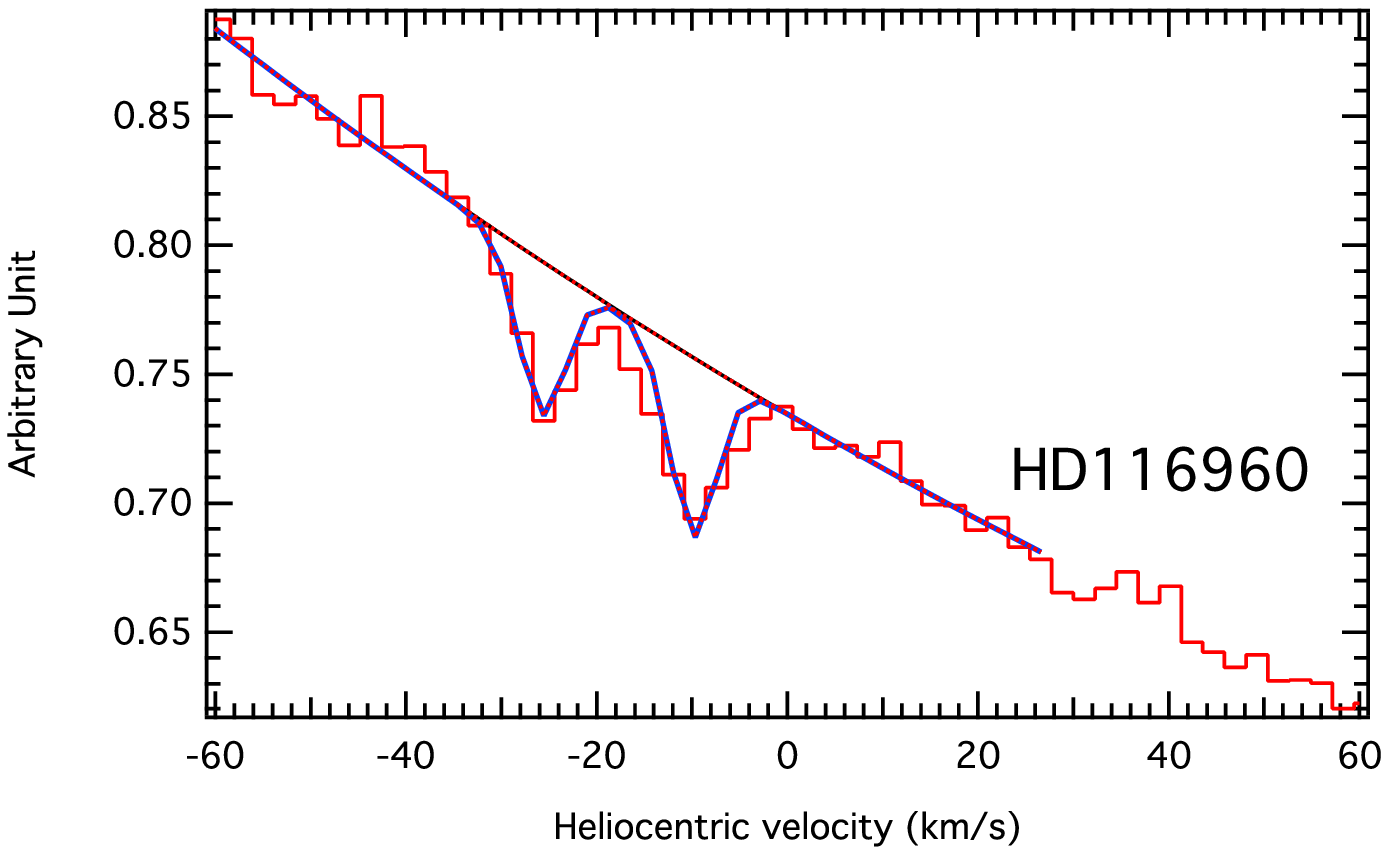}
\end{minipage}\hfill
\caption{Same as Fig. \ref{HD94194} {(in the article)}  but for interstellar CaII absorption for target stars: HD113365, HD115403, HD116379, and HD116960}
\label{HD113365ca}
\end{figure*}


\end{appendix}
\end{document}